Iliakis Dimitrios

"<u>D-Dimond1</u>"

Prototype Antenna RGA-05

Democritus University of Thrace

Polytechnic School of Xanthi

Department of Electrical and Computer Engineering

Section of Space and Telecommunications

Microwaves Laboratory

Antennas

Iliakis Dimitrios

"D-Dimond1"

Prototype Antenna RGA-05

Examining Committee

Associate Professor of High Voltages Michael Danikas

Assistant Professor of Digital Remote Sensing Dimitrios Diamantidis

Assistant Professor of Antennas Petros Zimourtopoulos

[Supervision]

Diploma Thesis #38

Xanthi

October 2006

*To Eleni,*
*who will be my future woman in one month*
*for her patience and her support,*
*for her cardiac love*

*To our child,*
*who is to be given birth,*
*with the errand of God,*
*the Easter of 2007*

*To my family*

*To Mr and Mrs Zimourtopoulos*

# Contents

Introduction

**Introduction**

At the present thesis a prototype antenna is studied. The essay is composed of the introduction, five chapters and eight annexes.

In chapter 1, firstly, the antenna's geometry was chosen with the supervisor's assistance. Then its dimensional characteristics and its geometrical description in space and at plane are outlined.

In chapter 2 an extensive theoretical analysis is carried out applying the acquired knowledge of the lessons "Antennas I : Analysis" and "Antennas II : Synthesis, Design" which I was taught at the Department of Electrical and Computer Engineering of the Democritus University of Thrace. From the theoretical analysis the final mathematical formulas of the antenna's radiation pattern are extracted. The radiation pattern defines how the antenna radiates in space. Next, the radiation pattern is depicted in space and at the three main planes utilizing two computer programs written in C language and their code is cited at the annexes A and B respectively.

In chapter 3 a computational study of the antenna is carried out utilizing the computer program RICHWIRE [4]. The antenna is divided in segments using the appropriate computer program written in C language (the code is cited at the Annex C). The results of the study are illustrated in appropriate figures which show the variations of the most important antenna's characteristic quantities as well as the radiation patterns at the three main planes. A study of the antenna's behavior as a function of the frequency follows at the range from 900 to 1300 [MHz]. Finally, comparative diagrams of the antenna's radiation patterns are quoted between the theoretical and computational study.

In chapter 4 the antenna's behavior as a function of frequency is studied at a more extensive range, approximately from 200 to 2850 [MHz].

Finally, in chapter 5 an improvement of the antenna's characteristics and its bandwidth, as defined at the current essay, is attempted by modifying its geometrical shape. Three models of geometrical modification are applied at five pre-selected frequency ranges. Then follow comparative results between the original and the improved antenna's dispositions at every examined frequency range, and eventually the final conclusions.

I would like earnestly to thank my supervisor assistant professor Mr Petros Zimourtopoulos for his guidance, patience and personal support which showed during the elaboration of the present thesis, as well as Mrs Nikolitsa Giannopoulou, for her priceless contribution and solidarity. Finally, I thank my fellow-students Giorgos Tsampazis and Giannis Giochtsis, who also elaborated their thesis with the same supervisor, for their constructive opinions and ideas exchange as well as for their psychological support.

# Chapter 1 : The D-Dimond1 antenna

## 1.1 : In General

In the present thesis an original antenna is studied. The main objective of this essay consists in the most complete possible study of the antenna's behavior, the comprehension of its particular characteristics and the investigation for their improvement, if this is possible. The farther objective of this essay is to derive practically beneficial, applicable and scientifically acceptable and accurate results through a methodical and systematic work.

At the beginning, the antenna's geometry had to be found. The supervisor set the requirements which had to be fulfilled. More specifically, the antenna's geometry: a) has to be original, b) has to be relatively easy in its construction, c) has to be plane and d) its constructional dimensions have to be limited practically inside a conceivable rectangle parallelogram with sides a and b the values of whom result indirectly from the feeding source position as follows: if the feed is defined as a point which abstains a1 and a2 equivalent from the sides of length b, that is to say: $a1 + a2 = a$, as well as b1, b2 from the sides of length a, that is to say: $b1 + b2 = b$, then the construction requires that $\max\{a1, a2\} \leq 15$ [cm] and $\max\{b1, b2\} \leq 30$ [cm].

The final geometry eventually resulted after the supervisor's proposition. However, successive "failed" proposed geometries were preceded because of their increased geometrical complexity as well as their disagreement with the requirements mentioned above. The geometrical representation of the antenna at the plane is shown in <u>Figure 1.2.1</u> and <u>Figure 1.2.2</u> while in space in <u>Figure 1.2.3</u>.

## 1.2 : Geometrical characteristics of the antenna

The D-Dimond1 antenna is composed from 7 straight segments, 6 parasitically and 1 active, at which the feeding source is connected. For this essay all straight segments are considered that they are positioned at the same plane and particularly at the yOz level. The feeding source of the antenna is considered that is found at the origin of the axes, that is to say is placed at the center of that straight segment which passes from the origin of the axes.

From <u>Figure 1.2.3</u> the dimensional characteristics of the antenna are becoming obvious. The biggest dimension at the z axis is represented by the length AB and is approximately 36.88 [cm]. Respectively, the biggest dimension at the y axis is represented by the length EX and is 13.50 [cm].

It is noted that the 7 straight segments which compose the present antenna become $\lambda/2$ dipoles at the frequency of 1111 [MHz].

The antenna's geometry was selected carefully in order to reveal as many symmetries as possible. This fact will considerably simplify the subsequent analytical and computational study. As it results from the figures, the antenna is symmetrical concerning both y and z axes that define its plane.

The polarization of the antenna arises easily because of its simple and symmetrical geometry and it will be shown in detail in the following chapter utilizing the theory of standing waves. The direction of the polarization is estimated by the resultant direction of the individual vectors of electric fields that are being developed along each antenna's segment. Also, it should be mentioned that the polarization is being determined only when the observation of the antenna from a point of its distant field is considered. For the D-Dimond1 antenna turns out that the polarization is linear along the y axis direction.

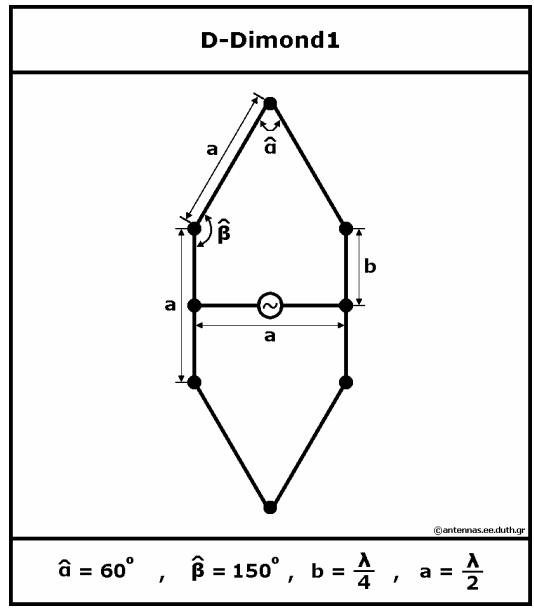

Figure 1 : Geometry at the plane and dimensional characteristics

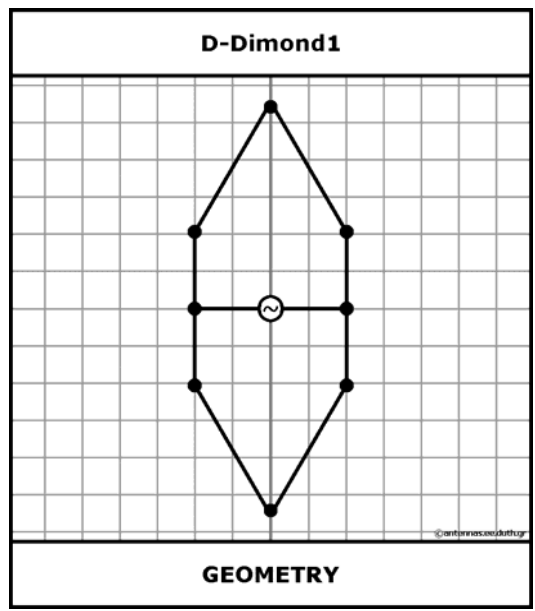

Figure 2 : Geometry at the plane with grid subdivided in λ/8

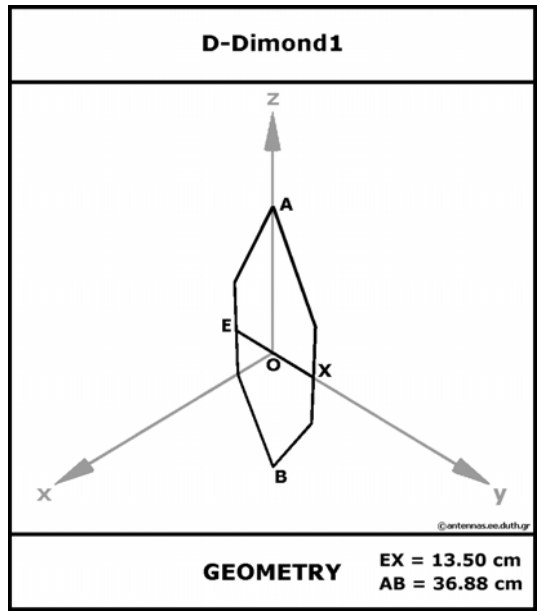

Figure 3 : Geometry in space

## 1.3 Origin of name

Initially it had been proposed by myself the name "Dim Diamond" which is composed by the synthetics "Dim" from Dimitris (Dimitris – my name) and from the "Diamond", describing somehow the geometry of the antenna. This name was proposed in the worldwide known forum of the radio amateurs, Rec Radio Amateur Antennas [7], from where the answers that I received were negative for the particular name.

Some user reported that the word "Diamond" is part of the brand name "Diamond Antenna Company". Therefore there could emerge copyright problems as somebody could think that the present antenna is for sale by this company.

A second user had intense objections for the first synthetic of the name, "Dim" because "it isn't sounds well", as he claimed, in the ears of the English-speaking, while it has also the meaning of "dim" (weak) in the terminology of E/M waves.

Finally, some other user "discovered" putting that name of the antenna in the search engine of Google a musical group in Brooklyn with the same name.

Consequently, the first name which was proposed abandoned. However there was a user named Reg Edwards who proposed a different possible name. Particularly he consulted me to alter the word "Diamond" in "Dimond" and to add also some kind of number so as to make the name less identifiable or even eligible as nomenclature from any other, eg the 1, so as a future prospect of development of the antenna into model 2 or 3 to be shown. Finally, "the addition of some number reveals a possible electromechanical background while it gives also better initial optical impression", as he declared.

Therefore, I followed the advice of Reg Edwards and I named the antenna D-Dimond1, where the additional prefix "D" is the initial letter of "Dimitris" (which is my name). The particular name does not refer in some product or commercial name after a short research that I carried out in the internet. At the same time, the "faint" impression of a "diamond" is somehow maintained as a reference in the geometry of the antenna.

# Chapter 2 : Theoretical study

## 2.1 : In General

In the present chapter the application of the theory of antennas [1] in the D-Dimond1will be attempted. It will be assumed that the antenna is located in the free space and the entire disposition will be confronted according to the theory of standing waves.

Initially the antenna will be divided in smaller straight segments. Then each straight segment will be considered as an independent dipole and for each one will be given the expression of its radiation pattern. Then, according to the principle of superimposition all the individual radiation patterns will be summed up. Finally, the total electric field of the antenna will be exported.

The radiation pattern for whatever straight antenna of thin wire, that is to say with wire's radius much smaller than the wavelength (practically radius smaller than 1/100 of wavelength), is given by:

$$(1): \overline{\mathbb{E}} = (\frac{1}{\lambda}) \begin{bmatrix} \ell_{\uparrow\theta} \\ \ell_{\uparrow\varphi} \end{bmatrix} e^{i\beta R_{k_r}} \dot{P}\dot{F}$$

where $\dot{PF}$ stands for the pattern factor of the radiation pattern and is given by the formula:

$$\dot{PF} = \int_{-L_A}^{L_T} I(\ell) e^{i\beta_{\ell_r}\ell} \, d\ell$$

For -dipole standing wave- antennas it is considered that the current distribution is the same with the one that results from open-circuited transmission lines of two parallel conductors, that is to say sinusoidal:

$$\dot{I}(\ell) = \dot{I}_0 \sin[\beta(h - |\ell|)]$$

where dipole's length L is defined as: $L = 2h$ and implies: $-h \leq \ell \leq +h$.

## 2.2 : Application of the standing wave theory

Initially, the theory of standing waves will be applied. The basic principles of this theory are summarized below:

1.    The current's direction reverses per segment with length $\lambda/2$.

2.    At the free terminals of the segments the current is nullified.

3.    Between two successive nodes of standing wave the current does not change its phase. Moreover, it will be assumed that Kirchhoff's laws are in effect.

At Figure 2.2.1 the following elements are indicated:

voltage nodes (spots with filling)

current nodes (spots without filling)

current's direction along each segment (with and without filling)

characteristic figure and direction of the feeding source

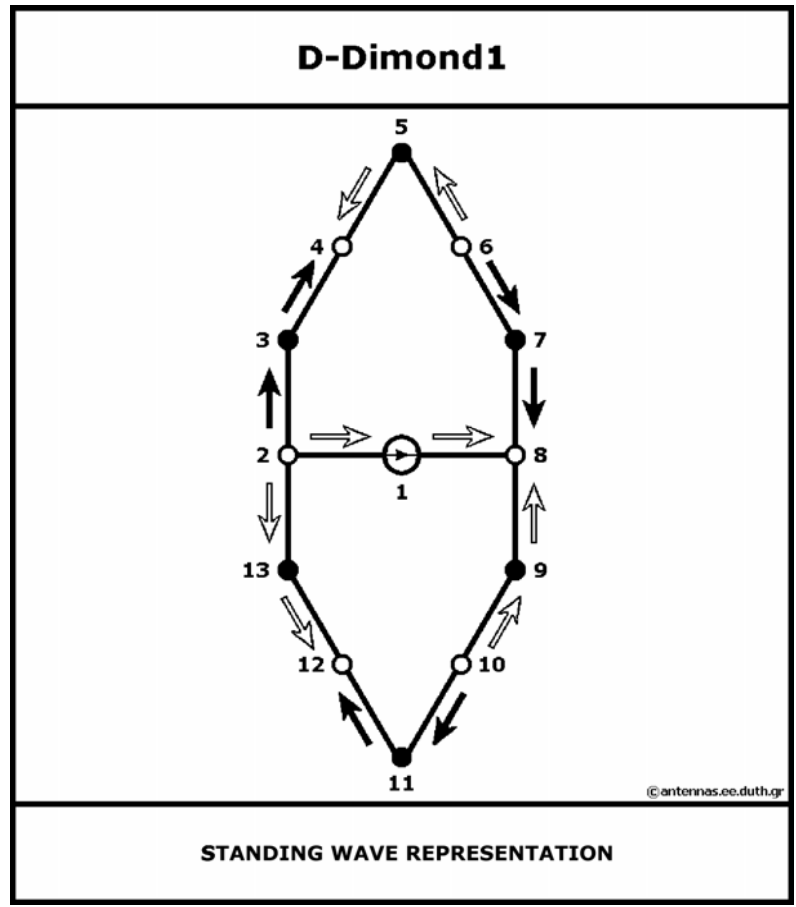

Figure 1 : D-Dimond1 antenna according to the theory of standing waves

In the Figure 2.2.1 13 nodes are numbered where along them run 7 current standing waves. Beginning from node 2 the 1ˢᵗ standing wave runs through the source via node 1 to node 8. Also, the second standing wave begins from node 2 and via node 3 ends up to node 4 and so forth. Generally, the current's direction is reversed at every segment with length λ/2 it crosses according to the theory of standing waves. Also, someone can easily verify that Kirchhoff 's first law is applied at each current node.

In the Figure 2.2.2 14 straight segments with length λ/4 are illustrated at which the antenna has been divided, numbered from one until fourteen. This is the minimum number of segments that it can be achieved. The straight segment 1 is defined by nodes 1 and 8. The straight segment 2 from nodes 7 and 8, and so forth.

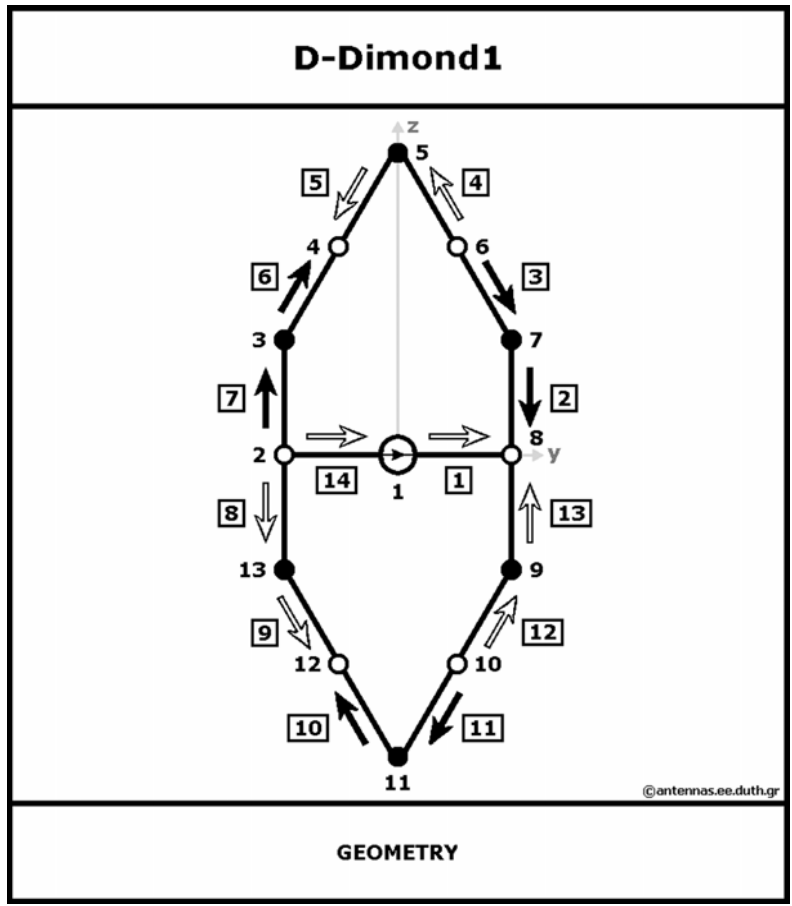

Figure 2 : D-Dimond1 antenna at the yOz plane

## 2.3 : Theoretical analysis

The aim of the theoretical analysis is to export the mathematical formula of the 2-dimensional radiation pattern (2D) of the antenna projected at the three main planes xOy, yOz, zOx as well as the 3-dimensional (3D) radiation pattern for the specific central frequency at 1111 [MHz].

In order to illustrate the 2D and 3D radiation pattern someone should firstly evaluate the mathematical expression of the total radiation pattern. This presupposes a specific geometrical representation of the antenna, that is to say its placement into a suitable system of coordinates. The simplification during the progress of this essay will be a criterion for the selec-tion of a suitable system of coordinates. Because of the plane geometry of D-Dimond 1 the antenna is placed at the yOz plane, as illustrated in <u>Figure 2.3.1</u>.

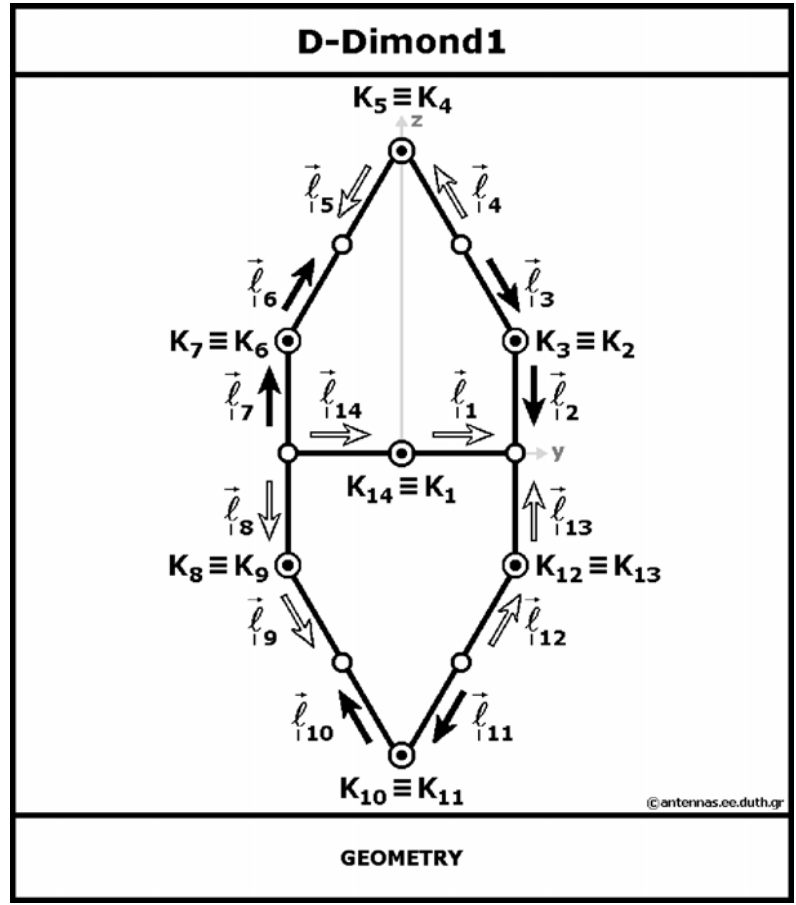

<u>Figure 1</u> : D-Dimond1 antenna at the yOz plane

So much in <u>Figure 2.3.1</u> as much as in <u>Figure 2.3.2</u> are displayed the centres of the straight segments, the calculation of which is explained analytically below, and also the unitary vectors which imply the direction of each one segment's current.

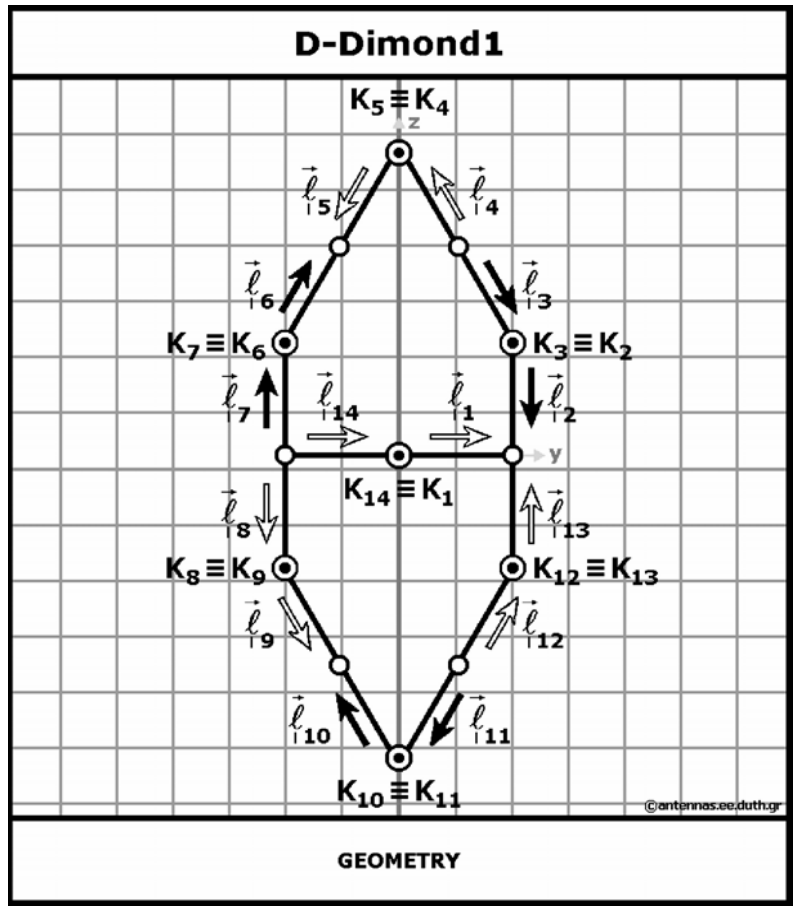

<u>Figure 2</u> : D-Dimond1 antenna at the yOz plane with grid subdivided in λ/8

The data for each of the fourteen straight segments of the antenna are given in <u>Table 2.3.1</u> and <u>Table 2.3.2</u>. The data's calculation process of three specific straight segments of the antenna will be shown below. The whole process for the antenna's remaining straight segments is completely similar.

**The data's calculation process for the segment 1**

α) The $|\ell_A|$ and $|\ell_T|$ are derived from the distribution of the standing wave at this segment (<u>Figure 2.2.1</u> and <u>2.2.2</u>), while their signs are derived from the position of the $\ell_A$ and $\ell_T$ in connection with $\ell = 0$ (<u>Figure 2.3.3</u>).

β) The $\vec{R}_A$ and $\vec{R}_T$ are derived from the specific geometry of the antenna and from the selection of the coordinate system.

γ) The $\vec{\ell}_1$ is estimated from the specific geometry of the antenna and from the standing wave's distribution along this segment (Figure 2.2.1). For this particular segment it is obtained:

$$\vec{R}_T = \frac{\lambda}{4}\vec{y}, \ \vec{R}_A = 0,$$

$$\ell_T = \frac{\lambda}{4}, \ \ell_A = 0$$

$$\vec{\ell}_1 = \frac{1}{\left|R_T - R_A\right|}\left(\vec{R}_T - \vec{R}_A\right) \Rightarrow \vec{\ell}_1 = \frac{1}{\left|\ell_T - \ell_A\right|}\frac{\lambda}{4}\vec{y} = \vec{y}$$

δ) For the calculation of $\vec{R}_\kappa$ the following formula is used:

$$\vec{R}_\kappa = \vec{R}_A - \ell_A\vec{\ell}_1 = \vec{R}_T - \ell_T\vec{\ell}_1 \Rightarrow$$

$$\vec{R}_\kappa = 0 - 0 = \frac{\lambda}{4}\vec{y} - \frac{\lambda}{4}\vec{y} = 0$$

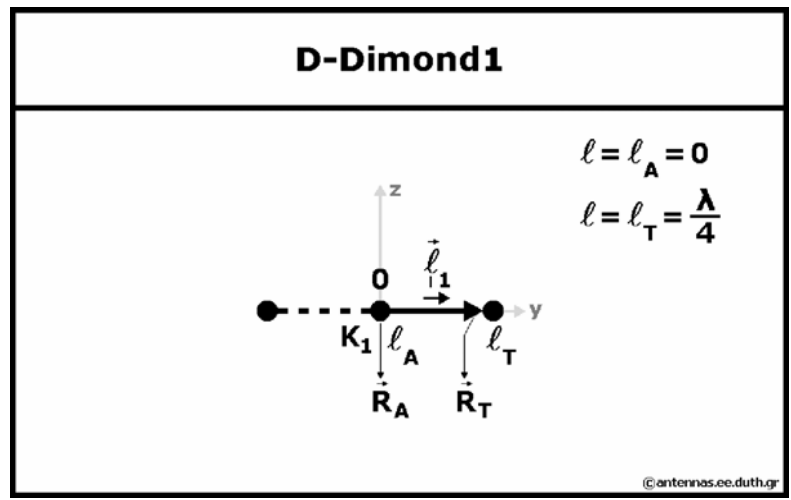

Figure 3 : Segment number 1

**The data's calculation process for segment 2**

α) and β) similarly with the first segment γ) The calculation of the $\vec{\ell}_2$ is based on the geometry of the antenna and on the standing wave's distribution at this segment (Figure 2.2.1). For this particular segment the following formulas are used:

$$\vec{R}_T = \frac{\lambda}{4}\vec{y}, \ \vec{R}_A = \frac{\lambda}{4}\vec{y} + \frac{\lambda}{4}\vec{z},$$

$$\ell_T = \frac{\lambda}{4}, \ \ell_A = 0$$

$$\vec{\ell}_2 = \frac{1}{|\vec{R}_T - \vec{R}_A|}\left(\vec{R}_T - \vec{R}_A\right) \Rightarrow \vec{\ell}_2 = \frac{1}{|\ell_T - \ell_A|}\left(\frac{\lambda}{4}\vec{y} - \frac{\lambda}{4}\vec{y} - \frac{\lambda}{4}\vec{z}\right) = \vec{z}$$

δ) For the calculation of $\vec{R}_\kappa$ the following formula is used:

$$\vec{R}_\kappa = \vec{R}_A - \ell_A\,\vec{\ell}_2 = \vec{R}_T - \ell_T\,\vec{\ell}_2 \Rightarrow$$

$$\vec{R}_\kappa = \frac{\lambda}{4}\vec{y} + \frac{\lambda}{4}\vec{z}$$

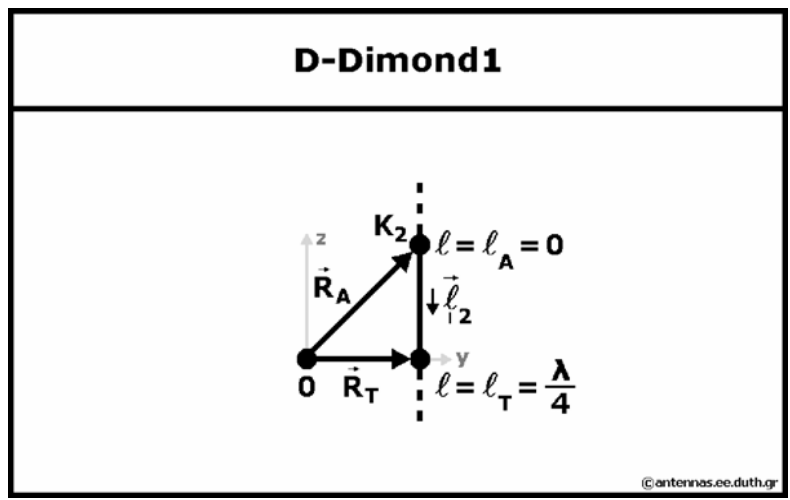

Figure 4 : Segment number 2

**The data's calculation process for segment 3**

α) and β) similarly with the first segment γ) The calculation of the $\vec{\ell}_3$ is based on the geometry of the antenna and on the standing wave's distribution at this segment (Figure 2.2.1). For this particular segment the following formulas are used:

$$\vec{R}_T = \frac{\lambda}{4}\vec{y} + \frac{\lambda}{4}\vec{z}, \ \vec{R}_A = \frac{\lambda}{8}\vec{y} + \frac{\lambda(1+\frac{\sqrt{3}}{2})}{4}\vec{z},$$

$$\ell_T = 0, \ \ell_A = -\frac{\lambda}{4}$$

$$\vec{\ell}_3 = \frac{1}{|R_T - R_A|}\left(\vec{R}_T - \vec{R}_A\right) \Rightarrow$$

$$\vec{\ell}_3 = \frac{1}{|\ell_T - \ell_A|}(\frac{\lambda}{4}\vec{y} + \frac{\lambda}{4}\vec{z} - \frac{\lambda}{8}\vec{y} - \frac{\lambda(1+\frac{\sqrt{3}}{2})}{4}\vec{z}) = \frac{1}{2}\vec{y} - \frac{\sqrt{3}}{2}\vec{z}$$

δ) For the calculation of $\vec{R}_\kappa$ the following formula is used:

$$\vec{R}_\kappa = \vec{R}_A - \ell_A \ \vec{\ell}_3 = \vec{R}_T - \ell_T \ \vec{\ell}_3 \Rightarrow$$

$$\vec{R}_\kappa = \frac{\lambda}{4}\vec{y} + \frac{\lambda}{4}\vec{z}$$

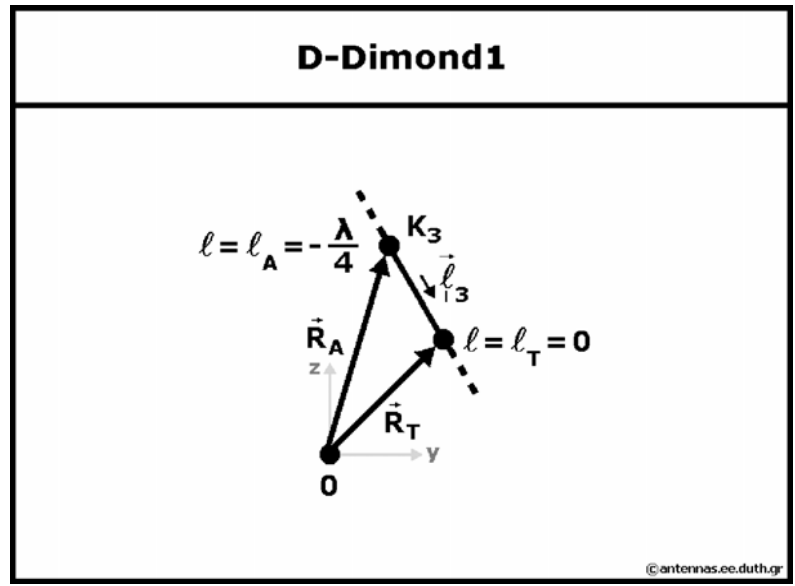

Figure 5 : Segment number 3

The data of the remaining segments are similarly calculated and are being given analytically in the following tables:

**GEOMETRY DATA I/II**

| # | $\ell$ | | $\vec{R}_A$ | | $\vec{R}_T$ | | $\ell$ | | $\vec{R}_\kappa$ | |
|---|---|---|---|---|---|---|---|---|---|---|
| | $\ell_A$ | $\ell_T$ | $\vec{y}$ | $\vec{z}$ | $\vec{y}$ | $\vec{z}$ | $\vec{y}$ | $\vec{z}$ | $\vec{y}_\kappa$ | $\vec{z}_\kappa$ |
| 1 | $0$ | $\dfrac{\lambda}{4}$ | $0$ | $0$ | $\dfrac{\lambda}{4}$ | $0$ | $1$ | $0$ | $0$ | $0$ |
| 2 | $0$ | $\dfrac{\lambda}{4}$ | $\dfrac{\lambda}{4}$ | $\dfrac{\lambda}{4}$ | $\dfrac{\lambda}{4}$ | $0$ | $0$ | $-1$ | $\dfrac{\lambda}{4}$ | $\dfrac{\lambda}{4}$ |
| 3 | $-\dfrac{\lambda}{4}$ | $0$ | $\dfrac{\lambda}{8}$ | $\dfrac{\lambda(1+\frac{\sqrt{3}}{2})}{4}$ | $\dfrac{\lambda}{4}$ | $\dfrac{\lambda}{4}$ | $\dfrac{1}{2}$ | $-\dfrac{\sqrt{3}}{2}$ | $\dfrac{\lambda}{4}$ | $\dfrac{\lambda}{4}$ |
| 4 | $-\dfrac{\lambda}{4}$ | $0$ | $\dfrac{\lambda}{8}$ | $\dfrac{\lambda(1+\frac{\sqrt{3}}{2})}{4}$ | $0$ | $\dfrac{\lambda(1+\sqrt{3})}{4}$ | $-\dfrac{1}{2}$ | $\dfrac{\sqrt{3}}{2}$ | $0$ | $\dfrac{\lambda(1+\sqrt{3})}{4}$ |
| 5 | $0$ | $\dfrac{\lambda}{4}$ | $0$ | $\dfrac{\lambda(1+\sqrt{3})}{4}$ | $-\dfrac{\lambda}{8}$ | $\dfrac{\lambda(1+\frac{\sqrt{3}}{2})}{4}$ | $-\dfrac{1}{2}$ | $-\dfrac{\sqrt{3}}{2}$ | $0$ | $\dfrac{\lambda(1+\sqrt{3})}{4}$ |
| 6 | $0$ | $\dfrac{\lambda}{4}$ | $-\dfrac{\lambda}{4}$ | $\dfrac{\lambda}{4}$ | $-\dfrac{\lambda}{8}$ | $\dfrac{\lambda(1+\frac{\sqrt{3}}{2})}{4}$ | $\dfrac{1}{2}$ | $\dfrac{\sqrt{3}}{2}$ | $-\dfrac{\lambda}{4}$ | $\dfrac{\lambda}{4}$ |
| 7 | $-\dfrac{\lambda}{4}$ | $0$ | $-\dfrac{\lambda}{4}$ | $0$ | $-\dfrac{\lambda}{4}$ | $\dfrac{\lambda}{4}$ | $0$ | $1$ | $-\dfrac{\lambda}{4}$ | $\dfrac{\lambda}{4}$ |
| 8 | $-\dfrac{\lambda}{4}$ | $0$ | $-\dfrac{\lambda}{4}$ | $0$ | $-\dfrac{\lambda}{4}$ | $-\dfrac{\lambda}{4}$ | $0$ | $-1$ | $-\dfrac{\lambda}{4}$ | $-\dfrac{\lambda}{4}$ |

| # | $\ell$ | | $\vec{R}_A$ | | $\vec{R}_T$ | | $\ell$ | | $\vec{R}_\kappa$ | |
|---|---|---|---|---|---|---|---|---|---|---|
| | $\ell_A$ | $\ell_T$ | $\vec{y}$ | $\vec{z}$ | $\vec{y}$ | $\vec{z}$ | $\vec{y}$ | $\vec{z}$ | $\vec{y}$ | $\vec{z}$ |
| 9 | $0$ | $\frac{\lambda}{4}$ | $-\frac{\lambda}{4}$ | $-\frac{\lambda}{4}$ | $-\frac{\lambda}{8}$ | $-\frac{\lambda(1+\frac{\sqrt{3}}{2})}{4}$ | $\frac{1}{2}$ | $-\frac{\sqrt{3}}{2}$ | $-\frac{\lambda}{4}$ | $-\frac{\lambda}{4}$ |
| 10 | $0$ | $\frac{\lambda}{4}$ | $0$ | $-\frac{\lambda(1+\sqrt{3})}{4}$ | $-\frac{\lambda}{8}$ | $-\frac{\lambda(1+\frac{\sqrt{3}}{2})}{4}$ | $-\frac{1}{2}$ | $\frac{\sqrt{3}}{2}$ | $0$ | $-\frac{\lambda(1+\sqrt{3})}{4}$ |
| 11 | $-\frac{\lambda}{4}$ | $0$ | $\frac{\lambda}{8}$ | $-\frac{\lambda(1+\frac{\sqrt{3}}{2})}{4}$ | $0$ | $-\frac{\lambda(1+\frac{\sqrt{3}}{2})}{4}$ | $-\frac{1}{2}$ | $-\frac{\sqrt{3}}{2}$ | $0$ | $-\frac{\lambda(1+\sqrt{3})}{4}$ |
| 12 | $-\frac{\lambda}{4}$ | $0$ | $\frac{\lambda}{8}$ | $-\frac{\lambda(1+\frac{\sqrt{3}}{2})}{4}$ | $\frac{\lambda}{4}$ | $-\frac{\lambda}{4}$ | $\frac{1}{2}$ | $\frac{\sqrt{3}}{2}$ | $\frac{\lambda}{4}$ | $-\frac{\lambda}{4}$ |
| 13 | $0$ | $\frac{\lambda}{4}$ | $\frac{\lambda}{4}$ | $-\frac{\lambda}{4}$ | $\frac{\lambda}{4}$ | $0$ | $0$ | $1$ | $\frac{\lambda}{4}$ | $-\frac{\lambda}{4}$ |
| 14 | $-\frac{\lambda}{4}$ | $0$ | $-\frac{\lambda}{4}$ | $0$ | $0$ | $0$ | $1$ | $0$ | $0$ | $0$ |

| # | $R_{\kappa_r}$ | $\ell_{|r}$ | $\ell_{|\theta}$ | $\ell_{|\varphi}$ |
|---|---|---|---|---|
| | | **GEOMETRY DATA II/II** | | |
| 1 | $0$ | $\sin\theta\sin\phi$ | $\cos\theta\sin\phi$ | $\cos\phi$ |
| 2 | $\frac{\lambda}{4}\sin\theta\sin\phi + \frac{\lambda}{4}\cos\theta$ | $-\cos\theta$ | $\sin\theta$ | $0$ |
| 3 | $\frac{\lambda}{4}\sin\theta\sin\phi + \frac{\lambda}{4}\cos\theta$ | $\frac{1}{2}\sin\theta\sin\phi - \frac{\sqrt{3}}{2}\cos\theta$ | $\frac{1}{2}\sin\theta\sin\phi + \frac{\sqrt{3}}{2}\cos\theta$ | $\frac{1}{2}\cos\phi$ |
| 4 | $\frac{\lambda}{4}(1+\sqrt{3})\cos\theta$ | $-\frac{1}{2}\sin\theta\sin\phi + \frac{\sqrt{3}}{2}\cos\theta$ | $-\frac{1}{2}\sin\theta\sin\phi - \frac{\sqrt{3}}{2}\cos\theta$ | $-\frac{1}{2}\cos\phi$ |
| 5 | $\frac{\lambda}{4}(1+\sqrt{3})\cos\theta$ | $-\frac{1}{2}\sin\theta\sin\phi - \frac{\sqrt{3}}{2}\cos\theta$ | $-\frac{1}{2}\sin\theta\sin\phi + \frac{\sqrt{3}}{2}\cos\theta$ | $-\frac{1}{2}\cos\phi$ |
| 6 | $-\frac{\lambda}{4}\sin\theta\sin\phi + \frac{\lambda}{4}\cos\theta$ | $\frac{1}{2}\sin\theta\sin\phi + \frac{\sqrt{3}}{2}\cos\theta$ | $\frac{1}{2}\sin\theta\sin\phi - \frac{\sqrt{3}}{2}\cos\theta$ | $\frac{1}{2}\cos\phi$ |
| 7 | $-\frac{\lambda}{4}\sin\theta\sin\phi + \frac{\lambda}{4}\cos\theta$ | $\cos\theta$ | $-\sin\theta$ | $0$ |

| # | $R_{\kappa_r}$ | $\ell_{|r}$ | $\ell_{|\theta}$ | $\ell_{|\varphi}$ |
|---|---|---|---|---|
| 8 | $-\frac{\lambda}{4}\sin\theta\sin\phi - \frac{\lambda}{4}\cos\theta$ | $-\cos\theta$ | $\sin\theta$ | $0$ |
| 9 | $-\frac{\lambda}{4}\sin\theta\sin\phi - \frac{\lambda}{4}\cos\theta$ | $\frac{1}{2}\sin\theta\sin\phi - \frac{\sqrt{3}}{2}\cos\theta$ | $\frac{1}{2}\sin\theta\sin\phi + \frac{\sqrt{3}}{2}\cos\theta$ | $\frac{1}{2}\cos\phi$ |
| 10 | $-\frac{\lambda}{4}(1+\sqrt{3})\cos\theta$ | $-\frac{1}{2}\sin\theta\sin\phi + \frac{\sqrt{3}}{2}\cos\theta$ | $-\frac{1}{2}\sin\theta\sin\phi - \frac{\sqrt{3}}{2}\cos\theta$ | $-\frac{1}{2}\cos\phi$ |
| 11 | $-\frac{\lambda}{4}(1+\sqrt{3})\cos\theta$ | $-\frac{1}{2}\sin\theta\sin\phi - \frac{\sqrt{3}}{2}\cos\theta$ | $-\frac{1}{2}\sin\theta\sin\phi + \frac{\sqrt{3}}{2}\cos\theta$ | $-\frac{1}{2}\cos\phi$ |
| 12 | $\frac{\lambda}{4}\sin\theta\sin\phi - \frac{\lambda}{4}\cos\theta$ | $\frac{1}{2}\sin\theta\sin\phi + \frac{\sqrt{3}}{2}\cos\theta$ | $\frac{1}{2}\sin\theta\sin\phi - \frac{\sqrt{3}}{2}\cos\theta$ | $\frac{1}{2}\cos\phi$ |
| 13 | $\frac{\lambda}{4}\sin\theta\sin\phi - \frac{\lambda}{4}\cos\theta$ | $\cos\theta$ | $-\sin\theta$ | $0$ |
| 14 | $0$ | $\sin\theta\sin\phi$ | $\cos\theta\sin\phi$ | $\cos\phi$ |

**Calculation of the radiation pattern's factor**

The expression of the radiation pattern's factor, $\dot{PF}$, at a segment where $\ell_A < \ell_T$, is calculated from a standing wave dipole $[-h, h]$ and is given by [6]:

$$(1): \dot{PF} = \dot{I}\left(\mathcal{P}\dot{\mathcal{F}}_{(\ell_T)} - \mathcal{P}\dot{\mathcal{F}}_{(\ell_A)}\right)$$

where:

$$\mathcal{P}\dot{\mathcal{F}}_{(\ell)} = \int \sin[\beta(h - s(\ell)\ell)]\, e^{i\beta\ell_r}\, d\ell$$

The preceding indefinite integral has the following analytical solution [5]:

$$(2): \mathcal{P}\dot{\mathcal{F}}_{(\ell)} = e^{i\beta\ell_r\ell}\, \frac{s(\ell)\cos[\beta(h - s(\ell)\ell)] + i\ell_r\sin[\beta(h - s(\ell)\ell)]}{\beta[s^2(\ell) - \ell_r^2]}$$

Remarks:

There are two cases of variations in the value of variable $\ell$ which result from Figure 2.3.1 and are based on the particular geometry of the antenna as well as the directions of each segment currents: At the 1st case $\ell_A < \ell < \ell_T = 0$, with $\ell_A = -\frac{\lambda}{4}$ and $\ell_T = 0$. Similarly, at the 2nd case $0 = \ell_A < \ell < \ell_T$, with $\ell_A = 0$ and $\ell_T = \frac{\lambda}{4}$.

The voltage's nodes are considered to be nullification points of the variable's $\ell$ value. Thus, the h constant will be equal with the distance between these points and the current's nodes. This distance, because of the specific geometry of the antenna, is constant for each segment and equals $\lambda/4$.

Taking into consideration all above it is obtained:

$$\beta\frac{\lambda}{4} = \frac{2\pi}{\lambda}\frac{\lambda}{4} = \frac{\pi}{2}$$

$$\sin(\beta\frac{\lambda}{4}) = \sin(\frac{\pi}{2}) = 1$$

$$\cos(\beta\frac{\lambda}{4}) = \cos(\frac{\pi}{2}) = 0$$

and focusing in the 1st case, 2.3. (1) gives:

$(3): \dot{PF} = \dot{I}_\delta \left[ \dot{\mathcal{PF}}_{(0)} - \dot{\mathcal{PF}}_{\left(-\frac{\lambda}{4}\right)} \right]$

Utilizing 2.3. (2) and 2.3. (3) it is obtained:

$$\dot{\mathcal{PF}}_{(0)} = \frac{-\cos\left[\beta(\frac{\lambda}{4}+0)\right] + i\,\ell_r \sin\left[\beta(\frac{\lambda}{4}+0)\right]}{\beta(1-\ell_r^2)} = \frac{\cos(\frac{\pi}{2}) + i\,\ell_r \sin(\frac{\pi}{2})}{\beta(1-\ell_r^2)} = i\,\frac{\ell_r}{\beta(1-\ell_r^2)}$$

$$\dot{\mathcal{PF}}_{\left(-\frac{\lambda}{4}\right)} = e^{-i\beta\ell_r\frac{\lambda}{4}}\,\frac{-\cos\left[\beta(\frac{\lambda}{4}-\frac{\lambda}{4})\right] + i\,\ell_r \sin\left[\beta(\frac{\lambda}{4}-\frac{\lambda}{4})\right]}{\beta(1-\ell_r^2)} = -\frac{e^{-i\beta\ell_r\frac{\lambda}{4}}}{\beta(1-\ell_r^2)}$$

Finally, 2.3. (3) becomes:

$$\dot{PF} = \dot{I}_\delta \left[ i\,\frac{\ell_r}{\beta(1-\ell_r^2)} + \frac{e^{-i\beta\ell_r\frac{\lambda}{4}}}{\beta(1-\ell_r^2)} \right] \Rightarrow$$

$$(4): \dot{PF} = \dot{I}_\delta\,\frac{\cos(\beta\ell_r\frac{\lambda}{4}) + i[\ell_r - \sin(\beta\ell_r\frac{\lambda}{4})]}{\beta(1-\ell_r^2)}$$

Remark

        The above formula of radiation's pattern factor has been derived as the variable $\ell$ varies along the interval $(-\frac{\lambda}{4}, 0)$. It will be considered from now on that 2.3. (4) will be valid without any proof for any other case of variable's $\ell$ variation along the pre-mentioned interval of values.

Focusing in the 2nd case, where variable's $\ell$ values vary along the interval $(0, \frac{\lambda}{4})$, 2.3.(1) becomes:

$$(5): \dot{PF} = \dot{I}_\delta \left[ \dot{\mathcal{PF}}_{\left(\frac{\lambda}{4}\right)} - \dot{\mathcal{PF}}_{(0)} \right]$$

$$\dot{\mathcal{PF}}_{\left(\frac{\lambda}{4}\right)} = e^{i\beta\ell_r\frac{\lambda}{4}} \frac{\cos[\beta(\frac{\lambda}{4} - \frac{\lambda}{4})] + i\ell_r\sin[\beta(\frac{\lambda}{4} - \frac{\lambda}{4})]}{\beta(1 - \ell_r^2)} = \frac{e^{i\beta\ell_r\frac{\lambda}{4}}}{\beta(1 - \ell_r^2)}$$

$$\dot{\mathcal{PF}}_{(0)} = \frac{\cos[\beta(\frac{\lambda}{4} - 0)] + i\ell_r\sin[\beta(\frac{\lambda}{4} - 0)]}{\beta(1 - \ell_r^2)} = \frac{\cos(\frac{\pi}{2}) + i\ell_r\sin(\frac{\pi}{2})}{\beta(1 - \ell_r^2)} = i\frac{\ell_r}{\beta(1 - \ell_r^2)}$$

Finally, 2.3.(5) gives:

$$\dot{PF} = \dot{I}_\delta\left[\frac{e^{i\beta\ell_r\frac{\lambda}{4}}}{\beta(1 - \ell_r^2)} - i\frac{\ell_r}{\beta(1 - \ell_r^2)}\right] \Rightarrow$$

$$(6): \dot{PF} = \dot{I}_\delta\frac{\cos(\beta\ell_r\frac{\lambda}{4}) - i[\ell_r - \sin(\beta\ell_r\frac{\lambda}{4})]}{\beta(1 - \ell_r^2)}$$

Remark

Following the same philosophy, the formula of radiation's pattern factor has been derived as the variable $\ell$ varies along the interval $(0, \frac{\lambda}{4})$. Similarly, it will be considered that 2.3. (6) will be in effect without proof for any other case of variable's $\ell$ variation along the pre-mentioned interval of values.

**Evaluation of the radiation pattern**

Since all the general formulas of the radiation pattern's factors that are in effect at each segment have been calculated, the radiation pattern for each segment henceforth can be evaluated. Utilizing the general formula 2.1. (1) as well as the data of tables 2.3.1 and 2.3.2 it is obtained for each segment:

For segment 1:

$$(7): \overline{\mathbb{E}}_1 = (\frac{\dot{I}_\delta}{2\pi})\frac{\cos(\frac{\pi}{2}\sin\theta\sin\phi)}{1 - (\sin\theta\sin\phi)^2} - i\frac{\sin\theta\sin\phi - \sin(\frac{\pi}{2}\sin\theta\sin\phi)}{1 - (\sin\theta\sin\phi)^2}\begin{bmatrix}\cos\theta\sin\phi \\ \cos\phi\end{bmatrix}$$

For segment 2:

$(8):\ \overline{\mathbb{E}}_2 = (\frac{\dot{I}\delta}{2\pi})(\frac{1}{\sin\theta})\left[\cos(\frac{\pi}{2}\cos\theta) + i[\cos\theta - \sin(\frac{\pi}{2}\cos\theta)]\right]e^{i\frac{\pi}{2}(\sin\theta\sin\phi + \cos\theta)}\ \vec{\theta}$

For segment 3:

$(9):\ \overline{\mathbb{E}}_3 = (\frac{\dot{I}\delta}{4\pi})\left[\frac{\cos[\frac{\pi}{4}(\sqrt{3}\cos\theta - \sin\theta\sin\phi)]}{1 - \frac{1}{4}(\sqrt{3}\cos\theta - \sin\theta\sin\phi)^2} - i\frac{\frac{1}{2}(\sqrt{3}\cos\theta - \sin\theta\sin\phi]}{1 - \frac{1}{4}(\sqrt{3}\cos\theta - \sin\theta\sin\phi)^2}\right.$

$\left.\frac{-\sin[\frac{\pi}{4}(\sqrt{3}\cos\theta - \sin\theta\sin\phi)}{1 - \frac{1}{4}(\sqrt{3}\cos\theta - \sin\theta\sin\phi)^2}\right]e^{i\frac{\pi}{2}(\sin\theta\sin\phi + \cos\theta)}\begin{bmatrix}\cos\theta\sin\phi + \sqrt{3}\sin\theta \\ \cos\phi\end{bmatrix}$

For segment 4:

$(10):\ \overline{\mathbb{E}}_4 = -(\frac{\dot{I}\delta}{4\pi})\left[\frac{\cos[\frac{\pi}{4}(\sqrt{3}\cos\theta - \sin\theta\sin\phi)]}{1 - \frac{1}{4}(\sqrt{3}\cos\theta - \sin\theta\sin\phi)^2} + i\frac{\frac{1}{2}(\sqrt{3}\cos\theta - \sin\theta\sin\phi)}{1 - \frac{1}{4}(\sqrt{3}\cos\theta - \sin\theta\sin\phi)^2}\right.$

$\left.\frac{-\sin[\frac{\pi}{4}(\sqrt{3}\cos\theta - \sin\theta\sin\phi)]}{1 - \frac{1}{4}(\sqrt{3}\cos\theta - \sin\theta\sin\phi)^2}\right]e^{i\frac{\pi}{2}[(1+\sqrt{3})\cos\theta]}\begin{bmatrix}\cos\theta\sin\phi + \sqrt{3}\sin\theta \\ \cos\phi\end{bmatrix}$

For segment 5:

$(11):\ \overline{\mathbb{E}}_5 = -(\frac{\dot{I}\delta}{4\pi})\left[\frac{\cos[\frac{\pi}{4}(\sin\theta\sin\phi + \sqrt{3}\cos\theta)]}{1 - \frac{1}{4}(\sin\theta\sin\phi + \sqrt{3}\cos\theta)^2} + i\frac{\frac{1}{2}(\sin\theta\sin\phi + \sqrt{3}\cos\theta)}{1 - \frac{1}{4}(\sin\theta\sin\phi + \sqrt{3}\cos\theta)^2}\right.$

$\left.\frac{-\sin[\frac{\pi}{4}(\sin\theta\sin\phi + \sqrt{3}\cos\theta)]}{1 - \frac{1}{4}(\sin\theta\sin\phi + \sqrt{3}\cos\theta)^2}\right]e^{i\frac{\pi}{2}[(1+\sqrt{3})\cos\theta]}\begin{bmatrix}\cos\theta\sin\phi - \sqrt{3}\sin\theta \\ \cos\phi\end{bmatrix}$

For segment 6:

$(12):\ \overline{\mathbb{E}}_6 = (\frac{\dot{I}\delta}{4\pi})\left[\frac{\cos[\frac{\pi}{4}(\sin\theta\sin\phi + \sqrt{3}\cos\theta)]}{1 - \frac{1}{4}(\sin\theta\sin\phi + \sqrt{3}\cos\theta)^2} - i\frac{\frac{1}{2}(\sin\theta\sin\phi + \sqrt{3}\cos\theta)}{1 - \frac{1}{4}(\sin\theta\sin\phi + \sqrt{3}\cos\theta)^2}\right.$

$\left.\frac{-\sin[\frac{\pi}{4}(\sin\theta\sin\phi + \sqrt{3}\cos\theta)]}{1 - \frac{1}{4}(\sin\theta\sin\phi + \sqrt{3}\cos\theta)^2}\right]e^{-i\frac{\pi}{2}(\sin\theta\sin\phi - \cos\theta)}\begin{bmatrix}\cos\theta\sin\phi - \sqrt{3}\sin\theta \\ \cos\phi\end{bmatrix}$

For segment 7:

$$(13): \overline{\mathbb{E}}_7 = -\left(\frac{\dot{I}\delta}{2\pi}\right)\left(\frac{1}{\sin\theta}\right)\left[\cos\left(\frac{\pi}{2}\cos\theta\right) + i[\cos\theta - \sin\left(\frac{\pi}{2}\cos\theta\right)]\right]e^{-i\frac{\pi}{2}(\sin\theta\sin\phi - \cos\theta)}\vec{\theta}_\parallel$$

For segment 8:

$$(14): \overline{\mathbb{E}}_8 = \left(\frac{\dot{I}\delta}{2\pi}\right)\left(\frac{1}{\sin\theta}\right)\left[\cos\left(\frac{\pi}{2}\cos\theta\right) - i[\cos\theta - \sin\left(\frac{\pi}{2}\cos\theta\right)]\right]e^{-i\frac{\pi}{2}(\sin\theta\sin\phi + \cos\theta)}\vec{\theta}_\parallel$$

For segment 9:

$$(15): \overline{\mathbb{E}}_9 = \left(\frac{\dot{I}\delta}{4\pi}\right)\left[\frac{\cos[\frac{\pi}{4}(\sin\theta\sin\phi - \sqrt{3}\cos\theta)]}{1 - \frac{1}{4}(\sin\theta\sin\phi - \sqrt{3}\cos\theta)^2} + i\frac{\frac{1}{2}(\sqrt{3}\cos\theta - \sin\theta\sin\phi)}{1 - \frac{1}{4}(\sqrt{3}\cos\theta - \sin\theta\sin\phi)^2}\right.$$

$$\left.\frac{-\sin[\frac{\pi}{4}(\sqrt{3}\cos\theta - \sin\theta\sin\phi)]}{1 - \frac{1}{4}(\sin\theta\sin\phi - \sqrt{3}\cos\theta)^2}\right]e^{-i\frac{\pi}{2}(\sin\theta\sin\phi + \cos\theta)}\begin{bmatrix}\cos\theta\sin\phi + \sqrt{3}\sin\theta \\ \cos\phi\end{bmatrix}$$

For segment 10:

$$(16): \overline{\mathbb{E}}_{10} = -\left(\frac{\dot{I}\delta}{4\pi}\right)\left[\frac{\cos[\frac{\pi}{4}(\sqrt{3}\cos\theta - \sin\theta\sin\phi)]}{1 - \frac{1}{4}(\sqrt{3}\cos\theta - \sin\theta\sin\phi)^2} - i\frac{\frac{1}{2}(\sqrt{3}\cos\theta - \sin\theta\sin\phi)}{1 - \frac{1}{4}(\sqrt{3}\cos\theta - \sin\theta\sin\phi)^2}\right.$$

$$\left.\frac{-\sin[\frac{\pi}{4}(\sqrt{3}\cos\theta - \sin\theta\sin\phi)]}{1 - \frac{1}{4}(\sqrt{3}\cos\theta - \sin\theta\sin\phi)^2}\right]e^{-i\frac{\pi}{2}[(1+\sqrt{3})\cos\theta]}\begin{bmatrix}\cos\theta\sin\phi + \sqrt{3}\sin\theta \\ \cos\phi\end{bmatrix}$$

For segment 11:

$$(17): \overline{\mathbb{E}}_{11} = -\left(\frac{\dot{I}\delta}{4\pi}\right)\left[\frac{\cos[\frac{\pi}{4}(\sin\theta\sin\phi + \sqrt{3}\cos\theta)]}{1 - \frac{1}{4}(\sin\theta\sin\phi + \sqrt{3}\cos\theta)^2} - i\frac{\frac{1}{2}(\sin\theta\sin\phi + \sqrt{3}\cos\theta)}{1 - \frac{1}{4}(\sin\theta\sin\phi + \sqrt{3}\cos\theta)^2}\right.$$

$$\left.\frac{-\sin[\frac{\pi}{4}(\sin\theta\sin\phi + \sqrt{3}\cos\theta)]}{1 - \frac{1}{4}(\sin\theta\sin\phi + \sqrt{3}\cos\theta)^2}\right]e^{-i\frac{\pi}{2}[(1+\sqrt{3})\cos\theta]}\begin{bmatrix}\cos\theta\sin\phi - \sqrt{3}\sin\theta \\ \cos\phi\end{bmatrix}$$

For segment 12:

$(18): \overline{\mathbb{E}}_{12} = (\frac{\dot{I}_\delta}{4\pi}) \left[ \frac{\cos[\frac{\pi}{4}(\sin\theta\sin\phi + \sqrt{3}\cos\theta)]}{1 - \frac{1}{4}(\sin\theta\sin\phi + \sqrt{3}\cos\theta)^2} + i\frac{\frac{1}{2}(\sin\theta\sin\phi + \sqrt{3}\cos\theta)}{1 - \frac{1}{4}(\sin\theta\sin\phi + \sqrt{3}\cos\theta)^2} \right.$

$\left. \frac{-\sin[\frac{\pi}{4}(\sin\theta\sin\phi + \sqrt{3}\cos\theta)]}{1 - \frac{1}{4}(\sin\theta\sin\phi + \sqrt{3}\cos\theta)^2} \right] e^{i\frac{\pi}{2}(\sin\theta\sin\phi - \cos\theta)} \begin{bmatrix} \cos\theta\sin\phi - \sqrt{3}\sin\theta \\ \cos\phi \end{bmatrix}$

For segment 13:

$(19): \overline{\mathbb{E}}_{13} = -(\frac{\dot{I}_\delta}{2\pi})(\frac{1}{\sin\theta}) \left[ \cos(\frac{\pi}{2}\cos\theta) - i[\cos\theta - \sin(\frac{\pi}{2}\cos\theta)] \right] e^{i\frac{\pi}{2}(\sin\theta\sin\phi - \cos\theta)} \vec{\theta}$

For segment 14:

$(20): \overline{\mathbb{E}}_{14} = (\frac{\dot{I}_\delta}{2\pi}) \frac{\cos(\frac{\pi}{2}\sin\theta\sin\phi)}{1 - (\sin\theta\sin\phi)^2} + i\frac{\sin\theta\sin\phi - \sin(\frac{\pi}{2}\sin\theta\sin\phi)}{1 - (\sin\theta\sin\phi)^2} \begin{bmatrix} \cos\theta\sin\phi \\ \cos\phi \end{bmatrix}$

The overall radiation pattern $\overline{\mathbb{E}}$ results from the vectorial sum of each "segmental" radiation pattern:

$$\overline{\mathbb{E}} = \overline{\mathbb{E}}_1 + \overline{\mathbb{E}}_2 + \overline{\mathbb{E}}_3 + \overline{\mathbb{E}}_4 + \overline{\mathbb{E}}_5 + \overline{\mathbb{E}}_6 + \overline{\mathbb{E}}_7 + \overline{\mathbb{E}}_8 + \overline{\mathbb{E}}_9 + \overline{\mathbb{E}}_{10} + \overline{\mathbb{E}}_{11} + \overline{\mathbb{E}}_{12} + \overline{\mathbb{E}}_{13} + \overline{\mathbb{E}}_{14}$$

The components $\dot{\mathbb{E}}_{o\lambda_\theta}$ και $\dot{\mathbb{E}}_{o\lambda_\phi}$ of the overall radiation pattern end up to be real except the contribution of $\dot{I}_\delta$ (Annex H). Thus, it results that:

$(21): \overline{\overline{\mathbb{E}}}_{o\lambda} = \begin{bmatrix} \mathbb{E}_{o\lambda} \vec{\theta} \\ \mathbb{E}_{o\lambda} \vec{\phi} \end{bmatrix}$

The final expressions of the two components of the overall radiation pattern are (Annex H):

(22): $\dot{\mathbb{E}}_{o\lambda_\theta} = (\frac{\dot{I}\delta}{2\pi})\left[(\frac{2}{\sin\theta})\left[\cos(\frac{\pi}{2}\cos\theta)\right]\left[\cos(a_2) - \cos(a_1)\right]\right.$

$$+ \frac{\cos(\frac{\pi}{2}b_1)}{1-b_1{}^2}c_1\left[\cos(a_1) - \cos(a_3)\right] + \frac{\cos(\frac{\pi}{2}b_2)}{1-b_2{}^2}c_2\left[\cos(a_2) - \cos(a_3)\right]$$

$$+ \frac{2\cos(\frac{\pi}{2}b_3)}{1-b_3{}^2}c_3 - (\frac{2}{\sin\theta})\left[\cos\theta - \sin(\frac{\pi}{2}\cos\theta)\right]\left[\sin(a_2) + \sin(a_1)\right]$$

$$+ \frac{b_1 - \sin(\frac{\pi}{2}b_1)}{1-b_1{}^2}c_1\left[\sin(a_3) - \sin(a_1)\right]$$

$$\left. + \frac{b_2 - \sin(\frac{\pi}{2}b_2)}{1-b_2{}^2}c_2\left[\sin(a_2) + \sin(a_3)\right]\right]$$

(23): $\dot{\mathbb{E}}_{o\lambda_\phi} = (\frac{\dot{I}\delta}{2\pi}\cos\phi)\left[\frac{\cos(\frac{\pi}{2}b_1)}{1-b_1{}^2}\left[\cos(a_1) - \cos(a_3)\right] + \frac{\cos(\frac{\pi}{2}b_2)}{1-b_2{}^2}\left[\cos(a_2) - \cos(a_3)\right]\right.$

$$+ \frac{2\cos(\frac{\pi}{2}b_3)}{1-b_3{}^2} + \frac{b_1 - \sin(\frac{\pi}{2}b_1)}{1-b_1{}^2}\left[\sin(a_3) - \sin(a_1)\right]$$

$$\left. + \frac{b_2 - \sin(\frac{\pi}{2}b_2)}{1-b_2{}^2}\left[\sin(a_2) + \sin(a_3)\right]\right]$$

where:

$a_1 = \frac{\pi}{2}(\sin\theta\sin\phi - \cos\theta)$

$a_2 = \frac{\pi}{2}(\sin\theta\sin\phi + \cos\theta)$

$b_1 = \frac{1}{2}(\sin\theta\sin\phi + \sqrt{3}\cos\theta)$

$c_1 = \cos\theta\sin\phi - \sqrt{3}\sin\theta$

$a_3 = \frac{\pi}{2}\left[(1 + \sqrt{3})\cos\theta\right]$

$b_2 = \frac{1}{2}(\sqrt{3}\cos\theta - \sin\theta\sin\phi)$

$c_2 = \cos\theta\sin\phi + \sqrt{3}\sin\theta$

$b_3 = \sin\theta\sin\phi$

$c_3 = \cos\theta\sin\phi$

Finally, the magnitude of the relation 2.3. (21) is given by the following expression:

$$(24): \mathbb{E}_{o\lambda} = \left|\overline{\overline{\mathbb{E}}}_{o\lambda}\right| = \left|\vec{\overline{\mathbb{E}}}_{o\lambda}\right| = \sqrt{\dot{\overline{\mathbb{E}}}_{o\lambda_\theta}{}^2 + \dot{\overline{\mathbb{E}}}_{o\lambda_\phi}{}^2}$$

At this point $\overline{\overline{\mathbb{E}}}_{o\lambda}$ can be calculated in space using the relations 2.3. (22), 2.3. (23) and 2.3. (24). The calculations will be carried out by a computer program written in C language (Annex A). The program flow as follows: firstly the magnitude of $\mathbb{E}_{o\lambda}$ per $5°$ is estimated, afterwards $\mathbb{E}_{max}$ is calculated and henceforth the normalized radiation pattern $\mathcal{E} = \dfrac{\mathbb{E}}{\mathbb{E}_{max}}$. Afterwards, the quantities $\mathcal{E}_X = \dfrac{\mathbb{E}}{\mathbb{E}_{max}} \sin\theta \cos\varphi$   $\mathcal{E}_Z = \dfrac{\mathbb{E}}{\mathbb{E}_{max}} \cos\theta$ $\mathcal{E}_y = \dfrac{\mathbb{E}}{\mathbb{E}_{max}} \sin\theta \sin\varphi$ are derived, which are the projections of the normalized radiation pattern onto the x, y and z axis. The program's output file, (E3OUT.TXT), is inserted as input file in RADPAT program and produces the extraction of the 3D radiation pattern which is illustrated in Figure 2.3.2 [2].

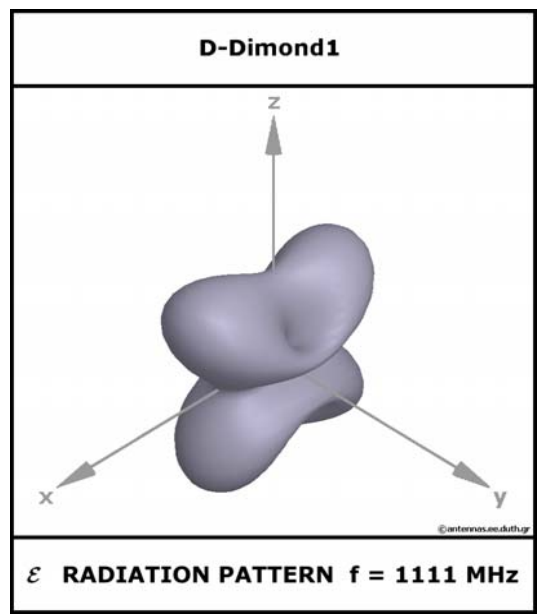

Figure 2 : The D-Dimond1 antenna's radiation pattern

Next, the extraction of the radiation pattern's expression ($\overline{\overline{\mathbb{E}}}_{o\lambda}$) at the three main planes follows. Therefore at the equations 2.3. (22) and 2.3. (23) the specific values that take the angle variables θ and φ at each one of the three planes is substituted. For the xOy plane $\hat{\theta} = 90°$, for the yOz plane $\hat{\varphi} = 90°$ or $\hat{\varphi} = 270°$ and for the zOx plane $\hat{\varphi} = 0°$ or $\hat{\varphi} = 180°$. The radiation pattern's evaluation at the three main planes is achieved with the assistance of a computer program written in C (Annex B). The program calculates the quantity's value per 1° and then normalizes it with its maximum value at each main plane. Below are cited the patterns at the three main planes in polar and cartesian coordinates [3].

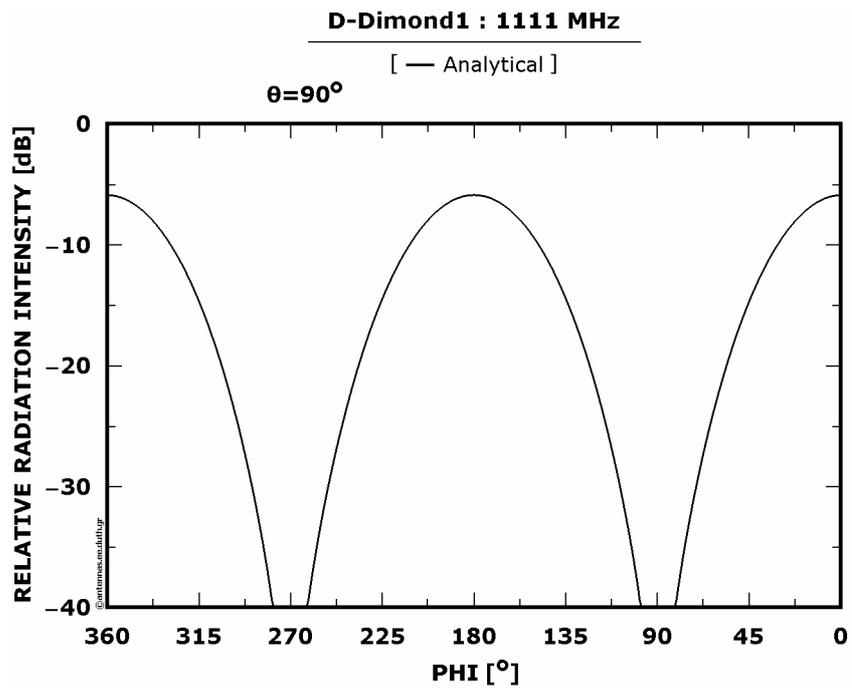

Figure 1 : Cartesian radiation pattern at xOy plane

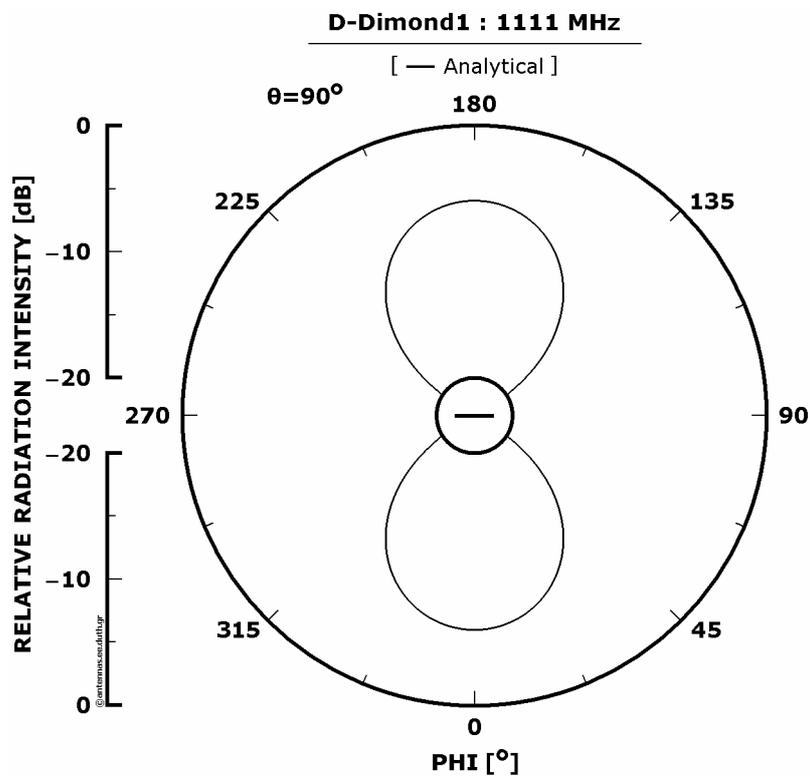

Figure 2 : Polar radiation pattern at xOy plane

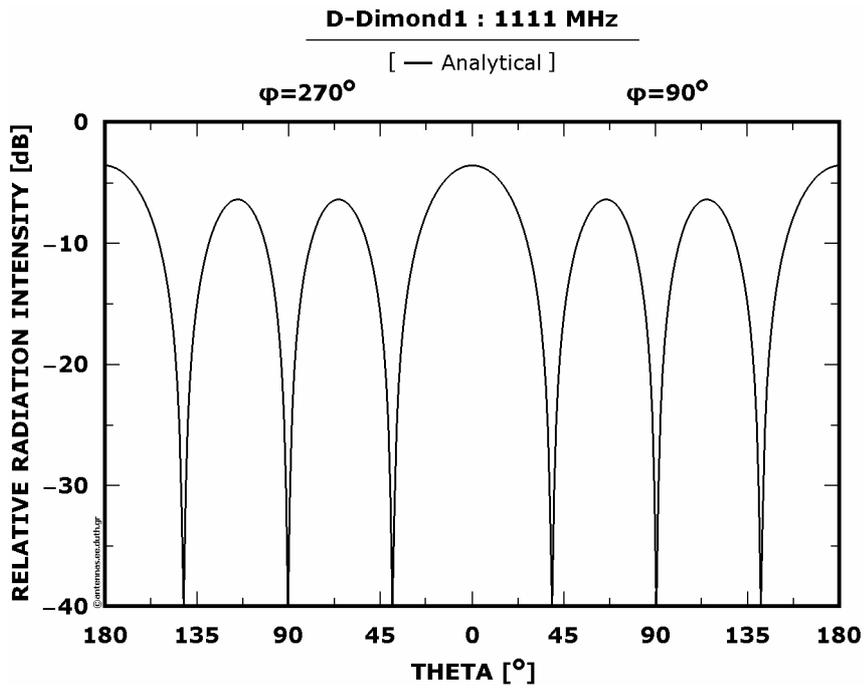

Figure 3 : Cartesian radiation pattern at yOz plane

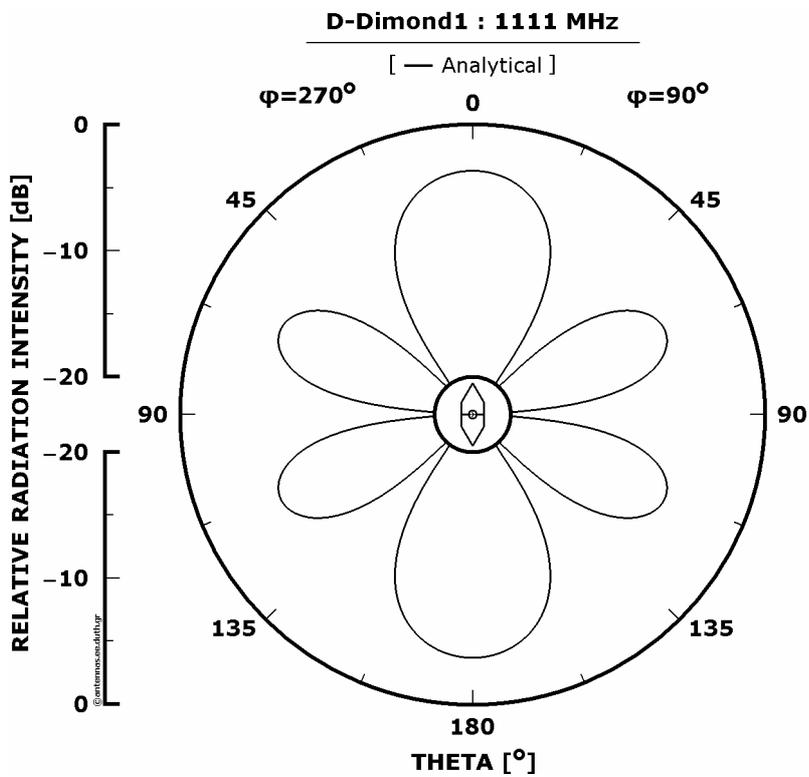

Figure 4 : Polar radiation pattern at yOz plane

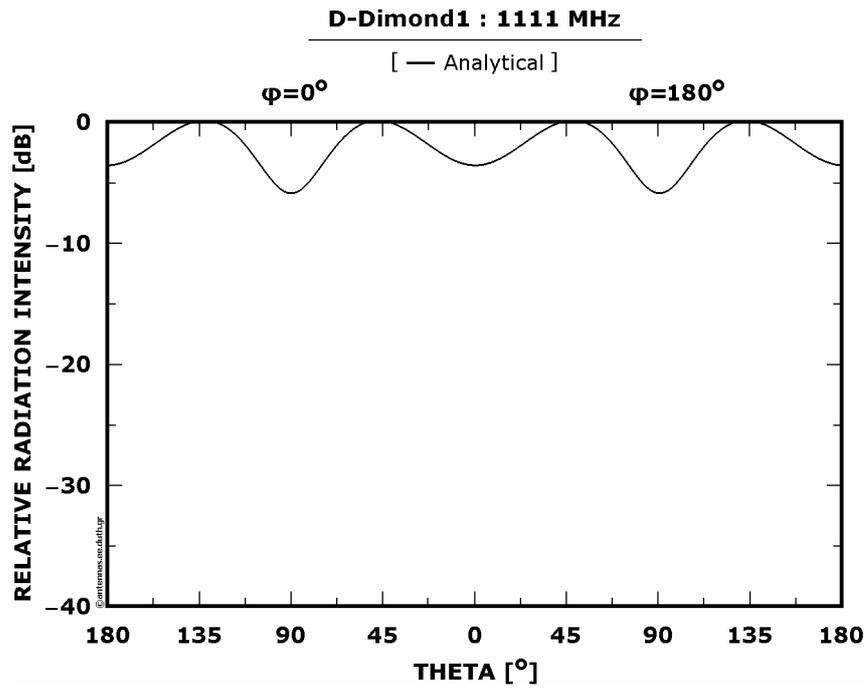

Figure 5 : Cartesian radiation pattern at zOx plane

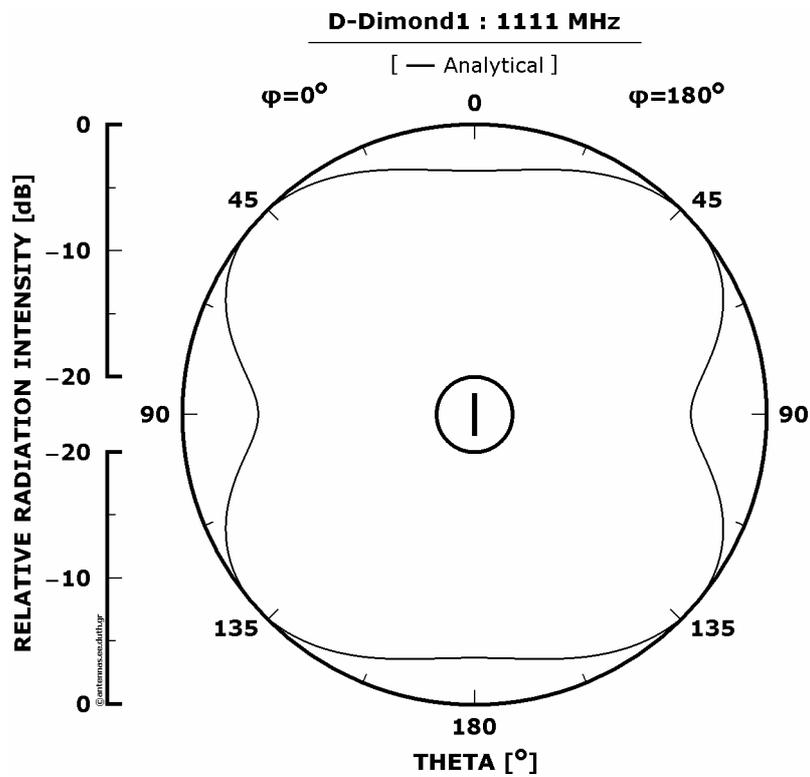

Figure 6 : Polar radiation pattern at zOx plane

# Chapter 3 : Computational study

## 3.1 : In General

It is generally admissible that before someone advances to the construction of an antenna a mathematical modelling and representation is necessary. This process is called simulation and assists so that some fundamental properties of the antenna can be determined before its realization and measurement. Also, simulation could play an important role at the optimization of the antenna right after its construction.

The knowledge of the current's distribution along the antenna's body is a critical factor for a successful simulation. Thus, various techniques have been developed in order that the current's distribution to be defined, the most important from which is the method of moments. This method handles the integral equations that result from the solution of the sources of EM fields (currents and charges). This method is effective and it is utilized mainly in EM problems that are related with linear materials, where the integral equations are degenerated in linear systems of equations.

**3.2 : Thesis methodology**

Despite of the effectiveness of the method of moments at the simplification of the integral equations, the handwritten solution of the linear systems of equations that result is almost impracticable and definitely outside of the aims of the present work. Therefore there are programs that have been developed which solve numerically and reliably these linear systems. One considerably known program is called THINWIRE, and its modified variant RICHWIRE, [4] which will be used at the present thesis. THINWIRE was initially developed in 1974 from J. H. Richmond written at FORTRAN language, which is the most suitable programming language to confront mathematical problems.

The simulation was carried out at the frequency of 1111 [MHz], which was selected as central operation frequency, where the wavelength is approximately 27 [cm]. The wire's radius from which the antenna should be constructed equals to λ [mm].

RICHWIRE is capable of dividing the antenna in more segments in order to achieve better precision to its results. The program functions with an input file which among else contains:

the coordinates of the nodes which compose the segments of the antenna

the way (orientation) that the nodes are connected between them in order to create the desirable segments

The problem of how many segments (thus, how many nodes too) the antenna should be divided emerges in order that RICHWIRE's input file will be created. It should be stressed out that there are certain requirements which are imported by the program itself:

1.      the length of each segment cannot exceed λ/4 (maximum distance between nodes equals λ/4)

2.      the length of each segment is not allowed to be smaller than the double wire's radius, that is to say 0.002 [m] (minimal distance between nodes equals to 0.002 [m])

3.      the feeding point of the antenna as well as the physical wires' connections should be points of division

Next step to the choice of number of segments is to examine the antenna's behaviour in a definite frequency rage. Thus, an examination will be carried out from 900 to 1300 [MHz], with a 10 [MHz] step.

Finally, the radiation patterns of D-Dimond1 that resulted from the process of the frequency scanning will be presented in space and at the three main planes.

The process of antenna's division is achieved with a suitable computer program (Annex C) written in C language where the number of sub-segments in which each segment of λ/4 can be divided is declared.

## 3.3 : Selection of the number of segments

Generally, as the number of segments increases, that is to say the length of each segment is decreased, the computational model's approximation of reality is improved. Thus, we are convicted that the simulation results "approach" closer to reality and particularly some reduction of their fluctuation's rate is expected as the number of segments is raising, that is to say there is an "ostensible" convergence. The observed convergence is described as "ostensible" because it is impossible to be depicted graphically by our limited-capability computers, as the calculation of infinite number of results is required. After a certain point the rate of the convergence is decreasing while the number of segments continues to increase therefore there isn't any practical improvement in accuracy. On the contrary, the increased number of segments leads to unnecessary computational resource's waste. The variation of the input impedance's real and imaginary part as a function of the number of segments is illustrated in Figure 3.3.1 which will be helpful in the effort of compromising the maximum practical convergence and the minimum number of segments.

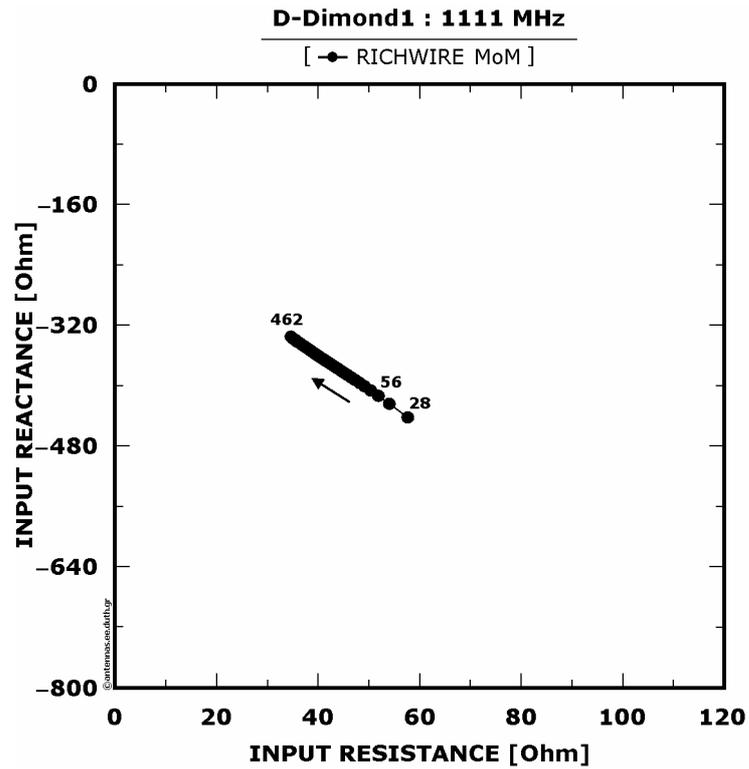

<u>Figure 1</u> : Input impedance as a function of the number of segments

Firstly, 28 segments have been selected while the division reaches its limit, that is 462 segments. It is observed that already from the 3$^{rd}$ division (56 segments) the "ostensible" convergence which is achieved is satisfactory, at least for the examined quantity. Below, the variations of the antenna's remainder quantities follow.

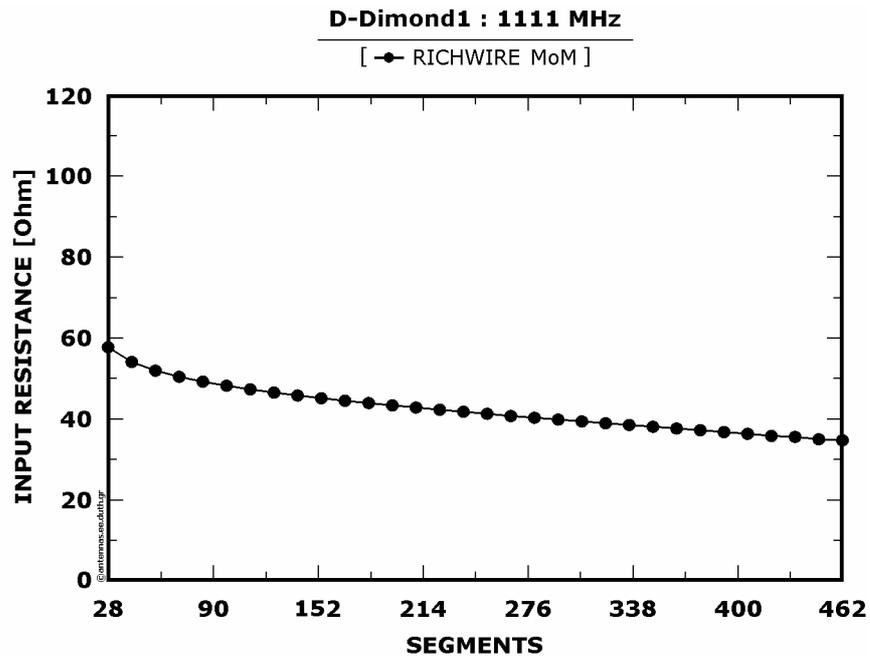

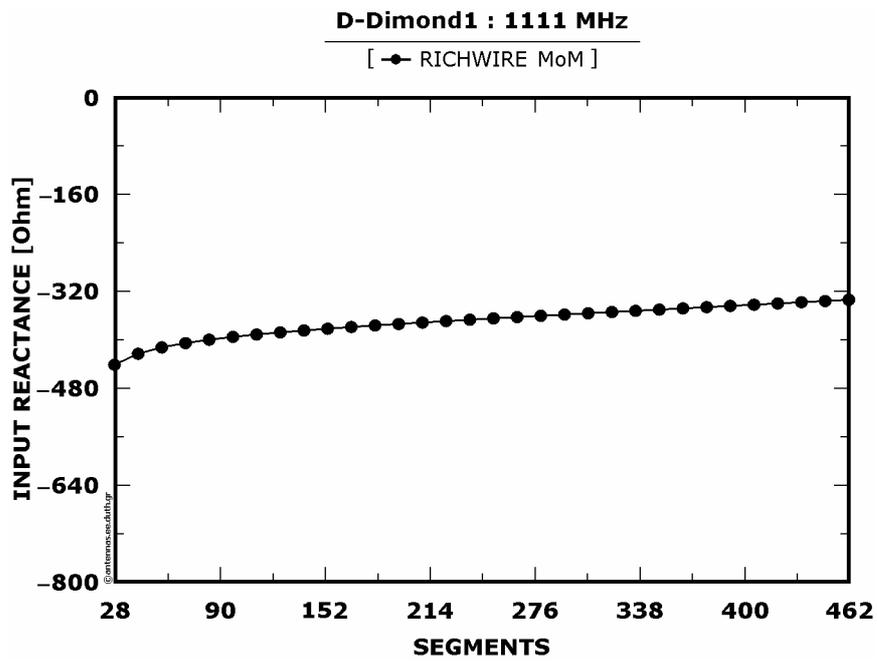

<u>Figure 2</u> : Real and imaginary part of the input impedance as a function of
the number of segments

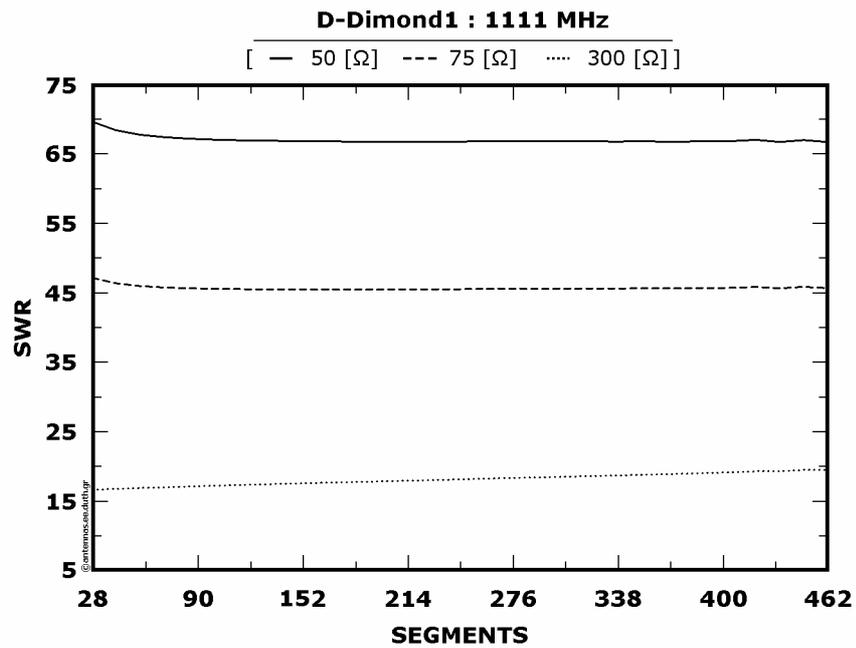

<u>Figure 3</u> : Standing wave ratios (SWRs) as a function of the number of segments

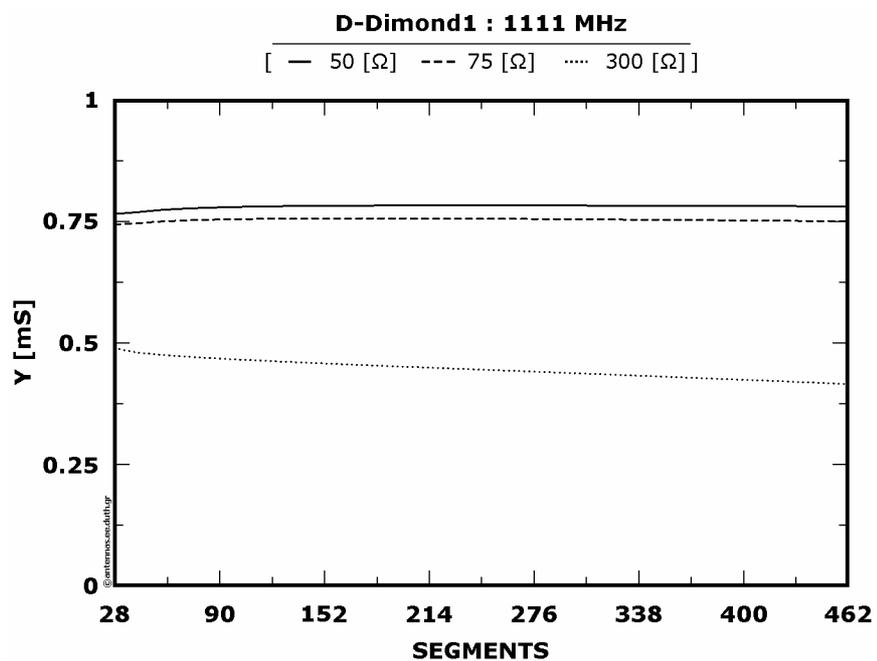

<u>Figure 4</u> : Normalized radiation intensities (Ys) as a function of the number of segments

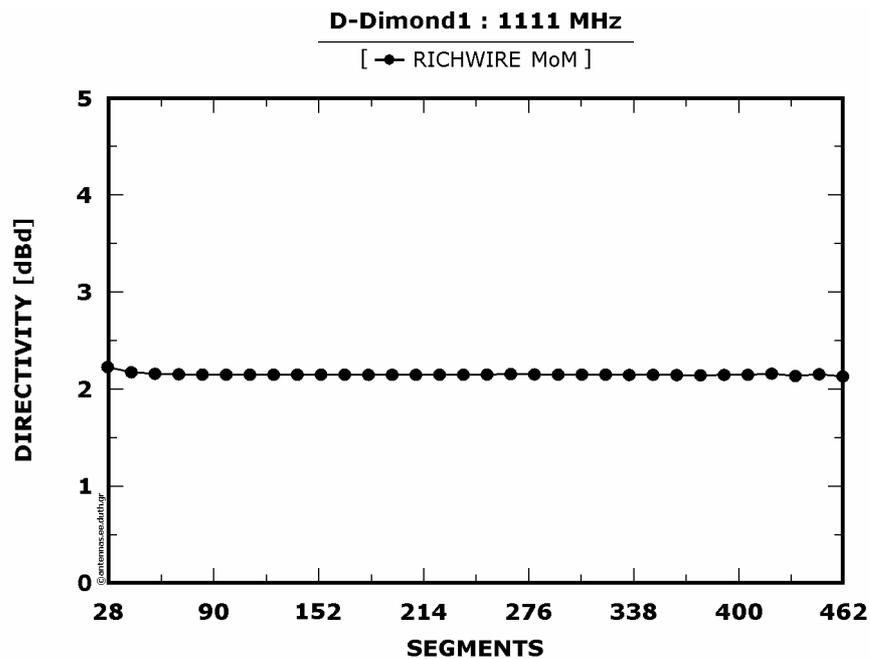

<u>Figure 5</u> : Directivity (with reference to dipole λ/2) as a function of the number of segments

After an observation of the above graphs it is realised that the values of the quantities are being stabilized from the start, that is, for a small number of divided segments. More specifically, the antenna's directivity and SWR which are shown in <u>Figure 3.3.5</u> and <u>Figure 3.3.</u>3 respectively are being stabilized at specific values in a short time. Regarding the input impedance that is illustrated in <u>Figure 3.3.1</u> its value's points begin to converge "ostensibly" after the third one. This point corresponds with division of D-Dimond1 in 56 segments. From now on the study will proceed with this particular number of segments.

**3.4 : Radiation patterns**

Below are cited the following figures: D-Dimond1's geometry in space, its radiation patterns at the three main planes and in space for the central operation frequency of 1111 [MHz] as well as for the edges of the examined frequency range (900 and 1300 [MHz]) and finally the relative intensity of radiation at the three main planes in polar and cartesian coordinates. Finally, there are comparison tables with the values of the antenna's quantities at the three specific frequencies that mentioned above.

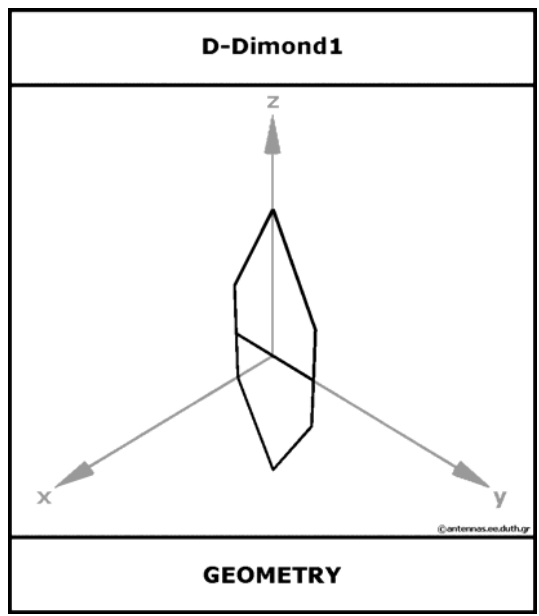

Figure 1 : Geometry of D-Dimond1 in space

### 3.4.1 : Patterns at 900 [MHz]

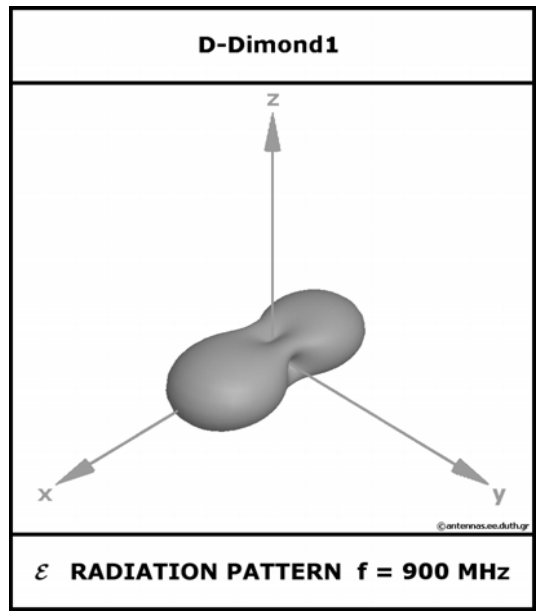

Figure 3 : The radiation pattern in space

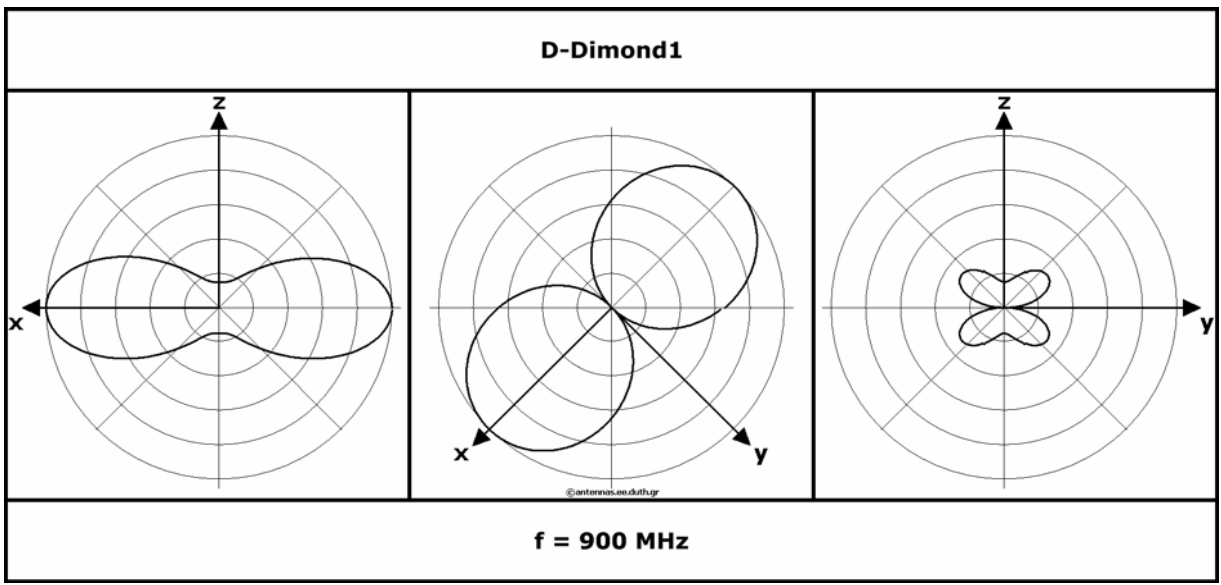

Figure 2 : The radiation pattern at the three main planes

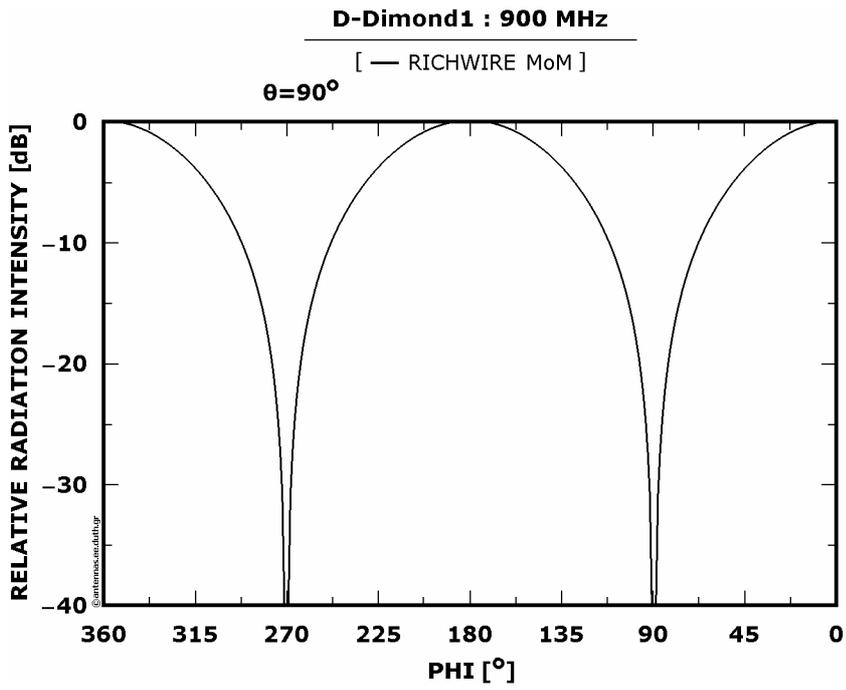

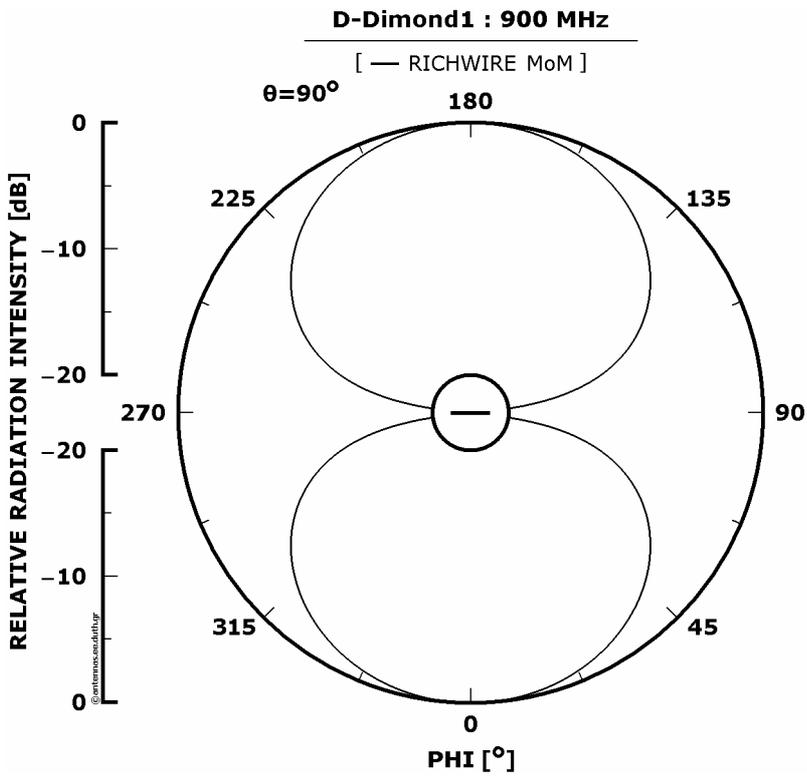

Figure 4 : Relative radiation intensities at xOy plane

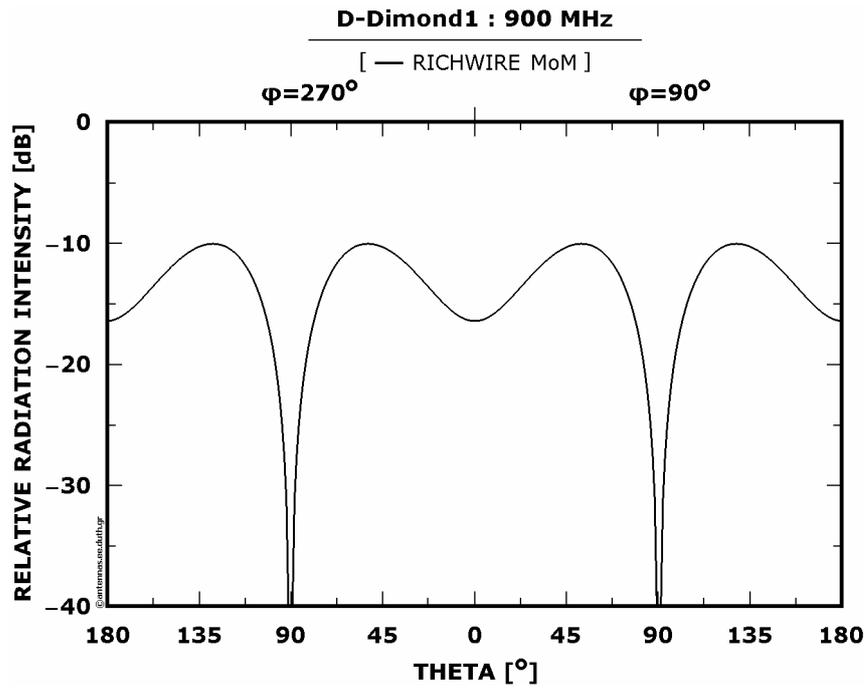

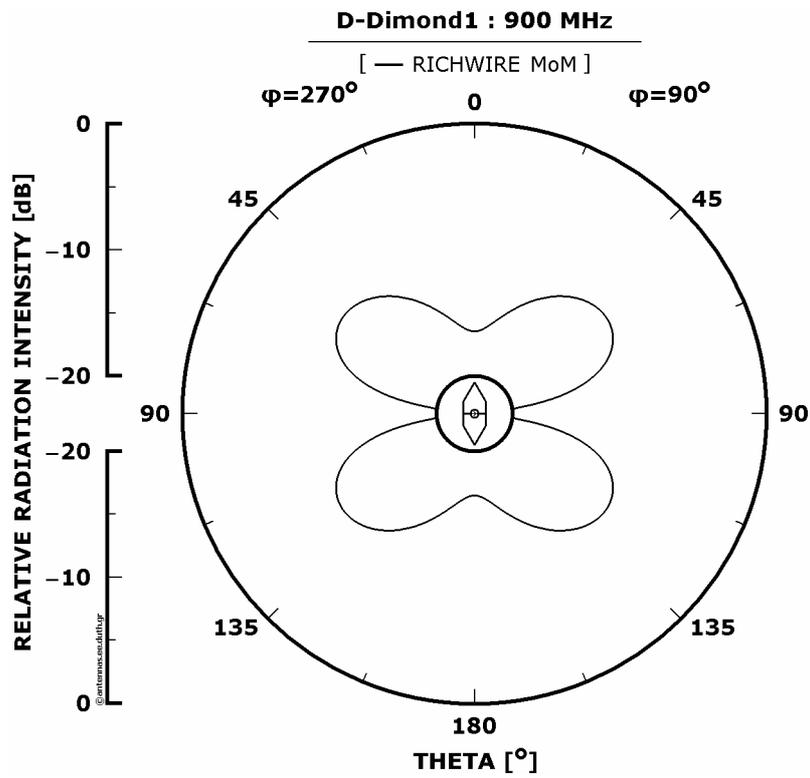

Figure 5 : Relative radiation intensities at yOz plane

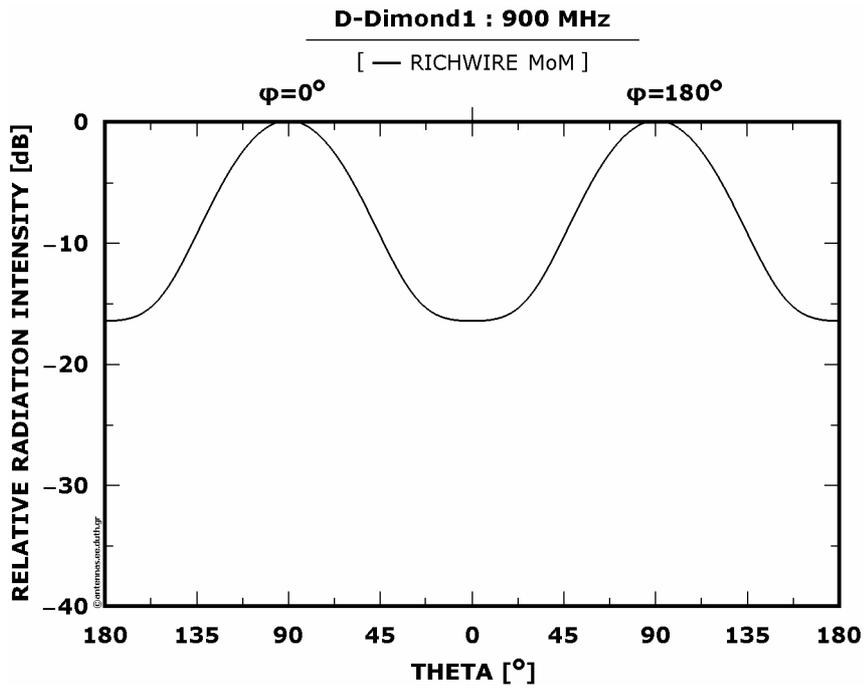

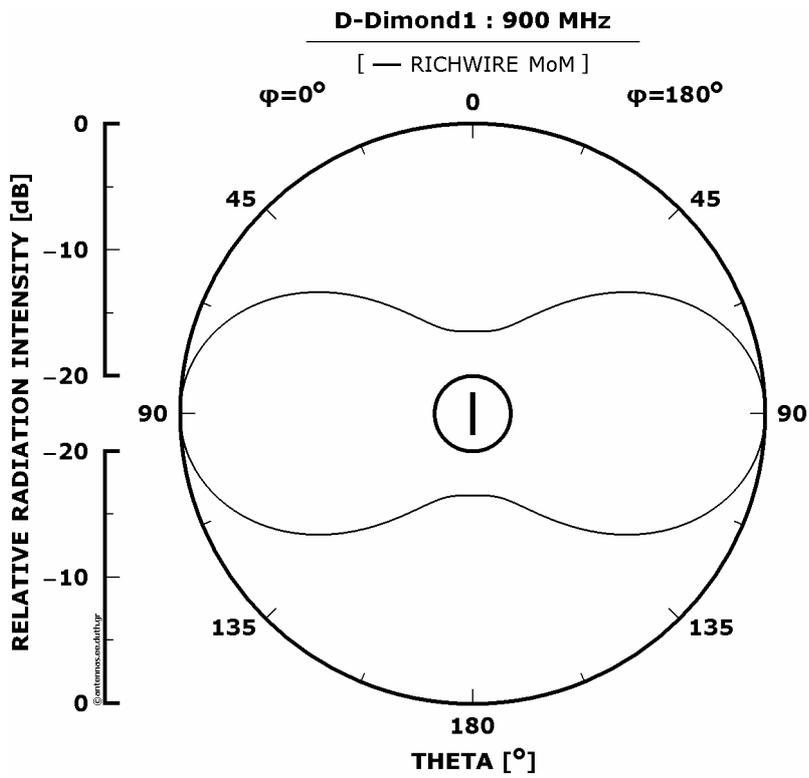

Figure 6 : Relative radiation intensities at zOx plane

### 3.4.2 : Patterns at 1111 [MHz]

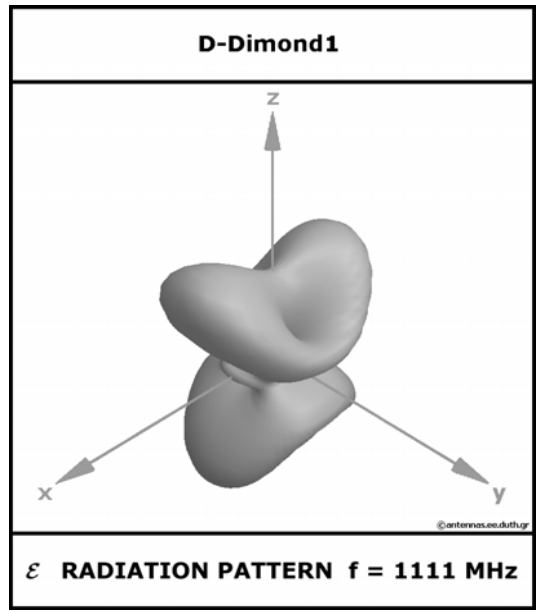

Figure 7 : The radiation pattern in space

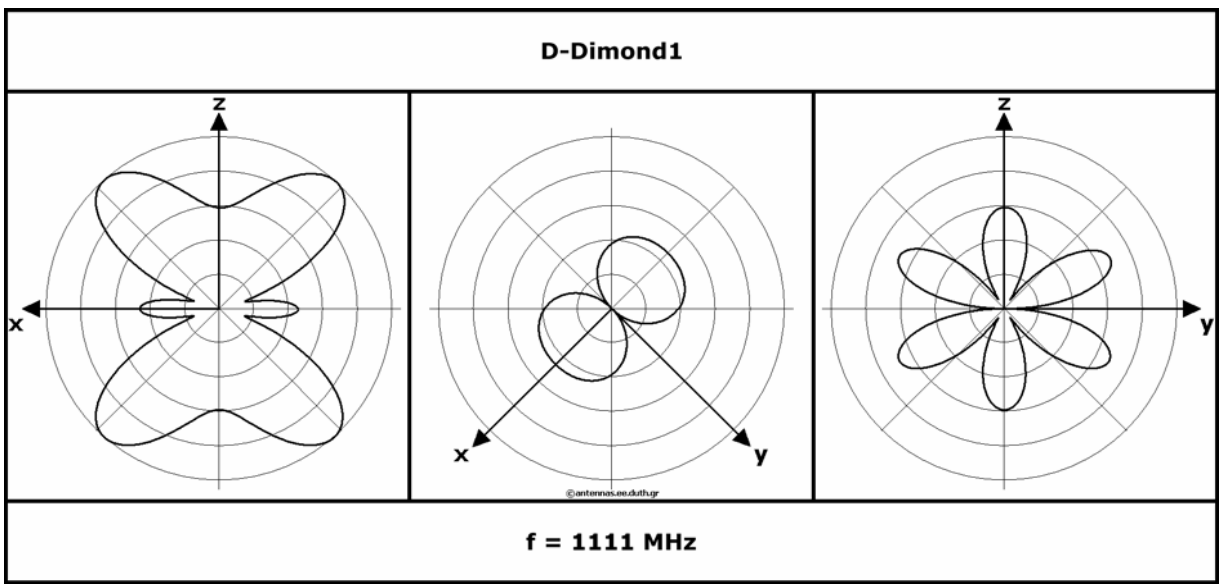

Figure 8 : The radiation pattern at the three main planes

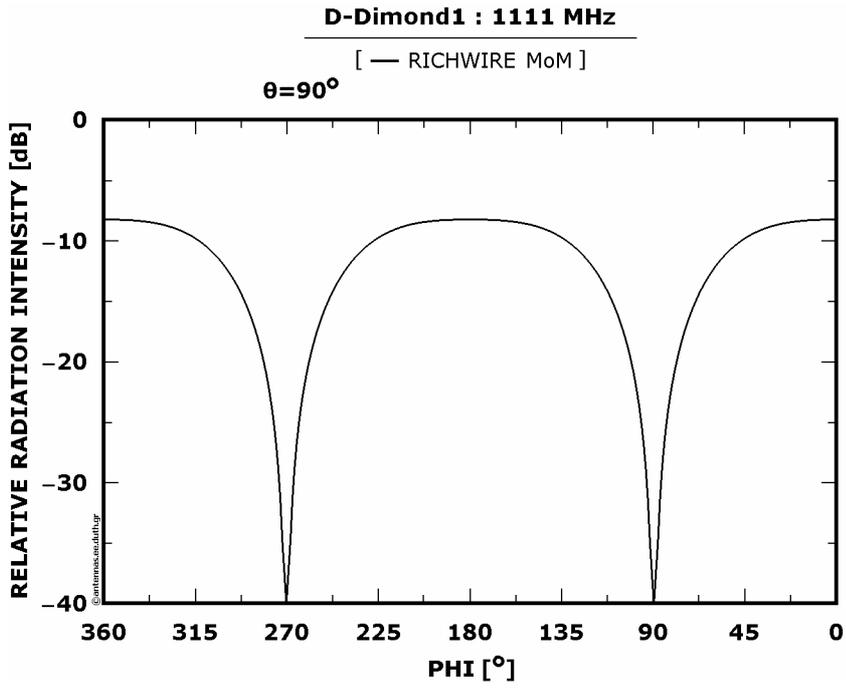

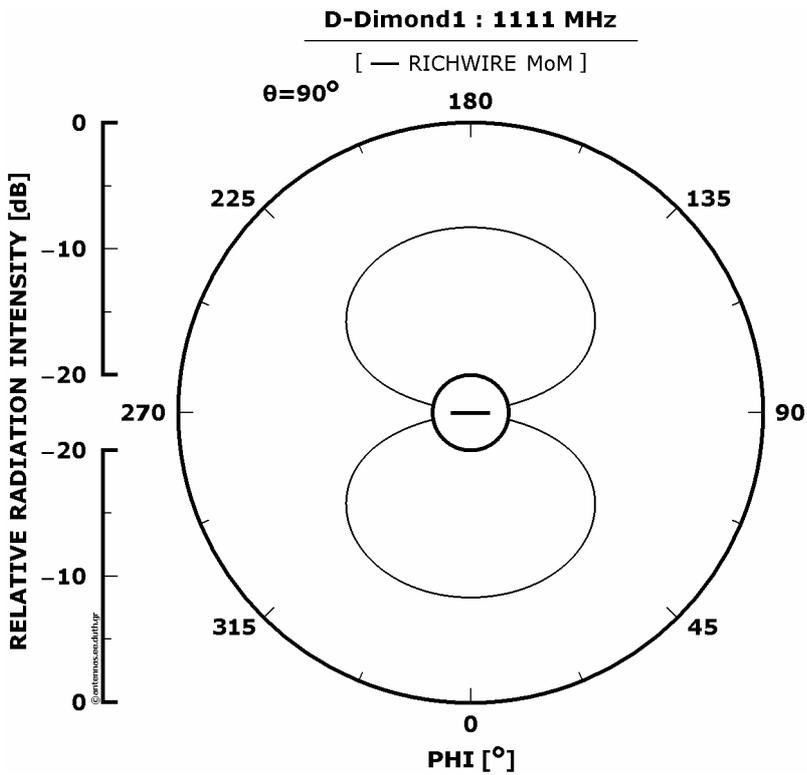

Figure 8 : Relative radiation intensities at xOy plane

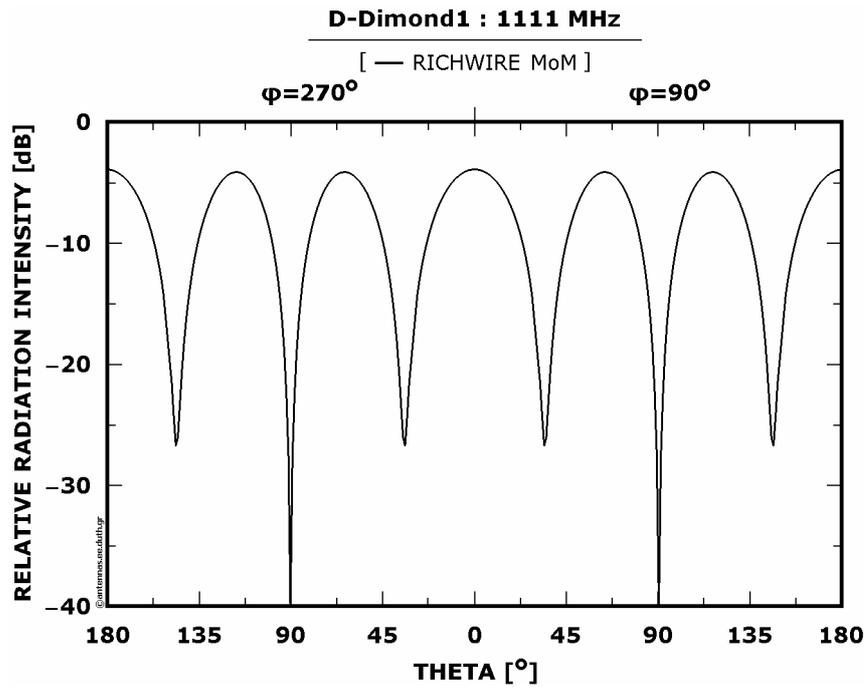

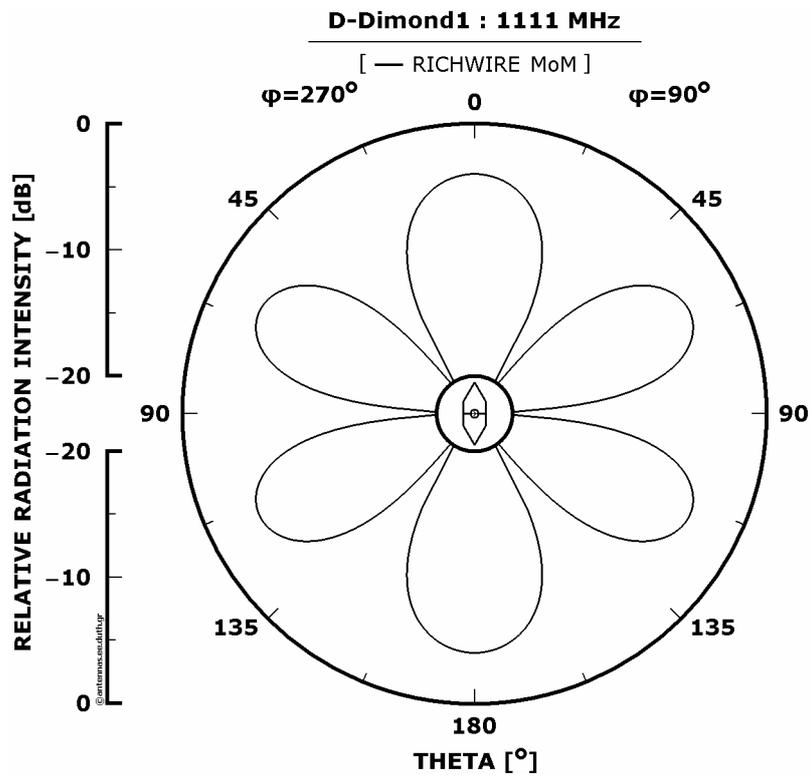

Figure 9 : Relative radiation intensities at yOz plane

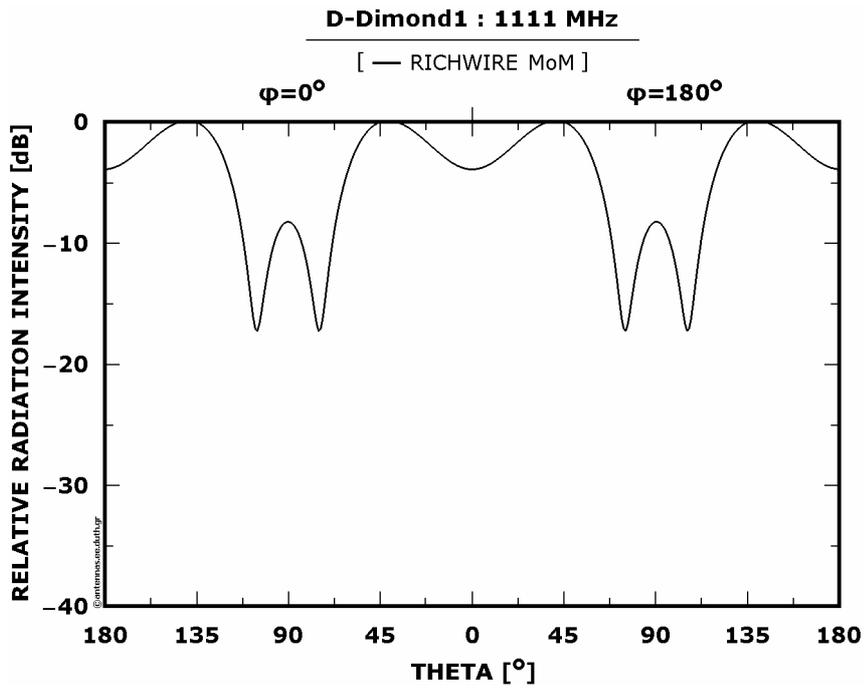

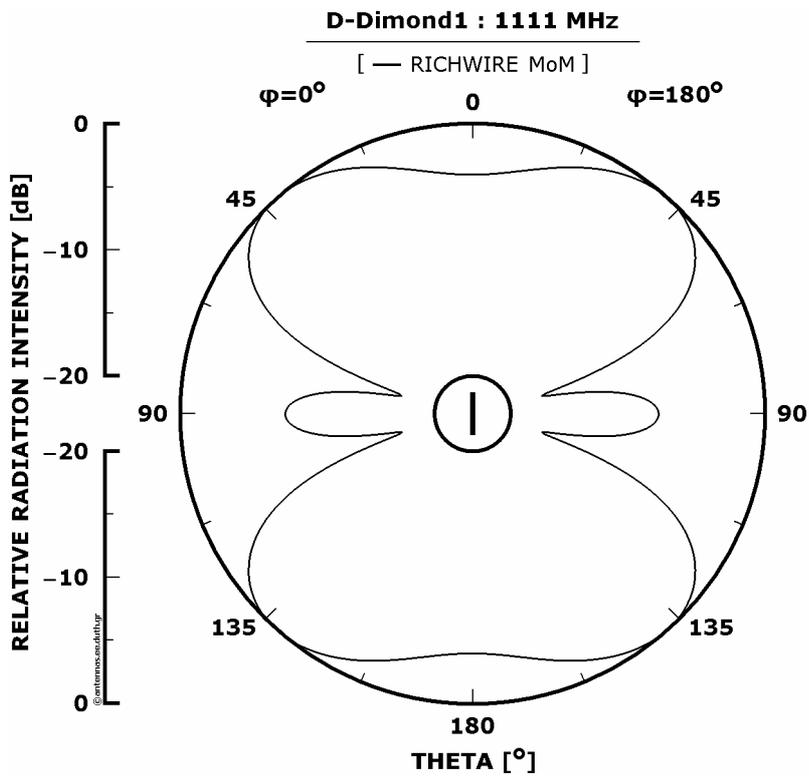

Figure 10 : Relative radiation intensities at zOx plane

### 3.4.3 : Patterns at 1300 [MHz]

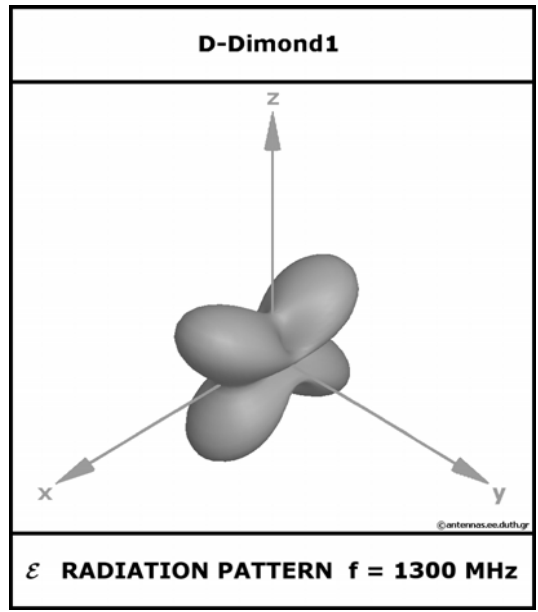

Figure 11 : The radiation pattern in space

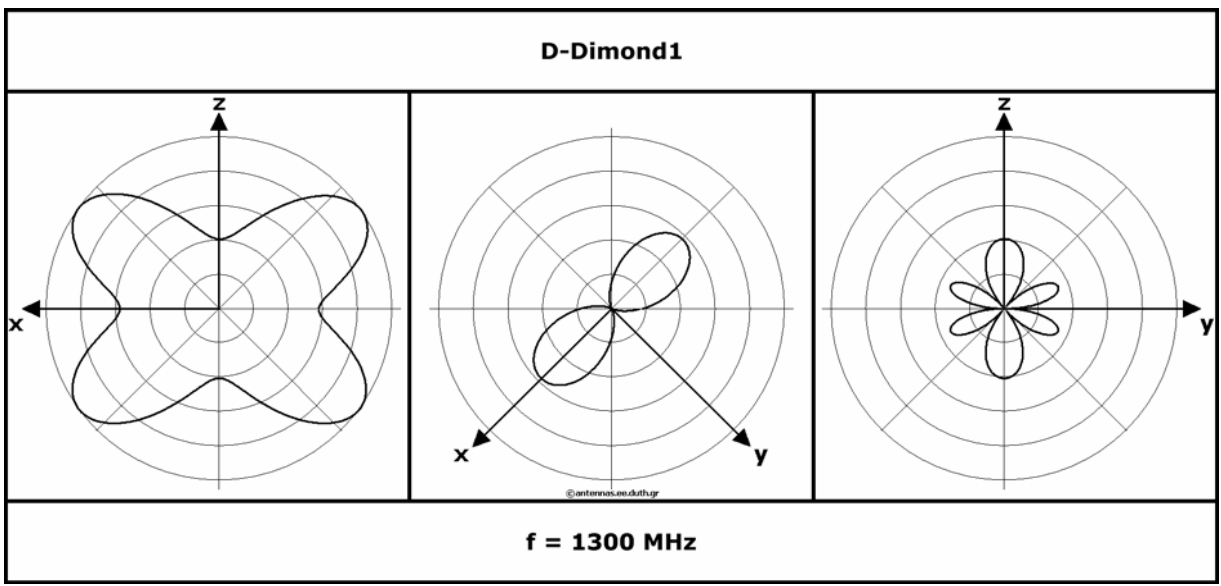

Figure 12 : The radiation pattern at the three main planes

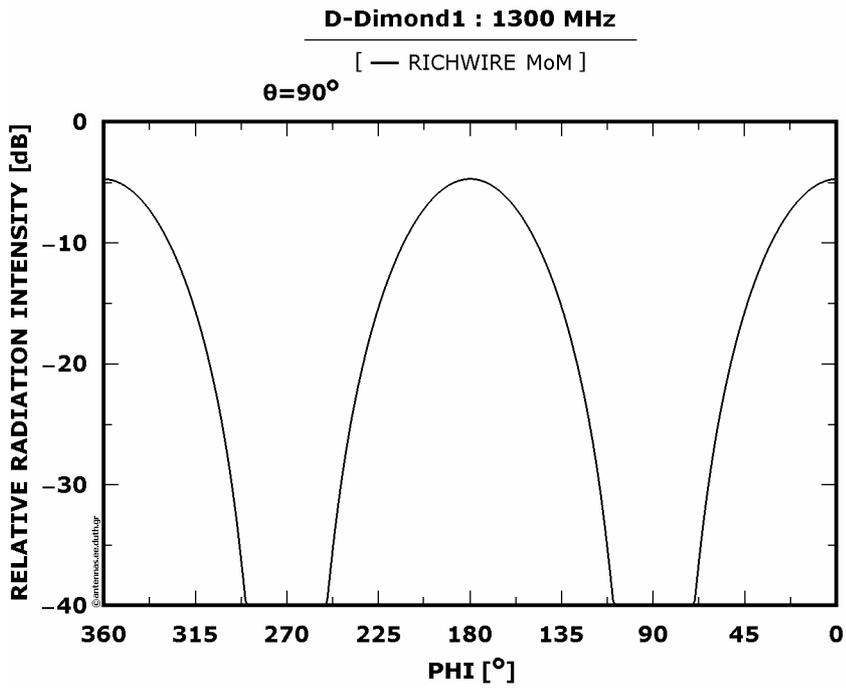

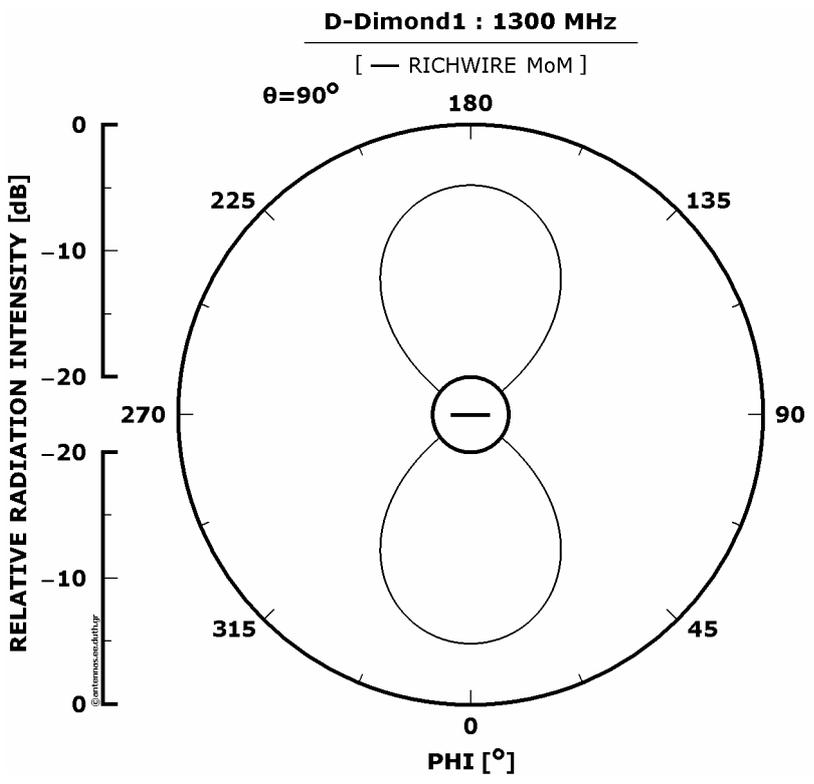

Figure 12 : Relative radiation intensities at xOy plane

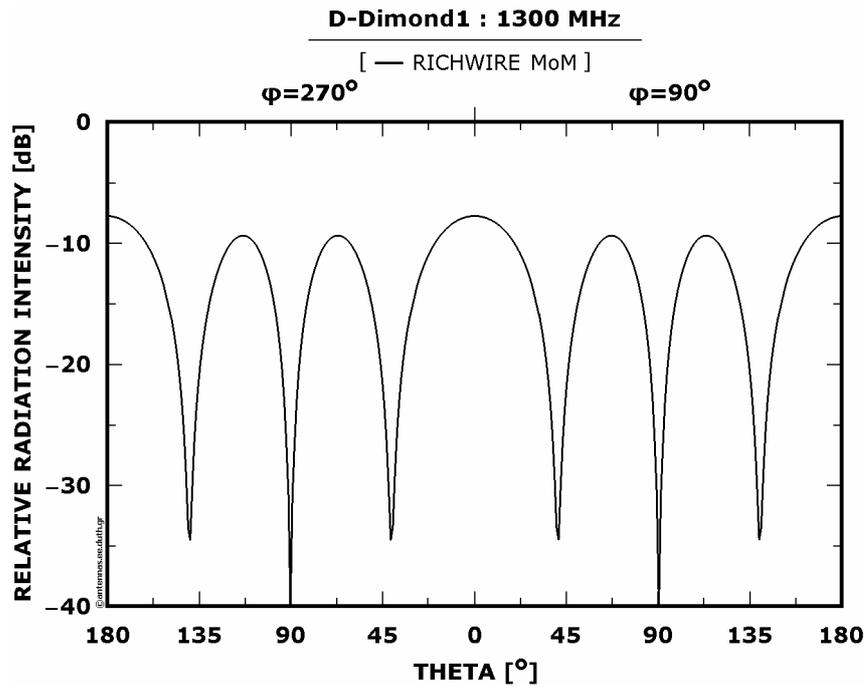

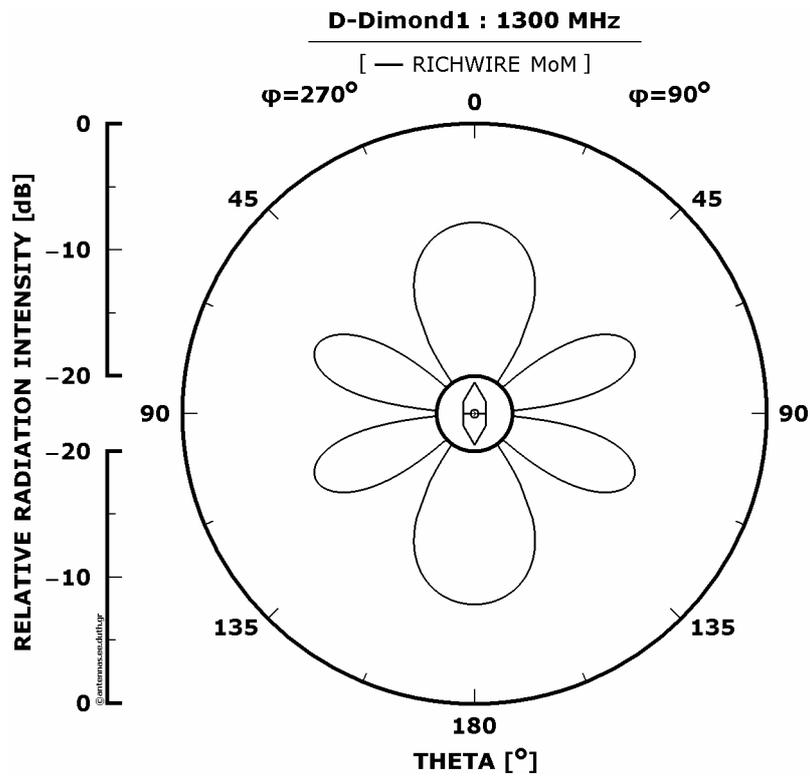

Figure 13 : Relative radiation intensities at yOz plane

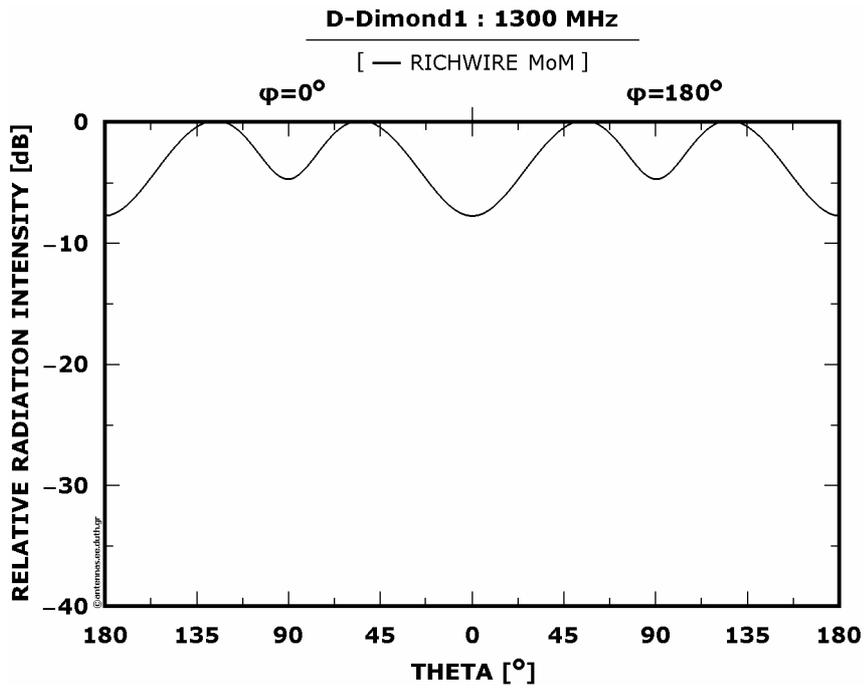

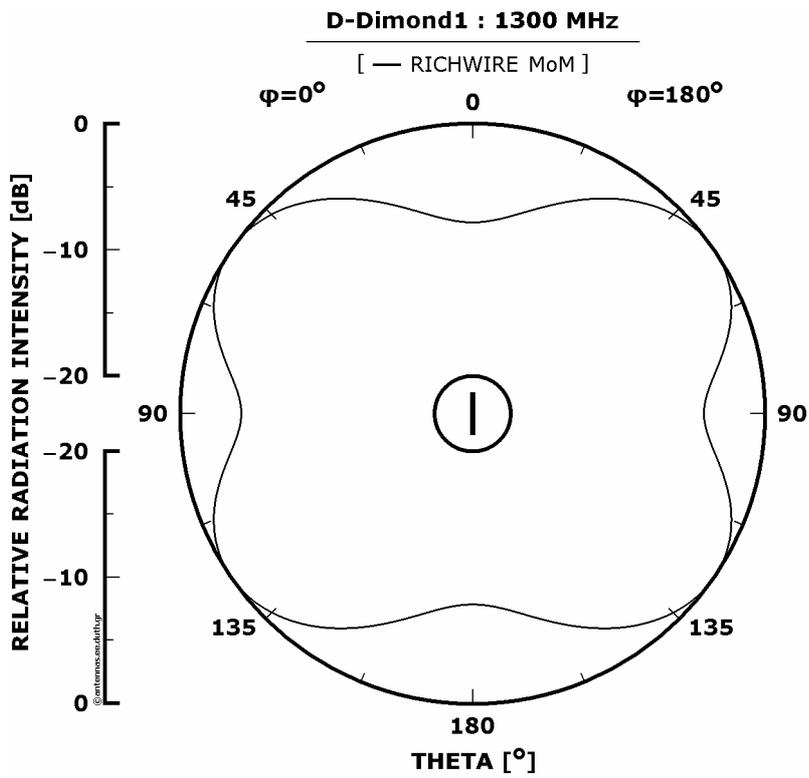

Figure 14 : Relative radiation intensities at zOx plane

**3.4.4 : Characteristic quantities**

Below tables are cited with the values of D-Dimond1's most essential characteristic quantities at 900, 1111 and 1300 [MHz] respectively:

Table 1 : Characteristic quantities of D-Dimond1 at 900 [MHz]

| $Z_{INP}$ [$\Omega$] | SWR(50 $\Omega$) | SWR(75 $\Omega$) | SWR(300 $\Omega$) | D [dBd] |
|---|---|---|---|---|
| $3421 - i967$ | 74 | 49 | 12 | 4.27 |

Table 2 : Characteristic quantities of D-Dimond1 at 1111 [MHz]

| $Z_{INP}$ [$\Omega$] | SWR(50 $\Omega$) | SWR(75 $\Omega$) | SWR(300 $\Omega$) | D [dBd] |
|---|---|---|---|---|
| $51.8 - i413$ | 68 | 46 | 17 | 2.15 |

Table 3 : Characteristic quantities of D-Dimond1 at 1300 [MHz]

| $Z_{INP}$ [$\Omega$] | SWR(50 $\Omega$) | SWR(75 $\Omega$) | SWR(300 $\Omega$) | D [dBd] |
|---|---|---|---|---|
| $377 - i29.8$ | 7.5 | 5 | 1.2 | 3.07 |

It becomes noticeable that SWRs' values are excessively high at the frequencies of 900 and 1111 [MHz]. Consequently, the utilization of the antenna at the pre-mentioned frequencies is impracticable. On the contrary, the situation is being improved at 1300 [MHz] and particularly for that SWR which is related with antenna's matching with transmission line of a 300 [$\Omega$] characteristic resistance, that is SWR(300). Moreover, the directivity's value for the specific frequency is considered to be satisfactory (3.07 [dBd]).

## 3.5 : Investigation in connection with frequency

In this section the behaviour of D-Dimond1, divided in 56 segments, will be examined as a function of frequency. The below graphs display the results of this investigation at the frequency range extending from 900 to 1300 [MHz], with a 10 [MHz] step.

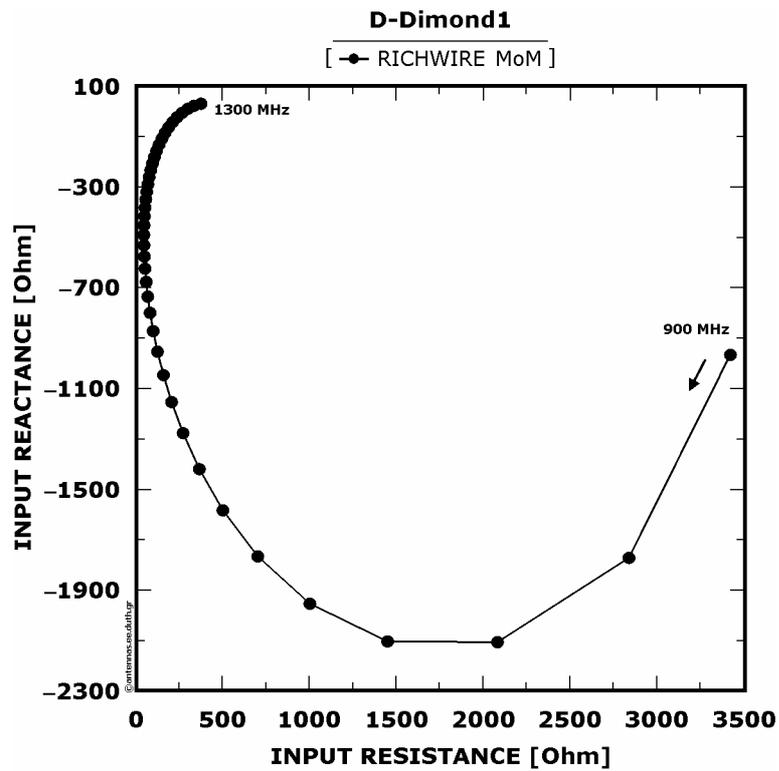

Figure 1 : Input impedance as a function of frequency

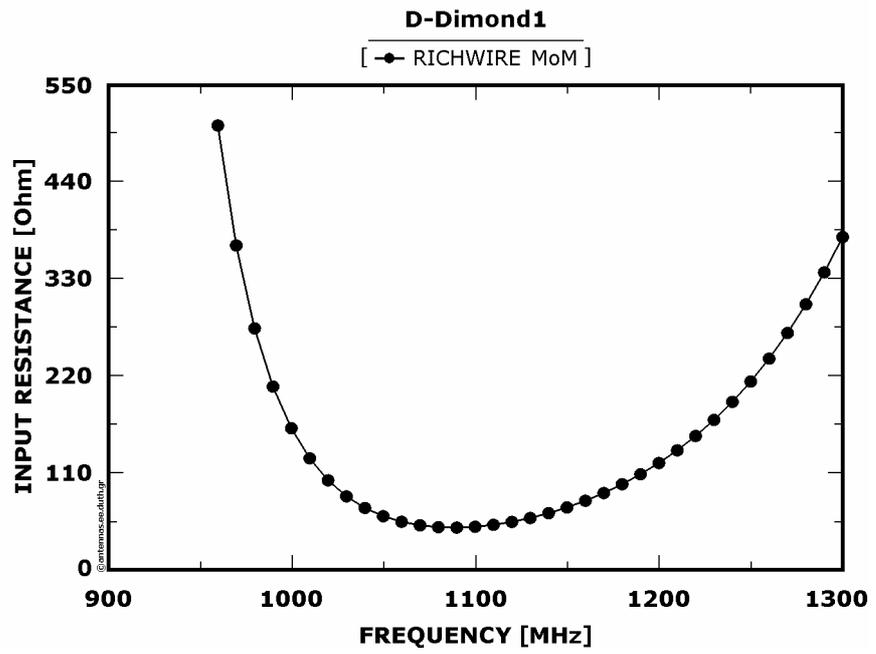

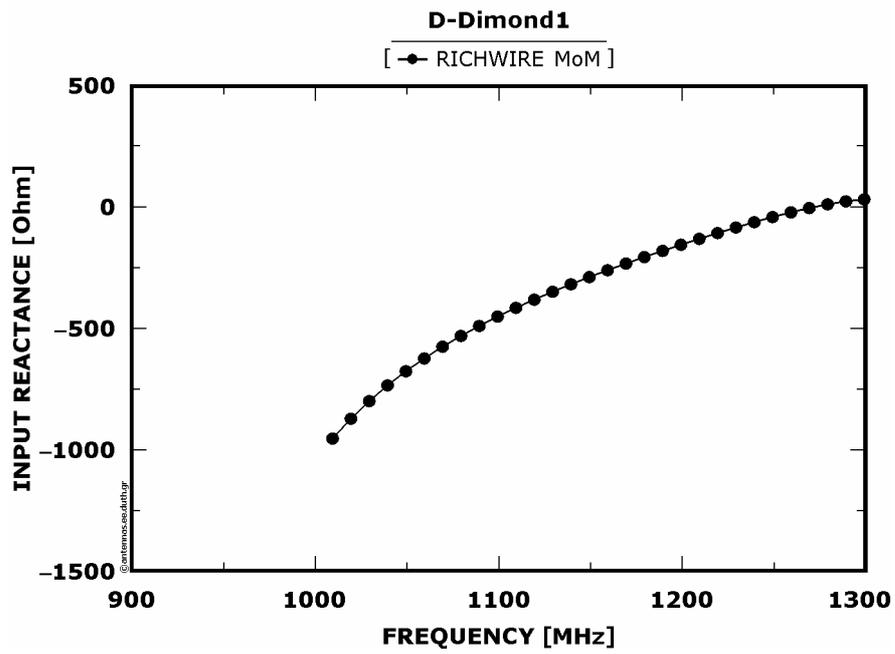

<u>Figure 2</u> : Real and imaginary part of input impedance as a function of frequency

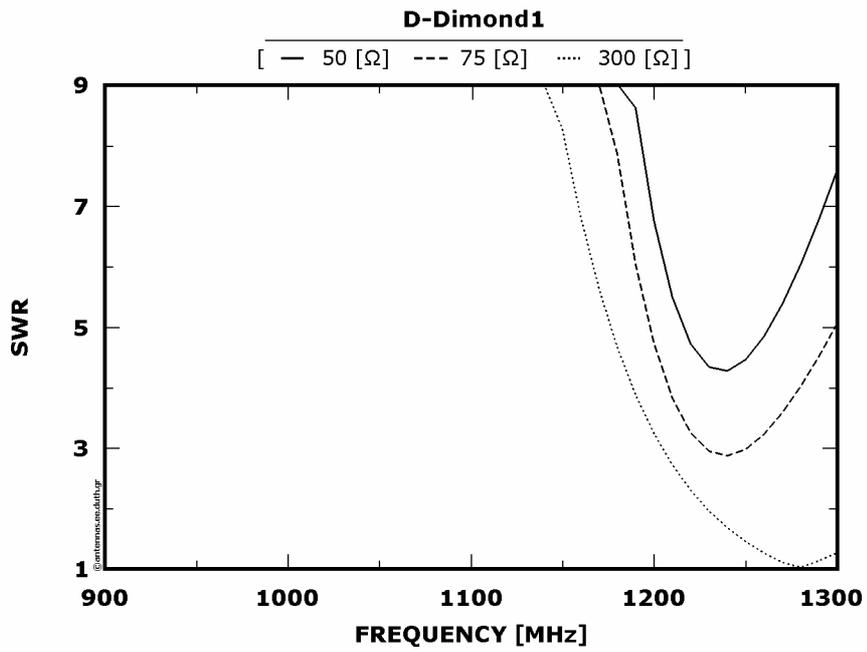

Figure 3 : SWRs as a function of frequency

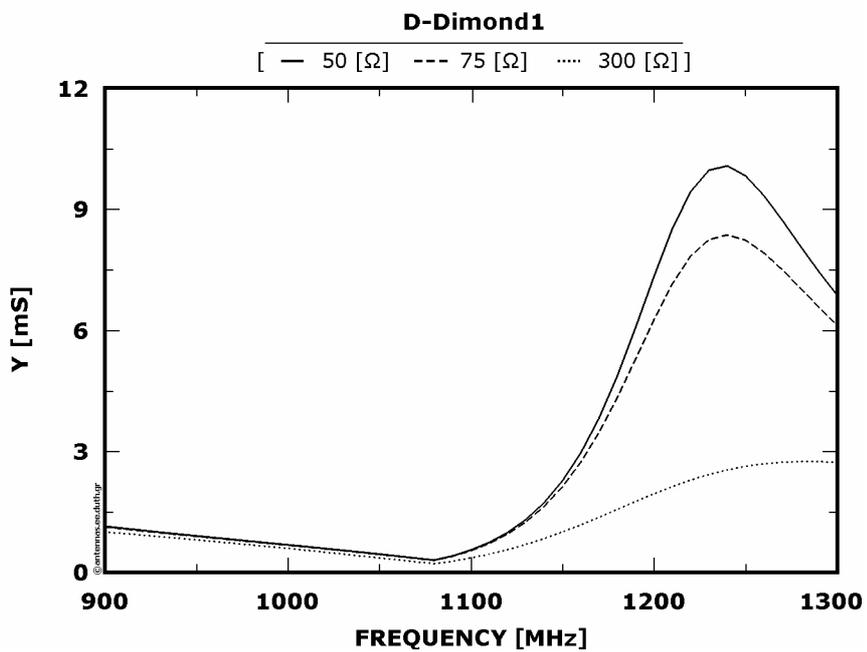

Figure 4 : Normalized radiation intensities as a function of frequency

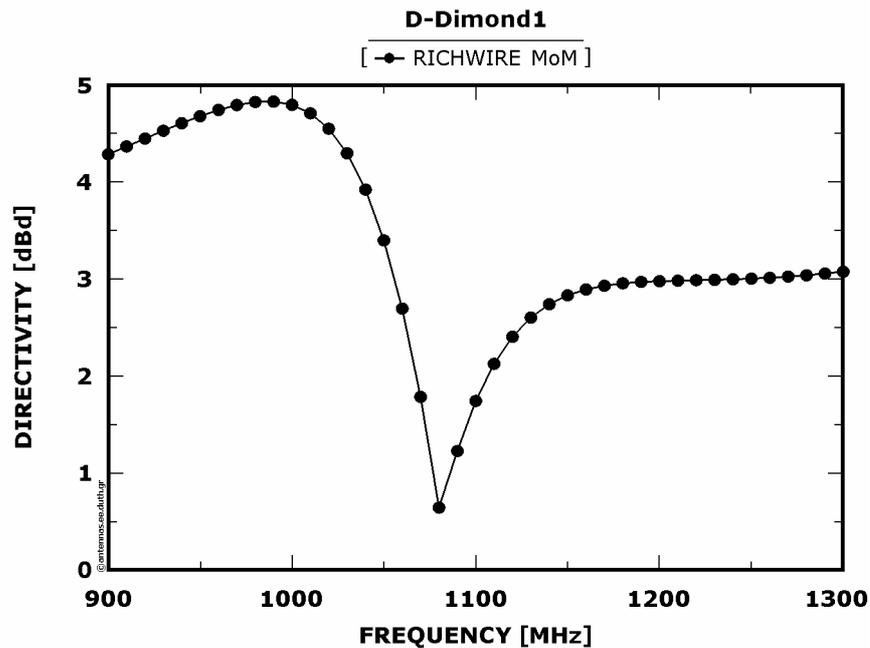

Figure 5 : Directivity (with reference to dipole λ/2) as a function of frequency

Note: At the previous graphs the values of antenna's quantities at the starting frequencies have been removed, where it is has been necessary, for the sake of discretion, as their values appeared to be excessively high.

According to Figure 3.5.3 SWR plots appear to range at quite high values and become acceptable at frequencies above 1200 [MHz].

Figure 3.5.5 illustrates the variation of directivity which ranges also at quite high values while approaching 5 [dBd], then it makes an abrupt "drop" approximately at 1080 [MHz] and is stabilised near 3 [dBd] as frequency increases.

At Figure 3.5.4 the plots of a slightly different quantity from the maximum radiation intensity, $U_{max}$, are illustrated. This particular quantity has been normalized and is symbolized as Y with dimensions of [mS]. Its mathematical expression is derived at Annex Z. Normalized radiation intensity is a useful indicator of the antenna's behaviour as it depends from both the directivity and SWR. More specifically, it is maximized at directivity's maximization (directly proportional) and at SWR's minimization (not exactly reversely proportional but a more complex dependence). With respect to Figure 3.5.5, maximisation of this quantity is expected at frequencies above 1200 [MHz].

It is worthwhile to define the meaning of an antenna's bandwidth which does not have an exclusive definition yet is defined according to the circumstances. The concept of bandwidth will be frequently mentioned at this thesis and generally can be defined as the frequency range at which an antenna behaves with a predetermined, acceptable way in connection with the variation of certain characteristics of its. More specifically, the 'bandwidth' term is defined as the frequency range at which an antenna's SWR takes values smaller or equal to two or three. However, the real frequency range at which an antenna will operate as a part of a communication system can be depended also from other factors, such as a minimal allowed fluctuation of directivity, therefore it can become much narrower than the calculated bandwidth according to the above definition.

A detailed table is cited below where the bandwidth is shown for each transmission line that is used for matching the antenna with its feeding source and for each maximum acceptable SWR value, which is predetermined from the system design (here is assumed to be 2 and 3). Furthermore, the directivity's values at the edges of the bandwidth range are indicated.

According to the Table 3.5.1 it results that for maximum acceptable SWR equals to 2 there is the possibility of connecting the antenna only with a transmission line of a 300 [Ω] characteristic resistance. Moreover the antenna should have to operate between 1230 and 1300 [MHz], that is a bandwidth of 70 [MHz]. The directivity's behaviour at the above frequency range is stationary. If the system design requires that maximum SWR value can reach 3 then there is also the possibility of using a 75 [Ω] transmission line, at frequencies between 1230 and 1250 [MHz], that is a narrow 20 [MHz] bandwidth. On the contrary using a 300 [Ω] transmission line offers a larger bandwidth, that is 90 [MHz], while the directivity's fluctuation is considered negligible.

<u>Table 1</u> : Frequency range from 900 to 1300 [MHz]

| Acceptable SWR | SWR(50) | | SWR(75) | | SWR(300) | |
|---|---|---|---|---|---|---|
| | BW [MHz] | D [dBd] | BW [MHz] | D [dBd] | BW [MHz] | D [dBd] |
| 2 | --- | --- | --- | --- | $1300 - 1230 = 70$ | 2.98 up to 3.07 |
| 3 | --- | --- | $1250 - 1230 = 20$ | 2.98 up to 3.0 | $1300 - 1210 = 90$ | 2.97 up to 3.07 |

Finally, as it was realised at the previous study, the excessively high SWR's values does not allow the operation of the antenna at the frequency range from 900 to 1300 except perhaps from a small region around 1250 [MHz]. Accordingly, searching for frequency ranges at which the antenna's behaviour will be more efficiently the variations of its characteristic quantities will be studied at a more extensive range of frequencies, an investigation which will be presented at the following chapter.

### 3.6 : Comparison between analytical and simulation results

At the $2^{nd}$ chapter, based at the antenna's theoretical dissertation, the mathematical formula of D-Dimond1's radiation pattern was derived and was depicted at the three main planes. At this section a comparison is carried out between the graphs which came from the analytical study and those ones that came from the computational study and were generated with the assistance of the RICHWIRE program, maintaining the minimum number of segments in which the antenna can be divided, that is 14.

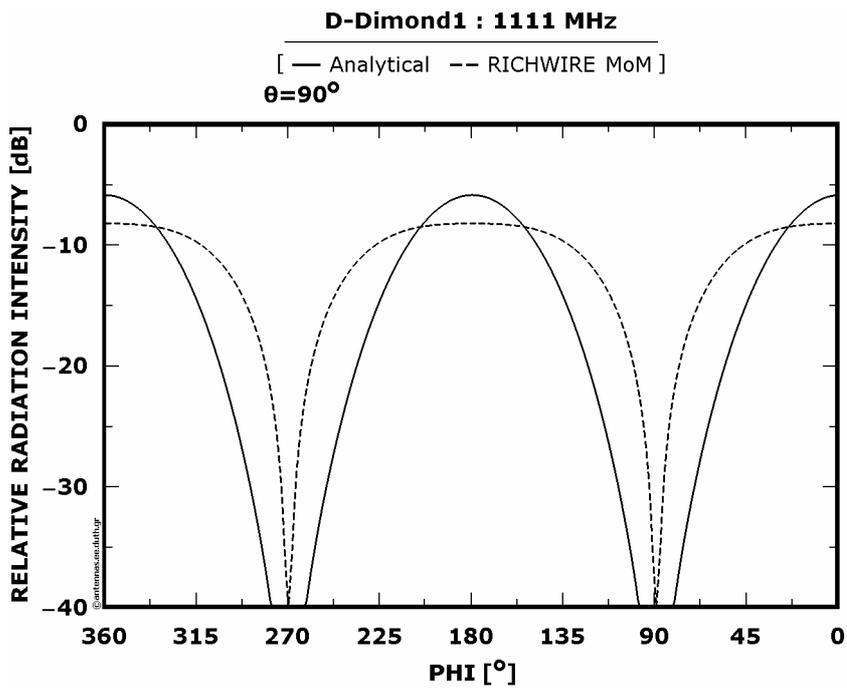

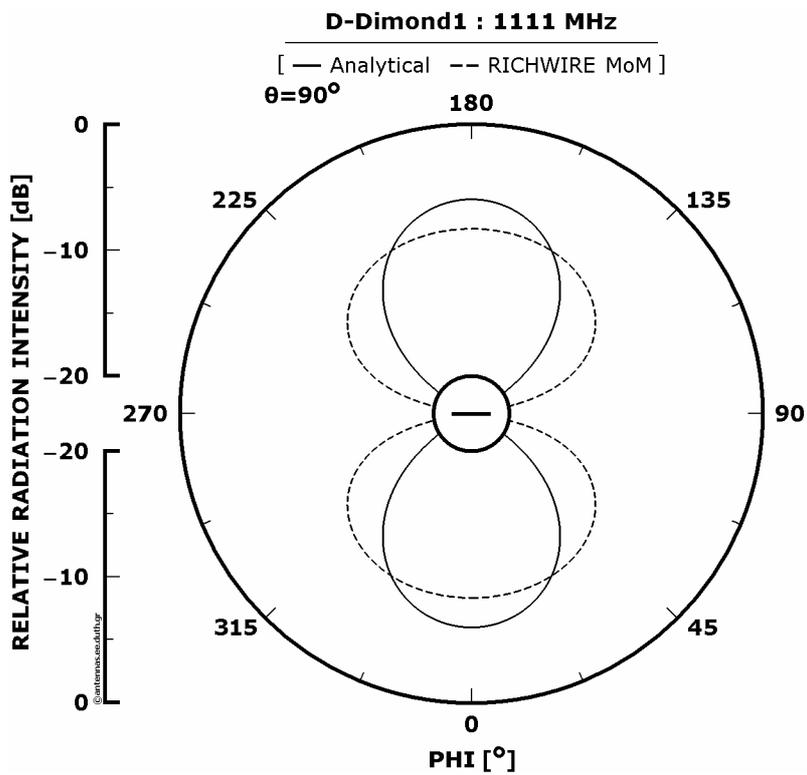

Figure 1 : Relative radiation intensities at xOy plane

**D-Dimond1 : 1111 MHz**

[ —— Analytical  — — RICHWIRE MoM ]

φ=270°                                          φ=90°

**D-Dimond1 : 1111 MHz**

[ —— Analytical  — — RICHWIRE MoM ]

φ=270°                    0                    φ=90°

Figure 2 : Relative radiation intensities at yOz plane

Figure 3 : Relative radiation intensities at zOx plane

According to the above figures it is deduced that the computational study - simulation verifies sufficiently the analytical study at least regarding to the relative radiation intensities at the xOy and yOz planes. The exception appears at the comparative presentation at zOx plane where certain divergences are observed.

# Chapter 4 : Study at a broader frequency range

## 4.1 : In General

The previous chapter's computational study indicated that antenna's characteristics were not satisfactory inside the selected frequency range of 900 to 1300 [MHz]. Specifically, SWRs' values were excessively high thus the antenna's operation at the particular range of frequencies is excluded. Nevertheless, the graphs according to the frequency showed that SWRs' values were decreasing from 1200 [MHz] and above. Consequently, a study at a broader frequency range has to be carried out. Moreover, the fact that the present study centres upon an original antenna, that is to say that it hasn't been studied by anyone up to now, justifies a further investigation regarding frequency so as to estimate its behaviour in a more complete way.

## 4.2 : Extensive frequency scan

Initially, a study was carried out centred upon the variation of antenna's characteristic quantities at a wide frequency range, touching upon the permissible operation frequency limits of the simulation program RICHWIRE. The frequency range that was investigated extents from 200 [MHz] to 2850 [MHz]. Multiple tunings of the antenna were found in at least five regions of frequencies and was attempted ex post facto further investigation on each one decreasing the step in order to achieve higher discretion and an in depth illustration of the rapid variations of the antenna's characteristics quantities, whenever it was required.

The tables below give the steps and the frequency ranges that were used:

| From [MHz] | To [MHz] | Step [MHz] | From [MHz] | To [MHz] | Step [MHz] |
|---|---|---|---|---|---|
| 200 | 900 | 100 | 1950 | 2000 | 25 |
| 900 | 1230 | 10 | 2000 | 2100 | 50 |
| 1230 | 1250 | 1 | 2100 | 2300 | 25 |
| 1250 | 1300 | 10 | 2300 | 2400 | 1 |
| 1300 | 1600 | 50 | 2400 | 2500 | 25 |
| 1625 | 1850 | 25 | 2500 | 2700 | 50 |
| 1850 | 1880 | 1 | 2700 | 2725 | 25 |
| 1880 | 1900 | 20 | 2725 | 2825 | 1 |
| 1900 | 1950 | 50 | 2825 | 2850 | 25 |

The total number of the frequency values that were taken was 335 which range from 200 to 2850 [MHz] with the variable step mentioned above. At the regions where the variations of the antenna's characteristic quantities were abrupt the step was decreasing so as to represent the variation with better discretion. On the contrary, at regions where weren't observed any satisfactory, interesting or non-stationary variations the step was increasing.

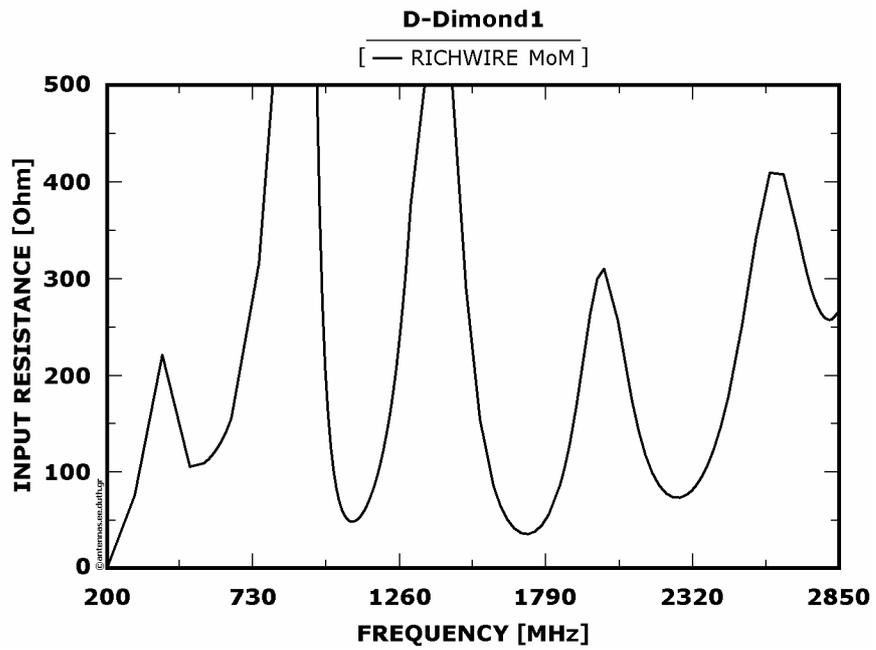

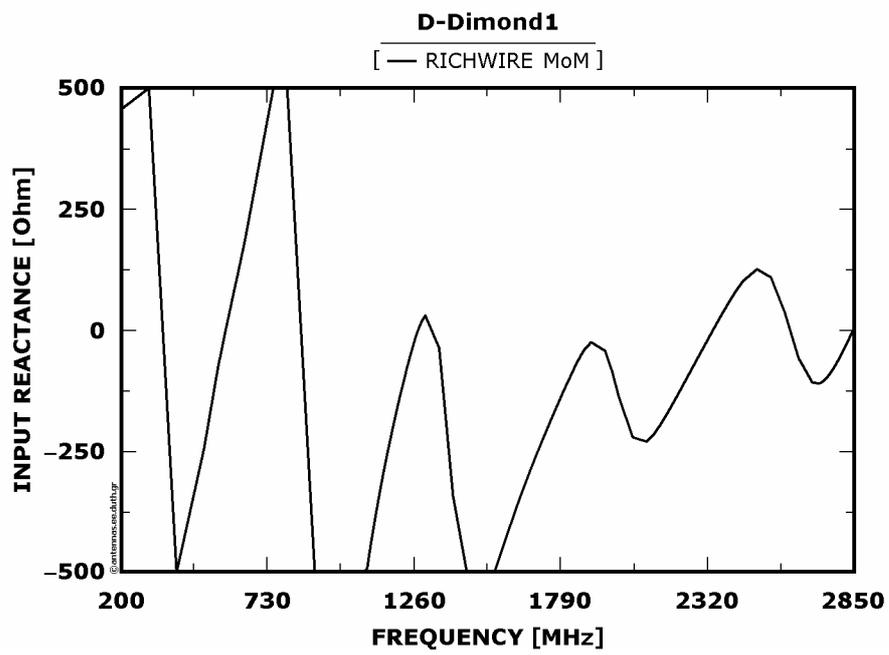

<u>Figure 1</u> : Real and imaginary part of the input impedance as a function of frequency

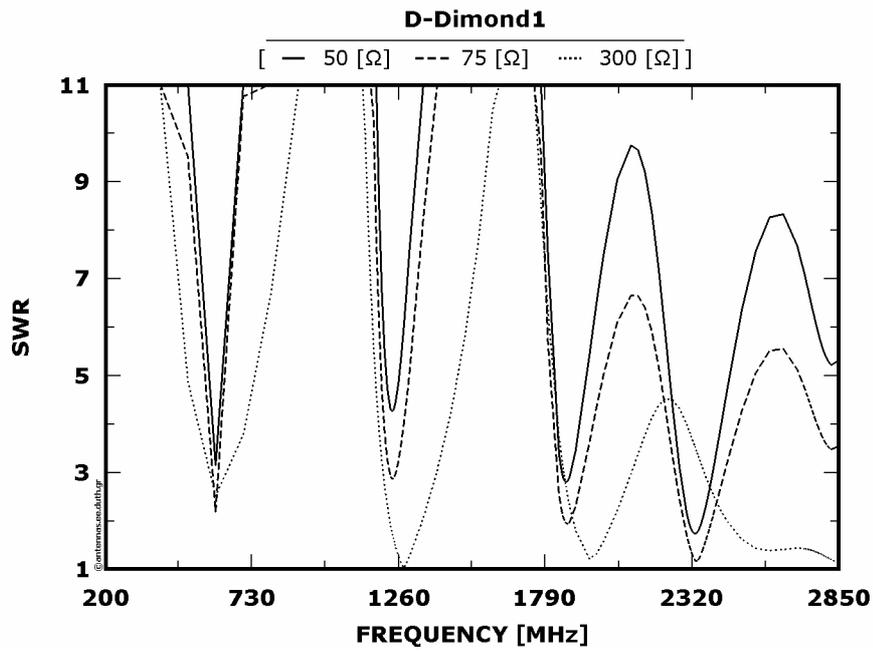

Figure 2 : SWRs as a function of frequency

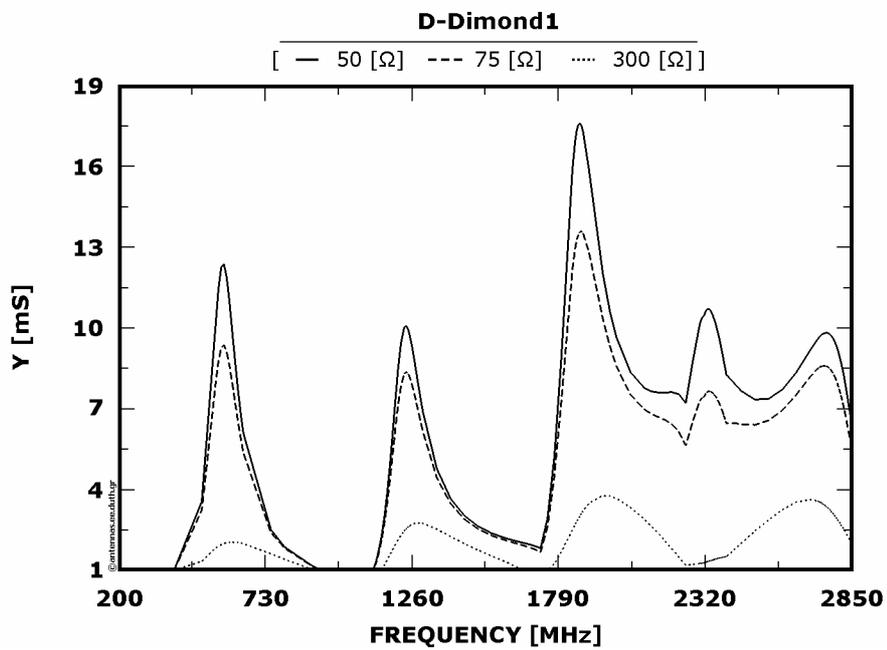

Figure 3 : Normalized radiation intensities as a function of frequency

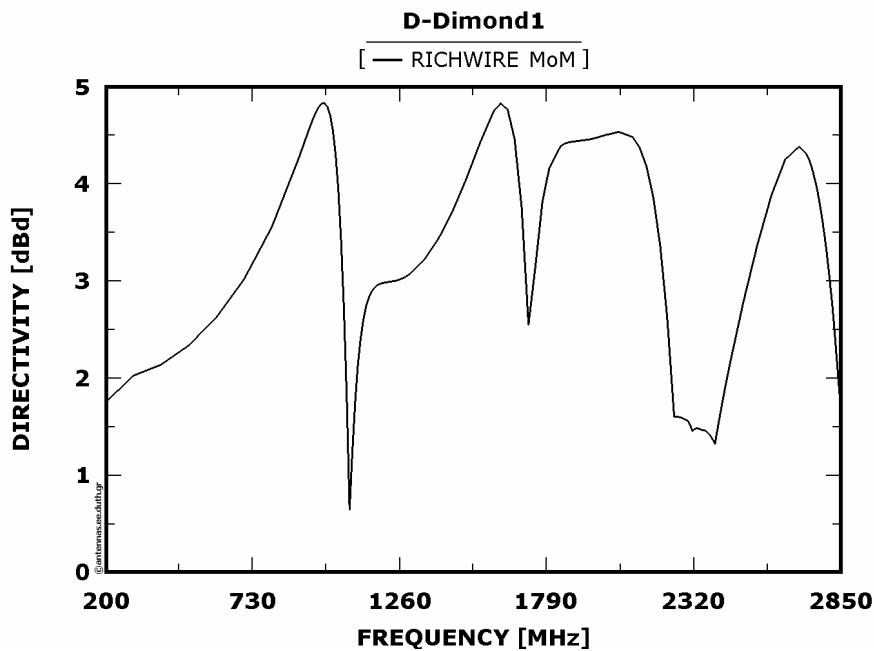

Figure 4 : Directivity (with reference to dipole λ/2) as a function of frequency

According to the above figures some interesting conclusions are deduced about D-Dimond1's behaviour in connection with frequency.

From Figure 4.2.1 the periodical interchange of resistance's maximums and minimums values is observed so as the variation of reactance's values on either side of nullification. The frequencies where input reactance is nullified turn to be interesting as antenna's tunings are revealed hence SWRs minimizations are occurred.

According to Figure 4.2.2 there are five frequency regions where SWRs' minimizations are occurred. SWR(50) and SWR(75) vary at the same way, that is to say their maximums and minimums values appear simultaneously, while SWR(300) follows differentiated variation as frequency increases. The lowest minimum value appears around the frequency of 2320 [MHz] concerning SWR(50) and SWR(75) and around the frequency of 1260 [MHz] concerning SWR(300) respectively . Finally, it is observed that SWRs are minimized for a broader range of frequencies as frequency rises.

Concerning Figure 4.2.4 successively maximum values of directivity are noticed at 990, 1625, 2050 and 2700 [MHz] and the highest maximum value reaches approximately 4.8 [dBd] at 1625 [MHz]. Interesting regions of frequencies are the ranges from 1800 to 2150 [MHz] where directivity's values are fixed above 4 [dBd], that is to say for a frequency range of 350 [MHz]. Moreover, a second interesting region commences from 2550 to 2810 [MHz] where directivity's values vary regularly above 3 [dBd], that is to say for a 300 [MHz] frequency range.

Finally, <u>Figure 4.2.3</u> displays the variations of normalized radiation intensities concerning resistances 50, 75 and 300 [$\Omega$], which are named arbitrarily as Y with dimensions [mS]. This quantity, as it has been noted, is directly proportional to the directivity, while its relation with the SWR is more complex than the reversely proportional and tends to maximize when SWR minimizes.

According to <u>Figure 4.2.3</u> five regions of function's Y maximization are clearly being distinguished. These particular five frequency ranges will be studied more extensively below one by one. The following frequency ranges have been selected:

| From [MHz] | To [MHz] |
|:---:|:---:|
| 400 | 800 |
| 1100 | 1500 |
| 1800 | 2200 |
| 2200 | 2500 |
| 2550 | 2850 |

At each one of the above ranges measurements have been made each 10 [MHz] while in certain sections of these regions, where the variations of the concerning quantities were too rapid, measurements have been made each 5 [MHz].

## 4.2.1 : Frequency range from 400 to 800 [MHz]

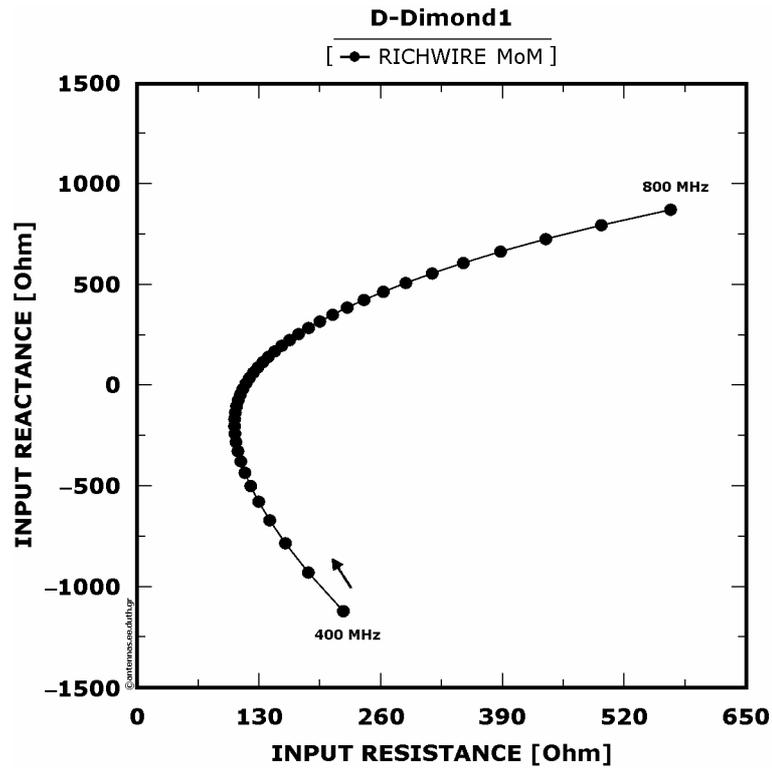

Figure 1 : Input impedance

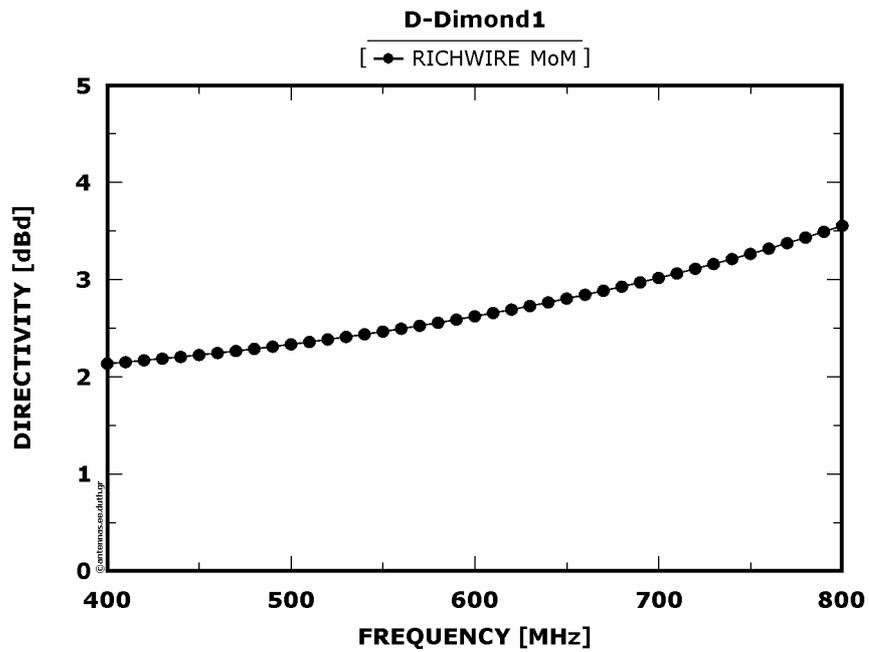

Figure 2 : Directivity (with reference to dipole λ/2) as a function of frequency

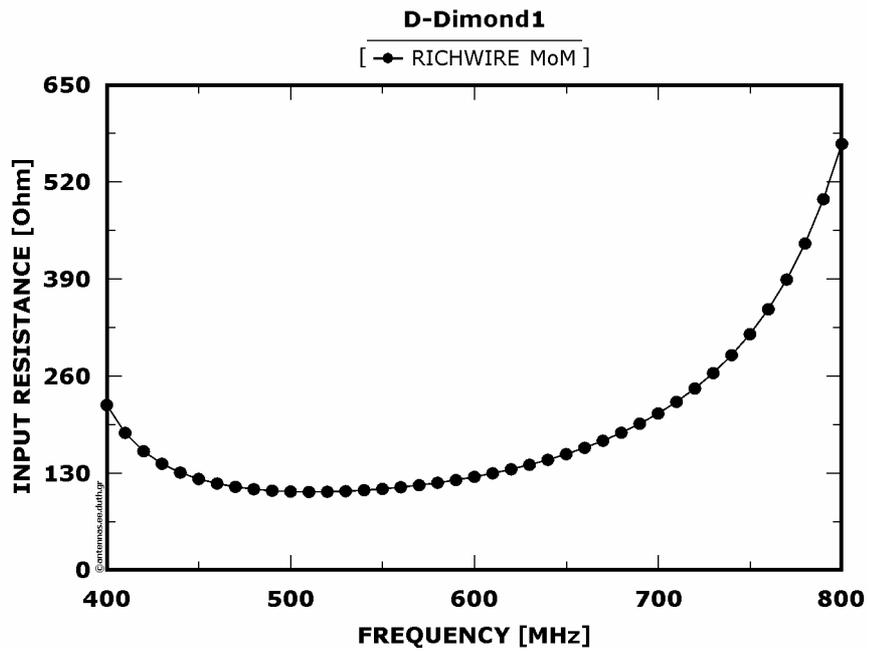

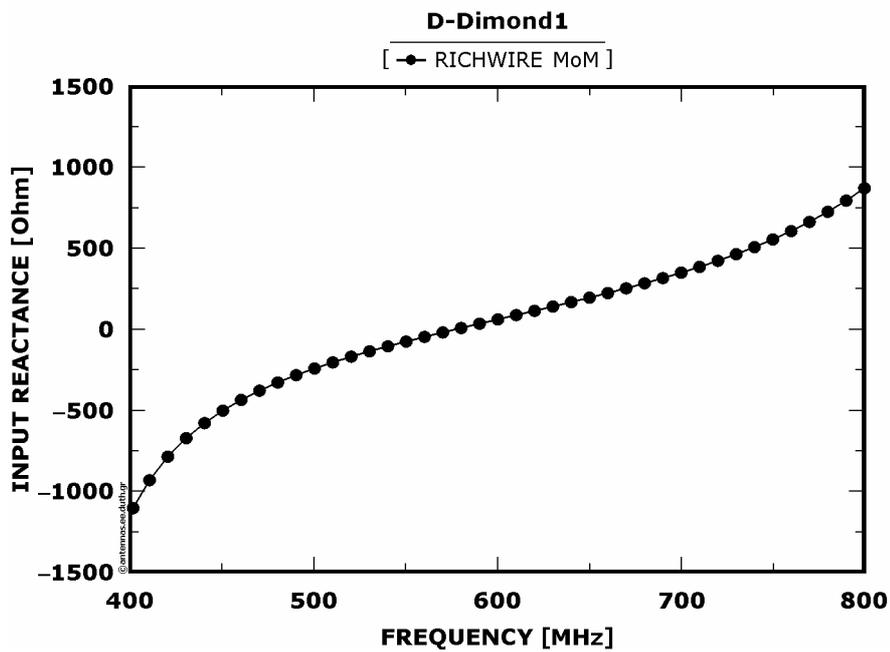

<u>Figure 3</u> : Real and imaginary part of input impedance as a function of frequency

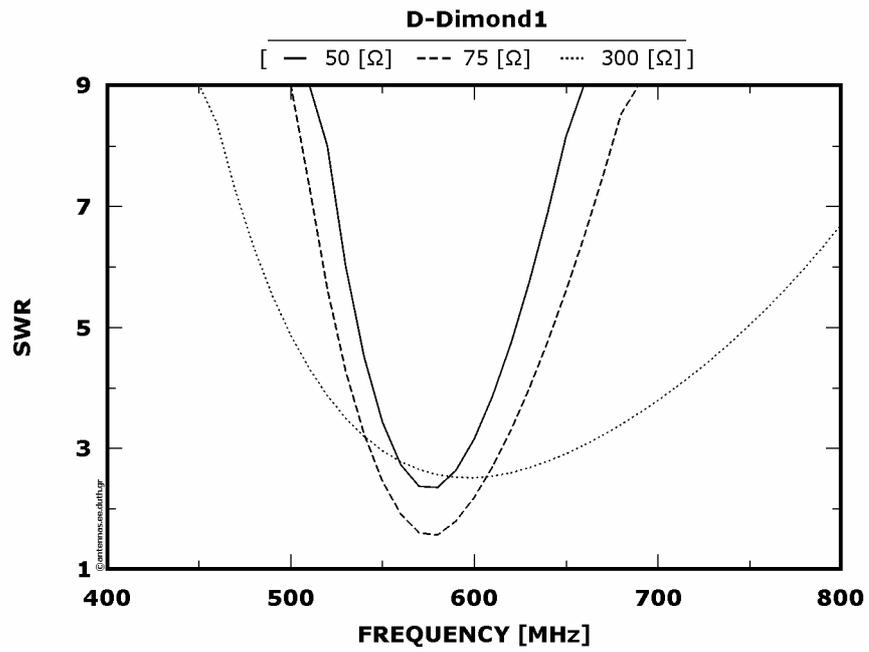

Figure 4 : SWRs as a function of frequency

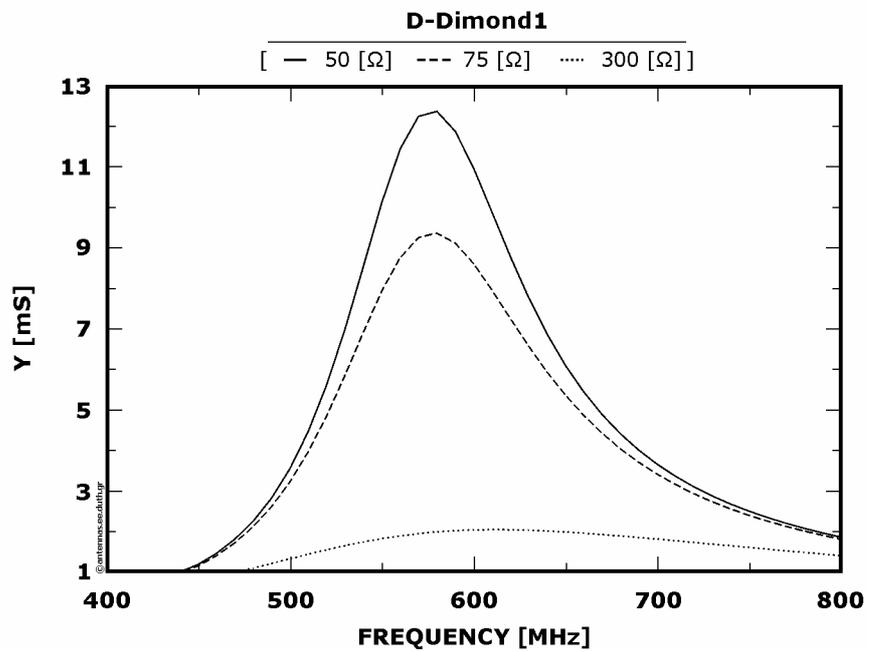

Figure 5 : Normalized radiation intensities as a function of frequency

**4.2.2 : Frequency range from 1100 to 1500 [MHz]**

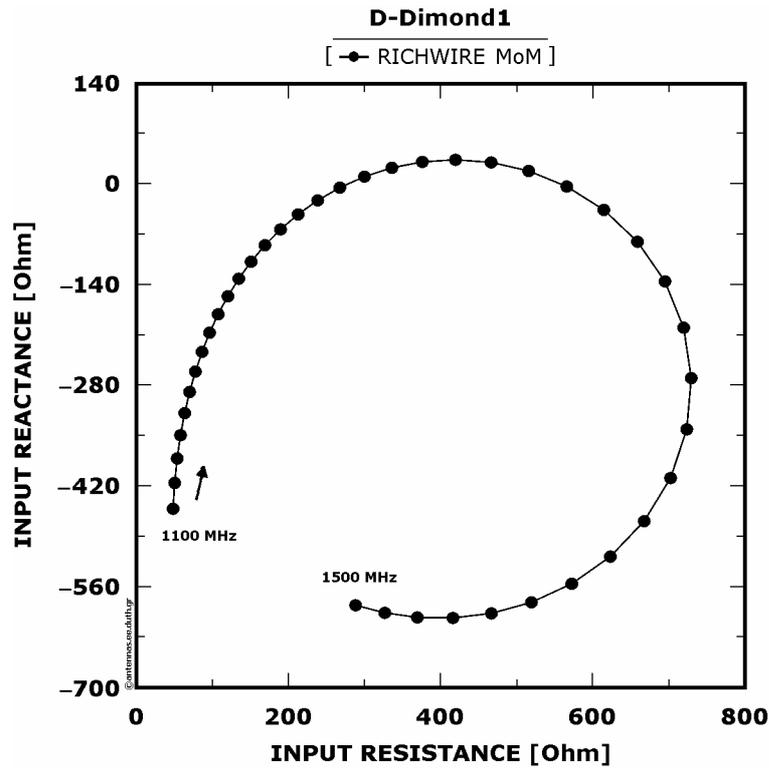

<u>Figure 1</u> : Input impedance

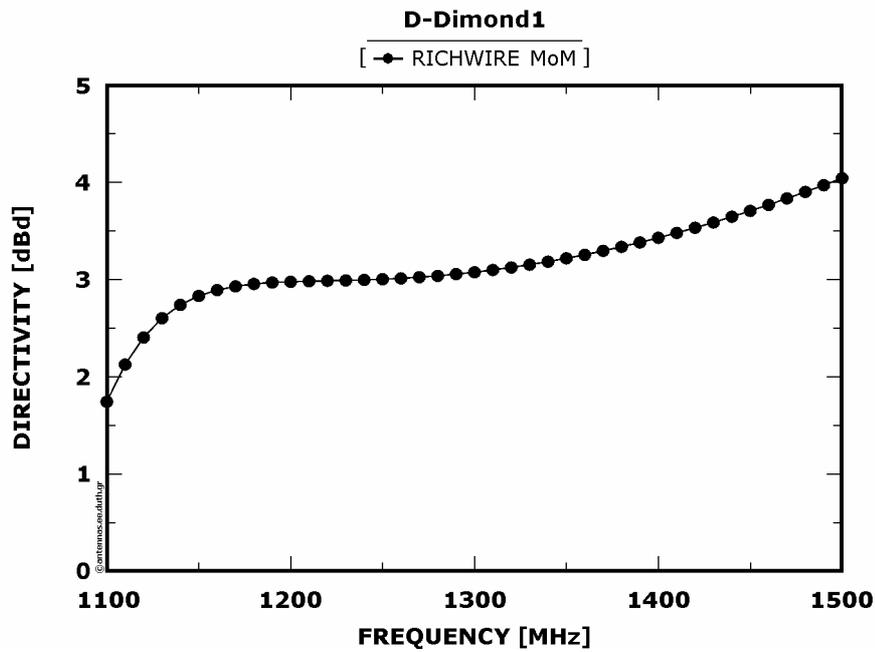

<u>Figure 2</u> : Directivity (with reference to dipole λ/2) as a function of frequency

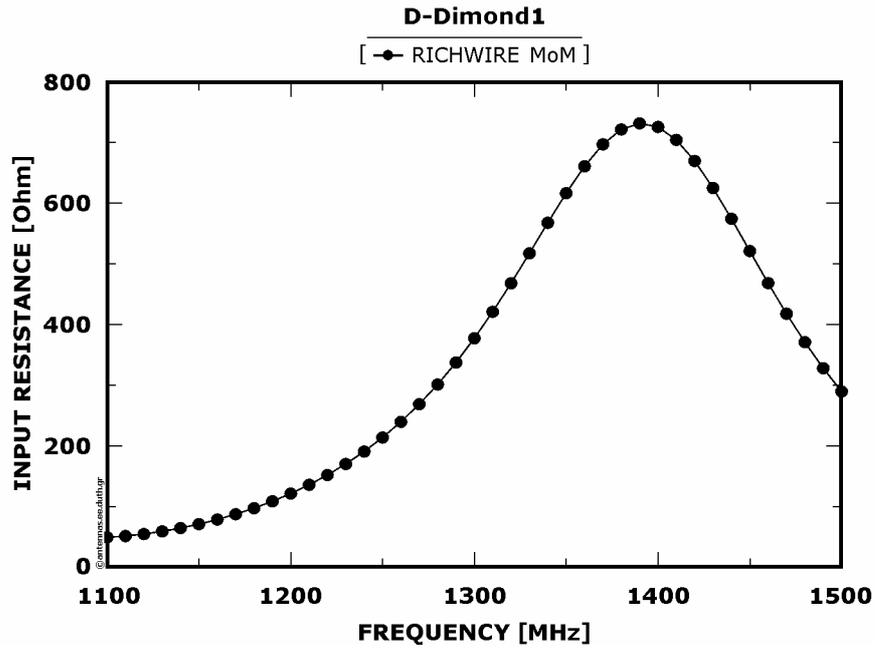

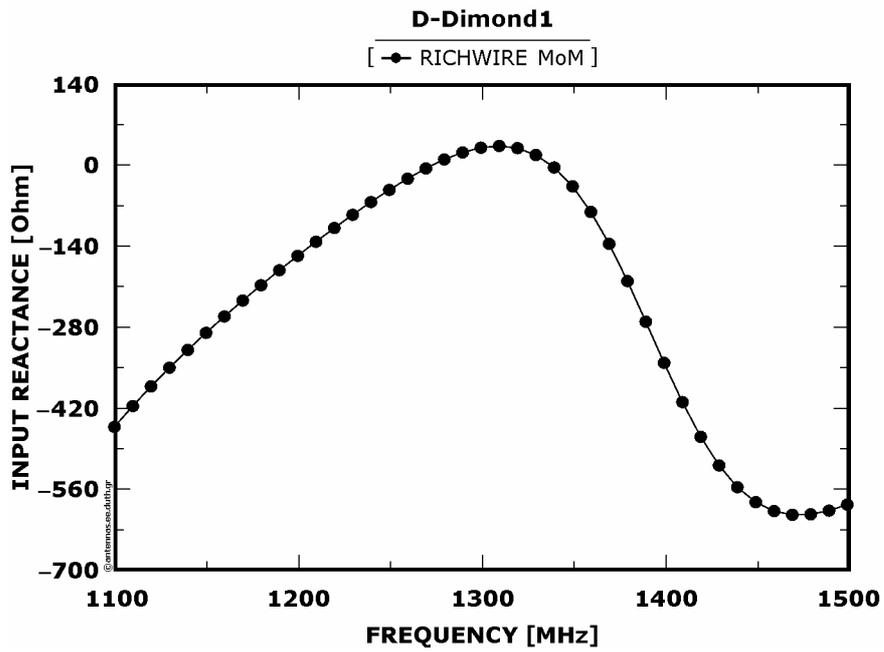

Figure 3 : Real and imaginary part of input impedance as a function of frequency

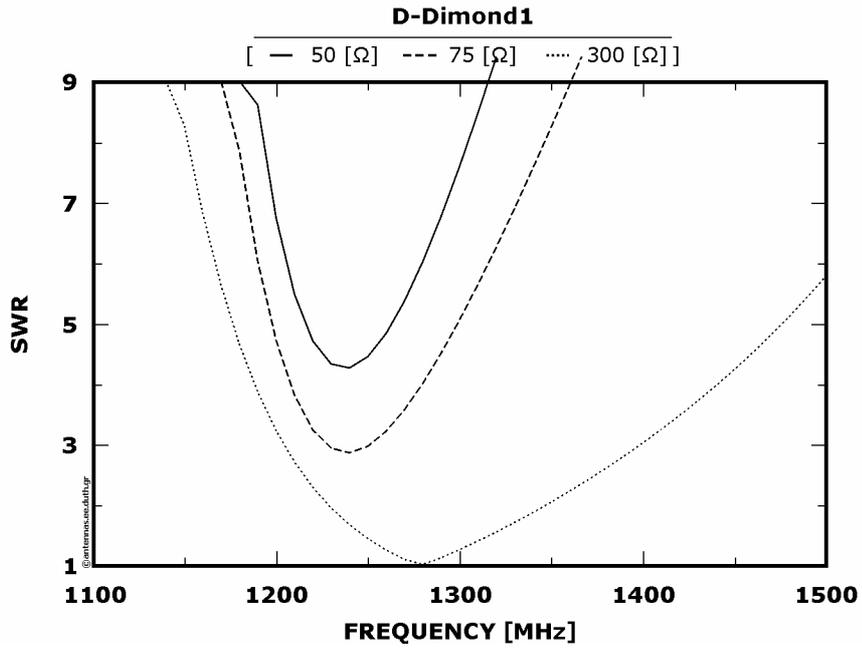

Figure 4 : SWRs as a function of frequency

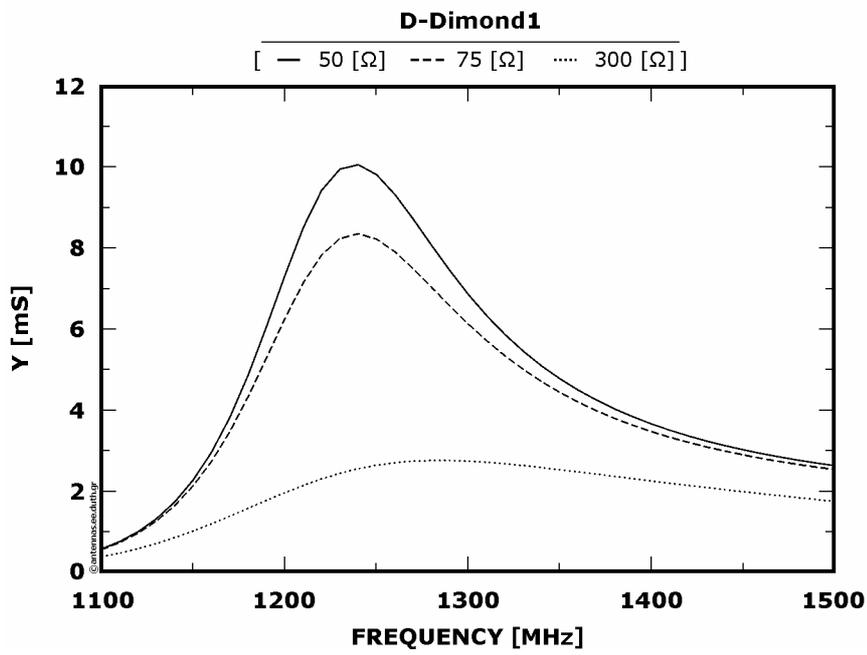

Figure 5 : Normalized radiation intensities as a function of frequency

**4.2.3 : Frequency range from 1800 to 2200 [MHz]**

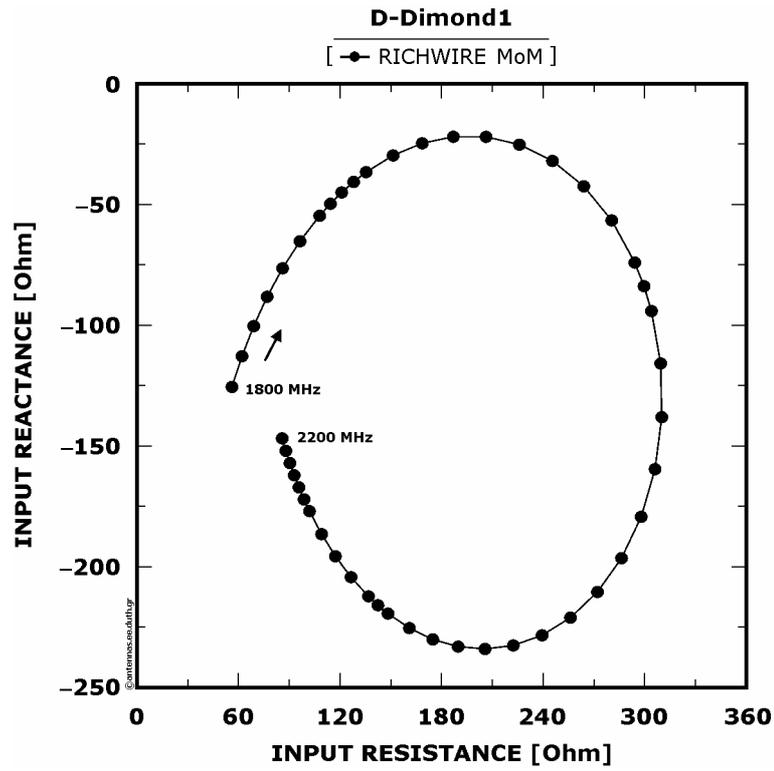

<u>Figure 1</u> : Input impedance

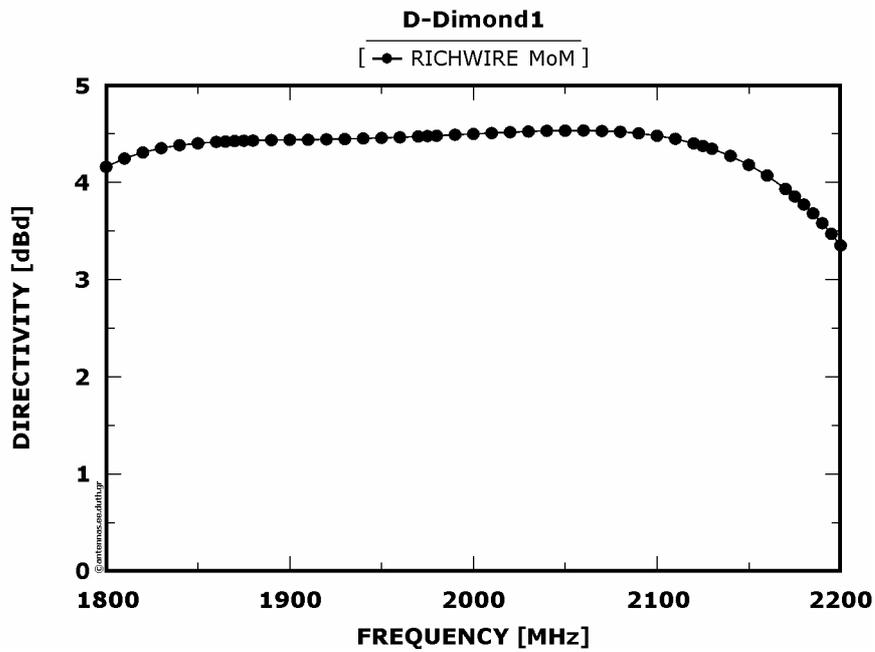

<u>Figure 2</u> : Directivity (with reference to dipole λ/2) as a function of frequency

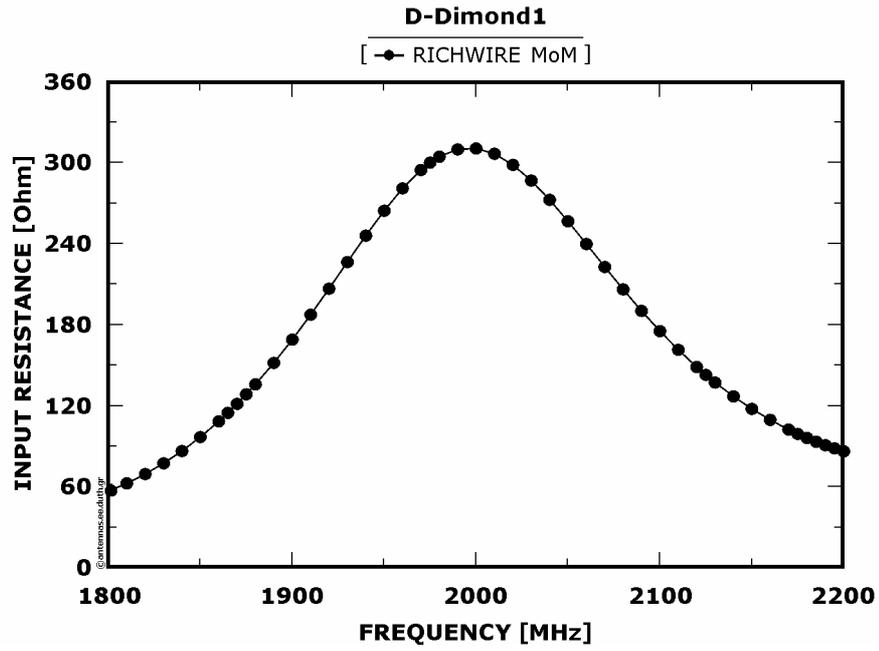

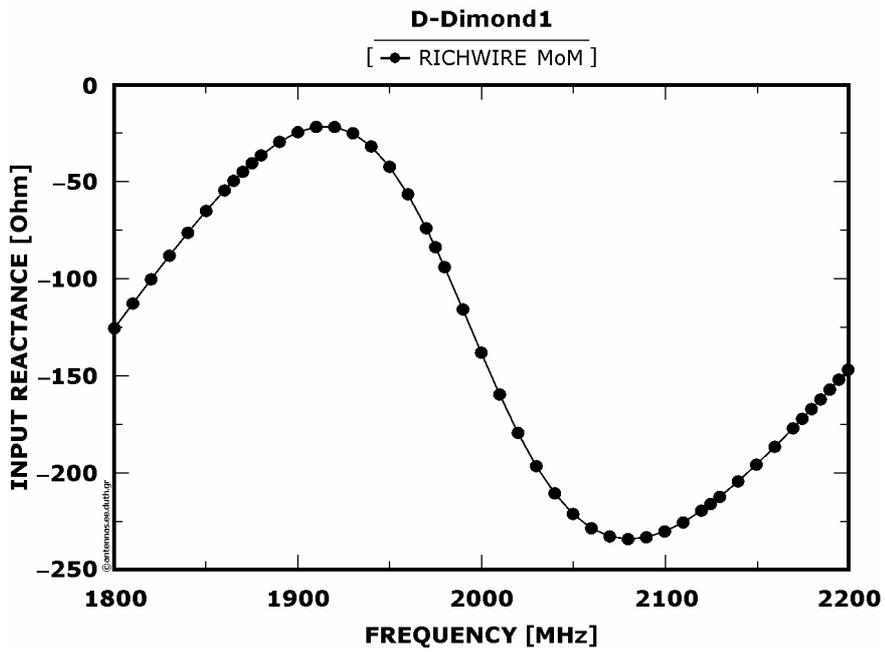

<u>Figure 3</u> : Real and imaginary part of input impedance as a function of frequency

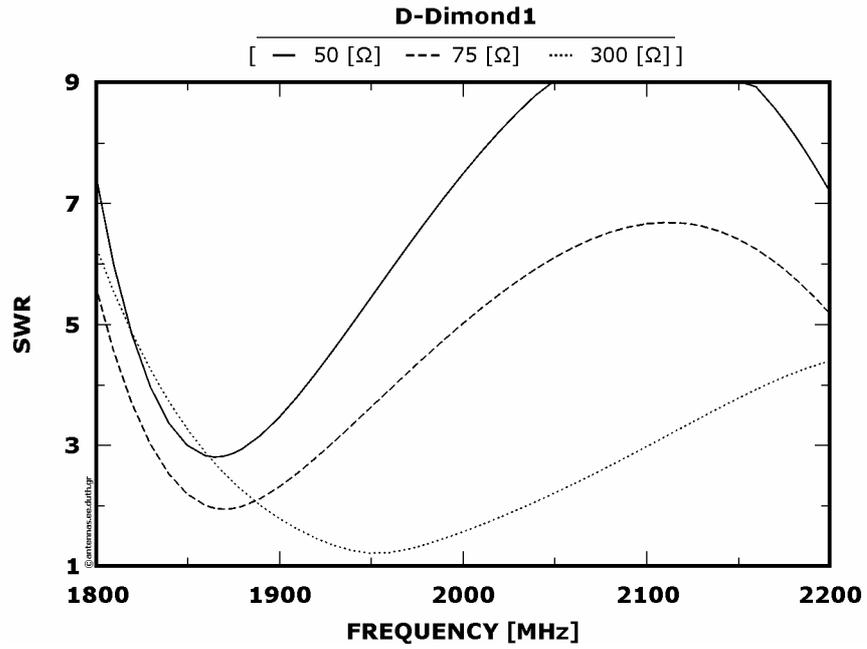

Figure 4 : SWRs as a function of frequency

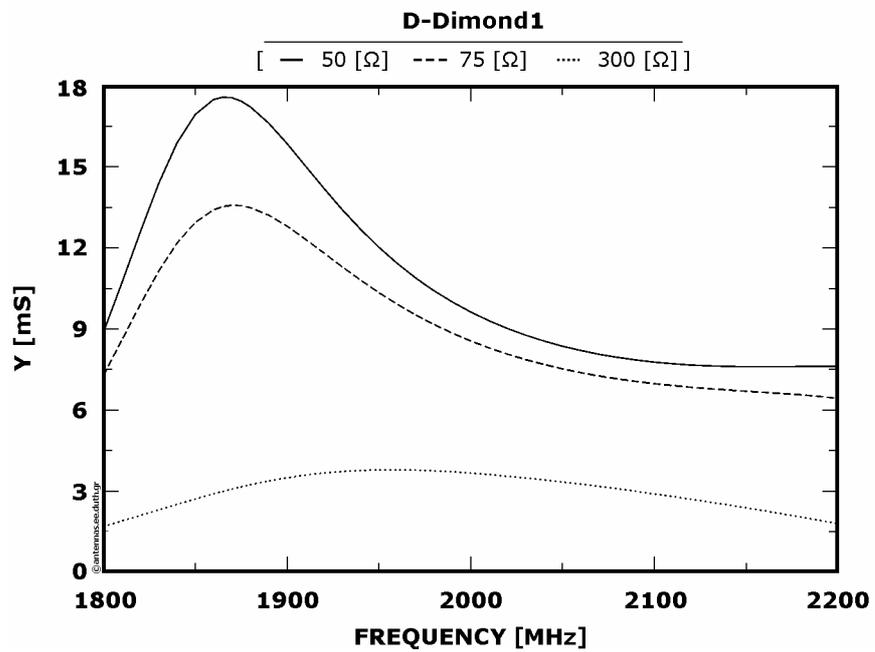

Figure 5 : Normalized radiation intensities as a function of frequency

**4.2.4 : Frequency range from 2200 to 2500 [MHz]**

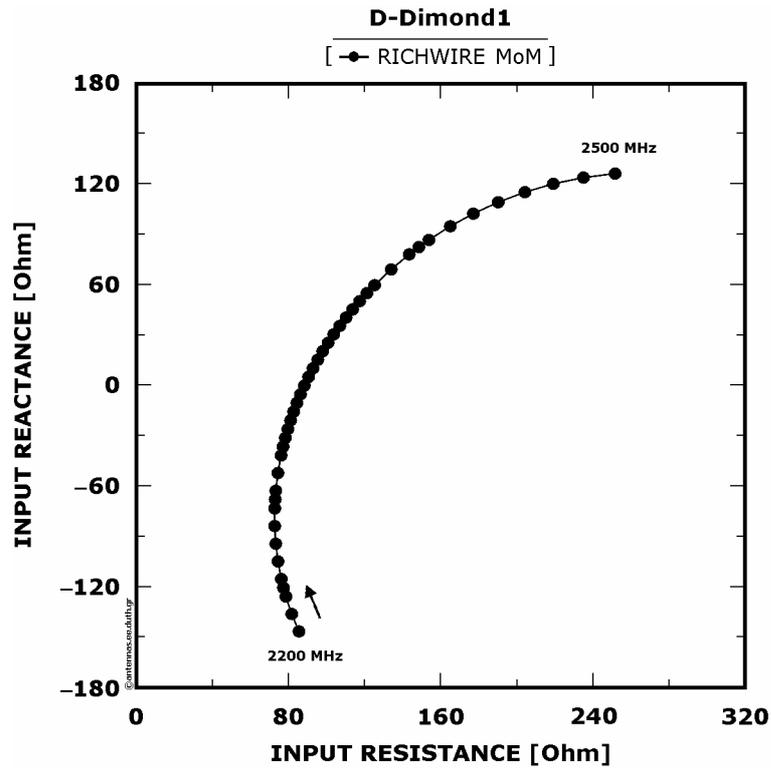

Figure 1 : Input impedance

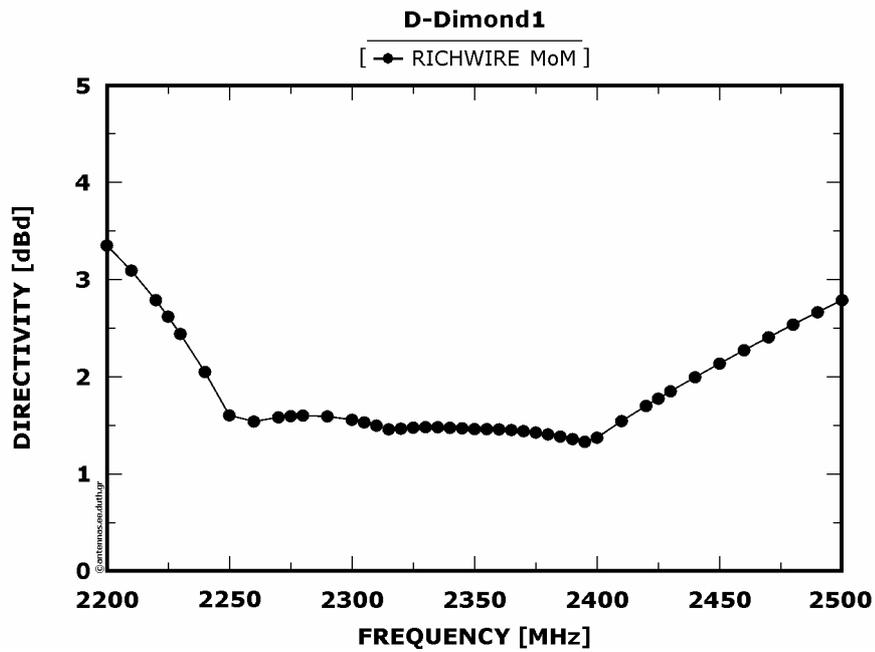

Figure 2 : Directivity (with reference to dipole λ/2) as a function of frequency

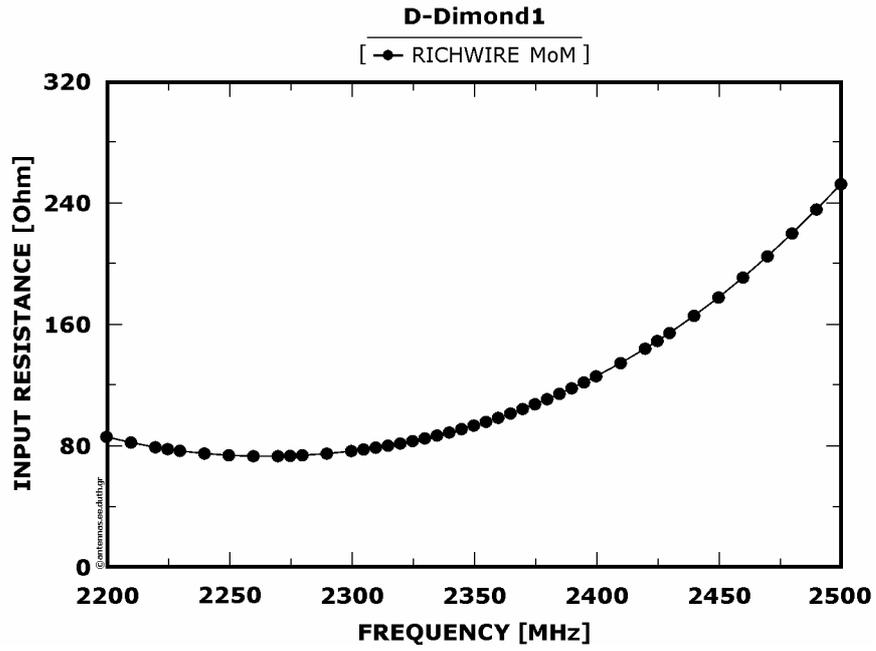

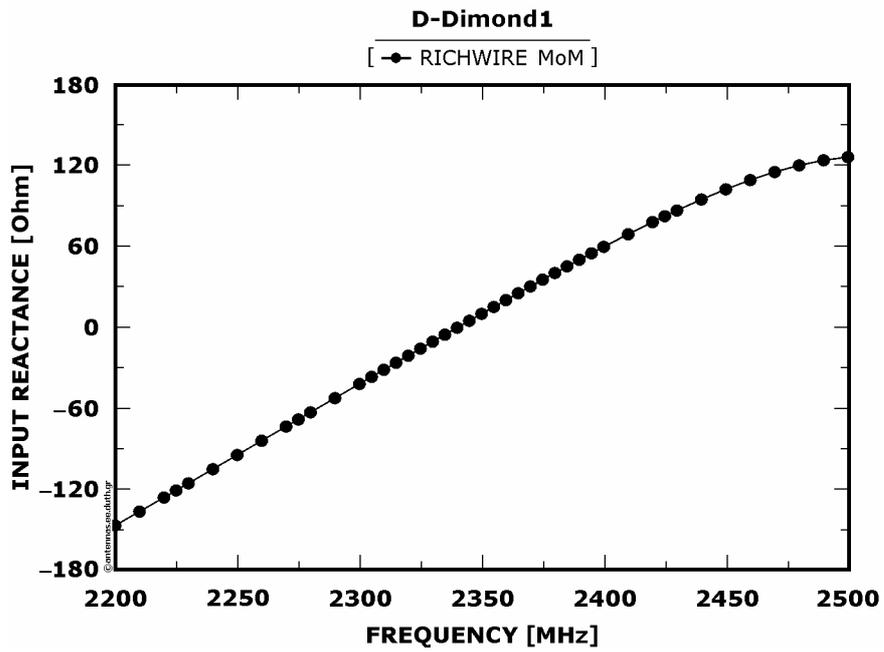

Figure 3 : Real and imaginary part of input impedance as a function of frequency

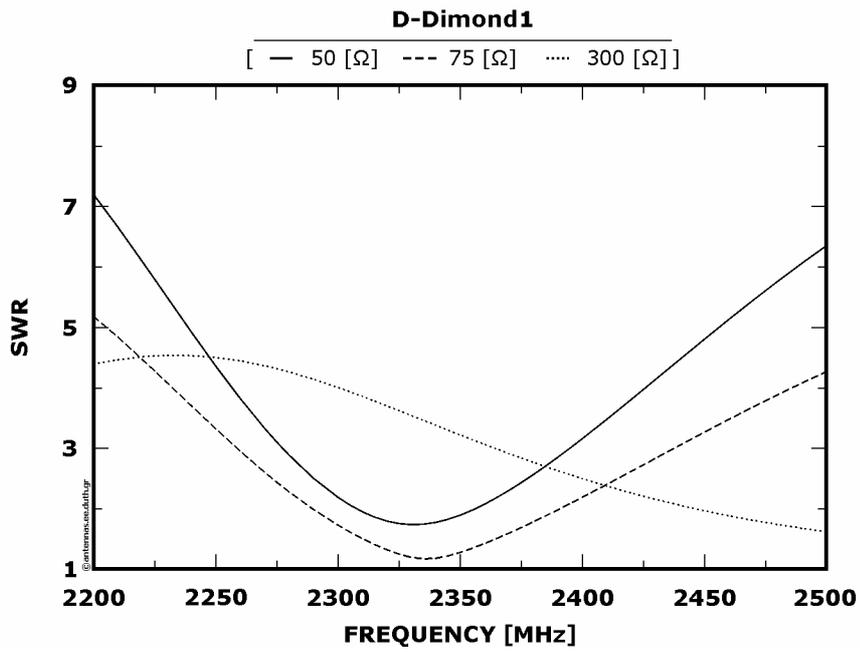

Figure 4 : SWRs as a function of frequency

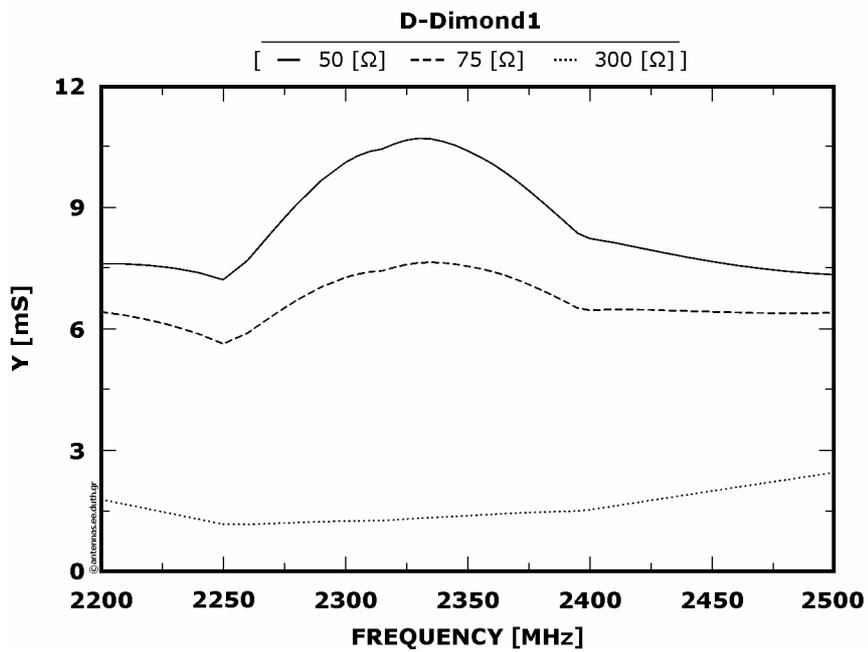

Figure 5 : Normalized radiation intensities as a function of frequency

### 4.2.5 : Frequency range from 2550 to 2850 [MHz]

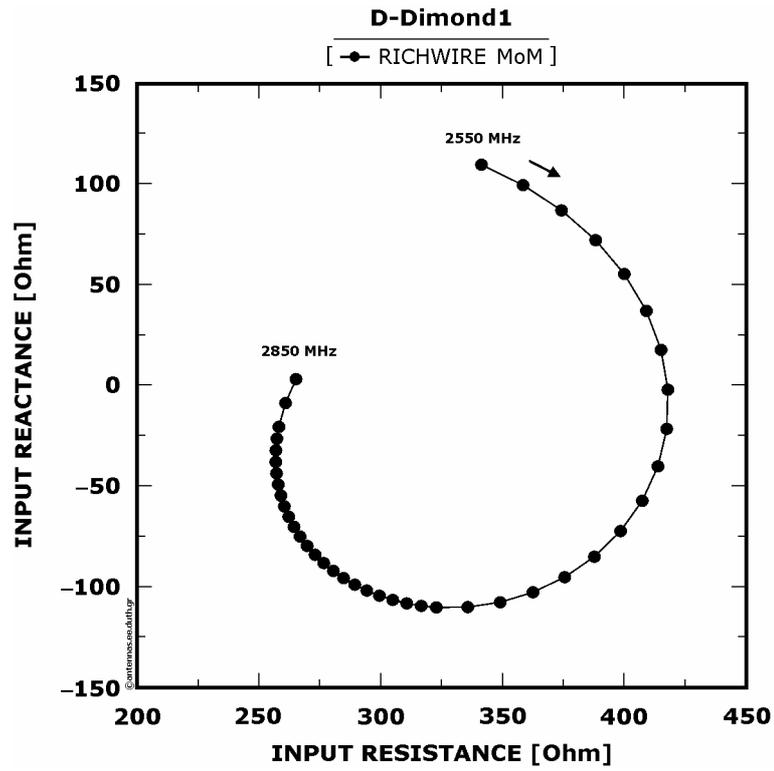

<u>Figure 1</u> : Input impedance

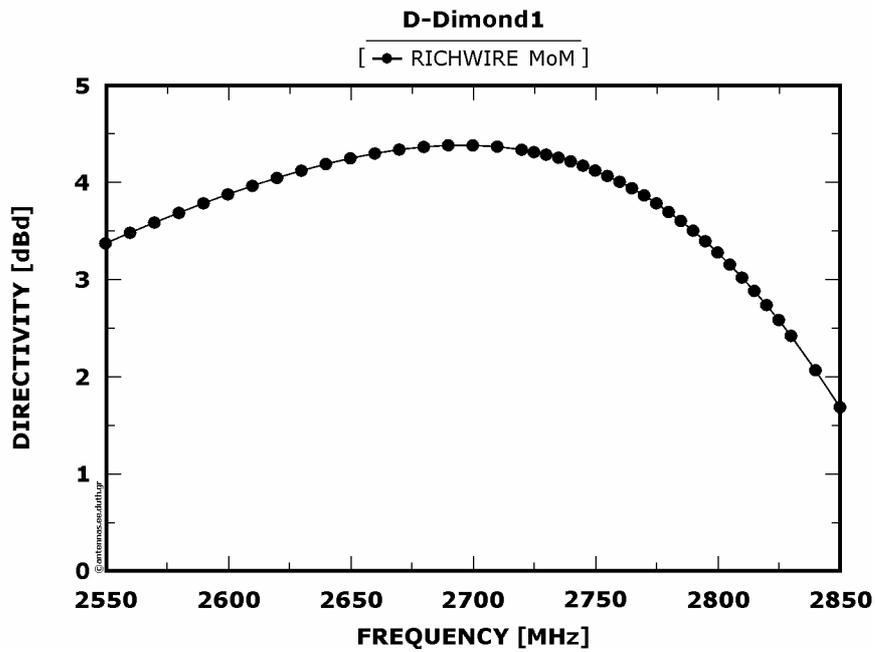

<u>Figure 2</u> : Directivity (with reference to dipole λ/2) as a function of frequency

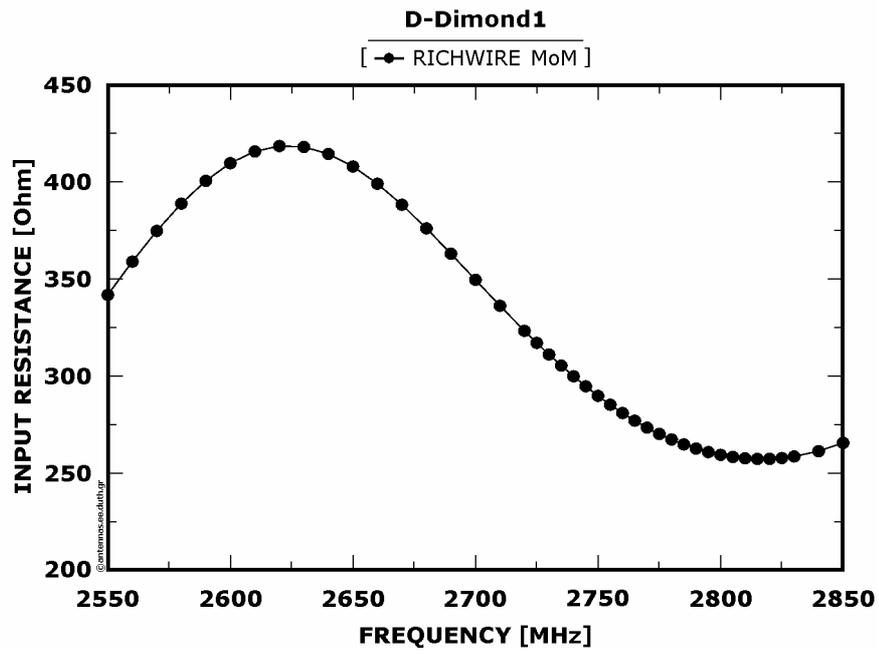

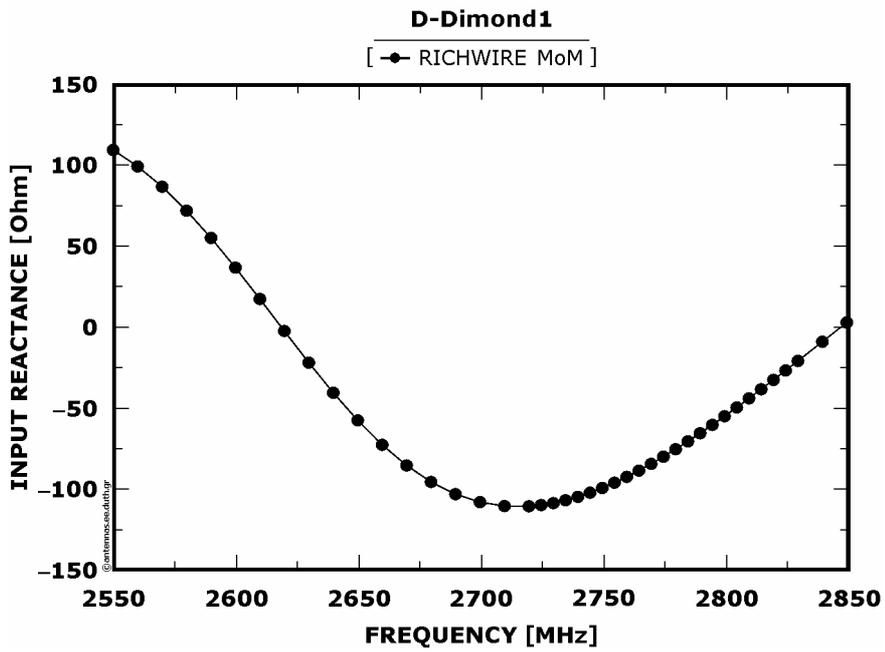

<u>Figure 3</u> : Real and imaginary part of input impedance as a function of frequency

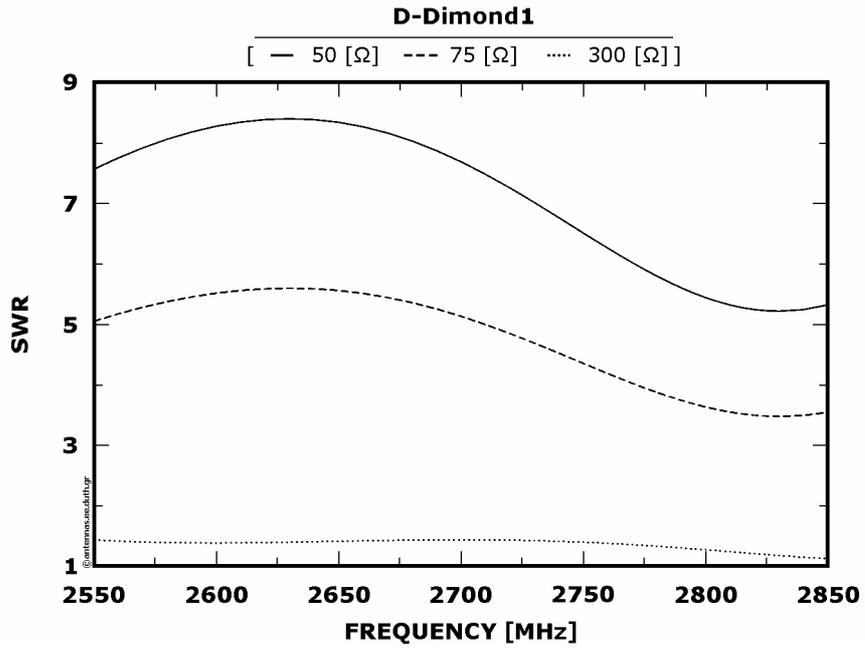

Figure 4 : SWRs as a function of frequency

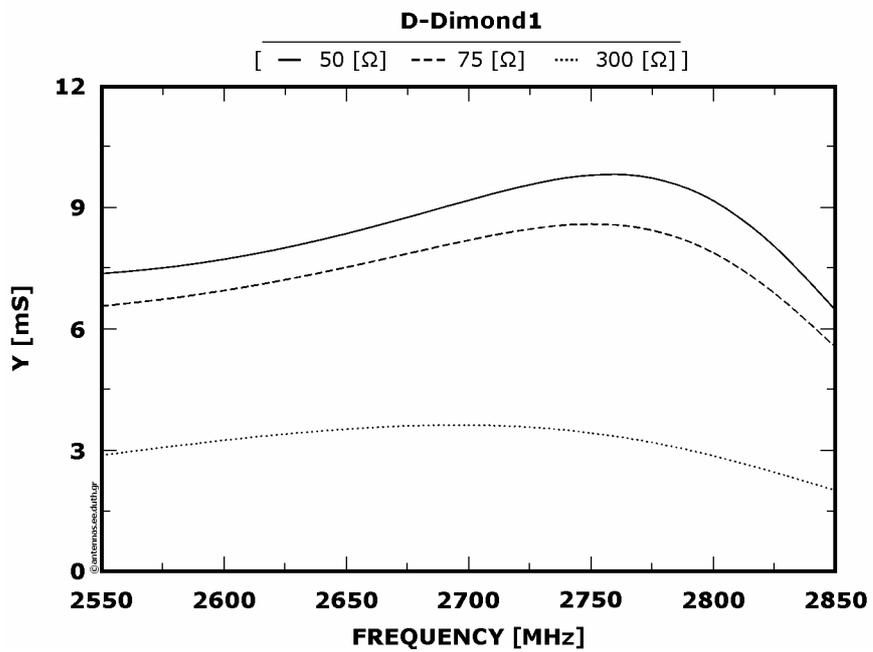

Figure 5 : Normalized radiation intensities as a function of frequency

**4.3 : Conclusions**

After the apposition of all essential graphs at the examined frequency ranges, an evaluation of the antenna's behaviour in connection with frequency has to be made. Below are cited analytical tables in which is shown the bandwidth depending on the transmission line which is used. The transmission lines match the antenna with its feeding source. Moreover, the antenna's maximum acceptable SWR is shown which is determined from the overall design of the communicating system in which the antenna is embodied. Finally, the directivity's values at the edges of the bandwidth are indicated.

According to Table 4.3.1 it is concluded that if the antenna is going to operate inside the examined bandwidth and if a SWR value higher than 2 isn't acceptable then a transmission line with a characteristic resistance of 75 [Ω] has to be used. The operation frequency will oscillate between 560 and 590 [MHz] and the directivity between the values of 2.49 and 2.58 [dBd]. Equivalently, if the overall design of the communicating system does not allow a SWR value above 3 then there is the capability of using transmission lines of 50,75 or 300 [Ω] which allow an operating bandwidth of 30,.70 and 110 [MHz] respectively.

As it appears from Table 4.3.2, if the system design requires that the maximum SWR value has to equal 2 then it should be used exclusively a transmission line of characteristic resistance 300 [Ω] and the antenna's operating frequency should range between 1230 and 1350 [MHz], offering a broad bandwidth, that is 120 [MHz]. The corresponding variation of directivity isn't significant and extents from 2.98 to 3.21 [dBd]. Respectively, for maximum SWR equals to 3 it should be used exclusively a 75 [Ω] transmission line, in a however narrower operational bandwidth, that is from 1230 to 1250 [MHz]. On the contrary, the usage of a 300 [Ω] transmission line increases even more the operational frequency range of the antenna, that is from 1210 to 1400 [MHz]. Moreover directivity is maintained relatively constant, oscillating between the values of 2.97 and 3.42 [dBd].

As it appears from Table 4.3.3, for maximum acceptable SWR equals to 2, a transmission line of characteristic resistance 75 or 300 [Ω] can be used. The 75 [Ω] transmission line offers a quite narrow operational bandwidth, that is 20 [MHz], ranging from 1860 to 1880 [MHz]. On the other hand, the usage of a 300 [Ω] transmission line offers a broader bandwidth ranging from 1890 to 2030 [MHz], that is 140 [MHz]. The directivity remains almost invariable at both cases taking quite high values, of the order of 4,4 [dBd]. If the system design allows maximum SWR value equals to 3 in that case all the transmission lines (50, 75 and 300 [$\Omega$]) can be utilized. The 50 [$\Omega$] transmission line offers a narrow operation frequency range while, on the other hand, a 300 [$\Omega$] transmission line offers a 240 [MHz] bandwidth, ranging from 1860 to 2100 [MHz]. The directivity maintains a negligible fluctuation, picking up high values.

Table 4.3.4 demonstrates that for a maximum acceptable SWR equals to 2 there is the possibility of using all the three transmission lines examined here (50,75 and 300 [$\Omega$]). The 50 [$\Omega$] transmission line offers a bandwidth of 50 [MHz], from 2310 to 2360 [MHz] and directivity picks up low values, ranging from 1.45 to 1.49 [dBd]. The transmission line of 75 [$\Omega$] allows the broadest bandwidth at this case, that is 100 [MHz], while the 300 [$\Omega$] transmission line offers a narrow bandwidth of 60 [MHz] and moreover a high directivity fluctuation, from 1.99 to 2.7 [dBd]. For a maximum SWR equals to 3 there is again the possibility of using all the three known transmission lines while there is a significant expansion of the antenna's operational bandwidth. Specifically, for transmission lines of 50, 75 and 300 [$\Omega$] the offering bandwidths are 110,180 and 140 [MHz] respectively, while directivity changes significantly in the case of using a 300 [$\Omega$] transmission line, that is from 1.45 to 2.7 [dBd].

According to Table 4.3.5 it appears that for the given frequency range of the antenna's operation only the usage of the 300 [$\Omega$] transmission line is possible, for both cases of the acceptable SWR. The bandwidth that is gained covers up the whole frequency range that is studied, that is to say from 2550 up 2850 [MHz], i.e. 300 [MHz], while directivity presents intense fluctuation starting from 3.36 [dBd] and reaching 4.37 [dBd], approaching at the end of the frequency range 1.68 [dBd], as shown in Figure 4.2.5.2.

Table 1 : Frequency range from 400 to 800 [MHz]

| Acceptable SWR | SWR(50) | | SWR(75) | | SWR(300) | |
|---|---|---|---|---|---|---|
| | BW [MHz] | D [dBd] | BW [MHz] | D [dBd] | BW [MHz] | D [dBd] |
| 2 | --- | --- | $590 - 560 = 30$ | 2.49 to 2.58 | --- | --- |
| 3 | $590 - 560 = 30$ | 2.49 to 2.58 | $610 - 540 = 70$ | 2.43 to 2.65 | $660 - 550 = 110$ | 2.46 to 2.84 |

Table 2 : Frequency range from 1100 to 1500 [MHz]

| Acceptable SWR | SWR(50) | | SWR(75) | | SWR(300) | |
|---|---|---|---|---|---|---|
| | BW [MHz] | D [dBd] | BW [MHz] | D [dBd] | BW [MHz] | D [dBd] |
| 2 | --- | --- | --- | --- | $1350 - 1230 = 120$ | 2.98 to 3.21 |
| 3 | --- | --- | $1250 - 1230 = 20$ | 2.98 to 3.0 | $1400 - 1210 = 190$ | 2.97 to 3.42 |

Table 3 : Frequency range from 1800 to 2200 [MHz]

| Acceptable SWR | SWR(50) | | SWR(75) | | SWR(300) | |
|---|---|---|---|---|---|---|
| | BW [MHz] | D [dBd] | BW [MHz] | D [dBd] | BW [MHz] | D [dBd] |
| 2 | --- | --- | $1880 - 1860 = 20$ | 4.41 to 4.42 | $2030 - 1890 = 140$ | 4.42 to 4.51 |
| 3 | $1880 - 1850 = 30$ | 4.39 to 4.42 | $1920 - 1830 = 90$ | 4.34 to 4.43 | $2100 - 1860 = 240$ | 4.41 to 4.47 |

Table 4 : Frequency range from 2200 to 2500 [MHz]

| Acceptable SWR | SWR(50) | | SWR(75) | | SWR(300) | |
|---|---|---|---|---|---|---|
| | BW [MHz] | D [dBd] | BW [MHz] | D [dBd] | BW [MHz] | D [dBd] |
| 2 | $2360 - 2310 = 50$ | 1.49 to 1.45 | $2390 - 2290 = 100$ | 1.59 to 1.35 | $2500 - 2440 = 60$ | 1.99 to 2.78 |
| 3 | $2390 - 2280 = 110$ | 1.60 to 1.33 | $2440 - 2260 = 180$ | 1.53 to 1.99 | $2500 - 2360 = 140$ | 1.45 to 2.78 |

Table 5 : Frequency range from 2550 to 2850 [MHz]

| Acceptable SWR | SWR(50) | | SWR(75) | | SWR(300) | |
|---|---|---|---|---|---|---|
| | BW [MHz] | D [dBd] | BW [MHz] | D [dBd] | BW [MHz] | D [dBd] |
| 2 | --- | --- | --- | --- | $2850 - 2550 = 300$ | 3.36 to 1.68 |
| 3 | --- | --- | --- | --- | $2850 - 2550 = 300$ | 3.36 to 1.68 |

# Chapter 5 : Improvement

## 5.1 : In General

In the present chapter an improvement of certain characteristics of the antenna is attempted by modifying its geometrical shape. Therefore, three geometrical models of modification were developed and studied.

According to the 1[st] model there is a displacement of the active dipole towards the positive z axis. The displacement step is $\lambda/128$ and the total displacement positions are 32. Thus, the maximum active dipole's lifting does not exceed $\lambda/4$ from its initial position. The displacement step has been chosen to be as short as possible while simultaneously it satisfies the condition which is set by our simulation program, RICHWIRE, which does not permit the distances between nodes to be smaller than the double of wire's radius, that is 2 [mm].

According to the 2[nd] model there was an attempt to elongate the two perpendicular monopoles that are located at the positive z axis. The elongation happens at the direction of z axis towards its positive end. Moreover, the elongation step is $\lambda/128$ thus the total positions are 32. Finally, the maximum achieved elongation is $\lambda/4$. The step was carefully chosen again as previously, that is to say a compromise between the minimum allowed distance between two successive nodes, which is strictly defined by RICHWIRE and simultaneously the maximum possible number of steps in order more discretion ability to be achieved.

At the 3$^{rd}$ model an elongation of the active dipole is attempted on either side of its edges and towards the y axis. The elongation step is chosen again to be λ/128, the total possible positions are 32 and the maximum horizontal elongation of the active dipole is λ.

Below are cited some figures with the corresponding antenna's geometries in space which result from the pre-mentioned models, illustrating two geometrical modifications for each one of them, for clarifying reasons.

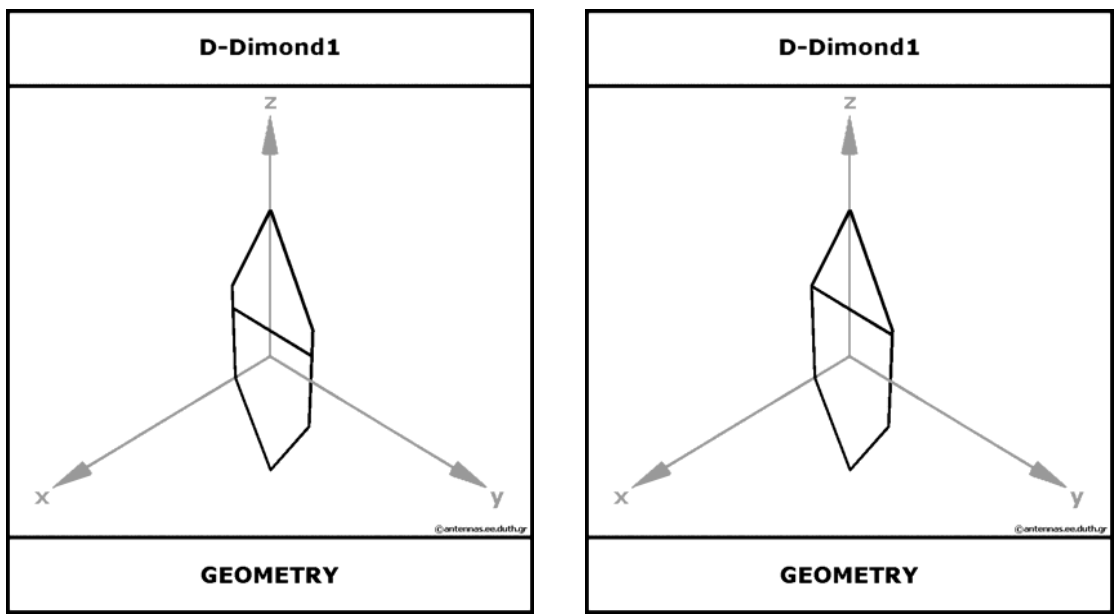

<u>Figure 1</u> : Displacement of the active dipole by λ/8 and λ/4 respectively

(1$^{st}$ model)

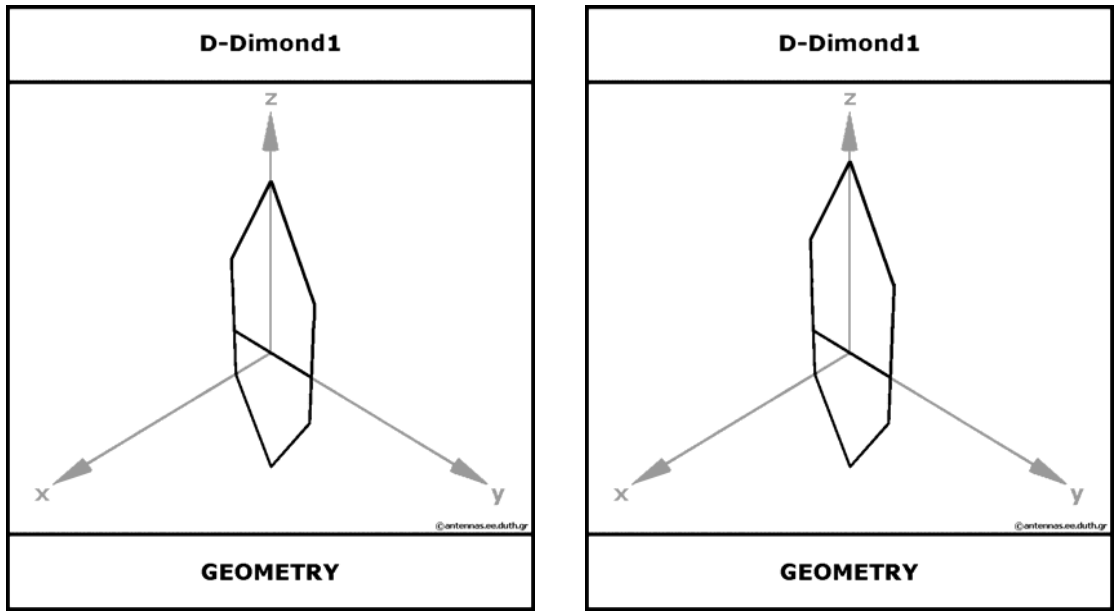

Figure 2 : Elongation of monopoles by λ/8 and λ/4 respectively
(2nd model)

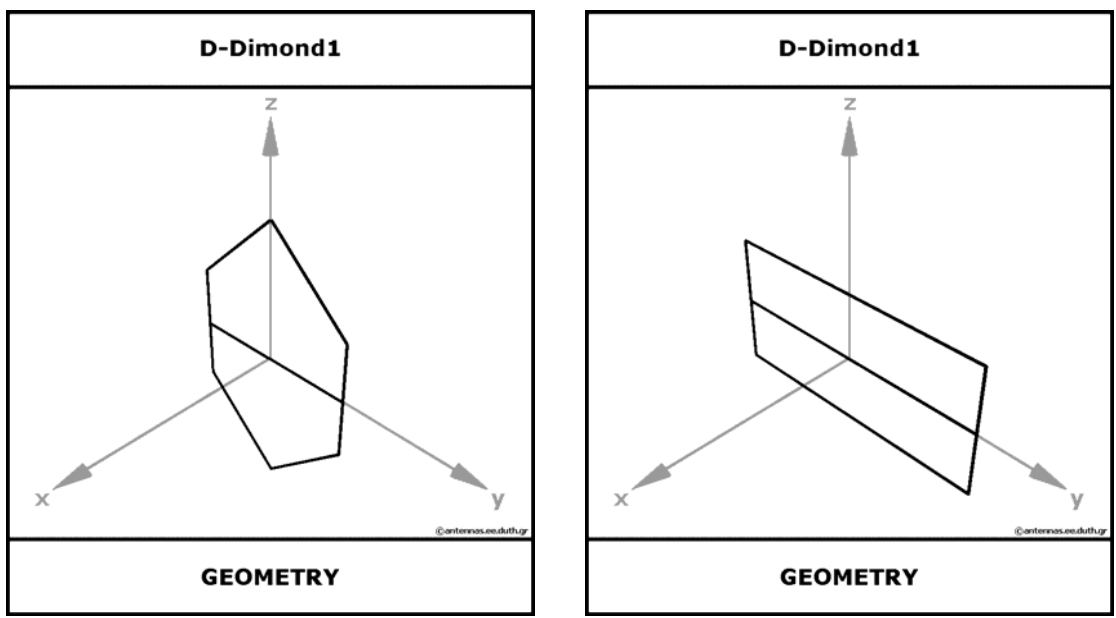

Figure 3 : Elongation at either side of active dipole's edges by λ/8 and λ/4 respectively
(3rd model)

**5.2 : Frequency selection**

Each one geometrical modification model is applied in order to investigate the variation of the antenna's characteristics at a specific frequency. Thus, a problem of the frequency selection at which the current thesis should examine the antenna's operation emerged. Therefore, for reasons that will be reported below, there have been selected the frequencies where the normalized radiation intensities, (Y [mS]), were maximized (mainly Ys in relation to 50 and 75 characteristic resistance). Thus, five specific frequencies from the equal in number frequency ranges that have been studied at the previous chapter were selected.

The philosophy of the reasons that contributed to the choice of these specific frequencies is the following: at these frequencies were presented the higher values of the maximum normalized radiation intensities, Y [mS]. The characteristic quantity Y was selected as auxiliary parameter for the frequencies' pick because it "responds" to the combination of minimal SWRs and maximum directivity, elements which are generally desirable for an antenna communication system. Thus, the investigation of how possible is the additional improvement of the above antenna's characteristics is studied, at the frequencies where the most desirable behavior is already has been observed.

Finally, according to this philosophy, there have been selected the following particular frequencies: 580, 1238, 1867, 2330 and 2760 [MHz]

].

### 5.2.1 : Frequency at 580 [MHz] – 1ˢᵗ model

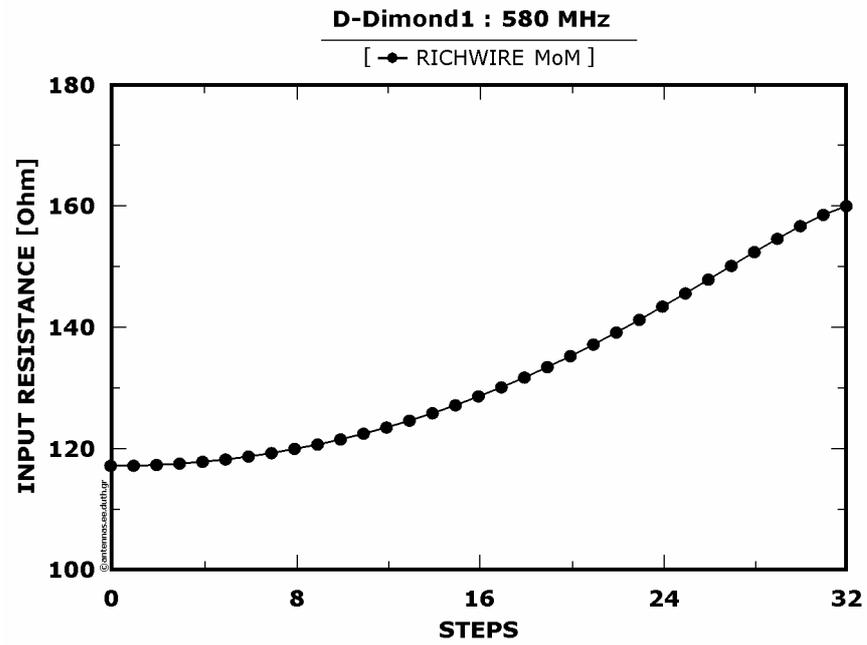

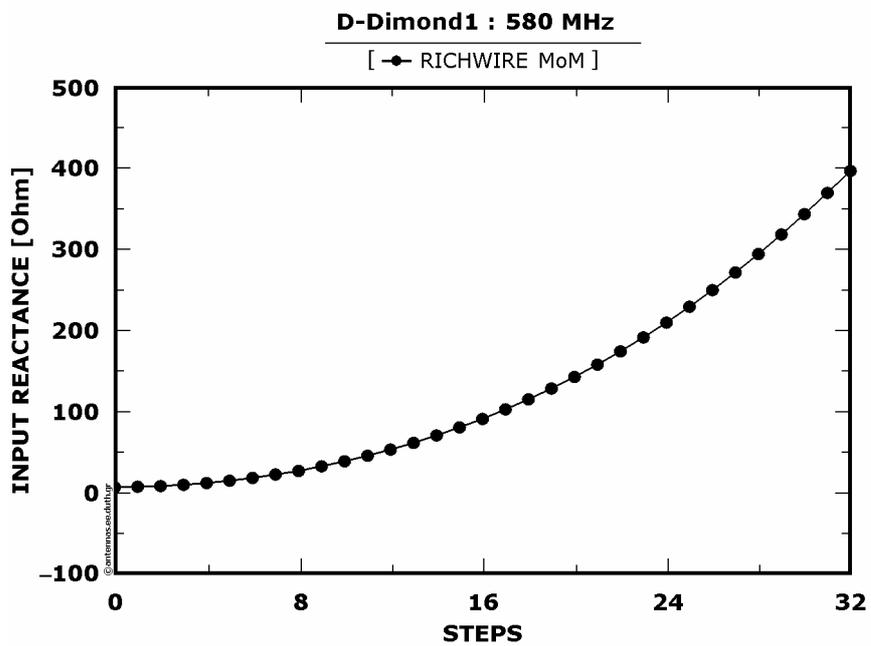

<u>Figure 1</u> : Real and imaginary part of the input impedance as a function of
the displacement steps

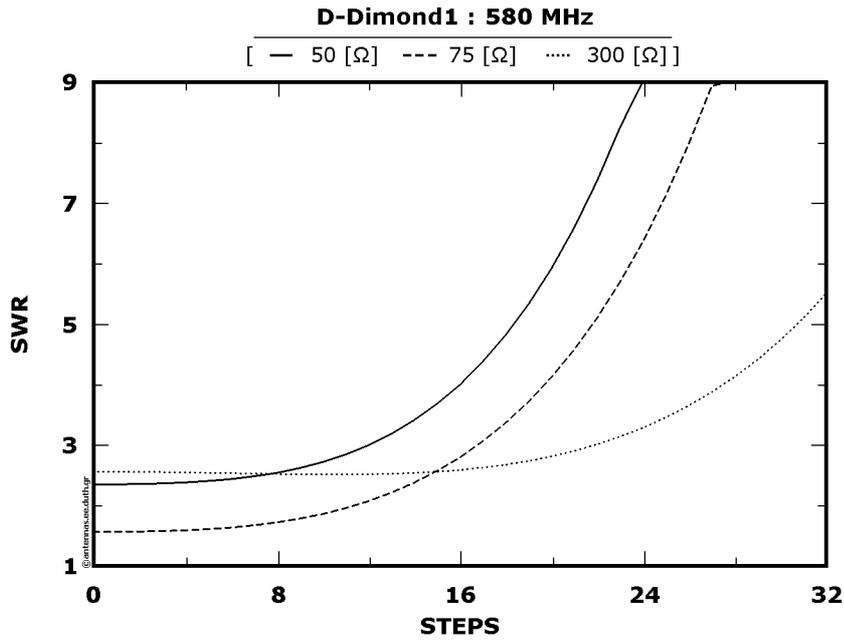

Figure 2 : SWRs as a function of the displacement steps

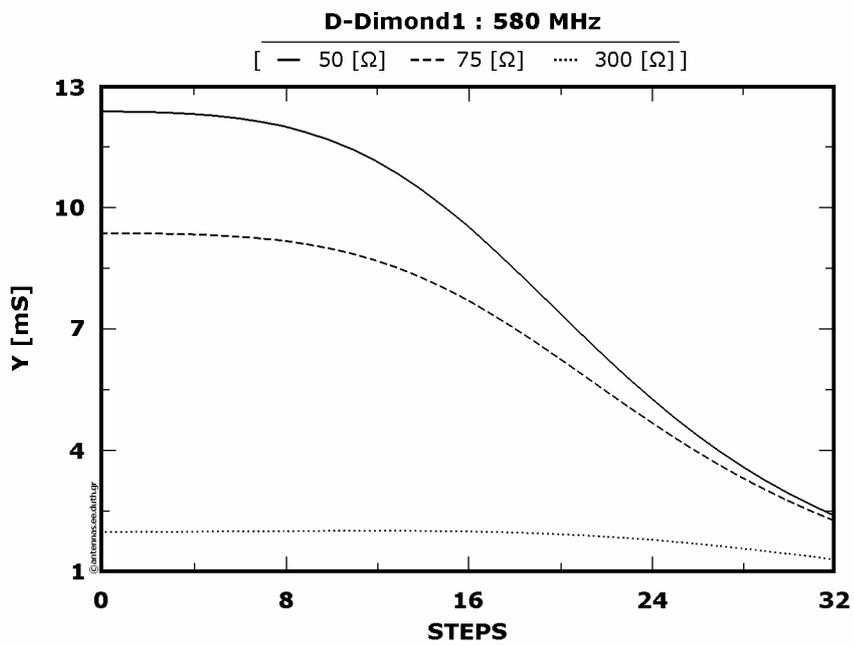

Figure 3 : Normalized radiation intensities as a function of the displacement steps

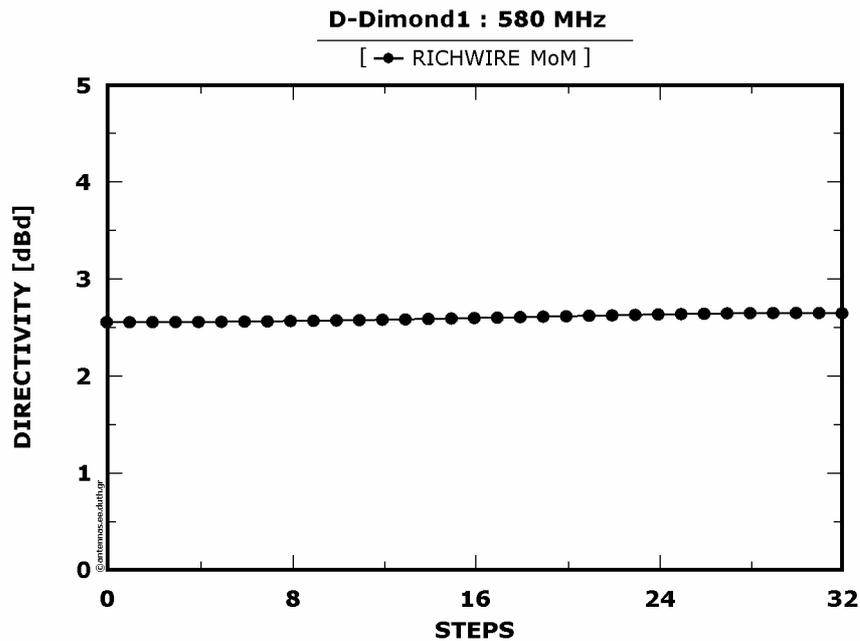

<u>Figure 4</u> : Directivity (with reference to dipole λ/2) as a function of the displacement steps

<u>Figure 5.2.1.2</u> shows that SWRs are increasing almost exponentially, except SWR(300) which remains practically constant with a value around 2.5 for the first λ/8 of the active dipole's displacement. Afterwards, SWR(300) commences to increase exponentially as well.

<u>Figure 5.2.1.4</u> shows that directivity remains almost unaffected from the displacement, assuming a value at around 2.5 [dBd].

Summarizing all the above, according to <u>Figure 5.2.1.3</u>, the almost exponential risings of SWRs in combination with a constant directivity produce almost corresponding exponential reductions of Ys, except of Y(300) which appears to be unaffected. Thus, it deduces that the particular improvement model does not succeed any desirable and interesting change of the antenna's characteristics at 580 [MHz]. Therefore the antenna's geometry is maintained as such.

**Model of perpendicular monopoles elongation (2ⁿᵈ model)**

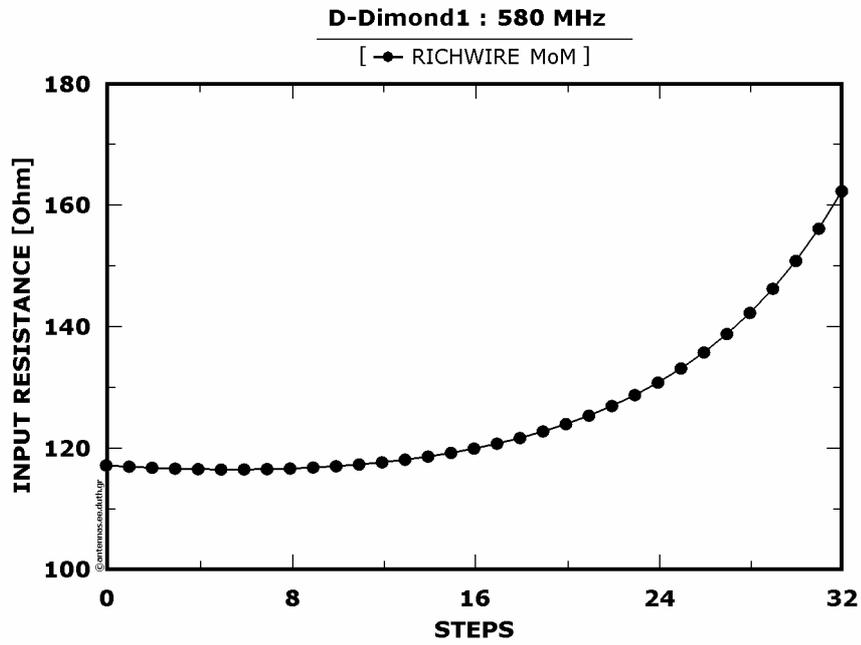

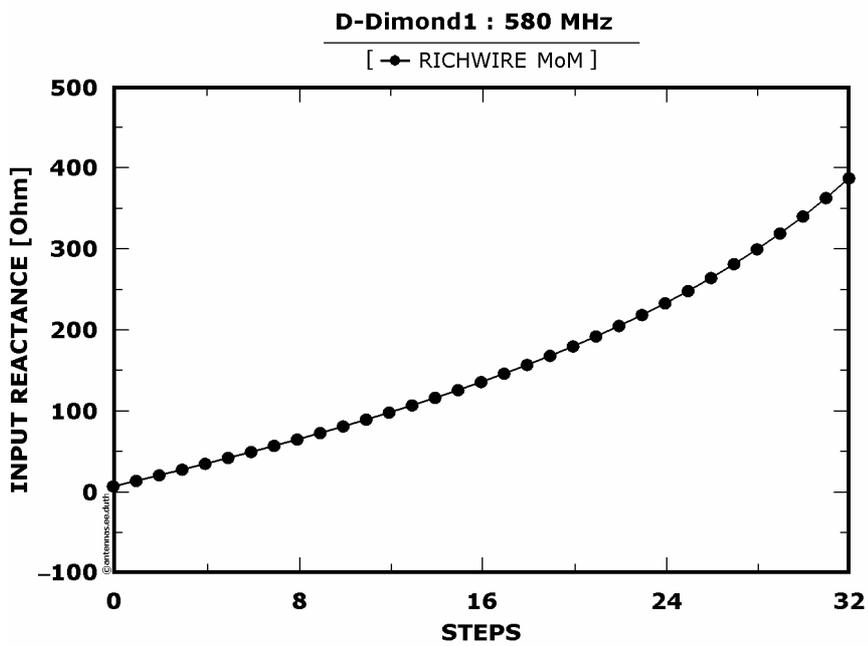

<u>Figure 5</u> : Real and imaginary part of the input impedance as a function of

the elongation steps

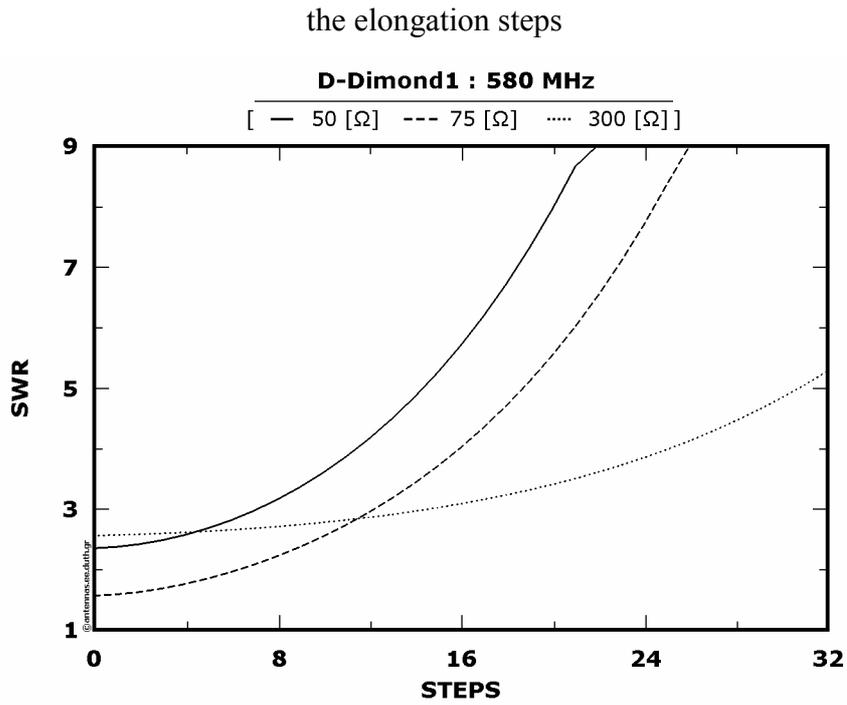

Figure 6 : SWRs as a function of the elongation steps

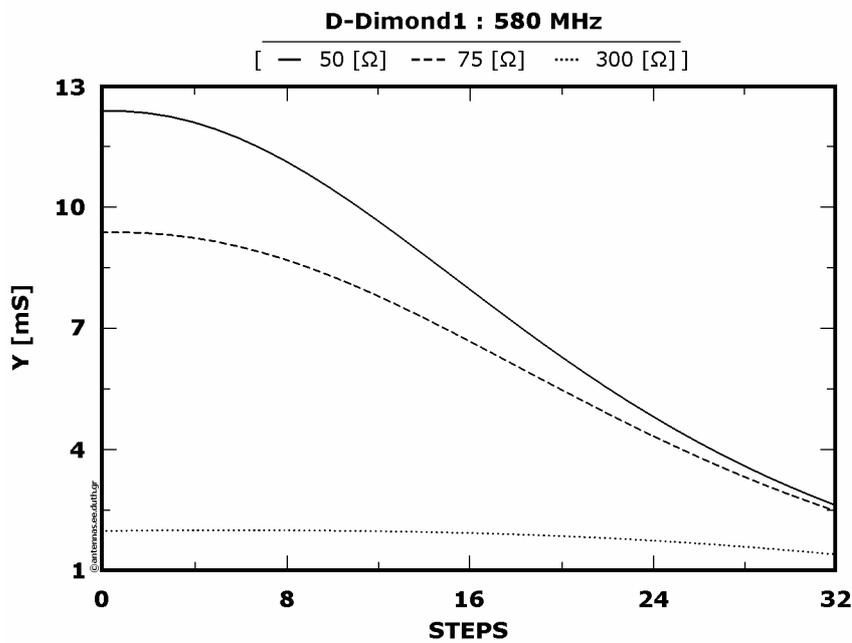

Figure 7 : Normalized radiation intensities as a function of the elongation steps

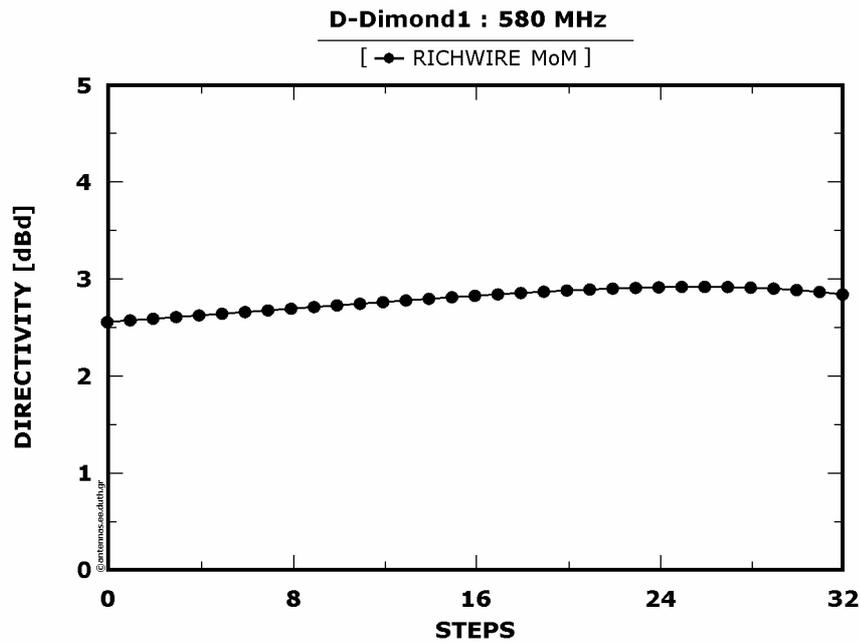

 : Directivity (with reference to dipole λ/2) as a function of the elongation steps

Figure 5.2.1.6 shows exponential increase of all the three SWR plots.

Figure 5.2.1.8 shows almost constant directivity as a function of the elongation steps.

Consequently, Ys are decreased almost exponential, as it results from Figure 5.2.1.7.

Summing all the above, the applied model is not judged satisfactory as it does not introduce any remarkable variations neither at directivity nor at SWRs. Therefore the next geometrical model will be applied without altering the original geometry of the antenna.

**Model of active dipole's elongation (3<sup>rd</sup> model)**

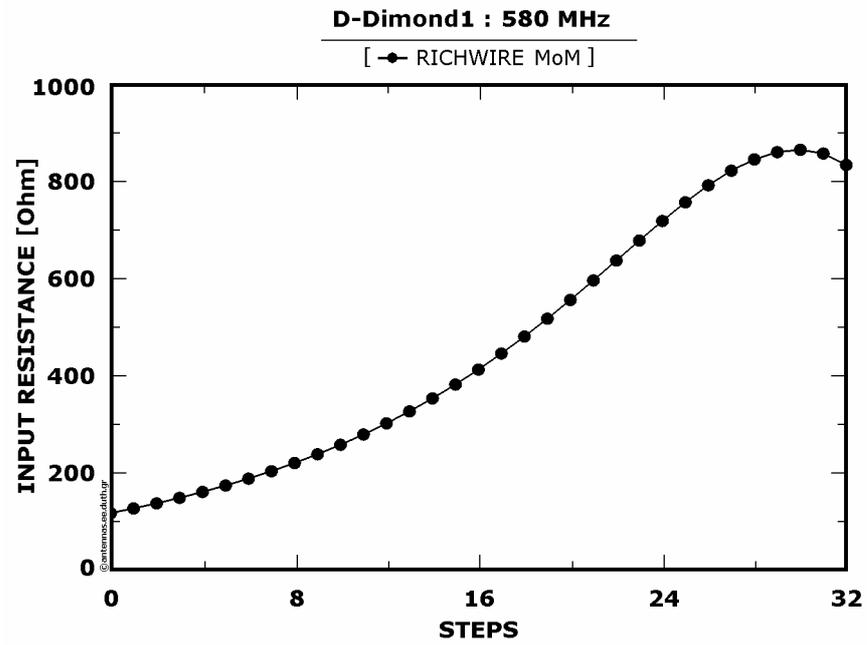

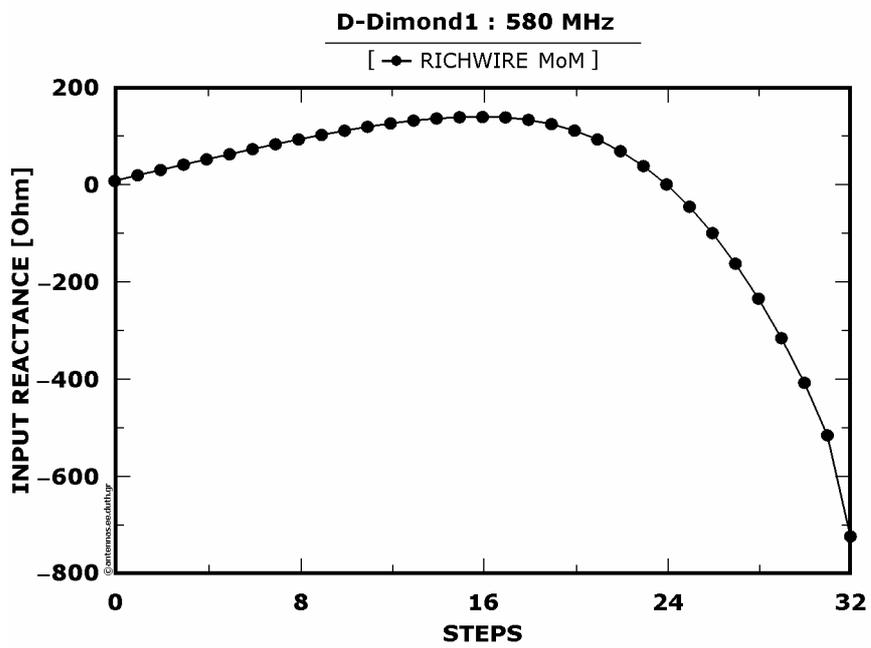

<u>Figure 9</u> : Real and imaginary part of the input impedance as a function of
the elongation steps

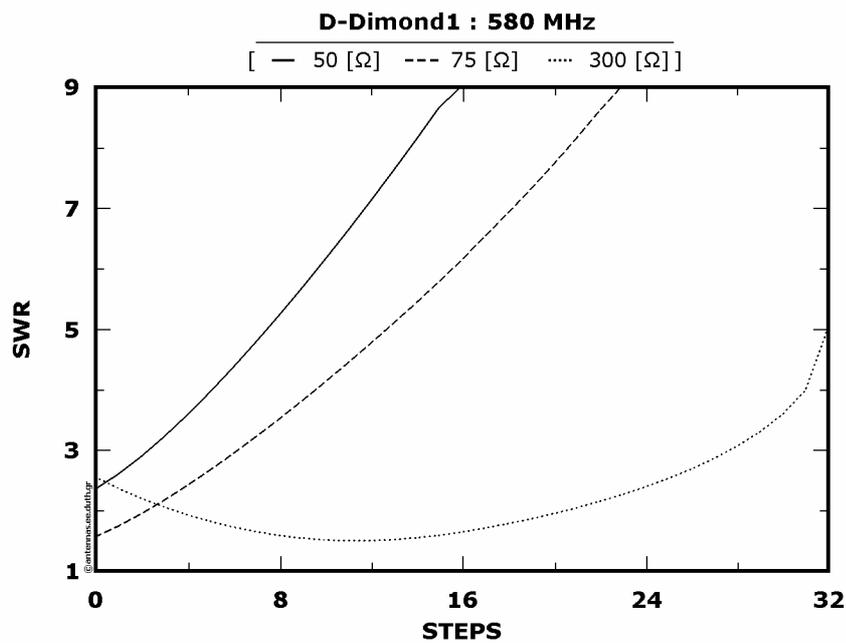

Figure 10 : SWRs as a function of the elongation steps

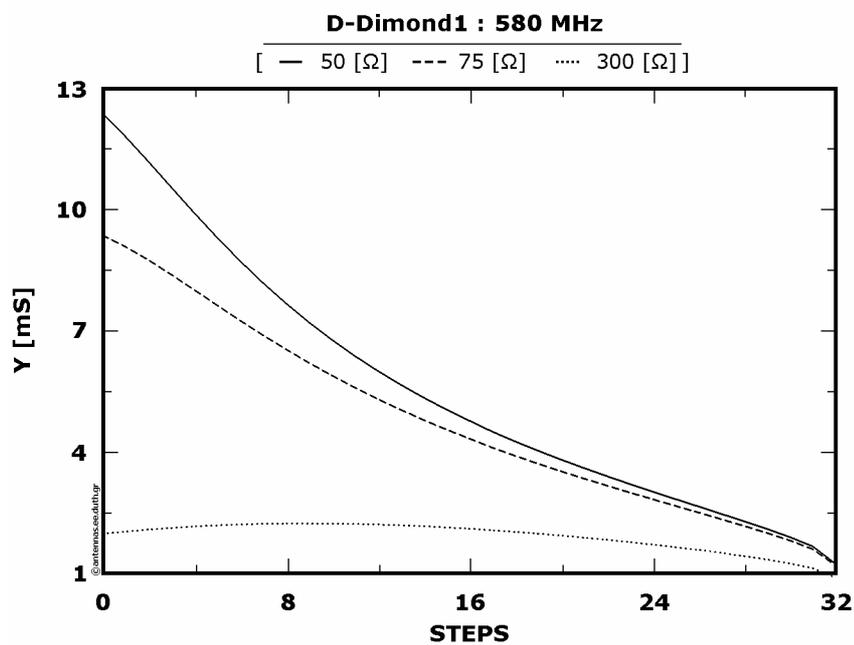

Figure 11 : Normalized radiation intensities as a function of the elongation steps

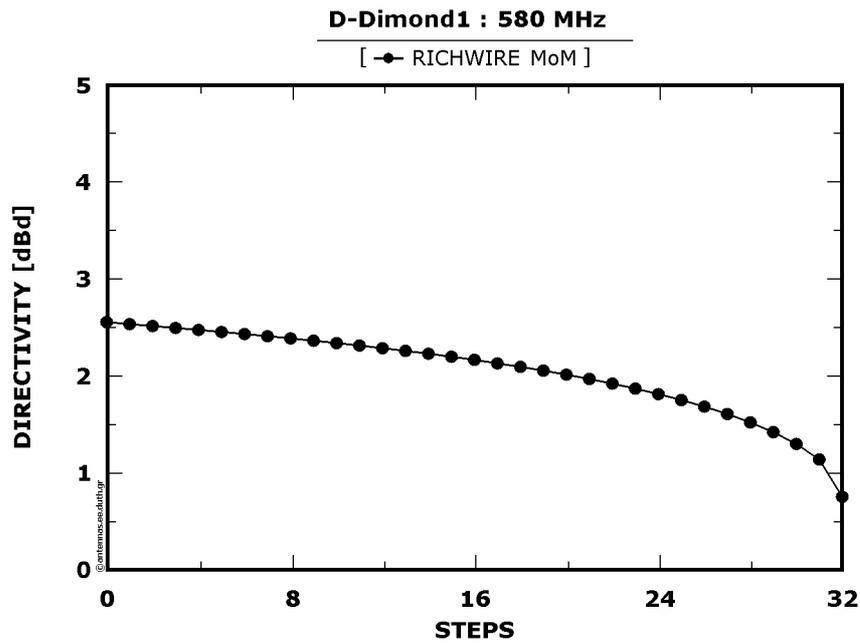

<u>Figure 12</u> : Directivity (with reference to dipole λ/2) as a function of the elongation steps

According to <u>Figure 5.2.1.10</u> there is an abrupt increase of SWRs with the exception of SWR(300) which diminishes roughly until the 12[th] elongation step.

The directivity decreases constantly as long as the elongation grows, as illustrated in <u>Figure 5.2.1.12</u>.

Thus, at <u>Figure 5.2.1.11</u> the decrease of both normalized radiation intensities Y(50) and Y(75) is displayed and also the small rising of Y(300) at the pre-mentioned range of elongation steps.

Judging from all the above after the application of the current model does not result satisfactory improvement for neither the directivity nor the SWRs. Consequently, the initial geometry of the antenna is maintained at 580 [MHz].

**5.2.2 : Frequency at 1238 [MHz] – Model of active dipole's displacement (1<sup>st</sup> model)**

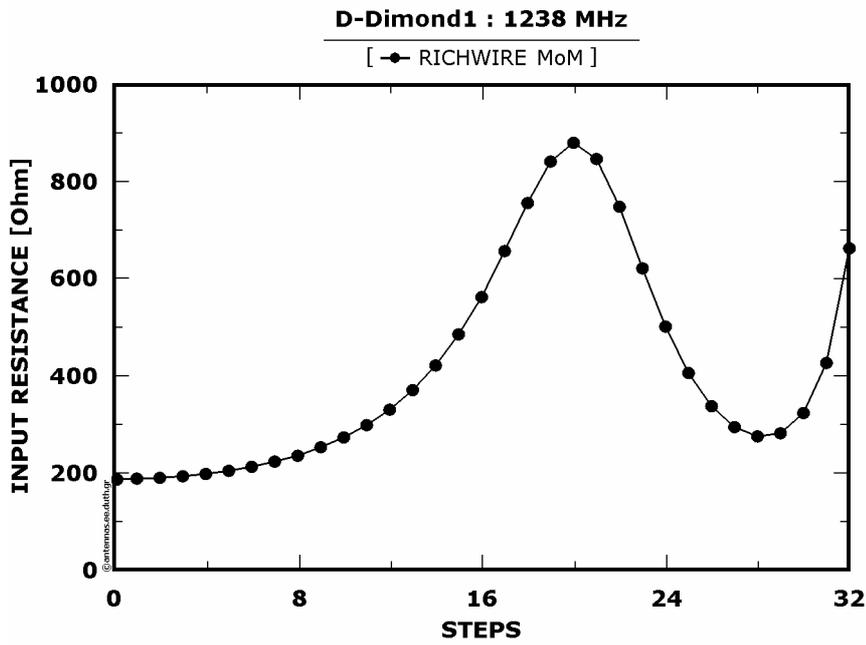

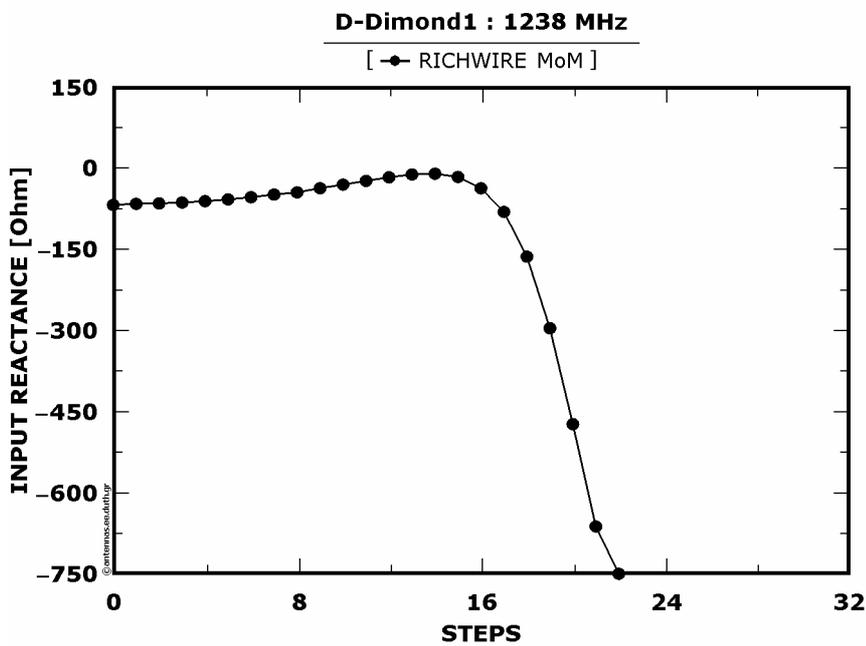

Figure 1 : Real and imaginary part of the input impedance as a function of
the displacement steps

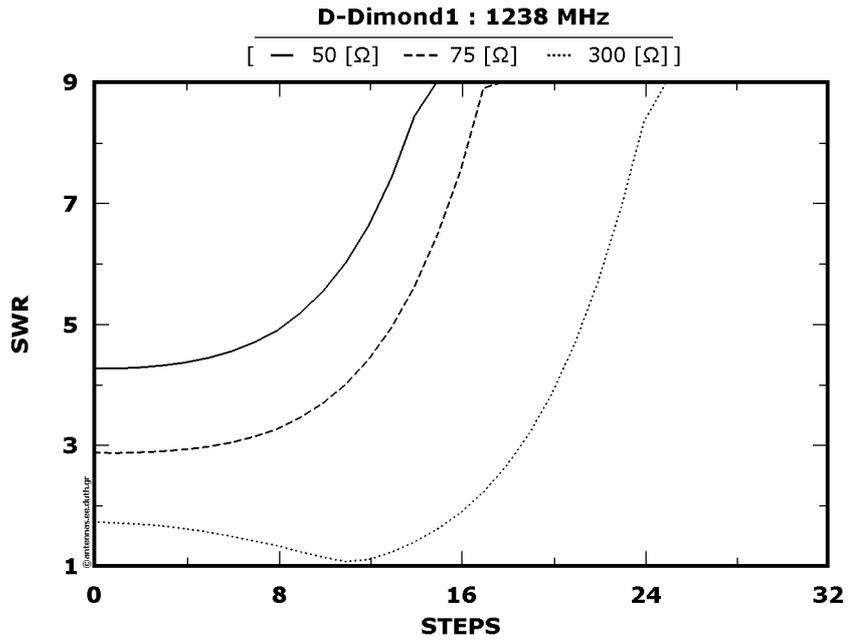

Figure 2 : SWRs as a function of the displacement steps

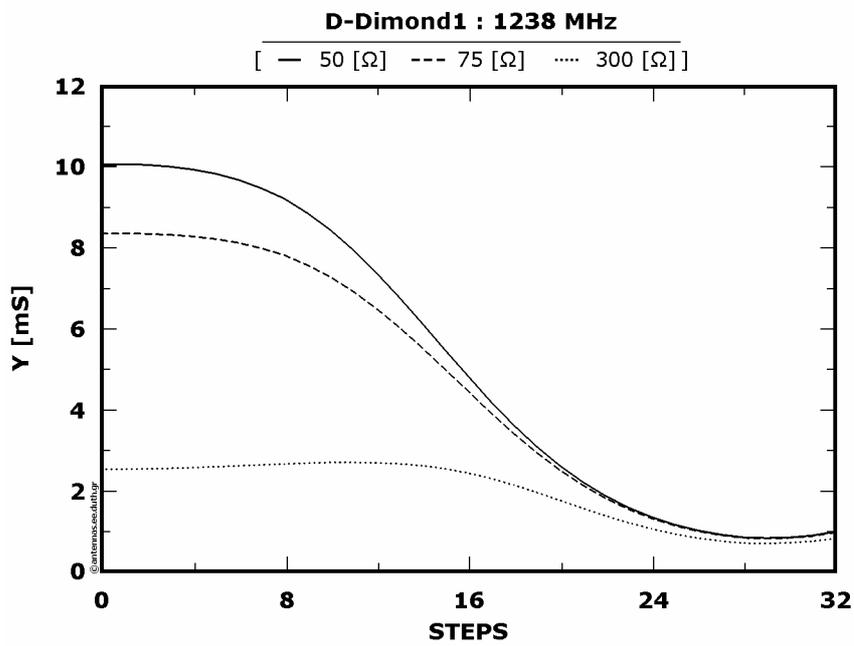

Figure 3 : Normalized radiation intensities as a function of the displacement steps

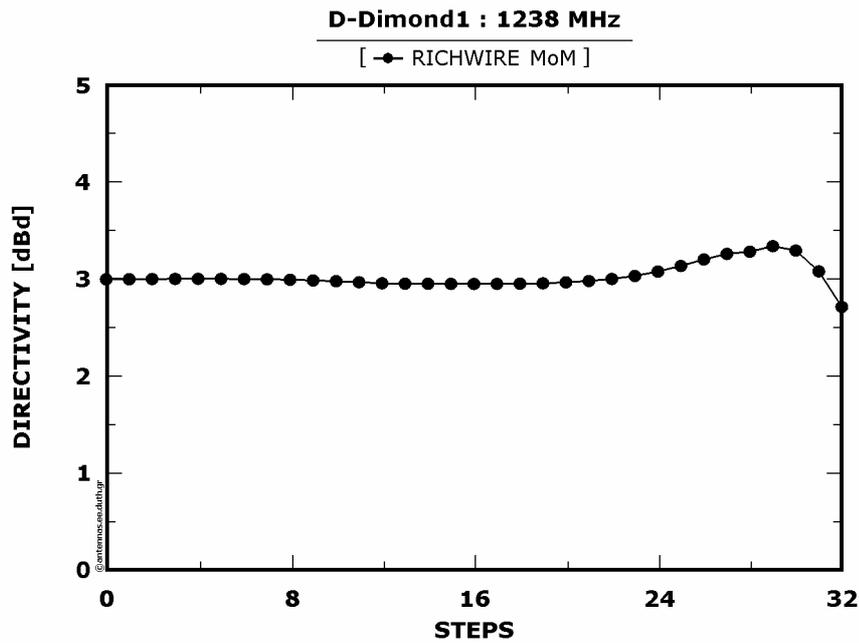

Figure 4 : Directivity (with reference to dipole λ/2) as a function of the displacement steps

According to Figure 5.2.2.2 abrupt increases at all the three SWR plots are observed exceeding the value of 9. The case at which SWR(300) is decreased taking almost the unitary value during the displacement steps between 4λ/64 and 6λ/64 is quite interesting.

From Figure 5.2.2.4 the invariable behavior of directivity is observed around the value of 3 [dBd].

Figure 5.2.2.3 displays the anticipated decrease of both Y(50) and Y(75) while a – small magnitude- maximization of Y(300) is occurred.

Deductively, the current model is not satisfying therefore any modification of the initial antenna's geometry isn't justifiable, even if SWR(300)decreases at some range of displacement steps.

**Model of perpendicular monopoles elongation (2<sup>nd</sup> model)**

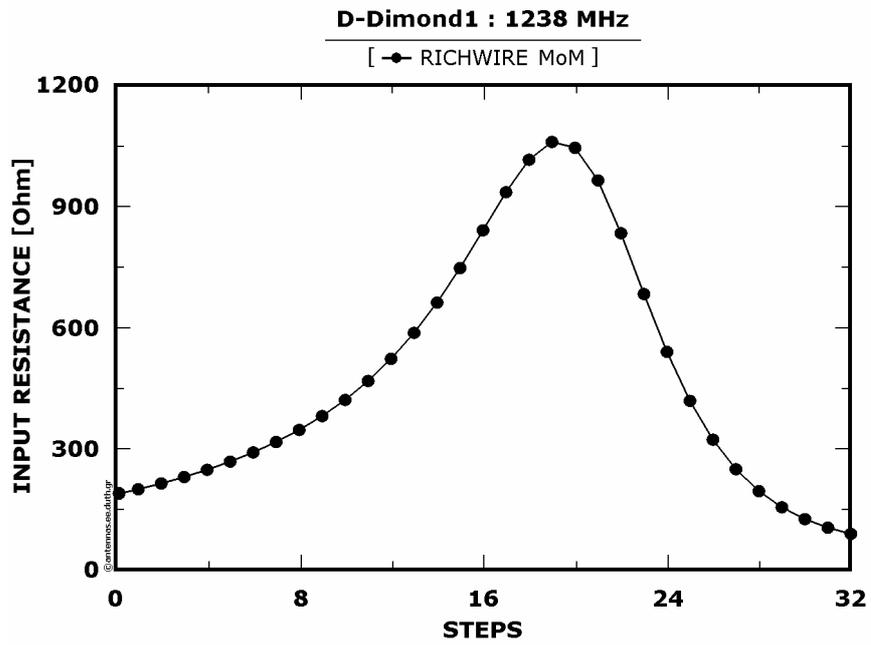

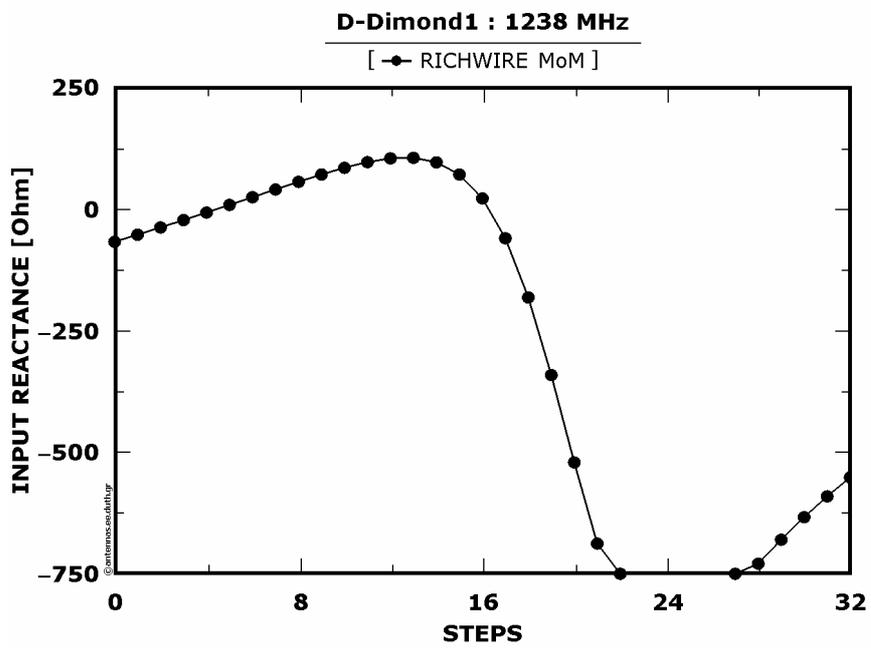

<u>Figure 5</u> : Real and imaginary part of the input impedance as a function of
the elongation steps

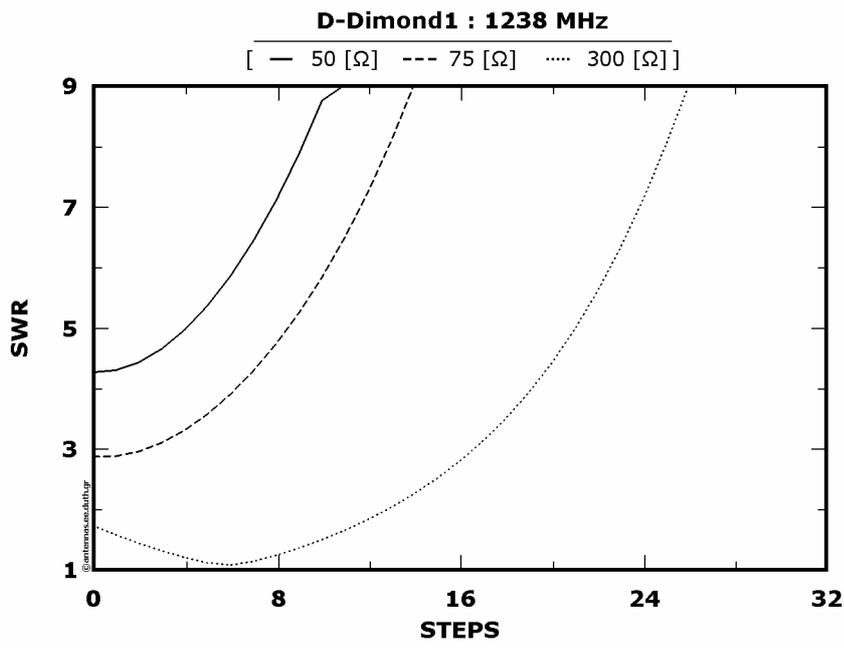

Figure 6 : SWRs as a function of the elongation steps

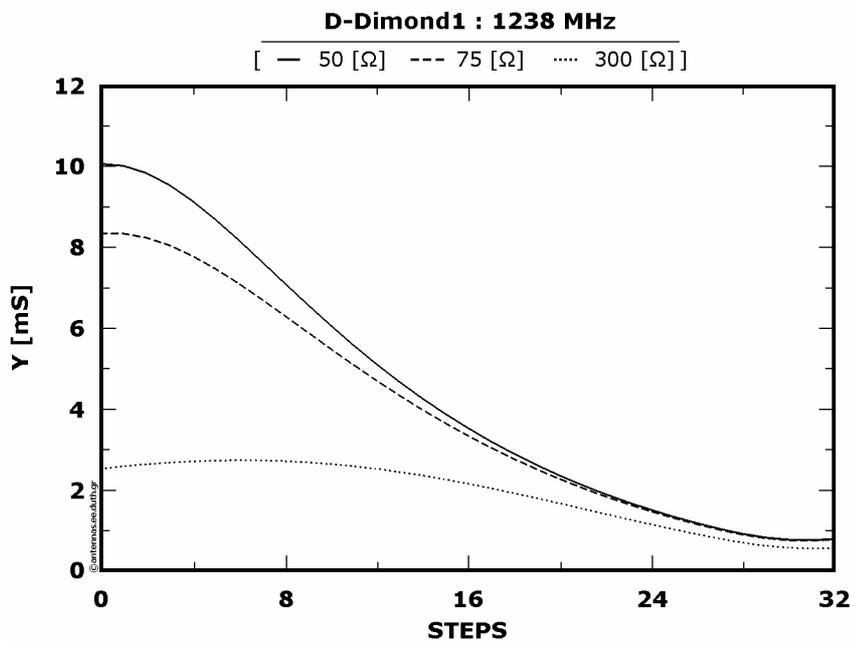

Figure 7 : Normalized radiation intensities as a function of
the elongation steps

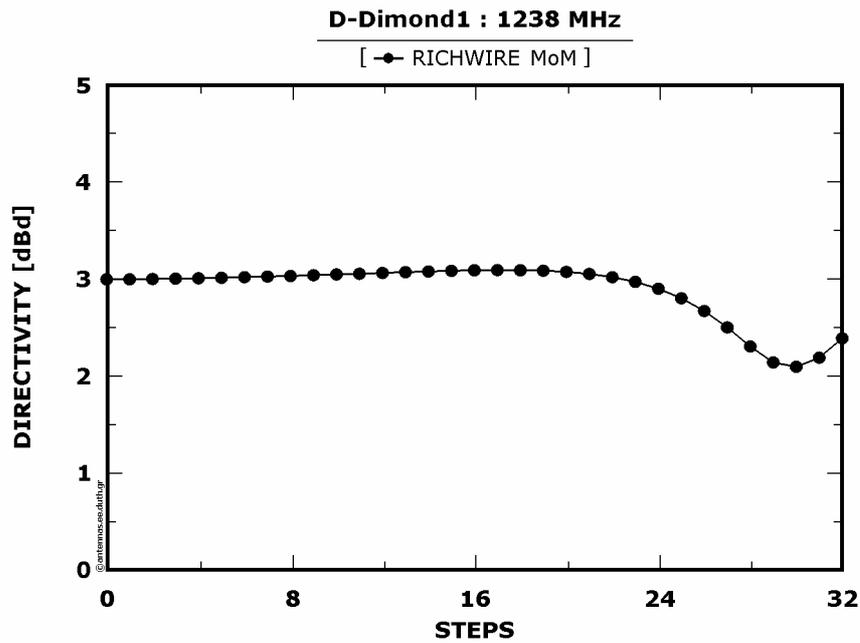

Figure 8 : Directivity (with reference to dipole λ/2) as a function of the elongation steps

From Figure 5.2.2.6 a -high inclined- exponential variation of SWR plots is observed. Exception constitutes again SWR(300) which produces a unitary minimal somewhere near the 6[th] step of elongation.

Directivity's variation does not present any interesting behavior, as it appears from Figure 5.2.2.8.

Consequently, from Figure 5.2.2.7 almost the same behavior of the normalized radiation intensities as in the previous model at 1238 [MHz] is observed while the maximum value that Y(300) takes isn't as much high as should be to justify any geometrical modification of the antenna regarding to the current model.

**Model of active dipole's elongation (3rd model)**

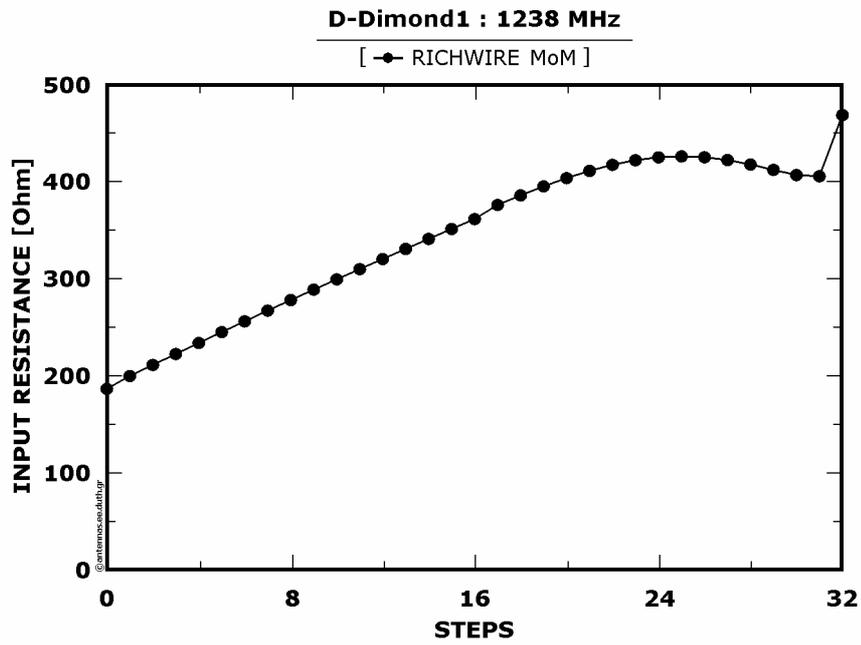

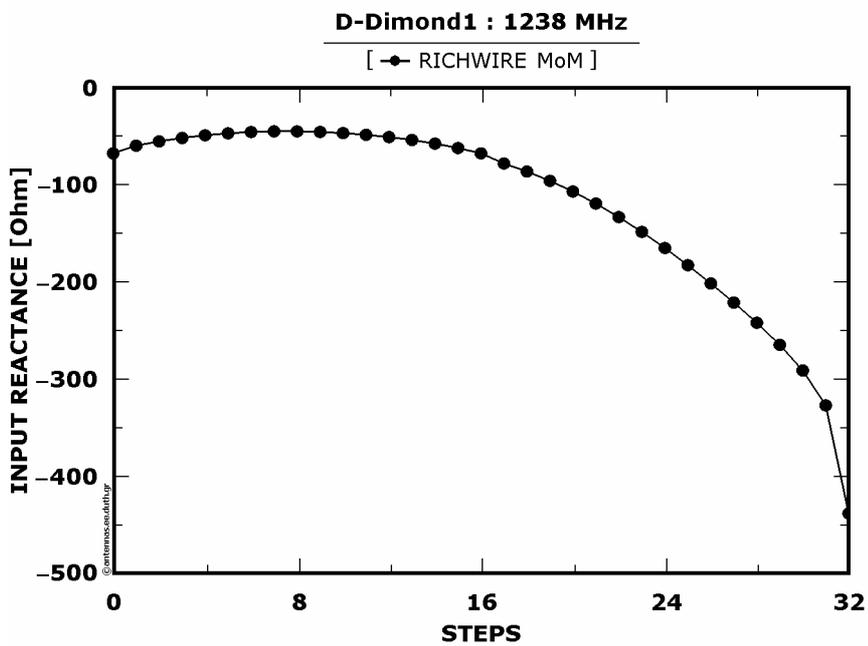

<u>Figure 9</u> : Real and imaginary part of the input impedance as a function of

the elongation steps

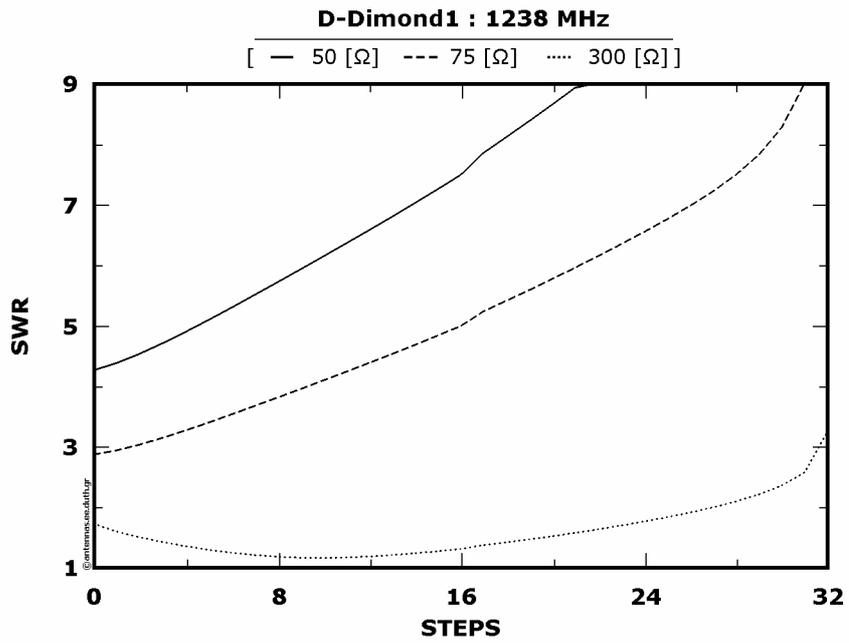

Figure 10 : SWRs as a function of the elongation steps

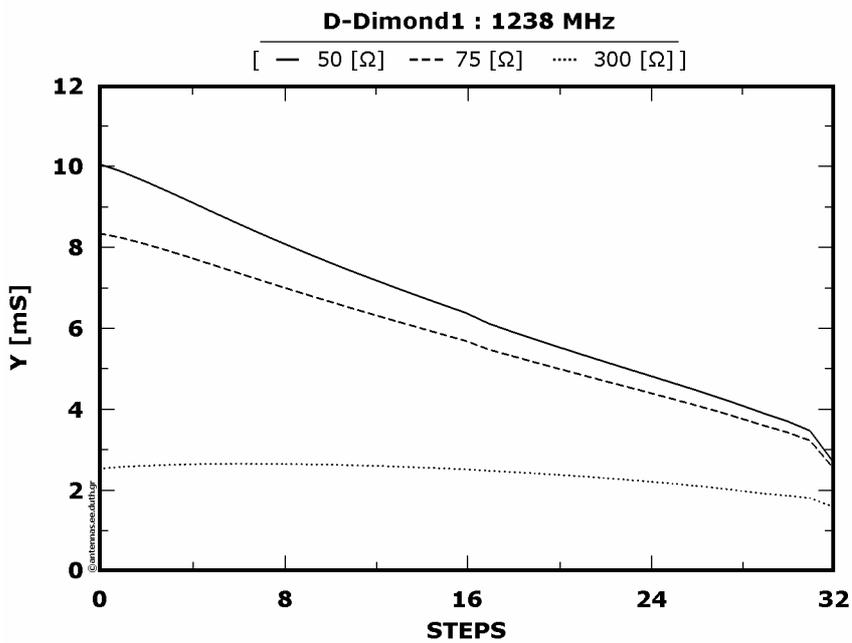

Figure 11 Normalized radiation intensities as a function of the elongation steps

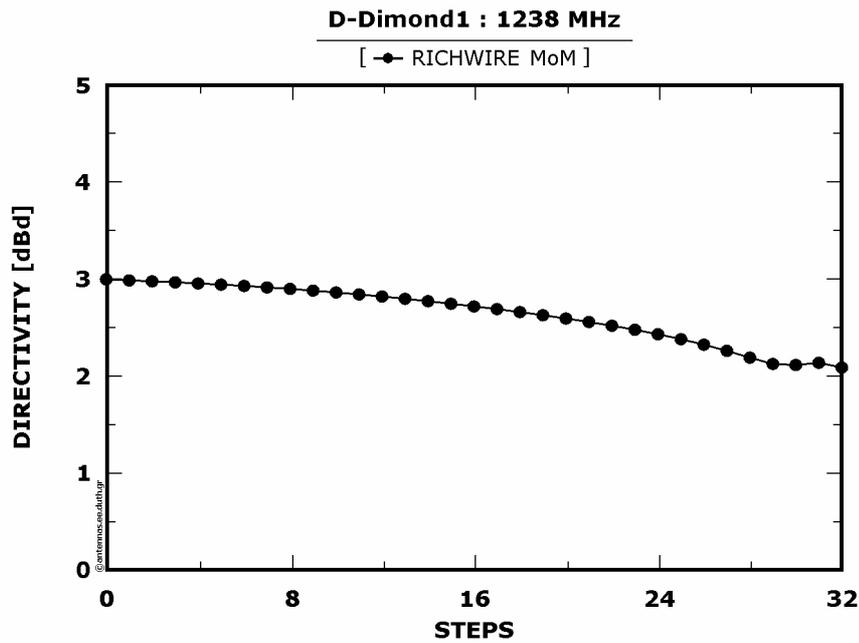

<u>Figure 12</u> : Directivity (with reference to dipole λ/2) as a function of the elongation steps

A similar behavior of SWR plots is observed from <u>Figure 5.2.2.10</u> as before, that is to say all the three ratios increase while a minimal for SWR(300) is observed around the 10<sup>th</sup> step of elongation.

Directivity is decreasing constantly with a low inclination, as illustrated in <u>Figure 5.2.2.12</u>.

Consequently, from <u>Figure 5.2.2.11</u> results a lowering of Ys except from Y(300) which is almost constant producing, as before, a slight maximum.

Deductively, none of each of the three preceding models is judged as satisfactory at 1238 [MHz] therefore the initial geometry of the antenna remains untouched.

## 5.2.3 : Frequency at 1867 [MHz] - Model of active dipole's displacement (1st model)

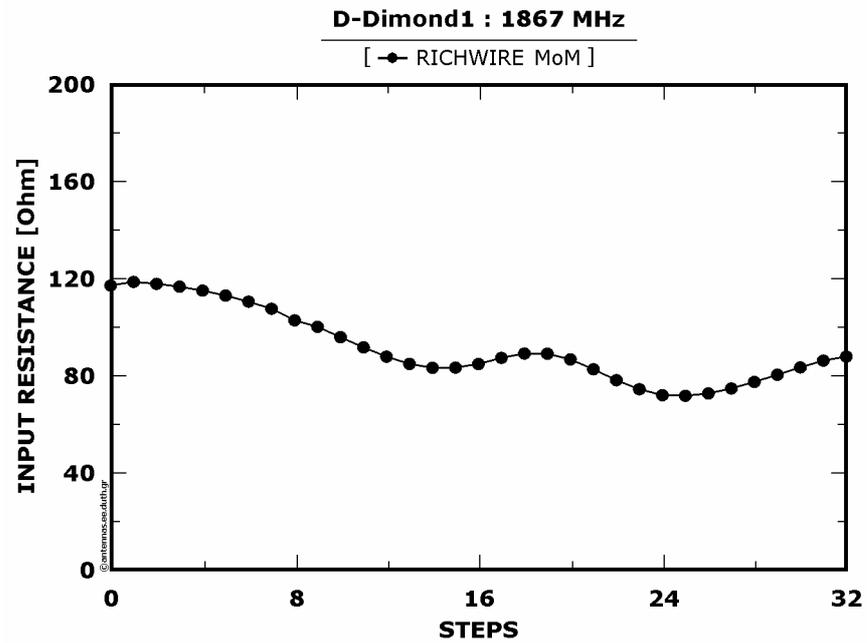

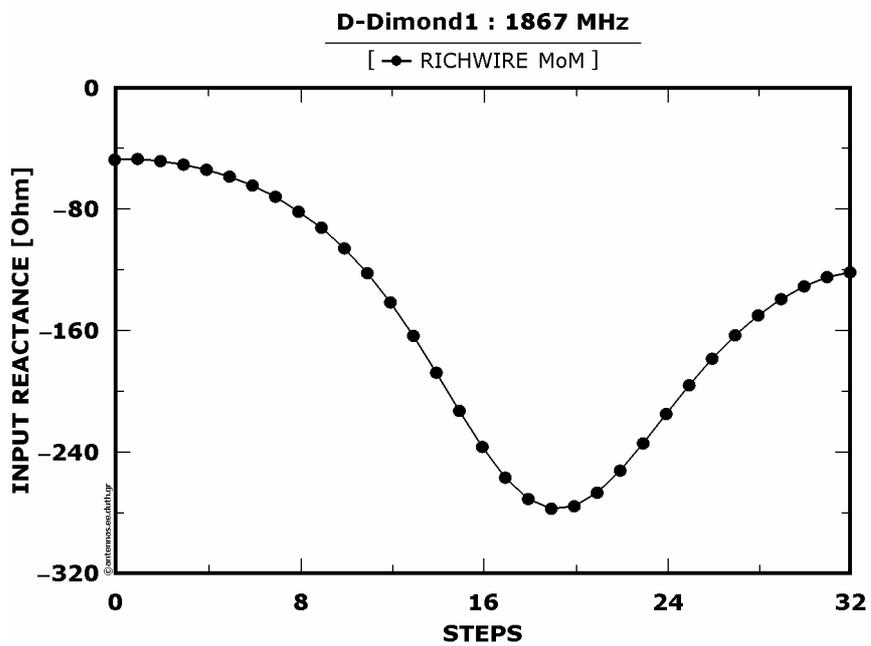

<u>Figure 1</u> : Real and imaginary part of the input impedance as a function of the displacement steps

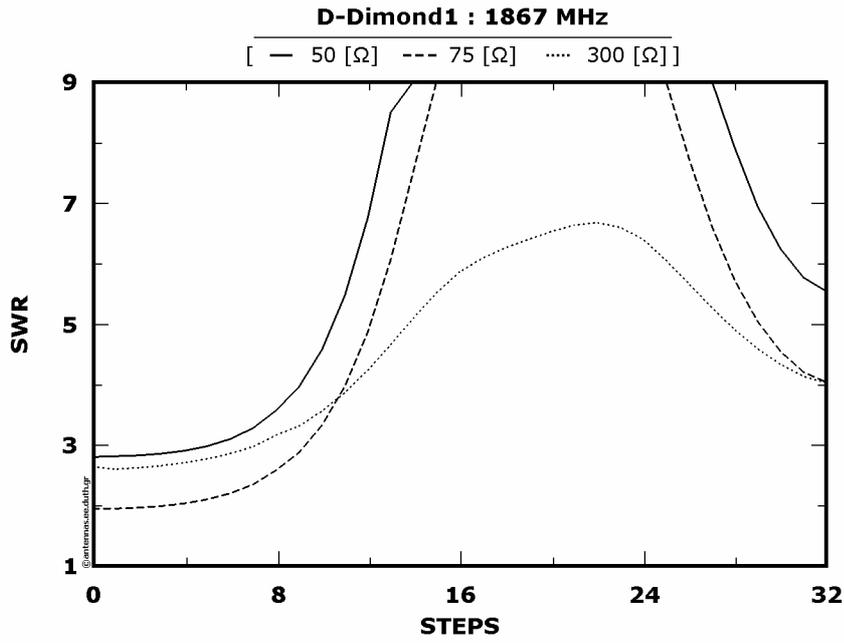

Figure 2 : SWRs as a function of the displacement steps

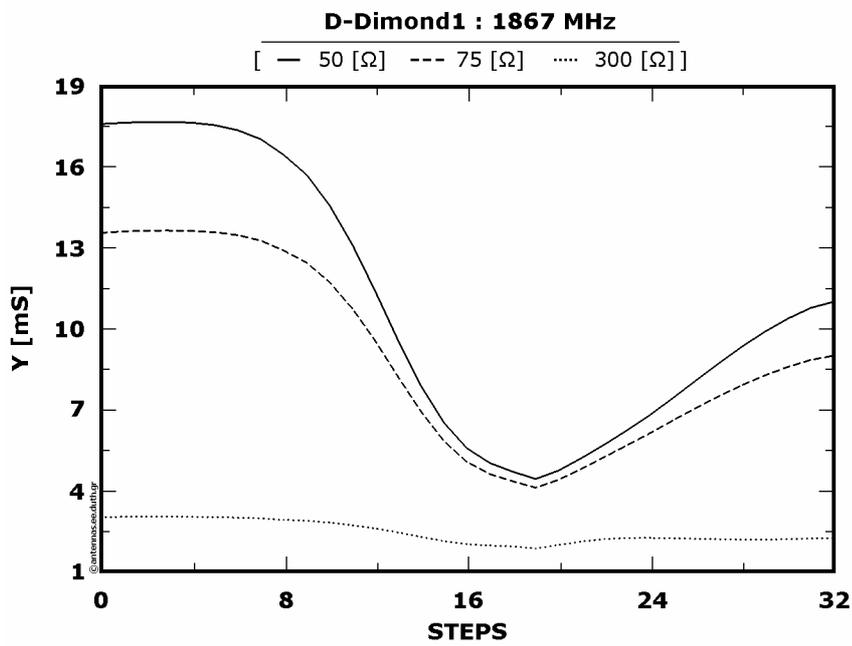

Figure 3 : Normalized radiation intensities as a function of the displacement steps

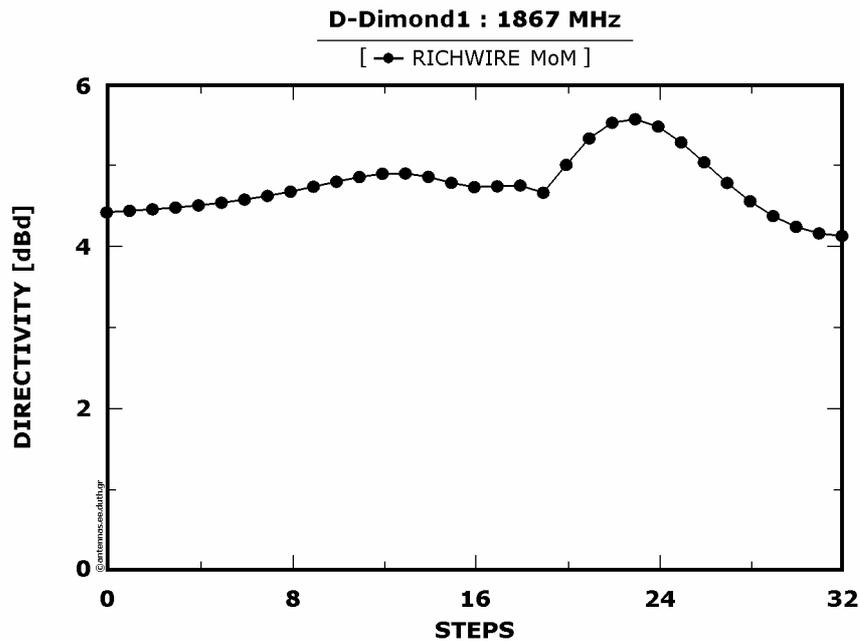

<u>Figure 4</u> : Directivity (with reference to dipole λ/2) as a function of the displacement steps

According to <u>Figure 5.2.3.2</u> a maximization of all SWR plots is observed and more specifically SWR(50) and SWR(75) exceed the value of 9 at a range of displacement steps.

Quite interesting appears to be the directivity's variation, as it is shown at <u>Figure 5.2.3.4</u>, where its value is getting over 5 [dBd] at range between 22[nd] and 26[th] step of displacement.

Summing all above and taking into consideration the <u>Figure 5.2.3.3</u> of normalized intensities radiations (Ys) it is deduced that there isn't any satisfactory combination of high directivity and low SWRs at the whole range of displacement. The improvement of directivity can be considered significant, however at the same time SWRs' values are disappointingly low.

Consequently, the current model cannot be applicable and therefore the initial geometry of the antenna remains intact.

**Model of perpendicular monopoles elongation (2nd model)**

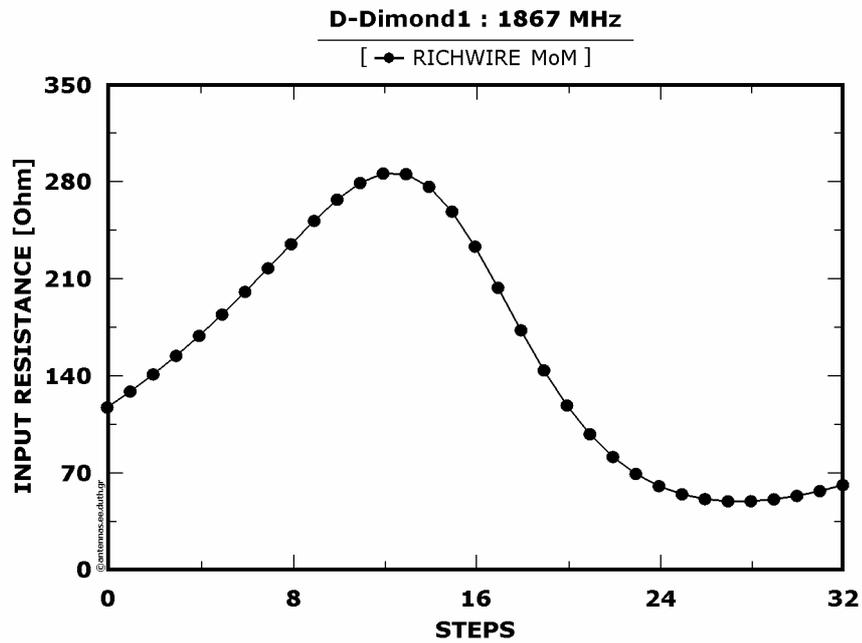

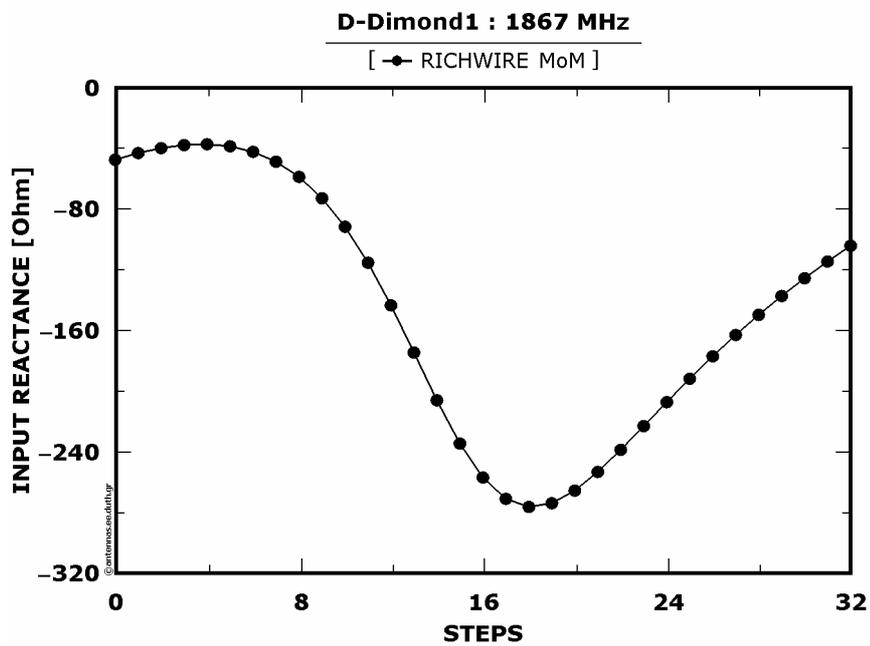

<u>Figure 5</u> : Real and imaginary part of the input impedance as a function of

the elongation steps

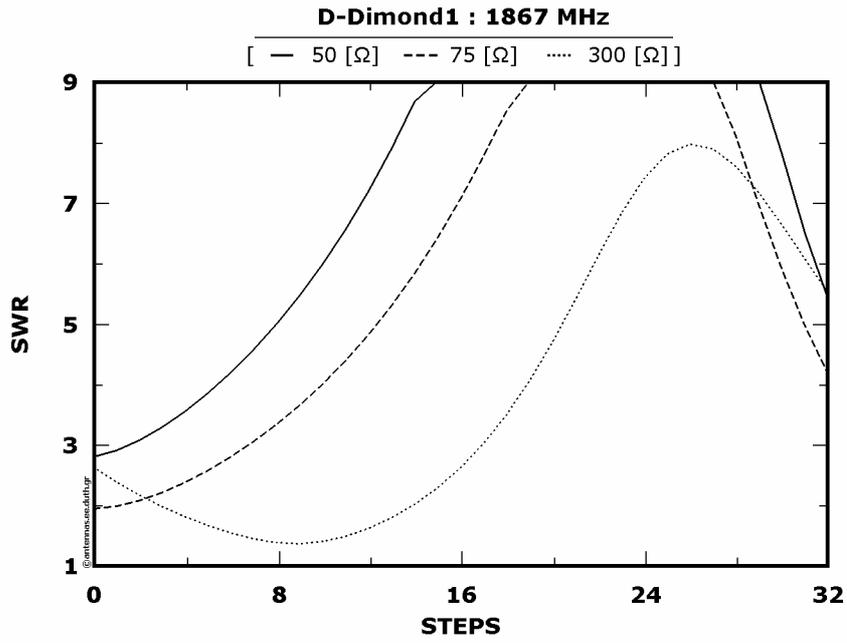

Figure 6 : SWRs as a function of the elongation steps

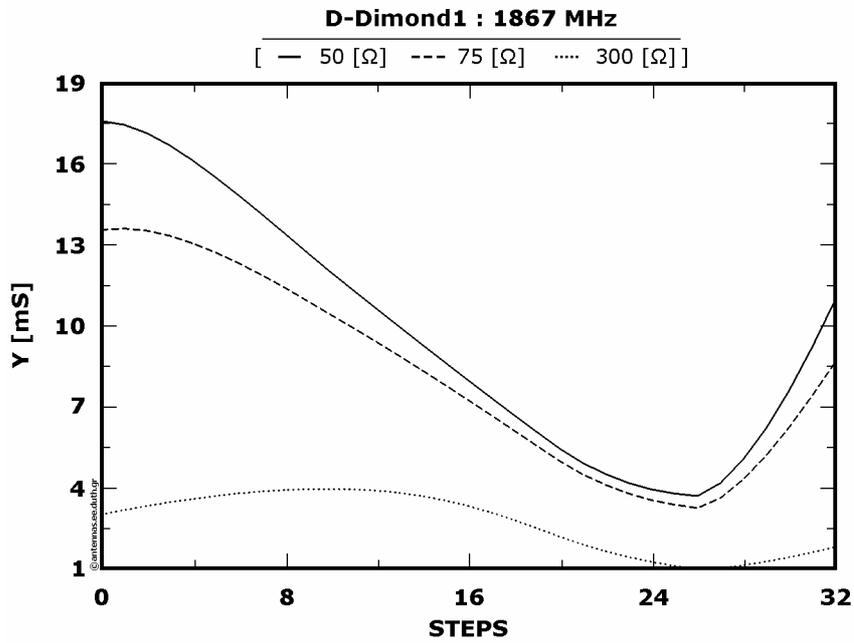

Figure 7 : Normalized radiation intensities as a function of the elongation steps

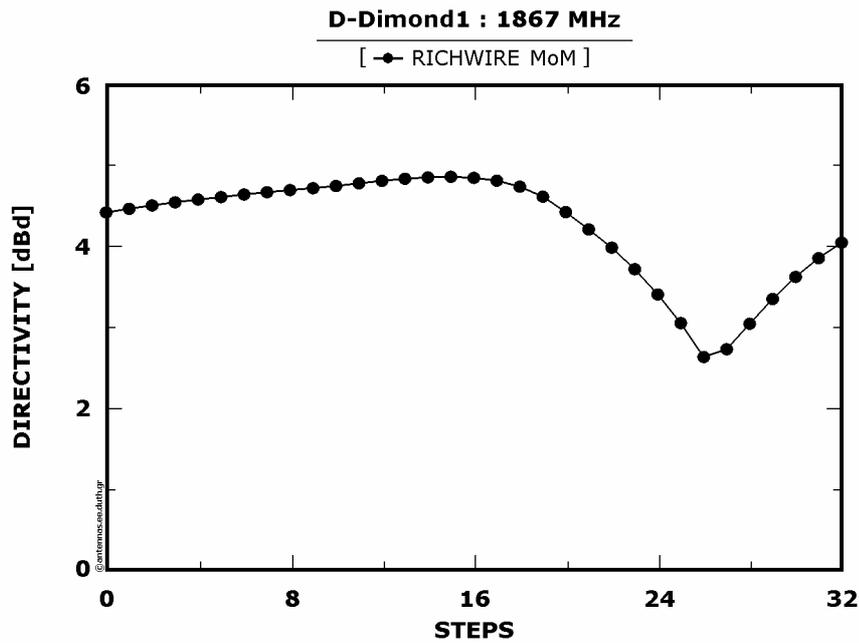

 : Directivity (with reference to dipole λ/2) as a function of the elongation steps

Figure 5.2.3.6 depicts the increase of all SWR plots. More specifically, SWR(50) and SWR(75) increase above the value of 9 for a range of elongation steps while SWR(300) is minimized at the 8<sup>th</sup> elongation step.

Directivity starts to increase with a small inclination approaching the value of 5 [dBd] and afterwards is minimized, as shown in Figure 5.2.3.8.

The normalized radiation intensities are decreasing in general excluding Y(300) which produces a small minimal, as shown in Figure 5.2.3.7.

Conclusively, the slight improvement which is observed with reference to 300 [Ω] resistance doesn't justify the application of the current model hence the geometrical modification of the antenna, as it isn't follow by equivalent improvements with reference to the rest of the resistances namely 50 and 75 [Ω].

**Model of active dipole's elongation (3<sup>rd</sup> model)**

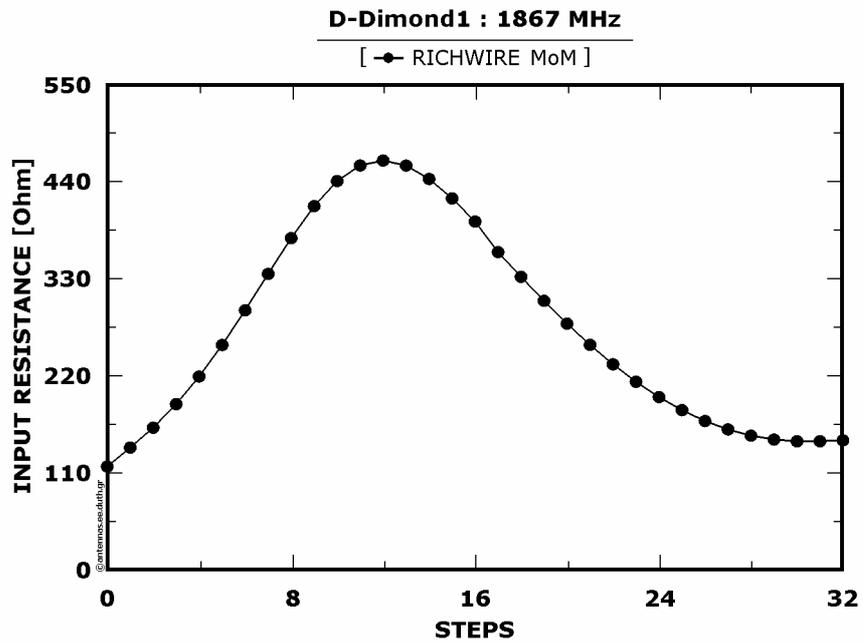

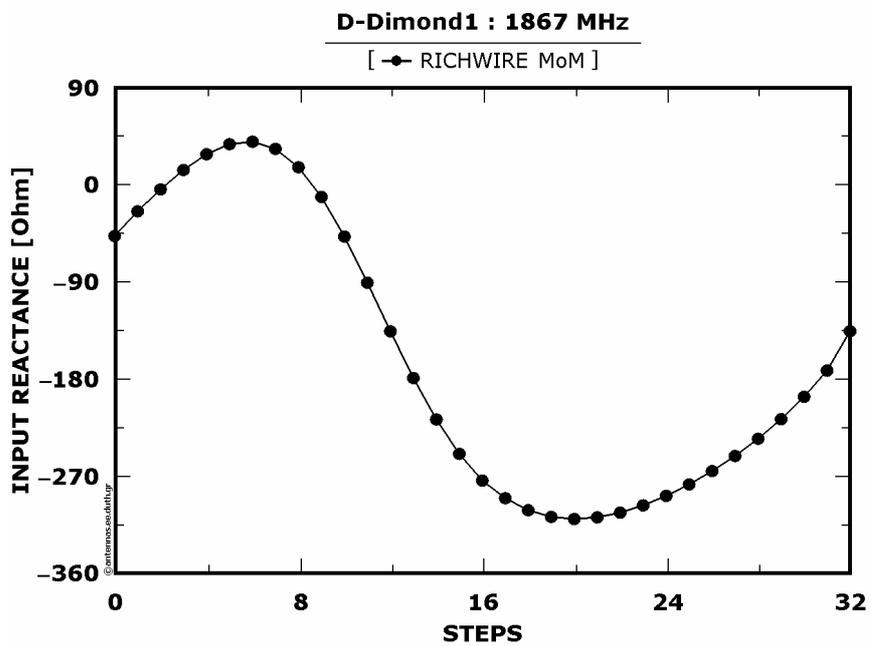

<u>Figure 10</u> : Real and imaginary part of the input impedance as a function of
the elongation steps

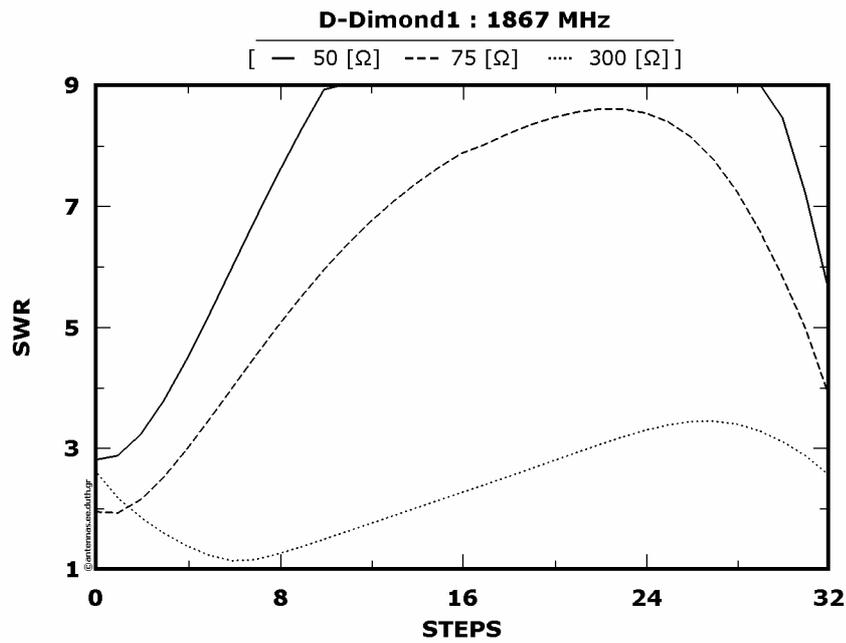

Figure 11 : SWRs as a function of the elongation steps

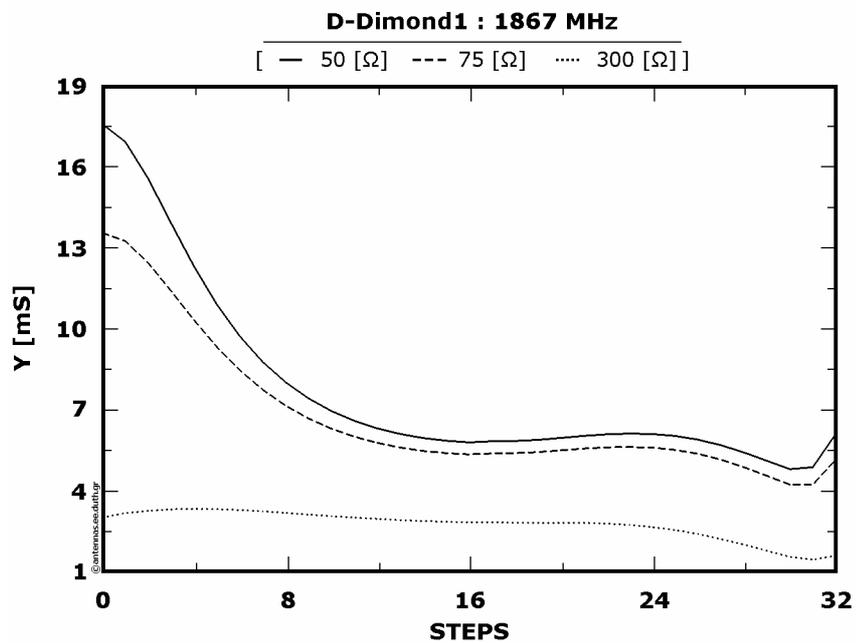

Figure 12 : Normalized radiation intensities as a function of the elongation steps

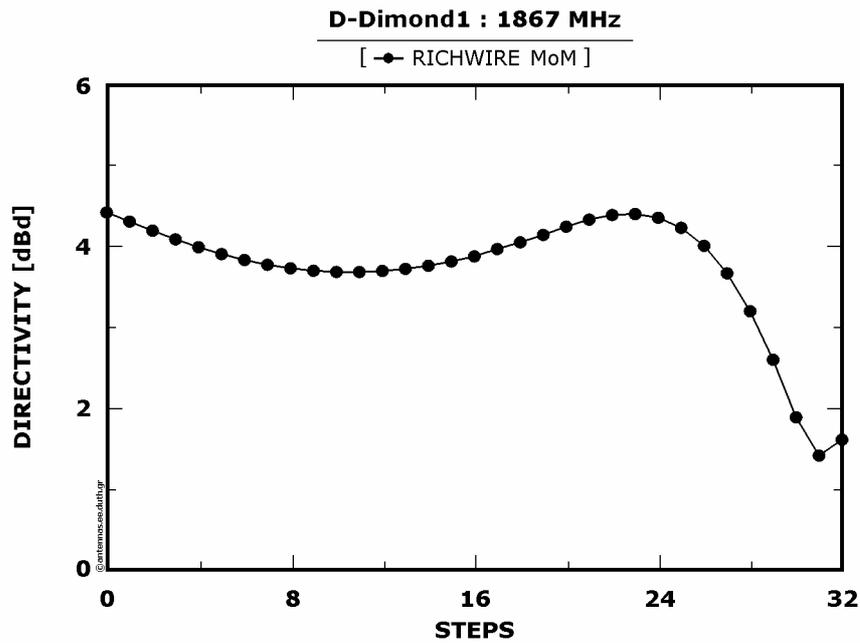

Figure 13 : Directivity (with reference to dipole λ/2) as a function of the elongation steps

According to Figure 5.2.3.10 all SWR plots generally increase and in particular SWR(50) is exceeding the value of 9 for a great range of elongation steps while SWR(300) produces a significant, almost unitary, minimal at the 6[th] elongation step.

The directivity's variation doesn't present any remarkable behavior, as shown in Figure 5.2.3.12.

Hence, in Figure 5.2.3.11 the normalized intensities of radiation are displayed where they generally decrease rapidly except Y(300) which remains almost invariable at a great range of elongation steps.

Deductively, so as the current model so as the other two pre-examined models are not being judged as satisfactory at the frequency of 1867 [MHz], as far as the improvement of the antenna's characteristics is concerned.

**5.2.4 : Frequency at 2330 [MHz] – Model of active dipole's displacement (1st model)**

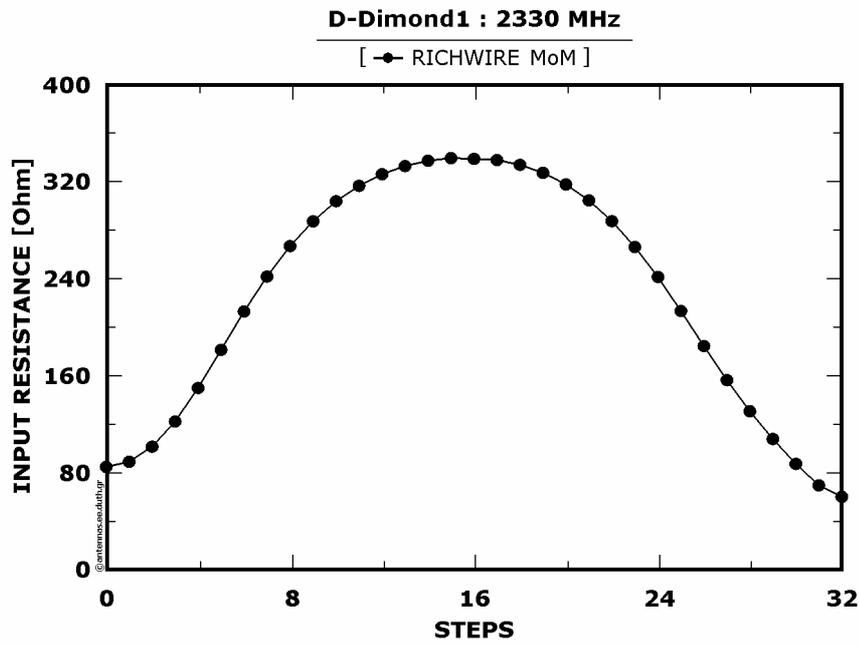

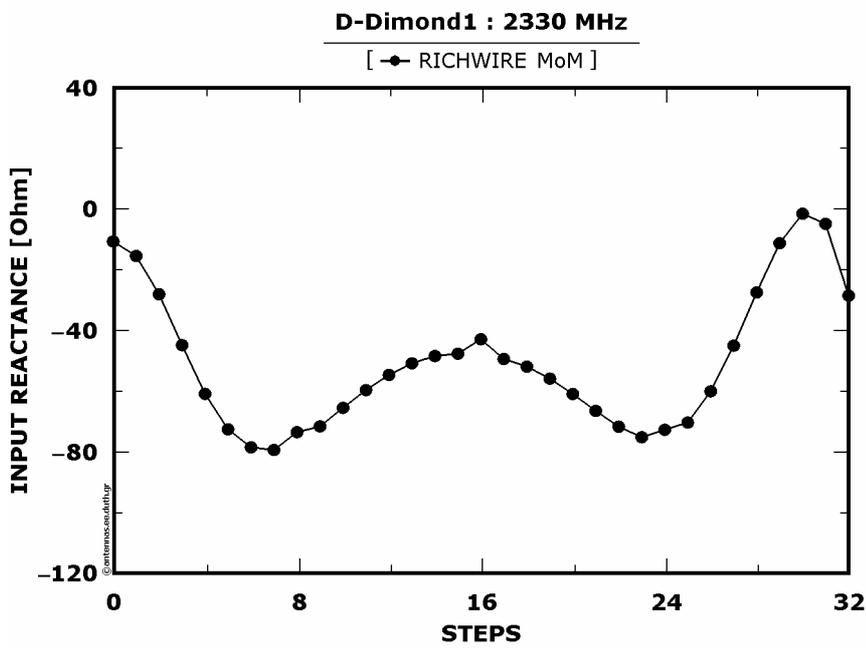

<u>Figure 1</u> : Real and imaginary part of the input impedance as a function of

the displacement steps

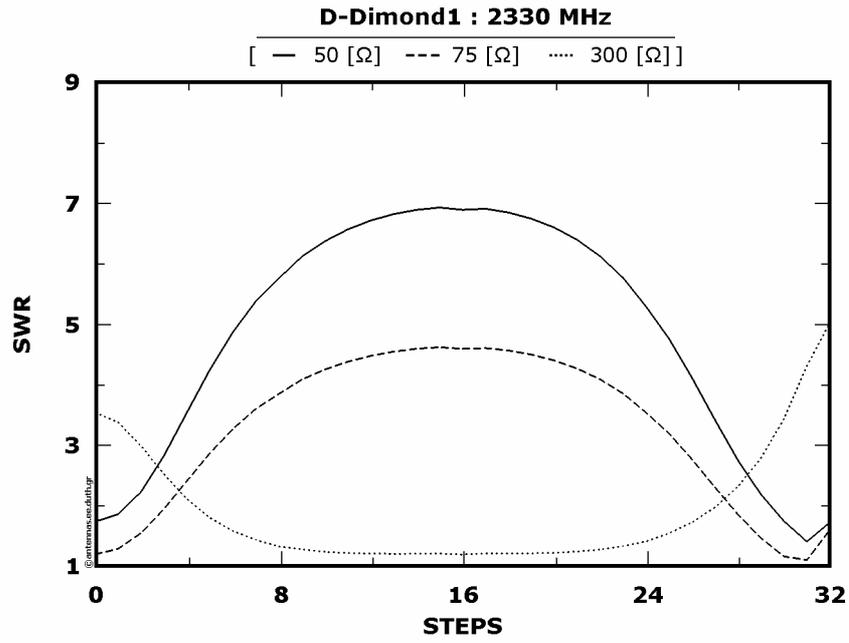

Figure 2 : SWRs as a function of the displacement steps

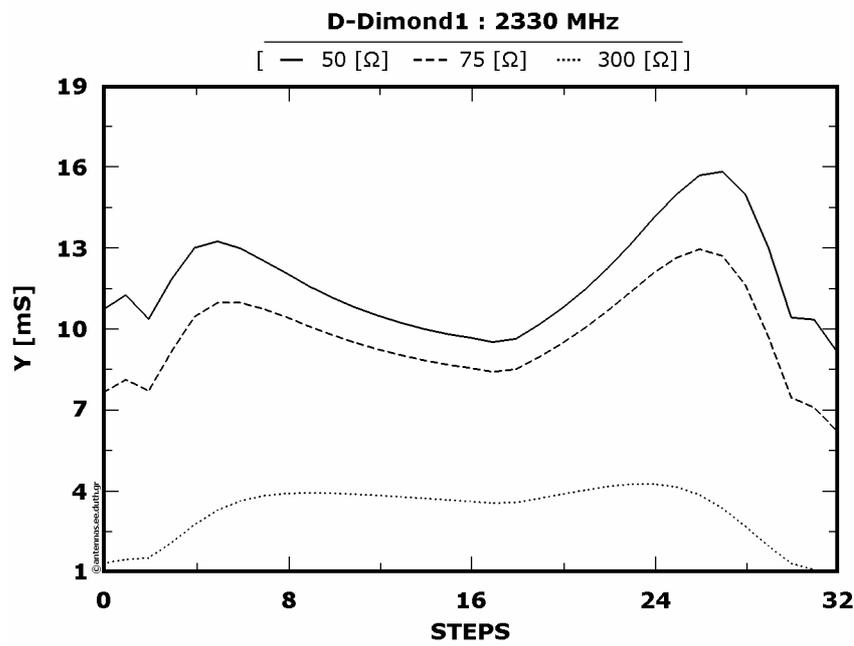

Figure 3 : Normalized radiation intensities as a function of the displacement steps

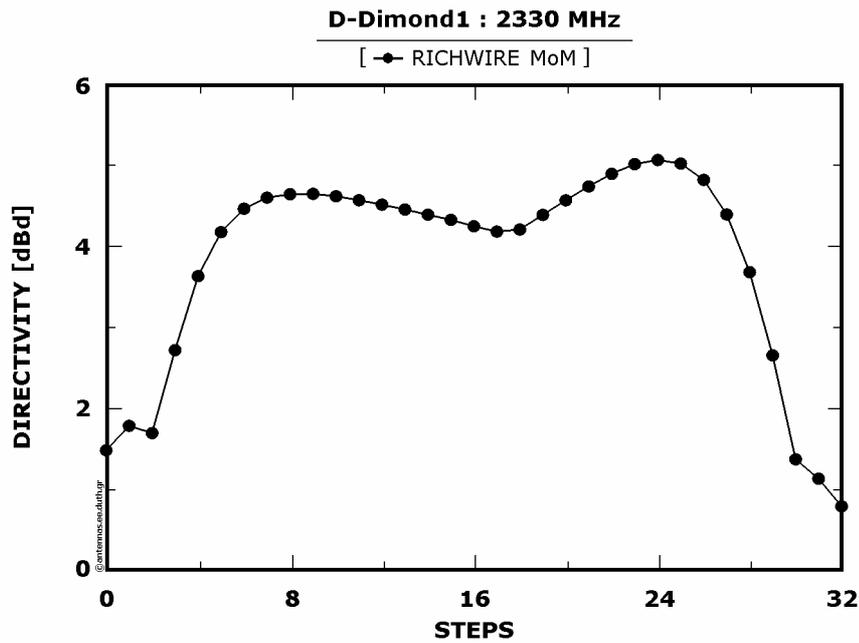

Figure 4 : Directivity (with reference to dipole λ/2) as a function of the displacement steps

According to Figure 5.2.4.2 SWR(300) decreases rapidly and its value remains unitary for a great range of displacements. Both SWR(50) and SWR(75) whereas initially increase they produce afterwards a considerable minimal about the 30[th] displacement step, that is to say where the antenna's tuning is happening, as it was depicted from Figure 5.2.4.1.

The directivity's variation appears to be equally interesting, as shown in Figure 5.2.4.4, whose value is over doubled and becomes maximum close to 5 [dBd] nearby the 24[th] displacement step.

From Figure 5.2.4.3 maximizations of the normalized radiation intensities are being observed. The maximum values of both Y(50) and Y(75) emerge at almost adjacent displacement steps. Consequently, the antenna's geometry is modified according to the 27[th] displacement step of the current model and the resultant geometry is maintained for the rest of the improvement process at 2330 [MHz].

**Model of perpendicular monopoles elongation (2$^{nd}$ model)**

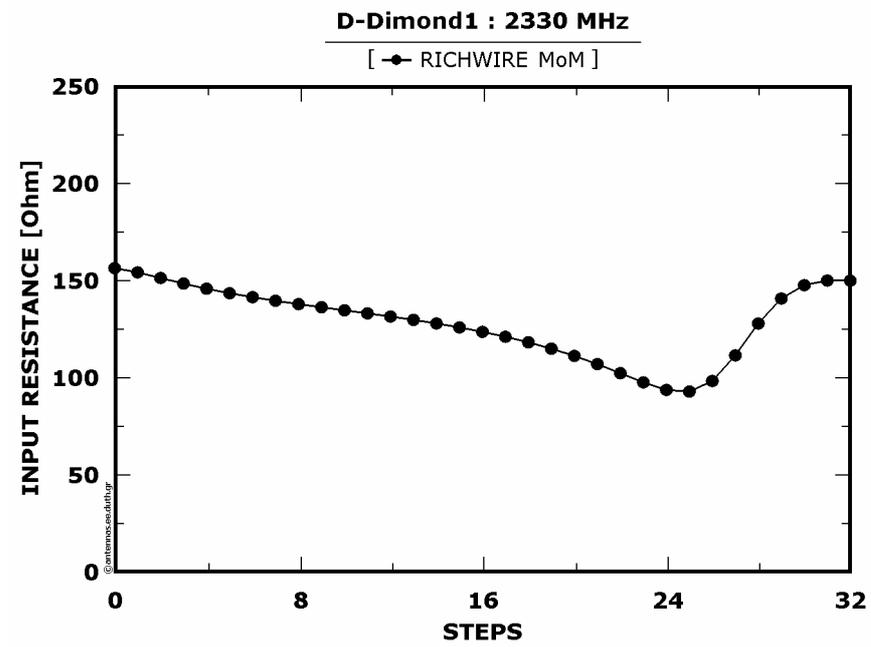

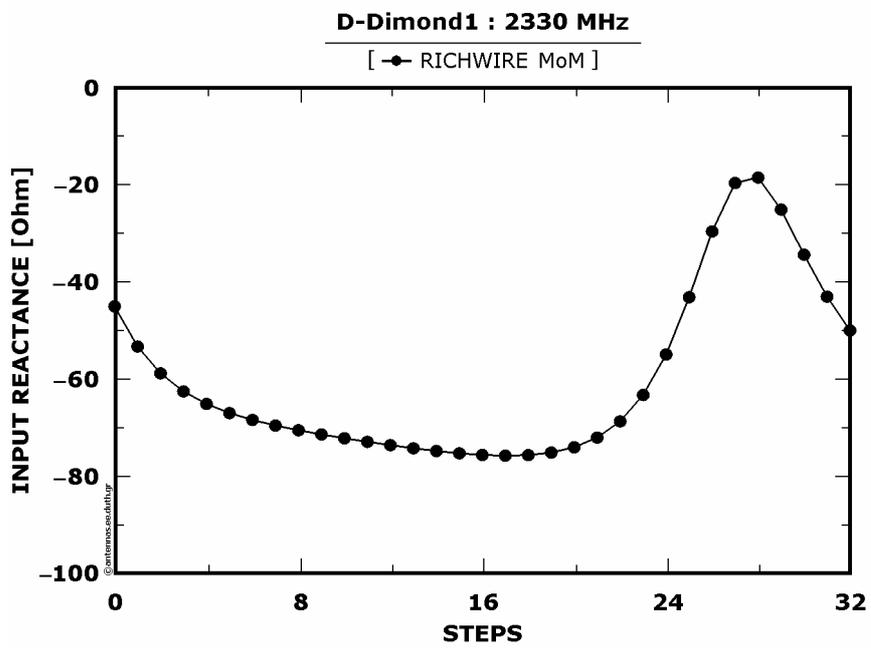

Figure 5 : Real and imaginary part of the input impedance as a function of the elongation steps

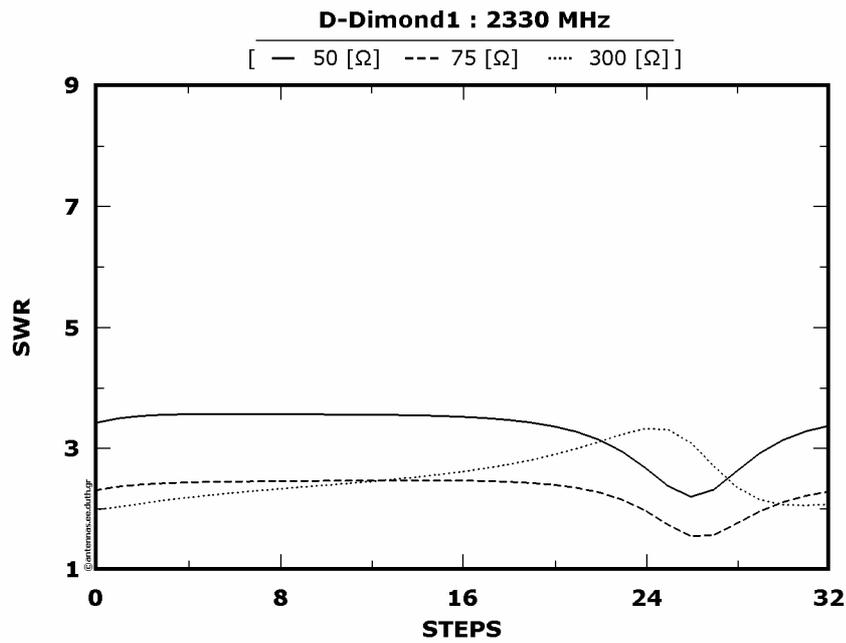

Figure 6 : SWRs as a function of the elongation steps

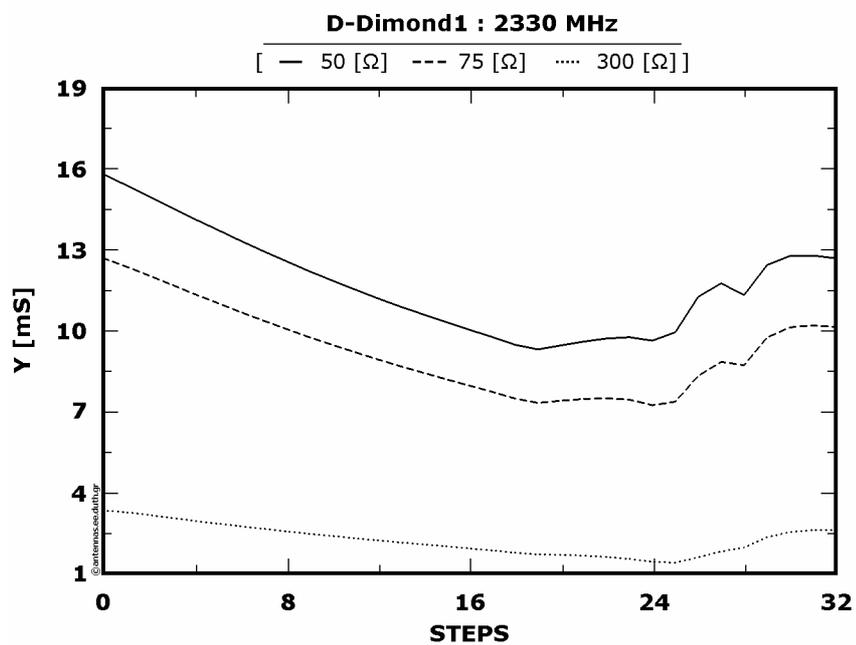

Figure 7 : Normalized radiation intensities as a function of the elongation steps

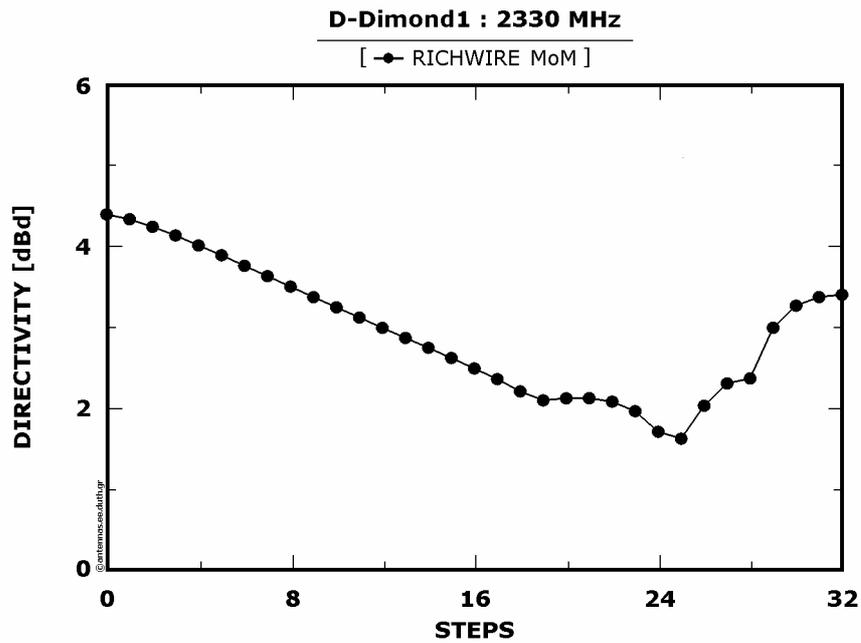

Figure 8 : Directivity (with reference to dipole λ/2) as a function of the elongation steps

SWRs' values fluctuate low as shown in Figure 5.2.4.6. More specifically, both SWR(50) and SWR(75) take a simultaneous "dive" at the 26th elongation step.

Directivity diminishes at the whole range of elongation steps, as illustrated in Figure 5.2.4.8.

According to the above variations the normalized radiation intensities don't produce any significant maximum at the whole range of elongation steps, as depicted in Figure 5.2.4.7.

Summing all above, directivity's quite low values do not assent to the modification of the antenna's geometry unless if a low SWR design is attempted regardless of the directivity.

**Model of active dipole's elongation (3<sup>rd</sup> model)**

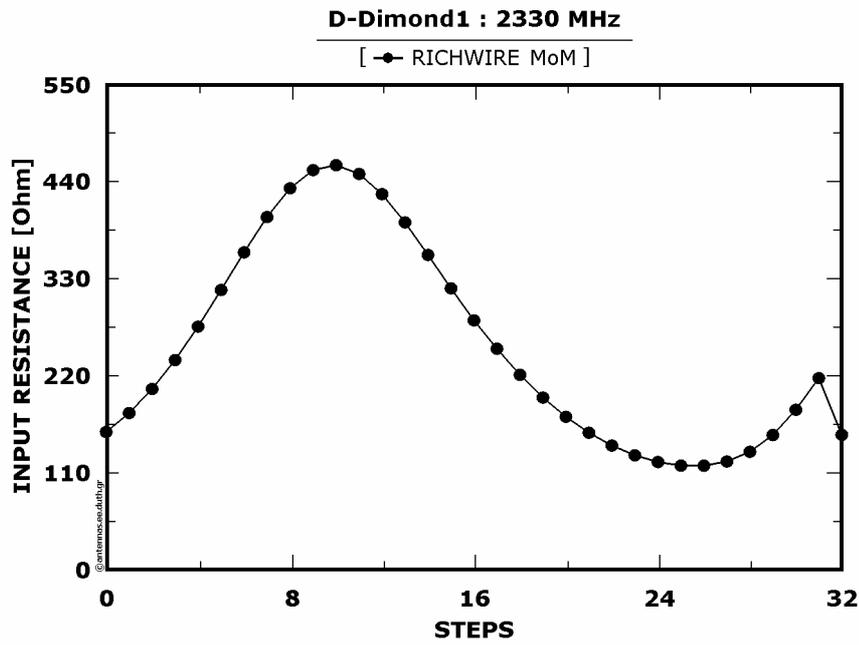

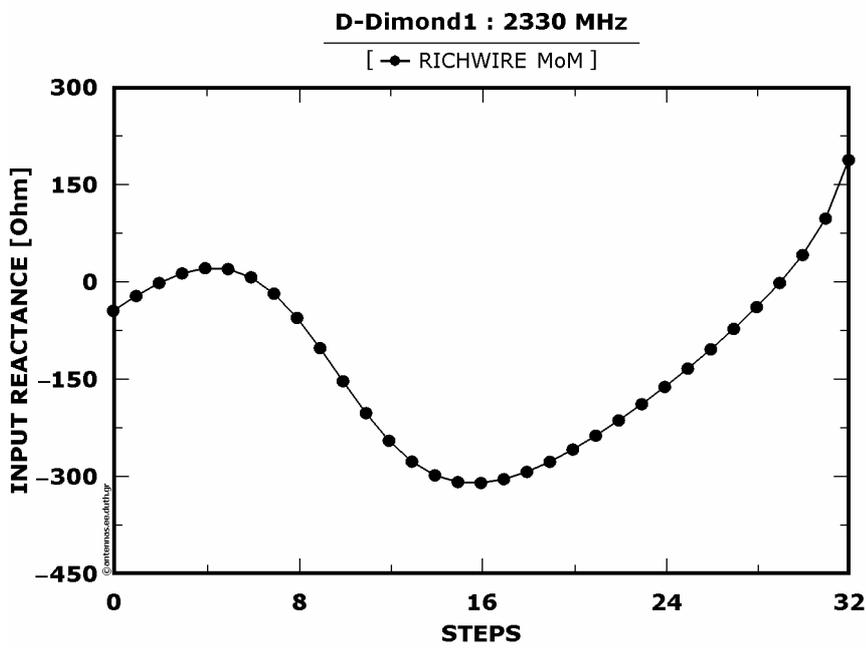

Figure 9 : Real and imaginary part of the input impedance as a function of
the elongation steps

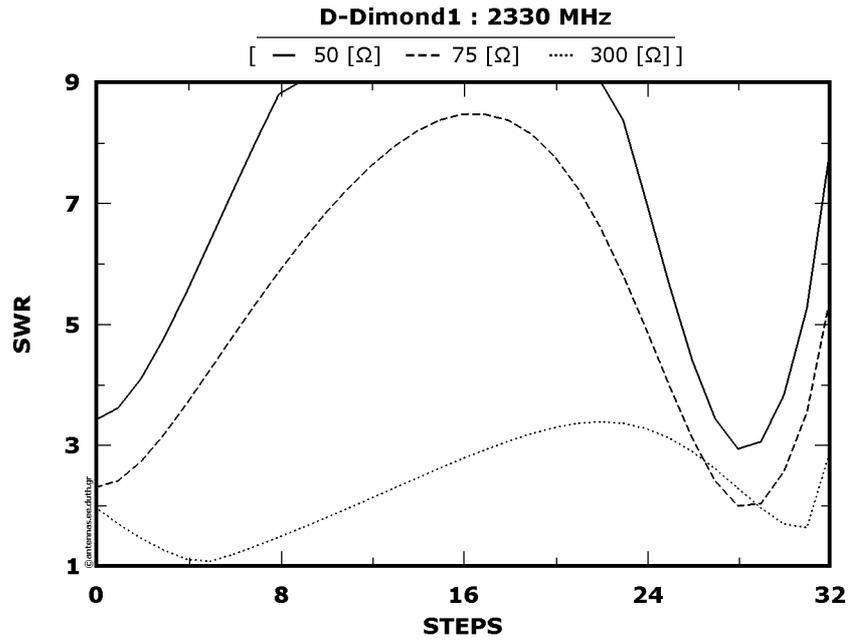

Figure 10 : SWRs as a function of the elongation steps

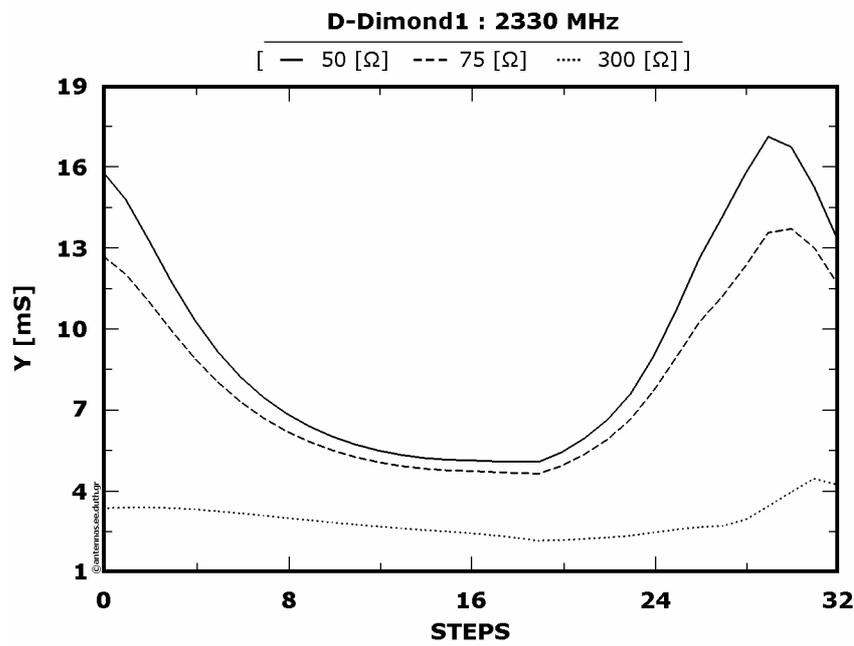

Figure 11 : Normalized radiation intensities as a function of the elongation steps

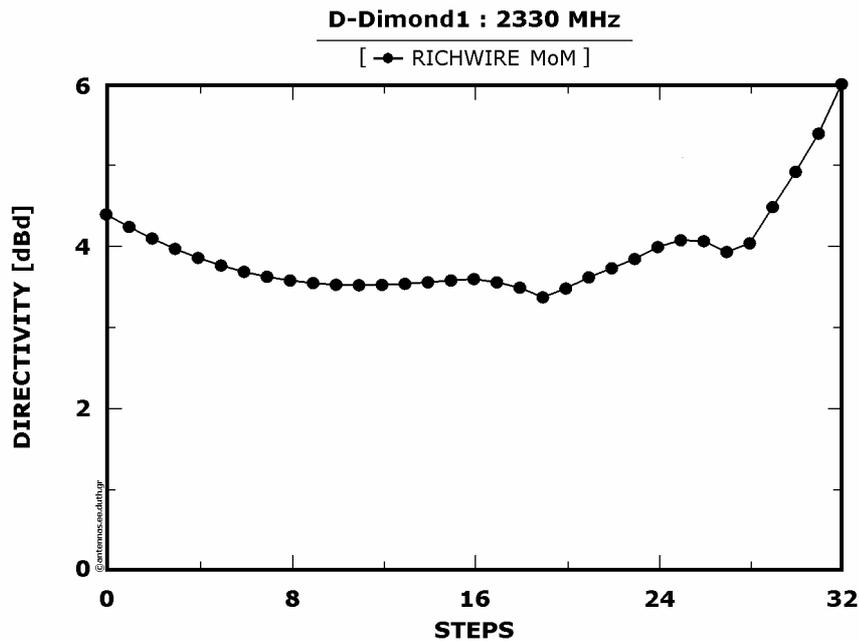

Figure 12 : Directivity (with reference to dipole λ/2) as a function of the elongation steps

According to Figure 5.2.4.10 minimal at all the three SWR plots are being observed. More specifically, both SWR(50) and SWR(75) are being minimized simultaneously approximately at the 28th elongation step.

Firstly the directivity's variation is presented quite firm yet increases afterwards as elongation steps advance, as shown in Figure 5.2.4.12, exceeding 5 [dBd].

According to all above, maximization of the normalized radiation intensities is being expected and actually emerges nearby the 29th elongation step for both Y(50) and Y(75), as shown in Figure 5.2.4.11.

Deductively, the geometrical modification of the former antenna's geometry according to the 29th elongation step of the current model is fully justifiable.

### 5.2.5 : Frequency at 2760 [MHz] – Model of active dipole's displacement (1st model)

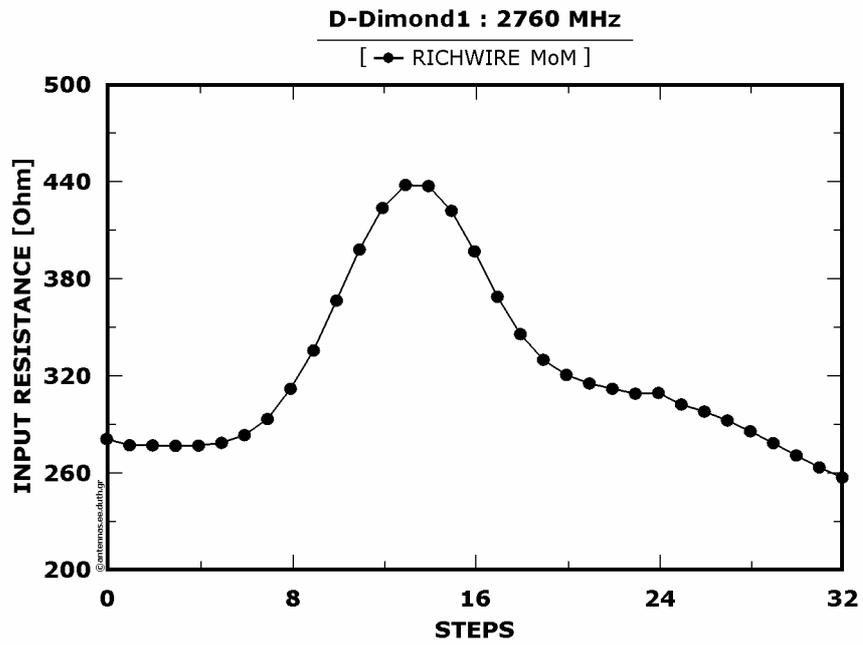

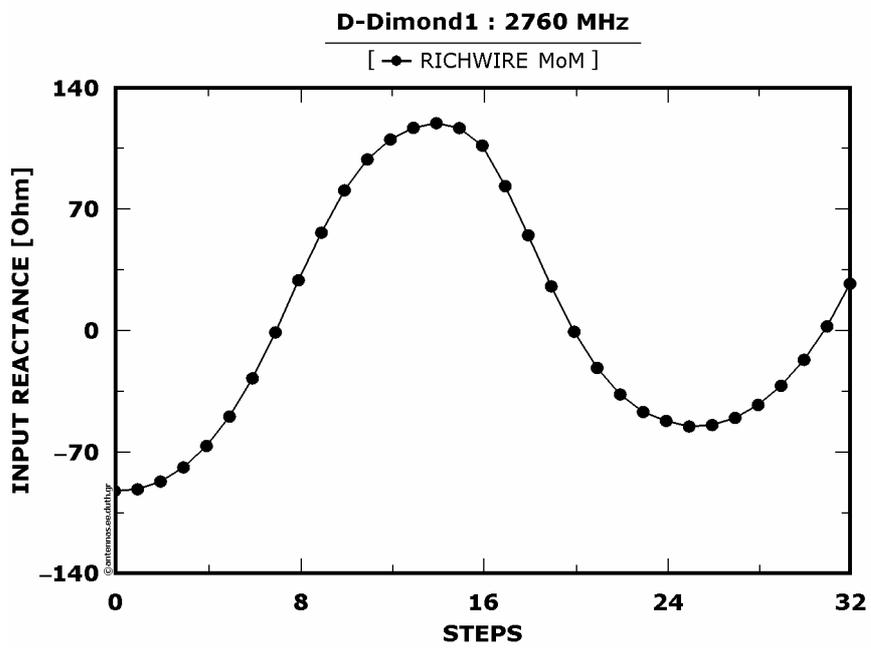

<u>Figure 1</u> : Real and imaginary part of the input impedance as a function of the displacement steps

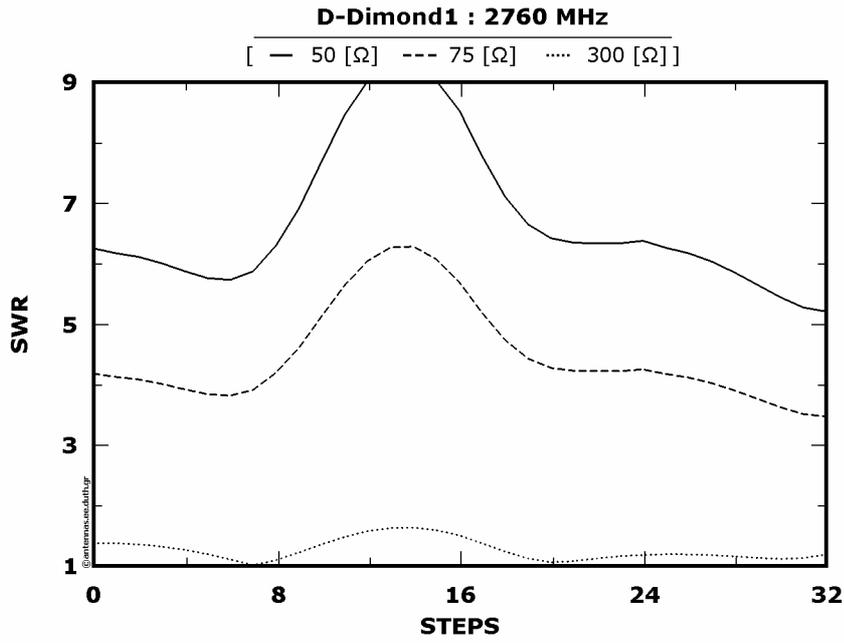

Figure 2 : SWRs as a function of the displacement steps

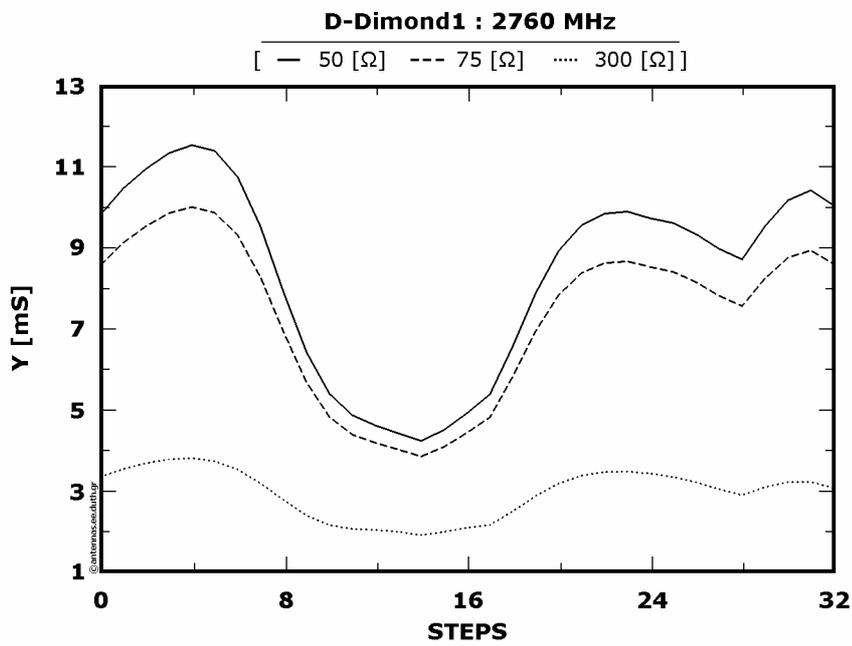

Figure 3 : Normalized radiation intensities as a function of displacement steps

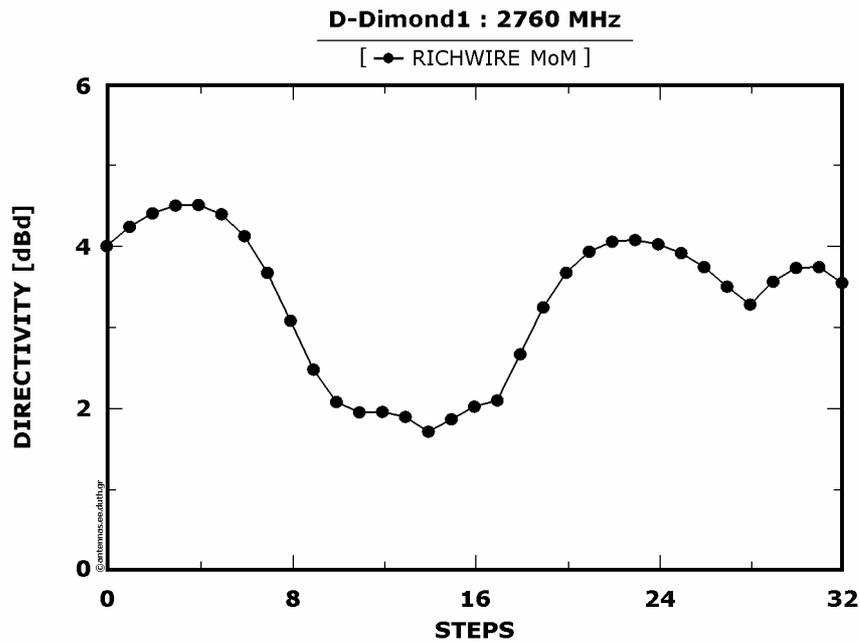

<u>Figure 4</u> : Directivity (with reference to dipole λ/2) as a function of the displacement steps

At <u>Figure 5.2.5.2</u> a similar variation of all the three SWR plots is being observed where there are two regions of minimums. Particularly, these regions are located for both SWR(50) and SWR(75), around the 6<sup>th</sup> and 32<sup>nd</sup> elongation step respectively. SWR(300)'s variation appears to be quite interesting as its value remains lower than 2 at the whole range of elongation.

From <u>Figure 5.2.5.4</u> the almost periodic directivity's variation is shown with its highest value around 4.5 [dBd] nearby the 4<sup>th</sup> elongation step.

Therefore, maximums of the normalized radiation intensities are being produced at the 4<sup>th</sup> elongation step, as shown in <u>Figure 5.2.5.3</u>, hence the antenna's geometrical modification according to the current model is justifiable.

**Model of perpendicular monopoles elongation (2<sup>nd</sup> model)**

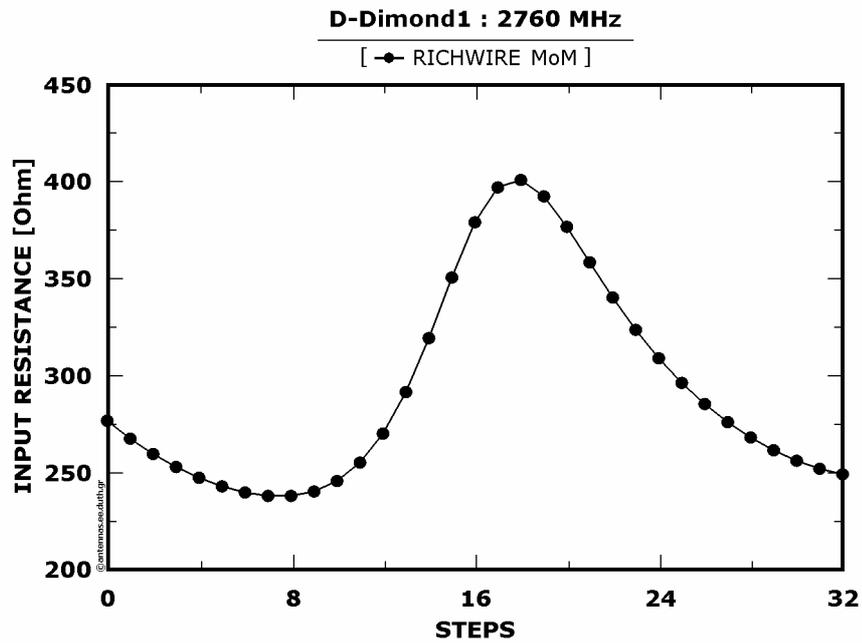

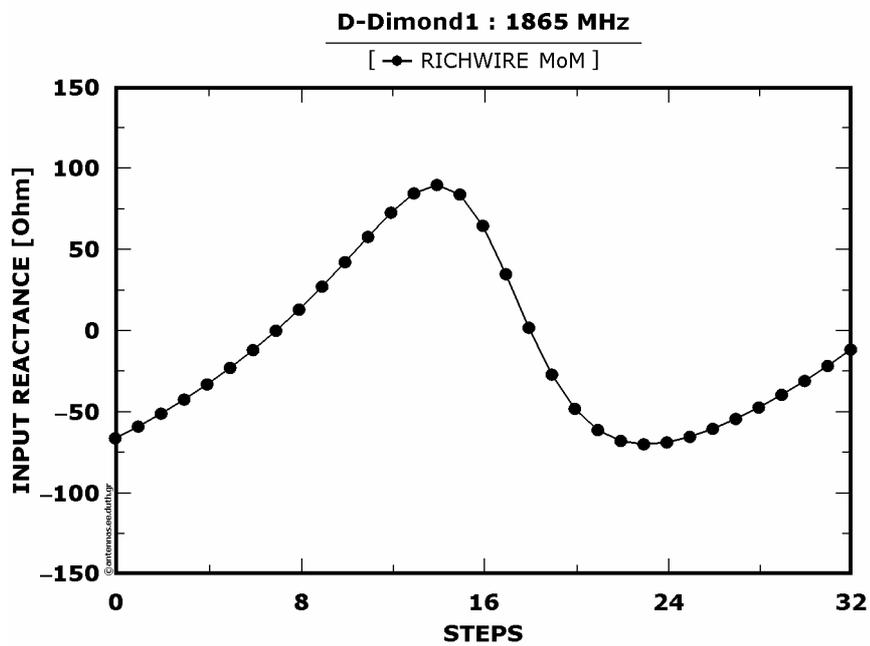

<u>Figure 5</u> : Real and imaginary part of the input impedance as a function of the elongation steps

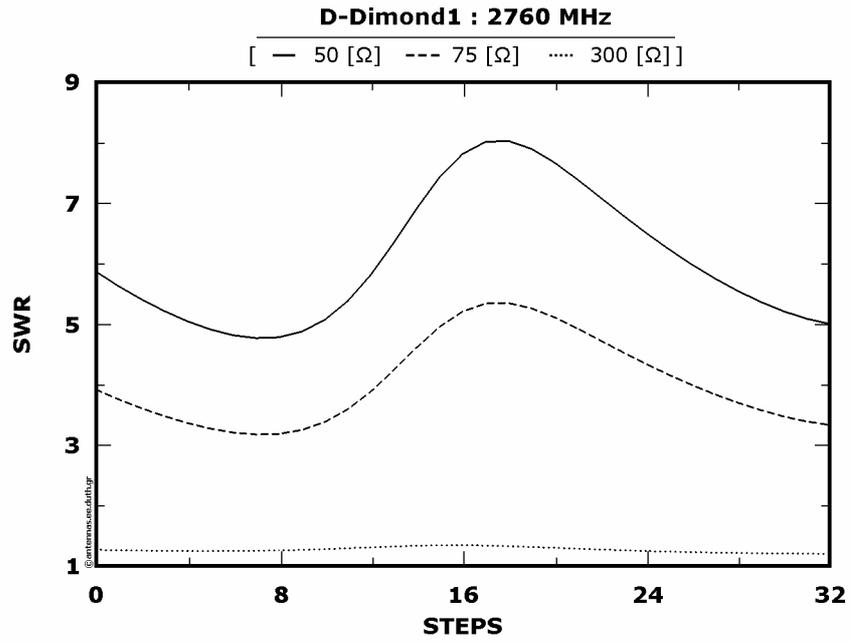

Figure 6 : SWRs as a function of the elongation steps

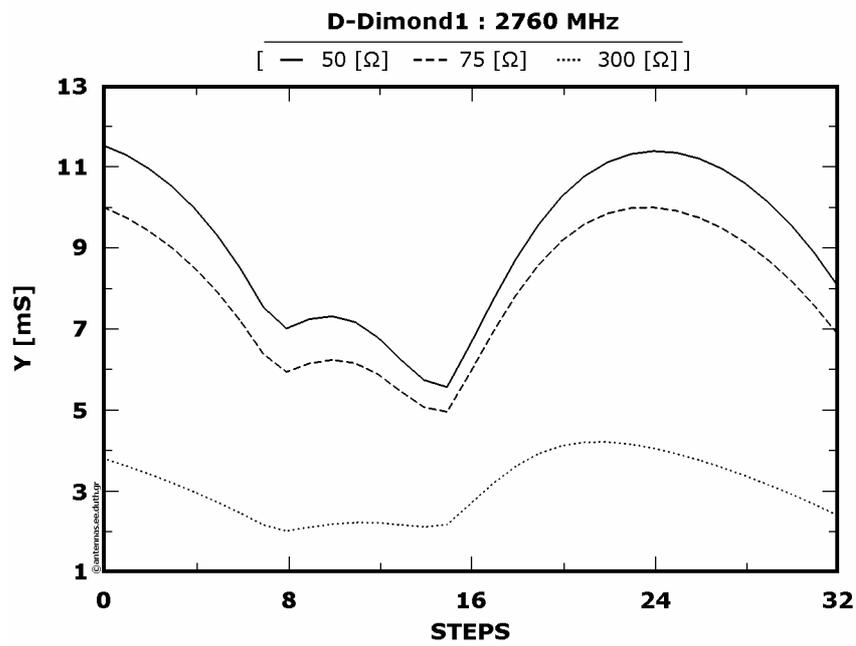

Figure 7 : Normalized radiation intensities as a function of the elongation steps

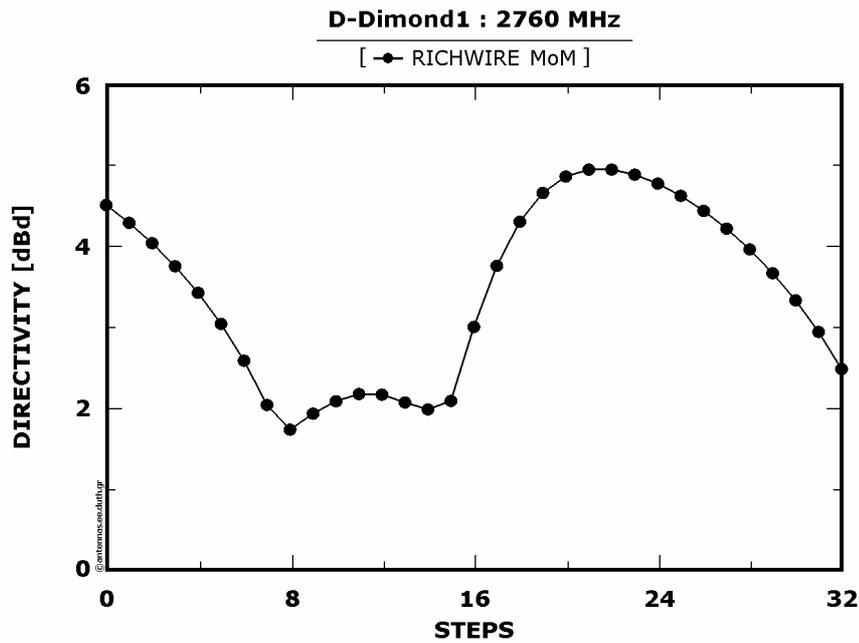

<u>Figure 8</u> : Directivity (with reference to dipole λ/2) as a function of the elongation steps

Both SWR(50) and SWR(75) have similar variations and produce two minimums nearby the 8th and 32nd elongation step respectively, as shown in <u>Figure 5.2.5.6</u>. On the other hand, SWR(300) remains almost invariant maintaining a unitary value at the whole range of elongation steps.

The directivity varies irregularly maintaining a value above 4 [dBd] at a quite broad range of elongation steps, as illustrated in <u>Figure 5.2.5.8</u>.

From <u>Figure 5.2.5.7</u> it is observed that both Y(50) and Y(75) do not exceed their initial values at the whole range of elongation steps while the maximum of Y(300) that it is produced is insignificant.

Consequently, any geometrical modification of the antenna according to the current model is not justifiable.

**Model of active dipole's elongation (3$^{rd}$ model)**

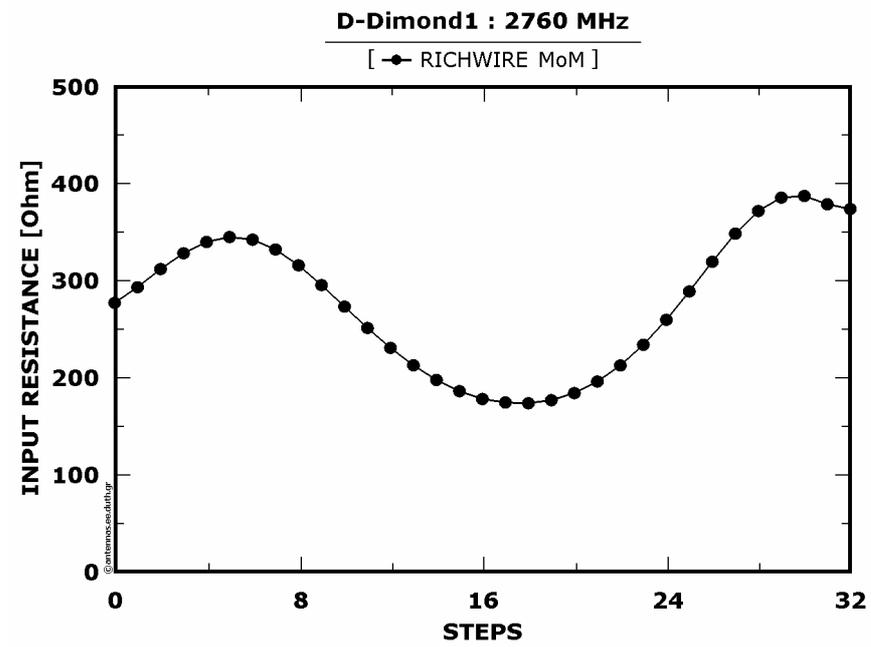

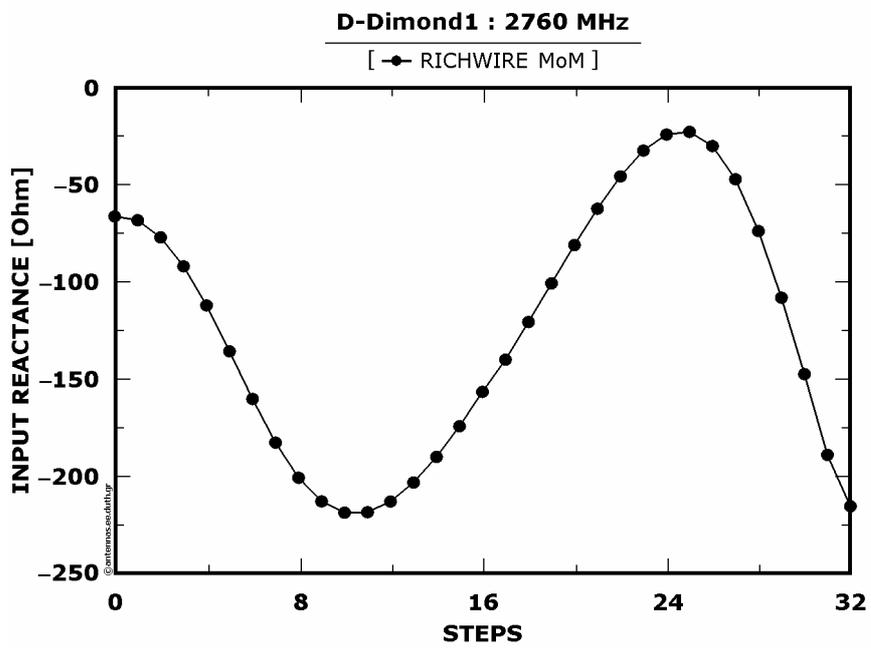

<u>Figure 9</u> : Real and imaginary part of the input impedance as a function of
the elongation steps

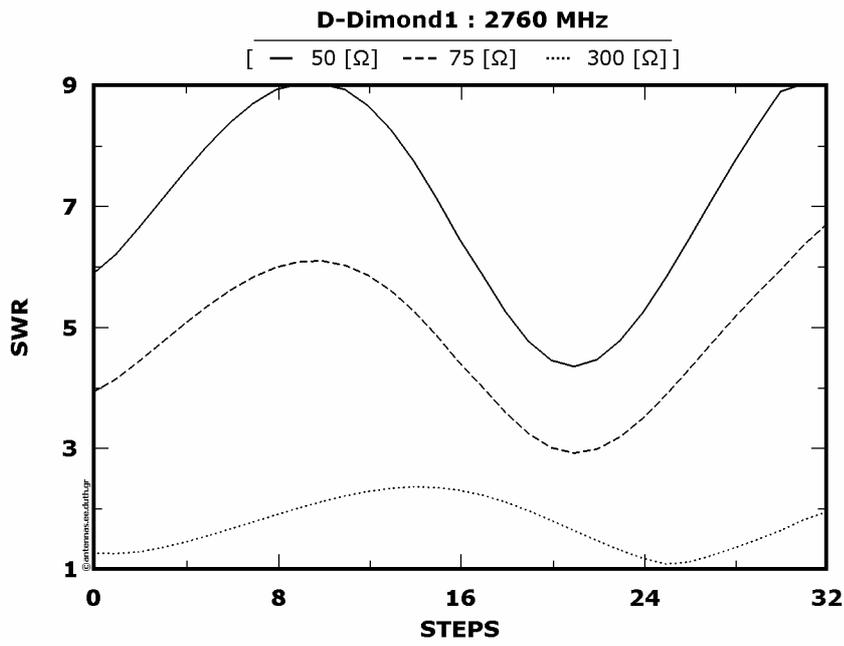

Figure 10 : SWRs as a function of the elongation steps

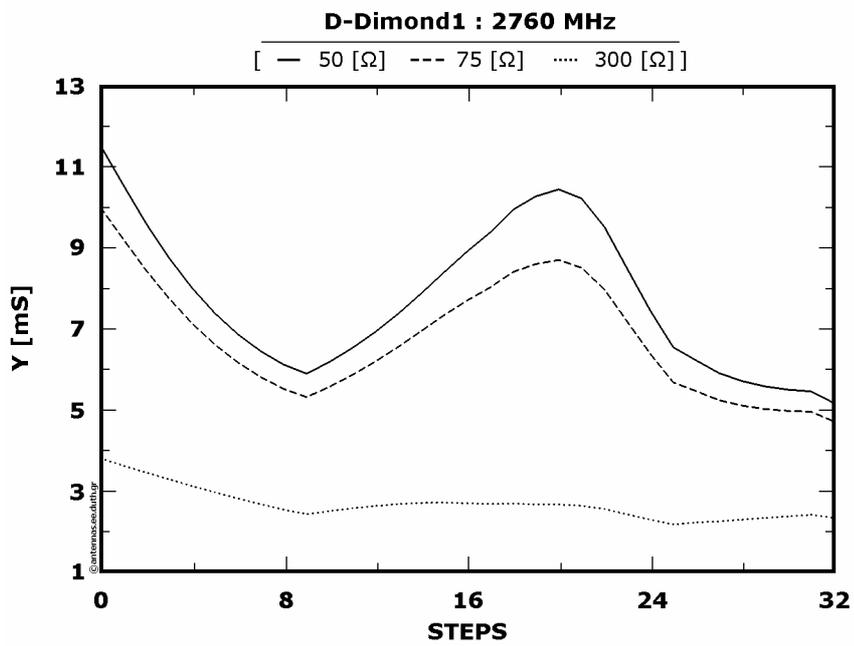

Figure 11 : Normalized radiation intensities as a function of the elongation steps

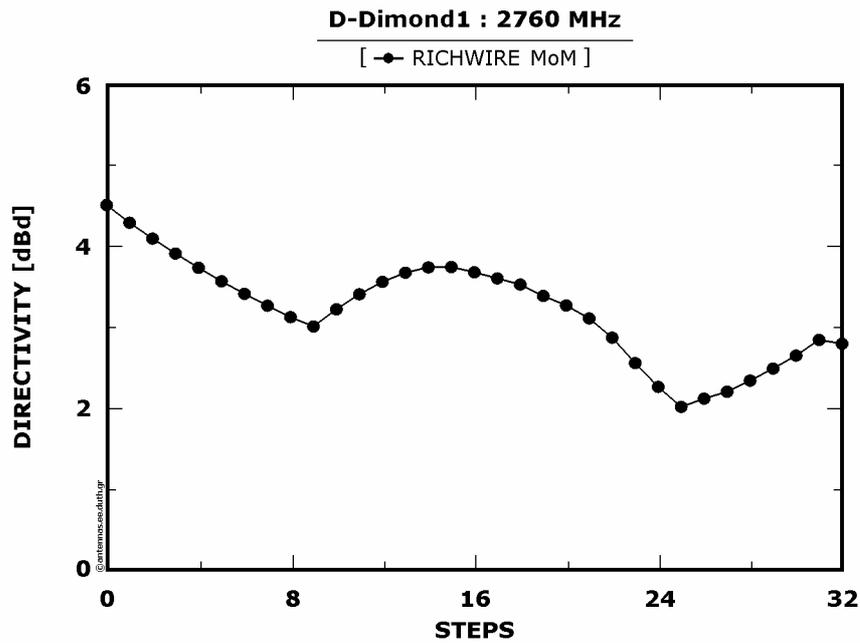

<u>Figure 12</u> : Directivity (with reference to dipole λ/2) as a function of the elongation steps

According to <u>Figure 5.2.5.10</u> a minimal which occurs simultaneously for both SWR(50) and SWR(75) is observed nearby the 22[th] step.

Directivity is generally decreasing during a periodic interchange of maximums and minimums, as shown in <u>Figure 5.2.5.12</u>.

Hence, none of the normalized radiation intensities exceeds its initial value, as shown in <u>Figure 5.2.5.11</u>, therefore any geometrical modification of the antenna according to the current model is not justifiable unless the design of the communicating system anticipates again low SWR regardless of the directivity, at 2760 [MHz].

**5.3 : Radiation Patterns**

After the application of the above three models of geometrical modification at five se-lected and specific frequencies which have been chosen from the broad frequency range from 200 to 2850 [MHz] it is worthwhile to adopt those geometries that have been modified and present a "better behavior" at the pre-mentioned frequencies, compared with the original ge-ometry of the antenna. Below are cited the 3-dimensional radiation patterns and the equivalent patterns at the three main planes of the modified improved dispositions. Moreover, below fol-low the improved geometry patterns in comparison with the initial geometry of the antenna.

**5.3.1 : Radiation patterns at 580 [MHz]**

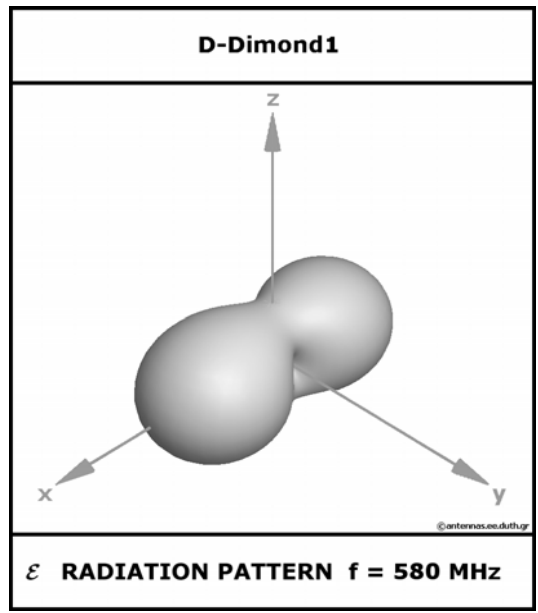

Figure 1 : Three-dimensional radiation pattern

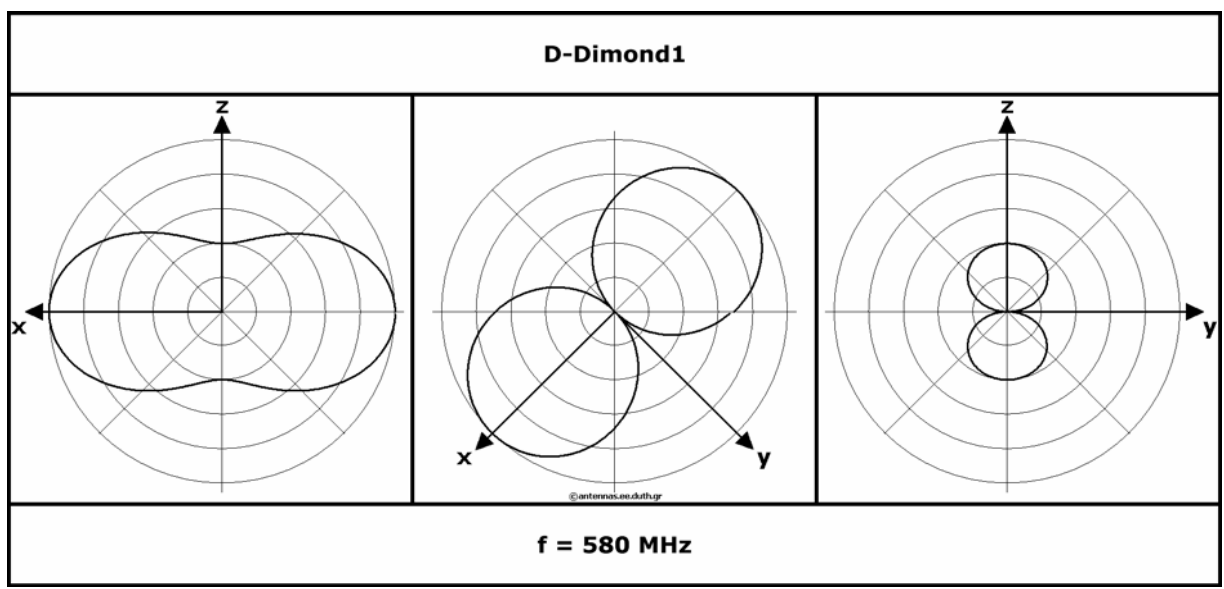

Figure 2 : Radiation pattern at the three main planes

**5.3.2 : Radiation patterns at 1238 [MHz]**

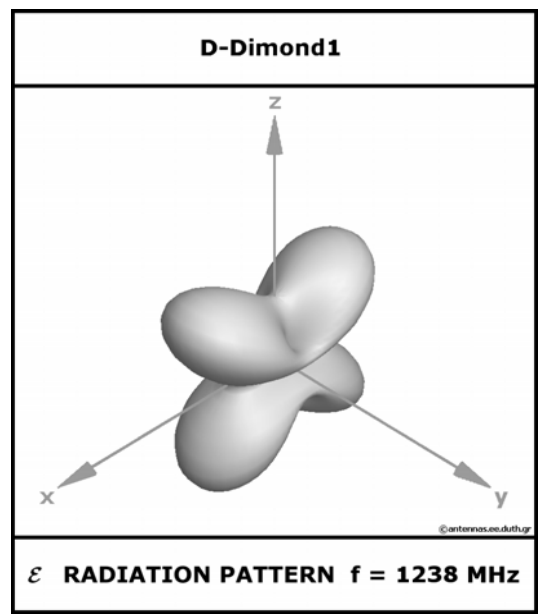

Figure 1 : Three-dimensional radiation pattern

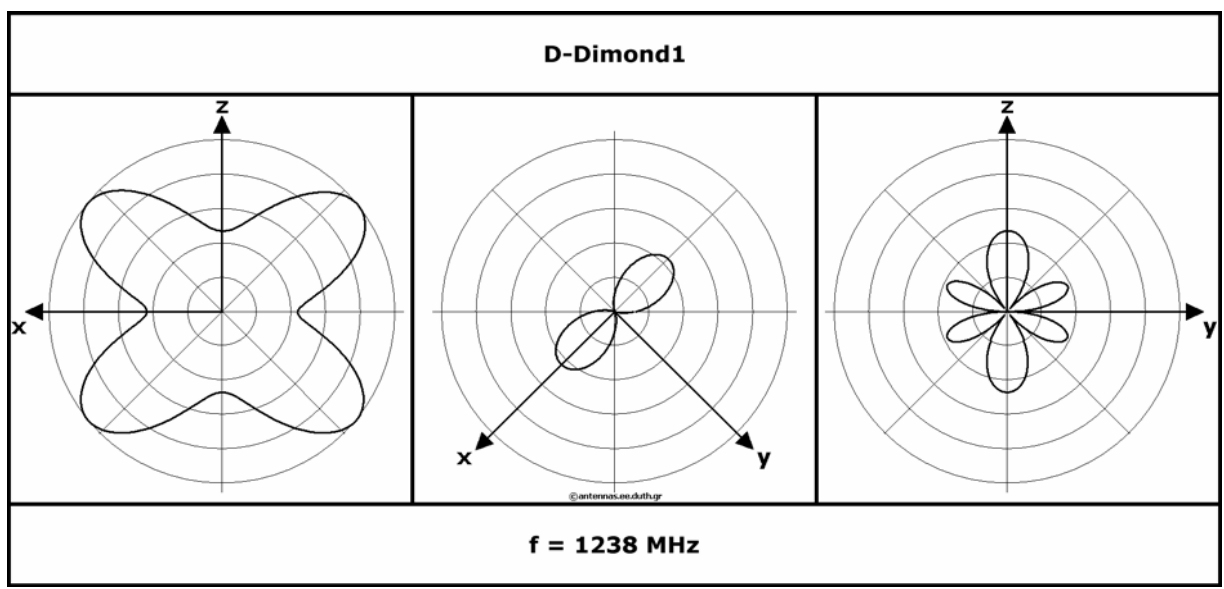

Figure 2 : Radiation pattern at the three main planes

### 5.3.3 : Radiation patterns at 1867 [MHz]

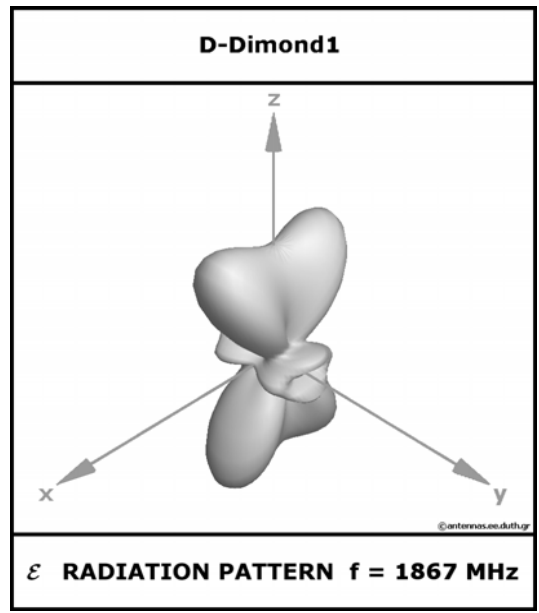

Figure 1 : Three-dimensional radiation pattern

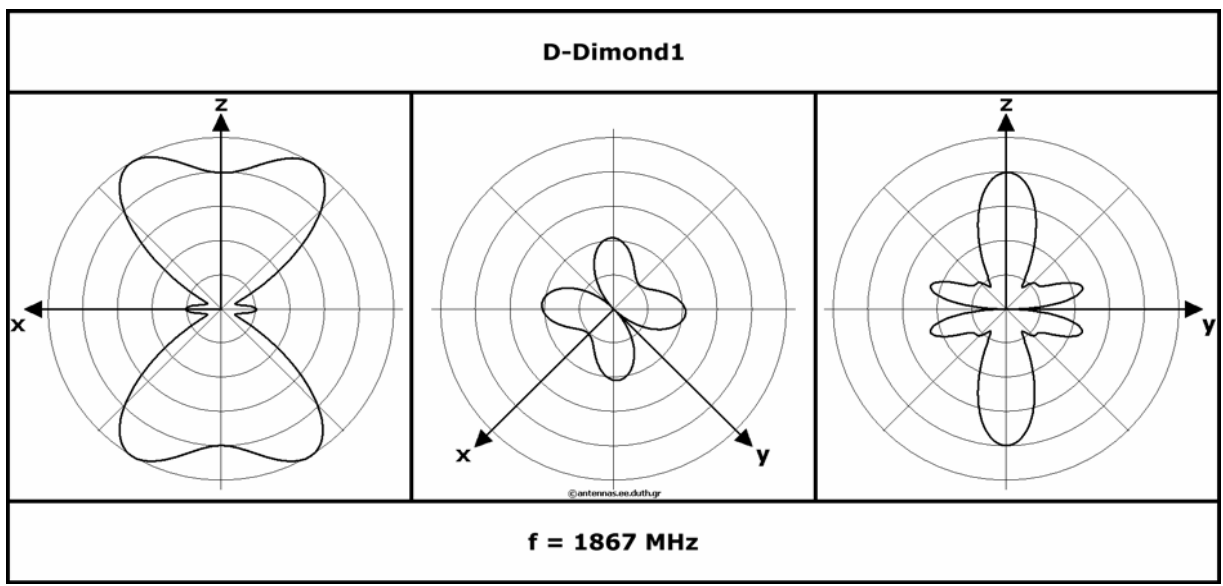

Figure 2 : Radiation pattern at the three main planes

**5.3.4 : Radiation patterns at 2330 [MHz]**

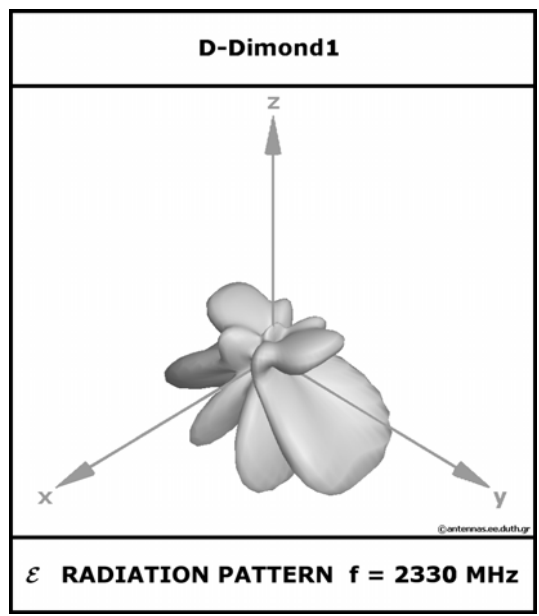

Figure 1 : Three-dimensional radiation pattern

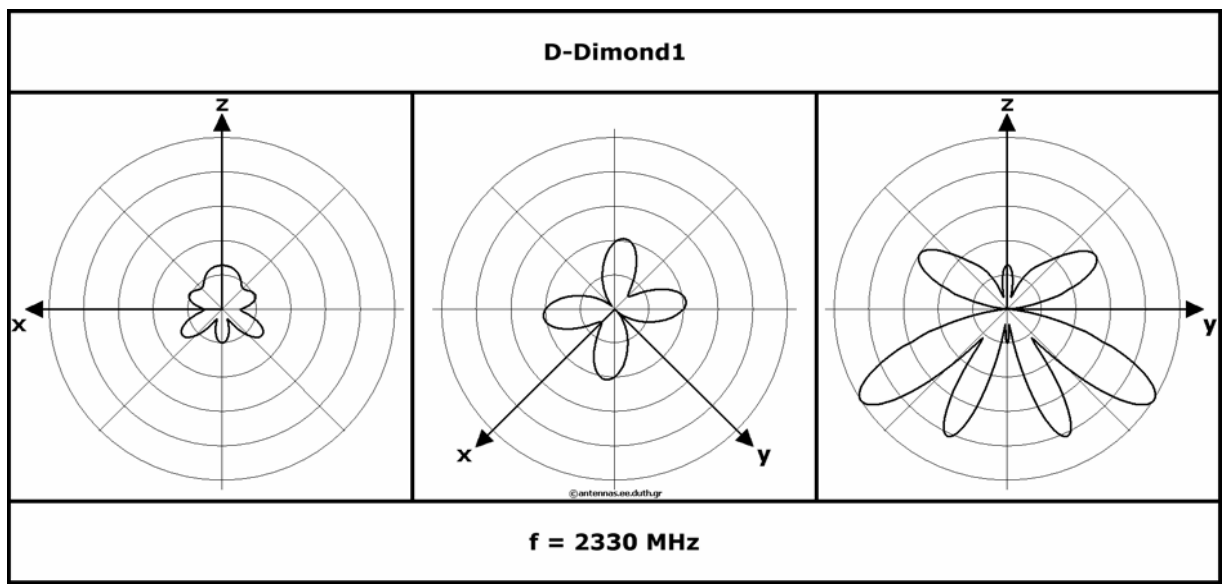

Figure 2 : Radiation pattern at the three main planes

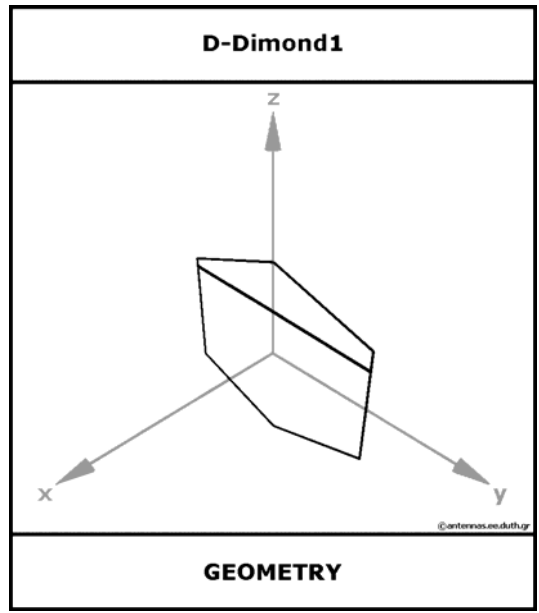

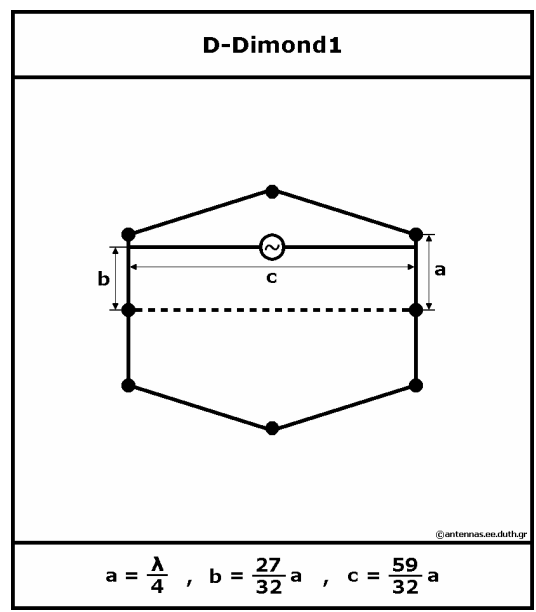

Figure 3 : Three-dimensional and two-dimensional antenna's geometry

**5.3.5 : Radiation patterns at 2760 [MHz]**

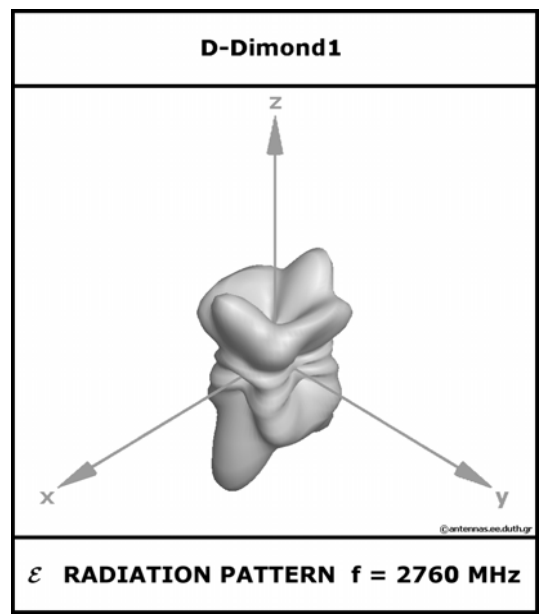

Figure 1 : Three-dimensional radiation pattern

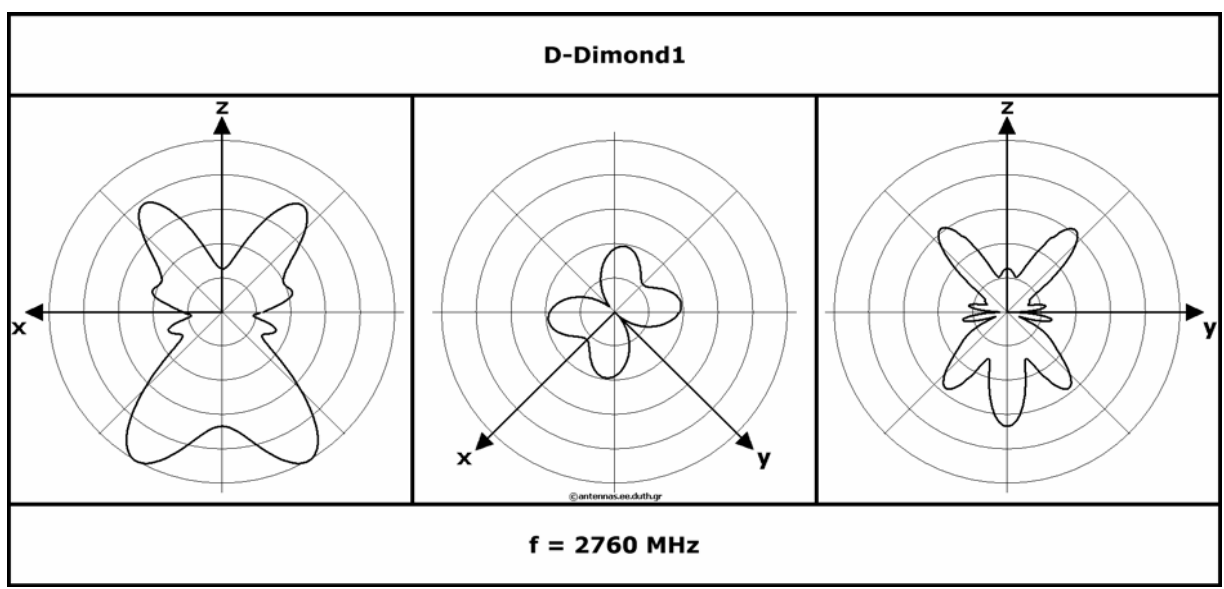

Figure 2 : Radiation pattern at the three main planes

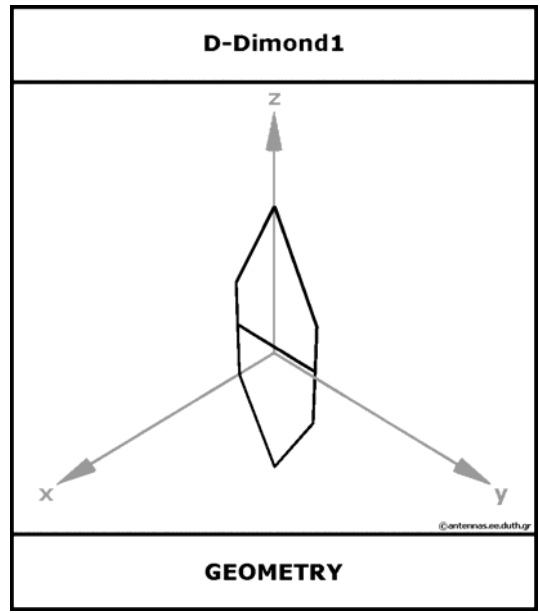

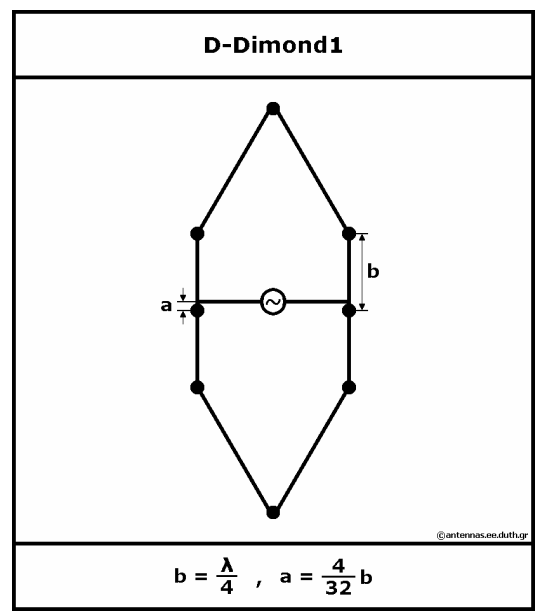

Figure 3 : Three-dimensional and two-dimensional antenna's geometry

**5.4 : Frequency scan**

　　　At the preceding section were studied the characteristic quantities of the modified – improved geometrical dispositions of the original antenna at one specific frequency. It is equally substantial to determine the behavior of these dispositions as a function of frequency in order to reveal some broadband characteristics, if there are any, that is, the capability of the "improved" antennas to maintain a desirable behaviour at a satisfactory frequency range.

　　　As it is observed from the preceding sections, two "improved" antenna's dispositions were emerged from the aggregate five frequency ranges. Therefore, the study regarding the frequency will be limited only at the following frequency ranges: from 2200 to 2500 [MHz] and from 2550 to 2850 [MHz] since the two "improved" models of D-Dimond1 operate at these specific frequency ranges.

**5.4.1 : Frequency range from 2200 to 2500 [MHz]**

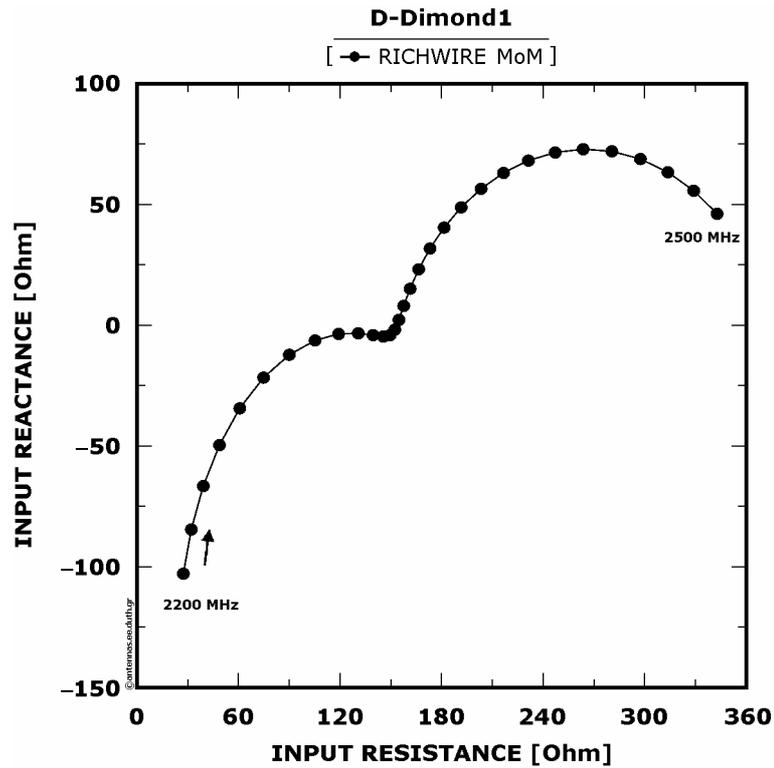

Figure 1 : Input impedance

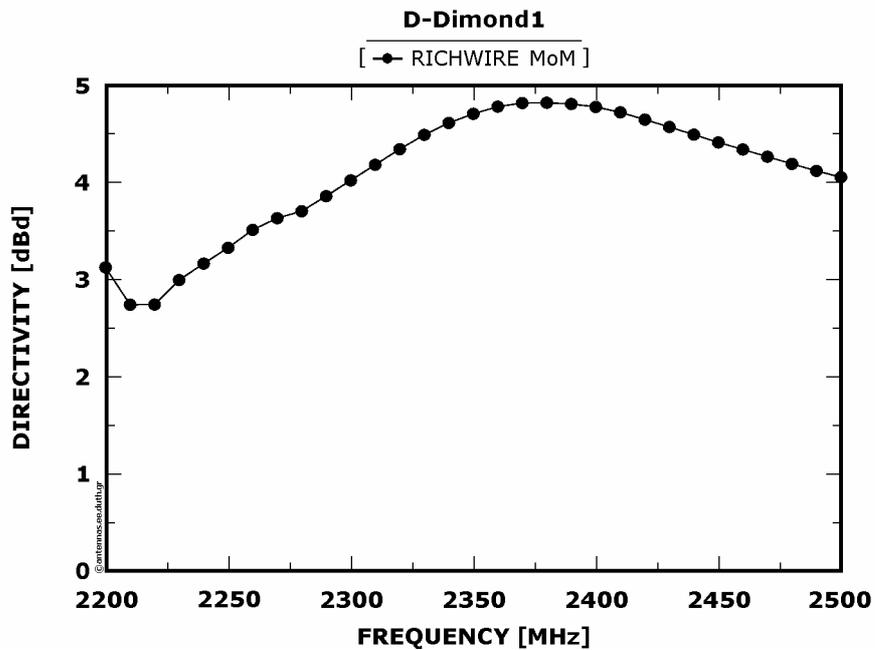

Figure 2 : Directivity (with reference to dipole λ/2) as a function of frequency

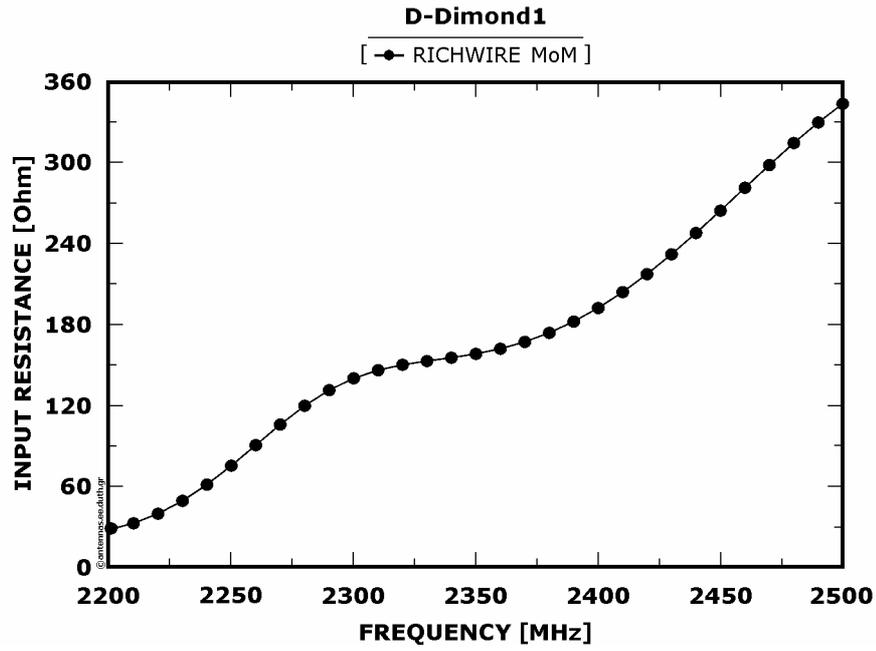

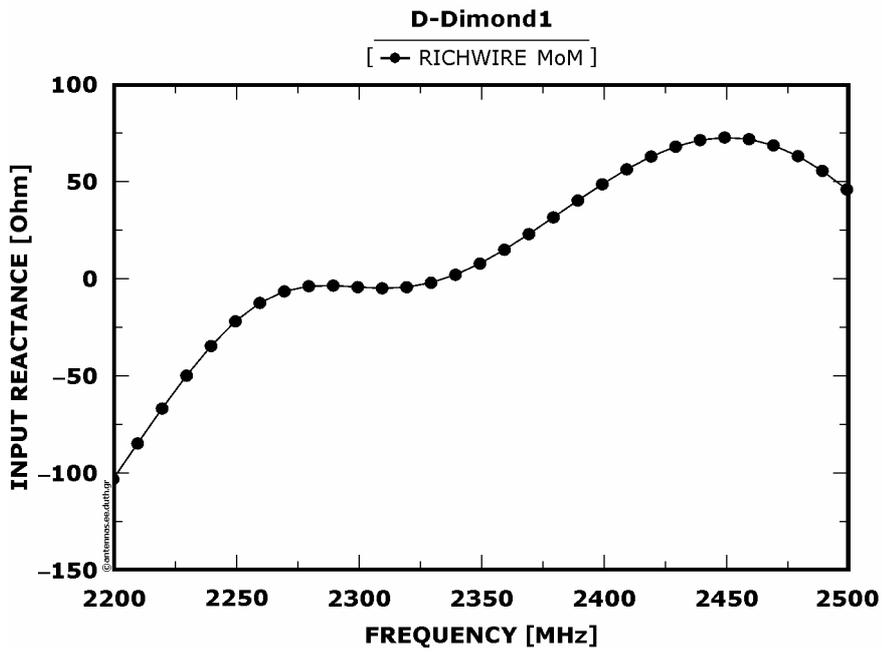

<u>Figure 3</u> : Real and imaginary part of the input impedance as a function of frequency

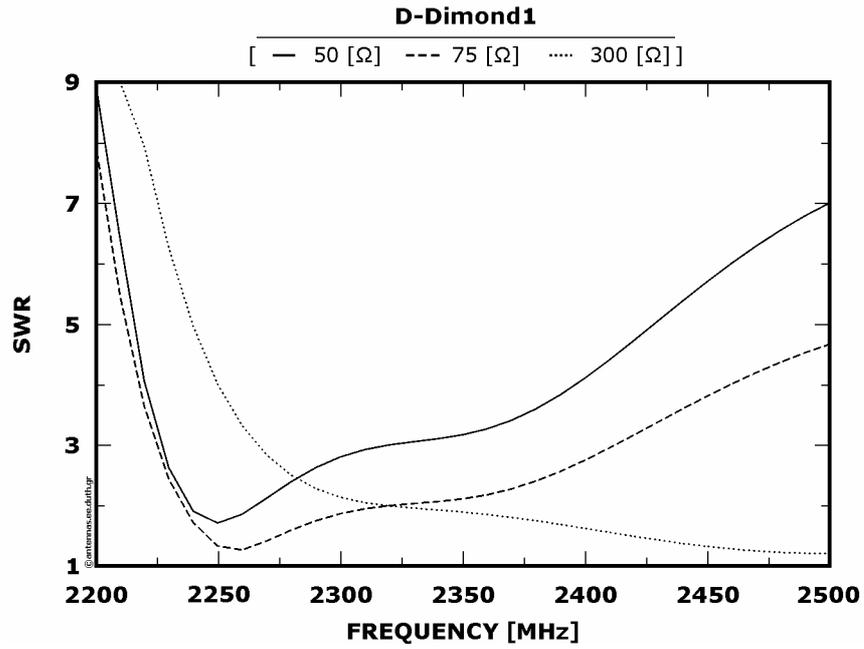

Figure 4 : SWRs as a function of frequency

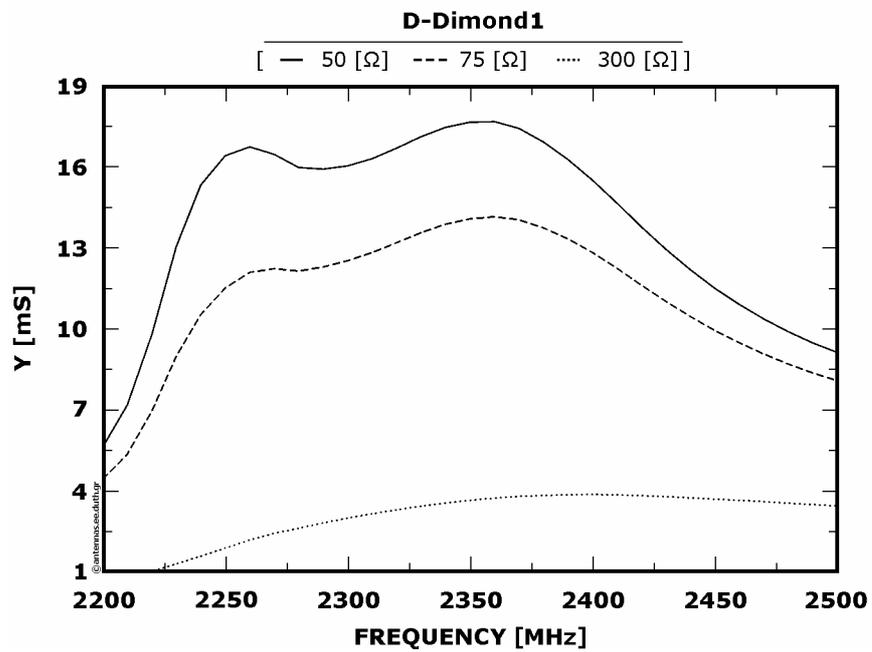

Figure 5 : Maximum normalized radiation intensities as a function of frequency

## 5.4.2 : Frequency range from 2550 to 2850 [MHz]

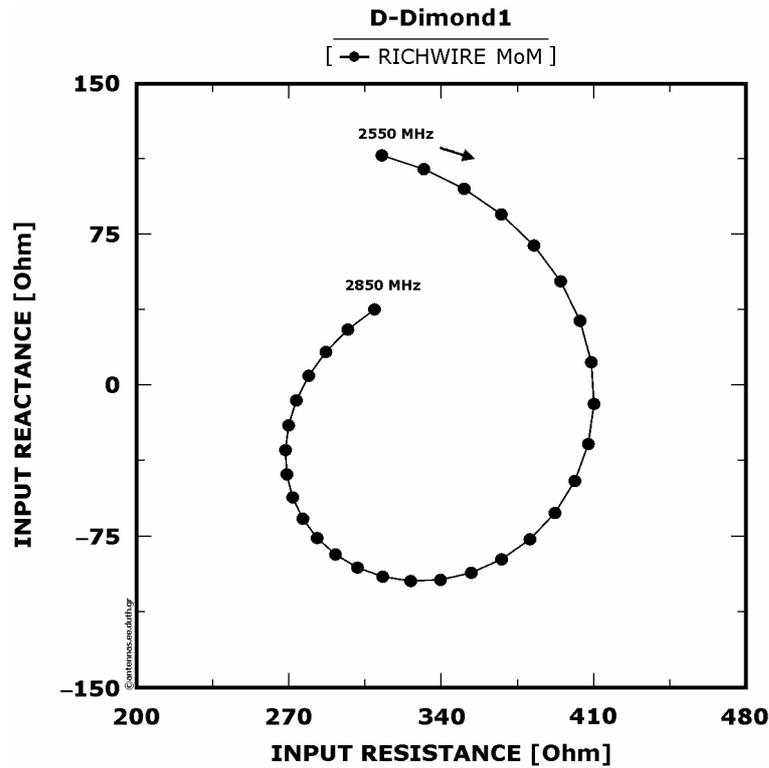

<u>Figure 1</u> : Input impedance

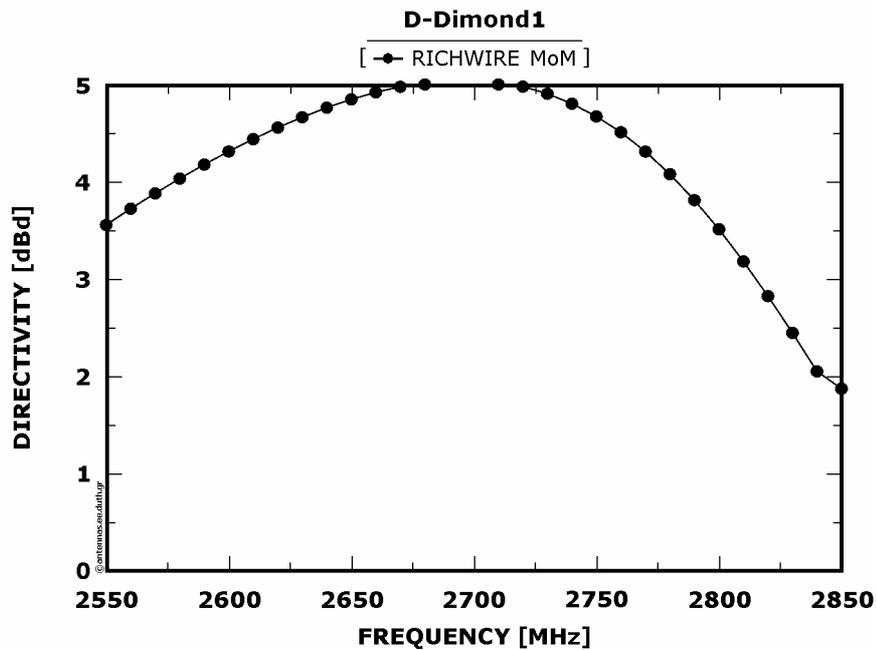

<u>Figure 2</u> : Directivity (with reference to dipole λ/2) as a function of frequency

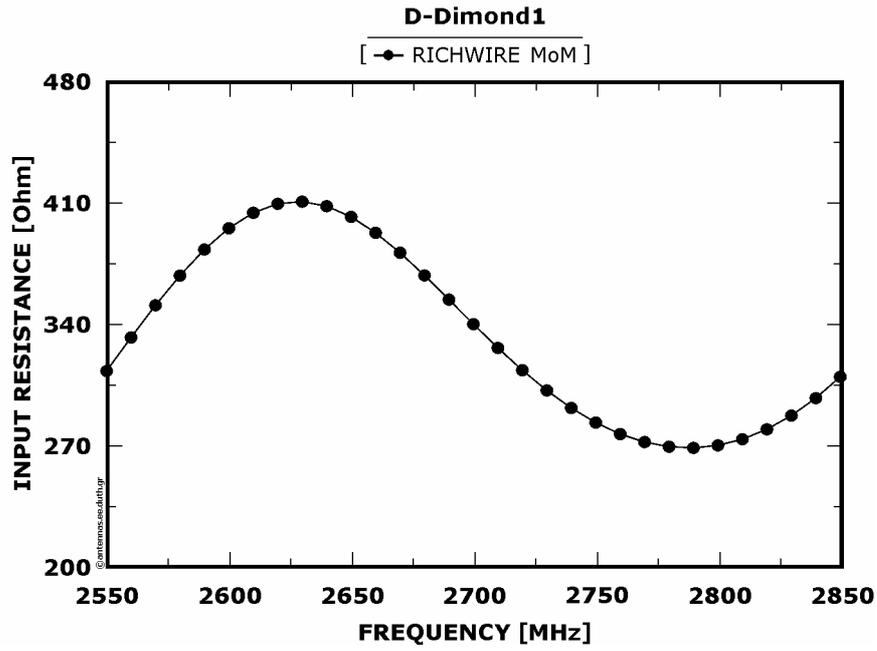

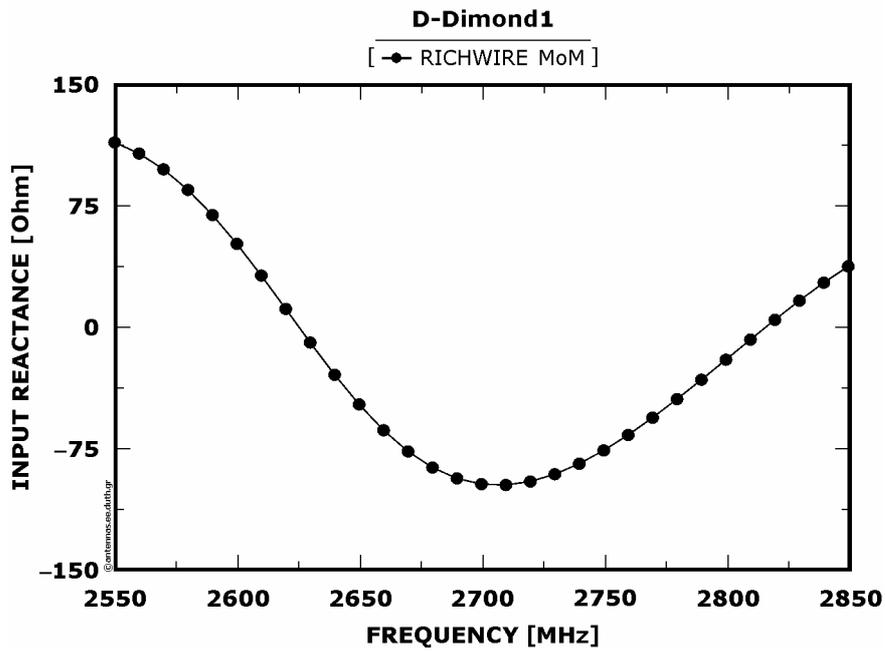

<u>Figure 3</u> : Real and imaginary part of the input impedance as a function of frequency

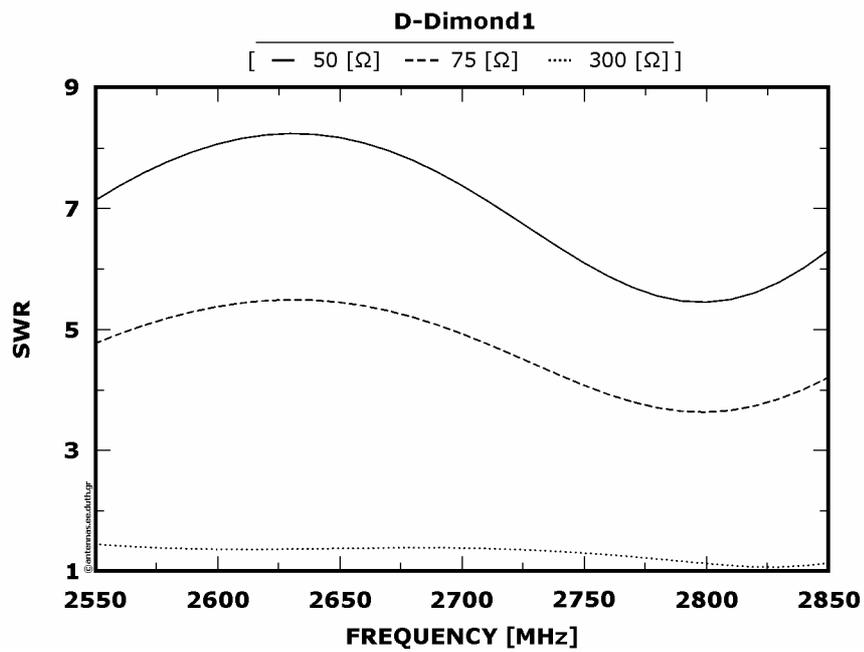

Figure 4 : SWRs as a function of frequency

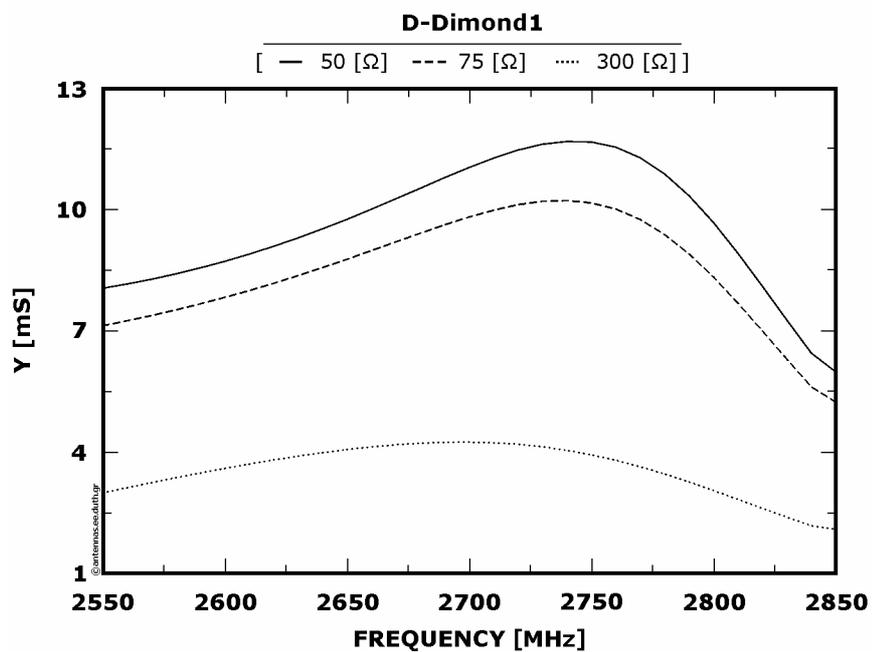

Figure 5 : Maximum normalized radiation intensities as a function of frequency

**5.5 : Comparisons**

Below are cited analytical tables which hold the values of the antenna's characteristic quantities at every examined frequency, before and after the antenna's improvement.

Table 1 : Characteristic quantities of D-Dimond1 at 580 [MHz]

| $Z_{INP}$ [$\Omega$] | SWR(50 $\Omega$) | SWR(75$\Omega$) | SWR(300$\Omega$) | D [dBd] |
|---|---|---|---|---|
| 117.1 + i6.5 | 2.3 | 1.5 | 2.5 | 2.55 |

Table 2 : Characteristic quantities of D-Dimond1 at 1238 [MHz]

| $Z_{INP}$ [$\Omega$] | SWR(50 $\Omega$) | SWR(75$\Omega$) | SWR(300$\Omega$) | D [dBd] |
|---|---|---|---|---|
| 186.3 − i68.3 | 4.2 | 2.8 | 1.7 | 2.99 |

Table 3 : Characteristic quantities of D-Dimond1 at 1867 [MHz]

| $Z_{INP}$ [$\Omega$] | SWR(50 $\Omega$) | SWR(75$\Omega$) | SWR(300$\Omega$) | D [dBd] |
|---|---|---|---|---|
| 117.0 − i47.9 | 2.8 | 1.9 | 2.6 | 4.41 |

Table 4 : Characteristic quantities of D-Dimond1 at 2330 [MHz]

(the 2nd line represent the counterpart values of the improved disposition)

| $Z_{INP}$ [Ω] | SWR(50 Ω) | SWR(75Ω) | SWR(300Ω) | D [dBd] |
|---------------|-----------|----------|-----------|---------|
| $84.9 - i\,10.8$ | 1.7 | 1.2 | 3.5 | 1.48 |
| $152.8 - i\,2.2$ | 3 | 2 | 1.9 | 4.48 |

Table 5 : Characteristic quantities of D-Dimond1 at 2760 [MHz]

(the 2nd line represent the counterpart values of the improved disposition)

| $Z_{INP}$ [Ω] | SWR(50 Ω) | SWR(75Ω) | SWR(300Ω) | D [dBd] |
|---------------|-----------|----------|-----------|---------|
| $280.9 - i\,92.4$ | 6.2 | 4.7 | 1.3 | 4.00 |
| $276.8 - i\,66.5$ | 5.8 | 3.9 | 1.2 | 4.50 |

Moreover, analytical figures which display the variation of the most significant characteristic quantities of the antenna are cited below. It is obvious that finally the improvement became feasible at two of the five pre-selected frequency ranges, that is 2200 to 2500 [MHz] and 2550 to 2850 [MHz]:

### 5.5.1 : Frequency range from 2200 to 2500 [MHz]

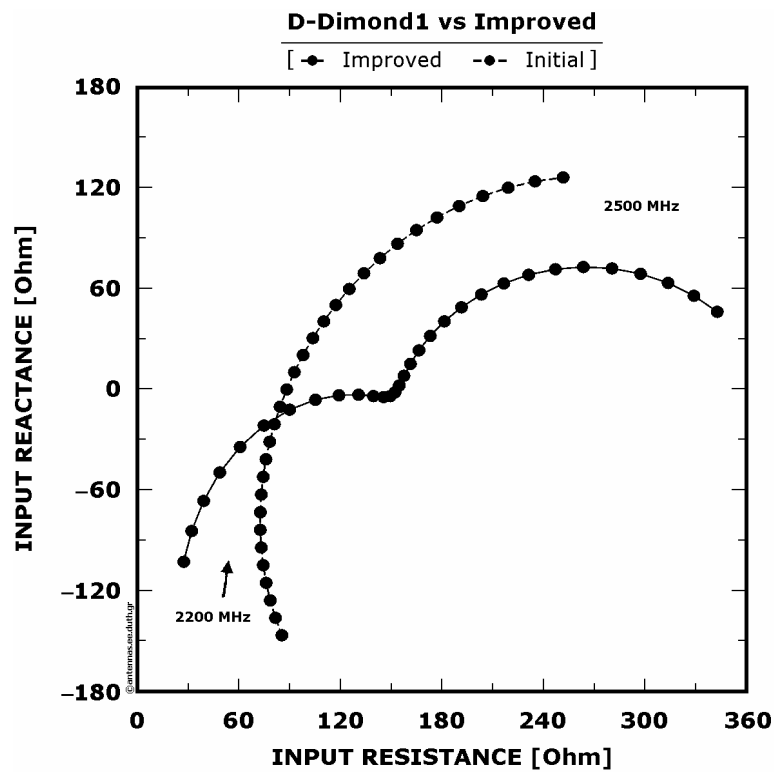

<u>Figure 1</u> : Input impedance of the initial and the improved disposition
as a function of frequency

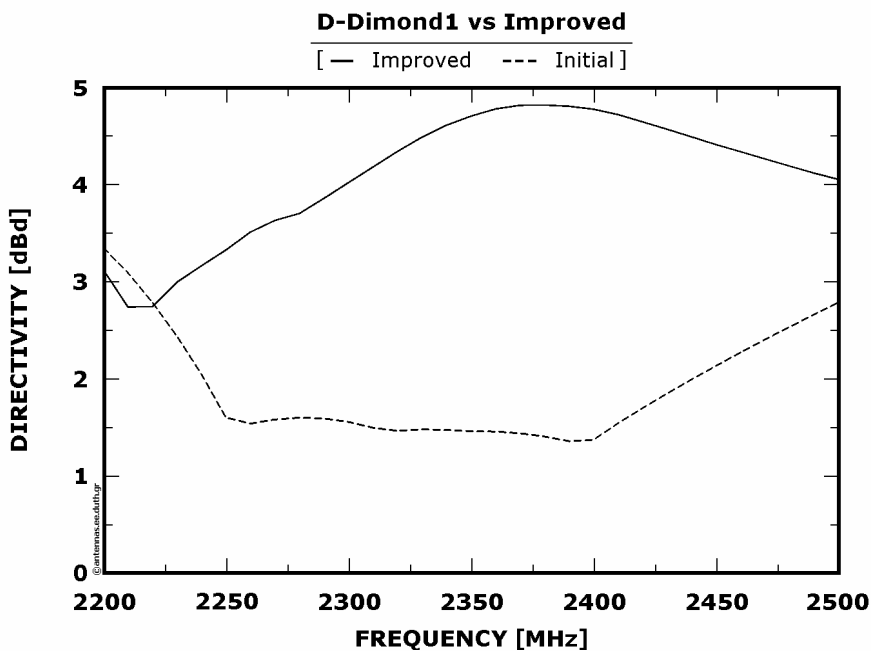

<u>Figure 2</u> : Directivity (with reference to dipole λ/2) of the initial and the improved disposition
as a function of frequency

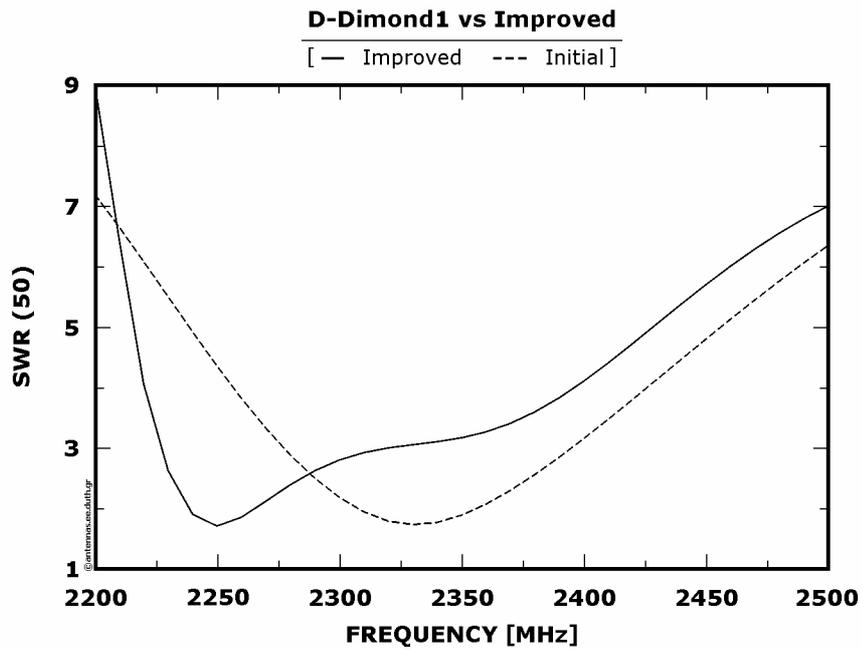

<u>Figure 3</u> : SWR (with reference to 50 [Ω]) at the initial and the improved disposition
as a function of frequency

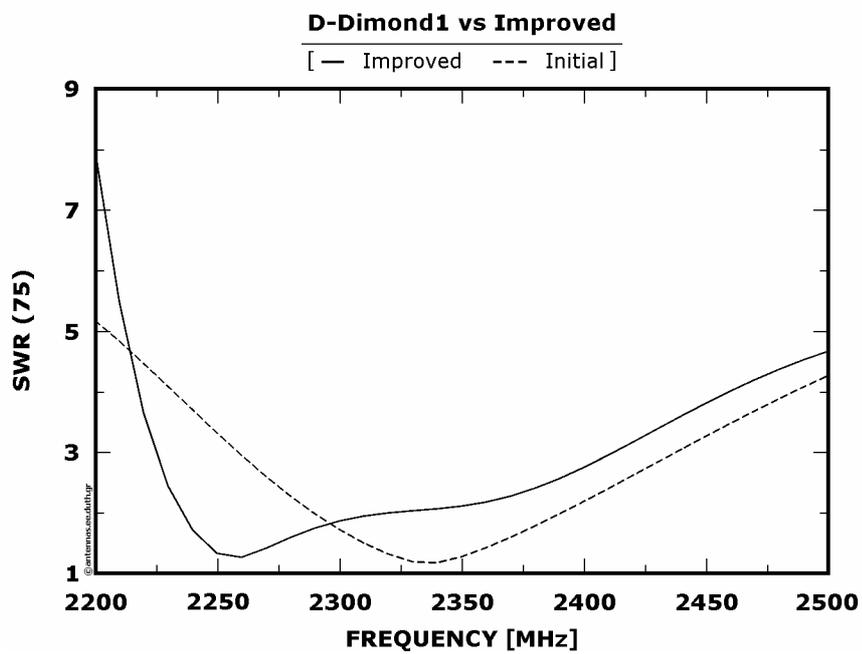

<u>Figure 4</u> : SWR (with reference to 75 [Ω]) at the initial and the improved disposition
as a function of frequency

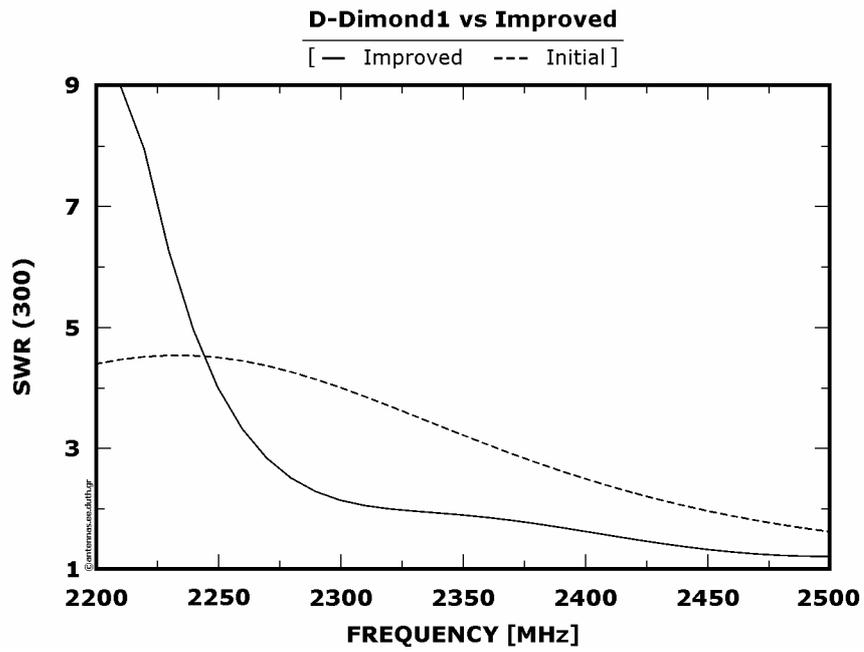

Figure 5 : SWR (with reference to 300 [Ω]) at the initial and the improved disposition as a function of frequency

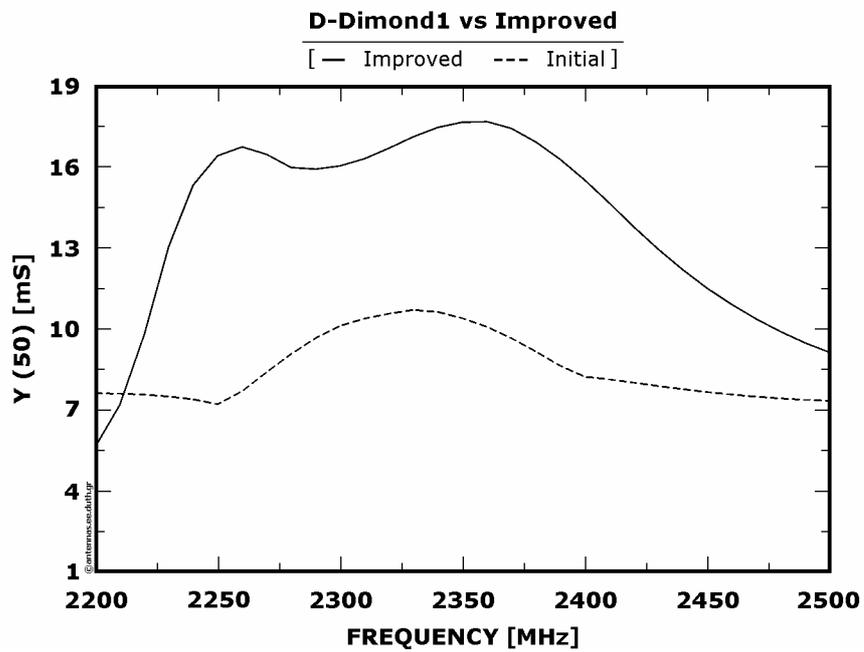

Figure 6 : Normalized radiation intensity (with reference to 50 [Ω]) at the initial and the improved disposition as a function of frequency

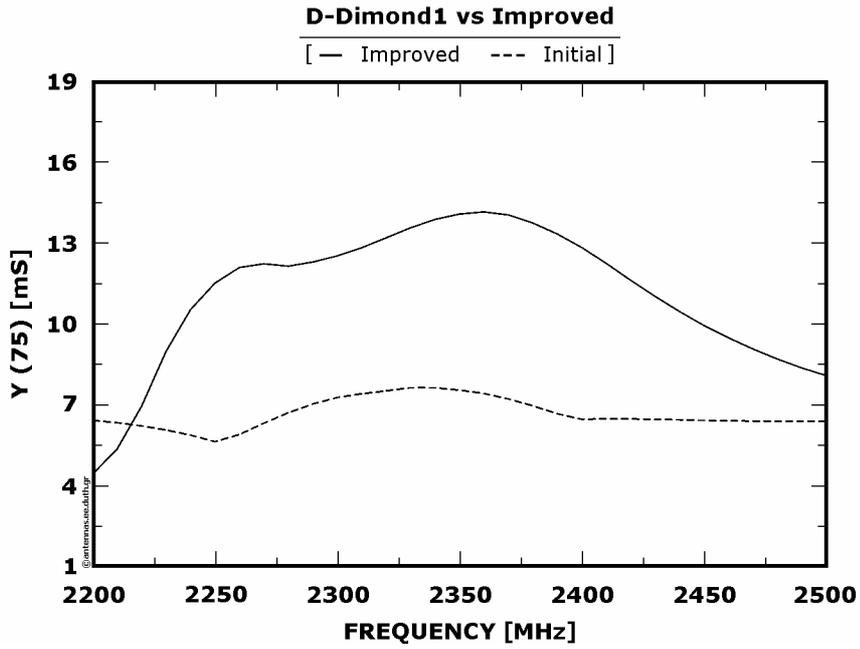

<u>Figure 7</u> : Normalized radiation intensity (with reference to 75 [Ω]) at the initial and the improved disposition as a function of frequency

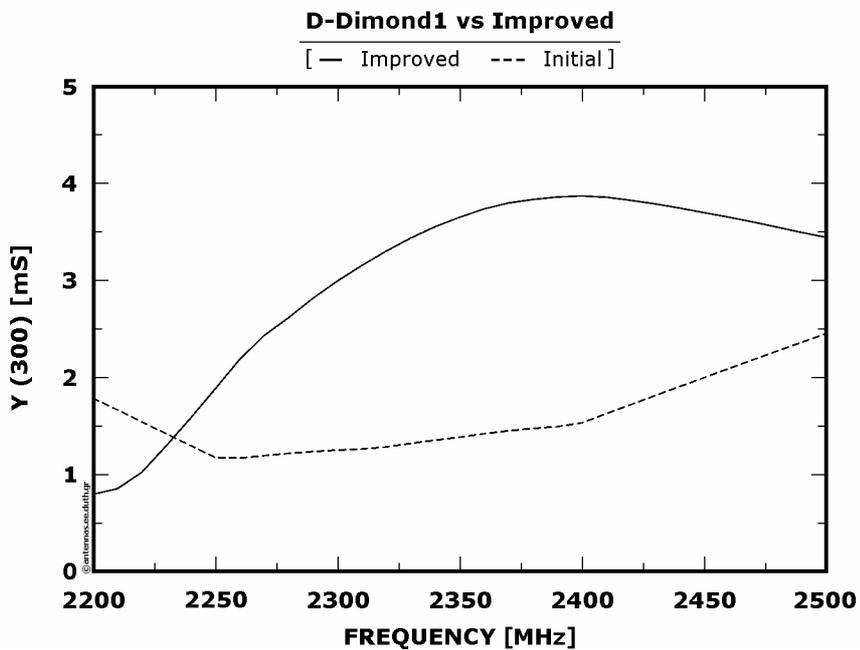

<u>Figure 8</u> : Normalized radiation intensity (with reference to 300 [Ω]) at the initial and the improved disposition as a function of frequency

**5.5.2 : Frequency range from 2550 to 2850 [MHz]**

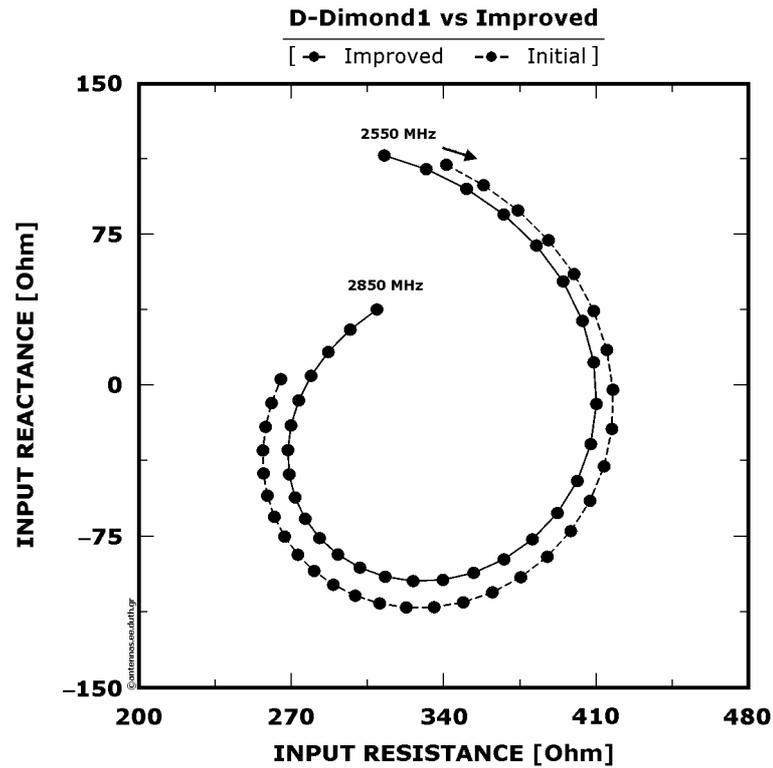

<u>Figure 1</u> : Input impedance of the initial and the improved disposition
as a function of frequency

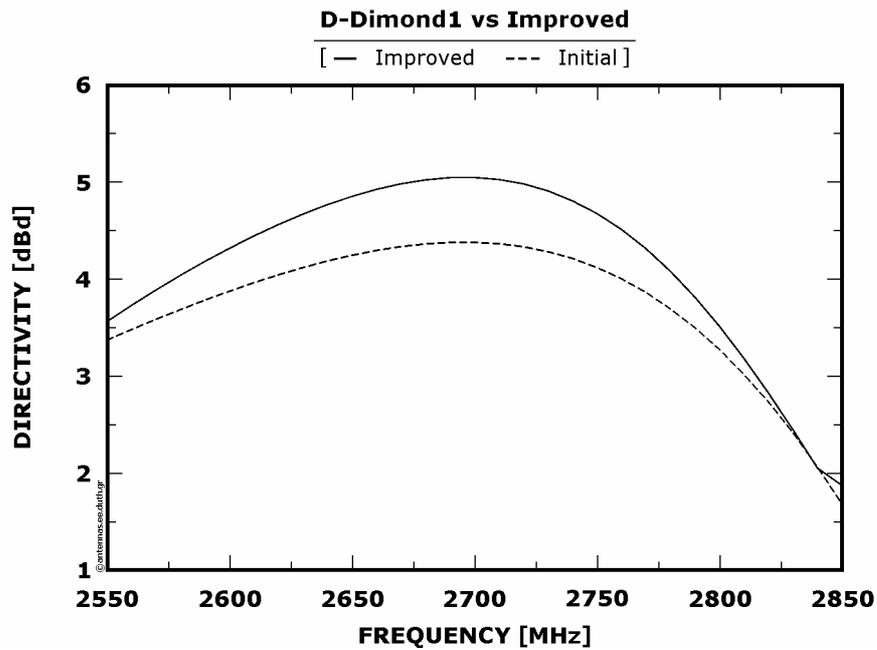

<u>Figure 2</u> : Directivity (with reference to dipole λ/2)of the initial and the improved disposition
as a function of frequency

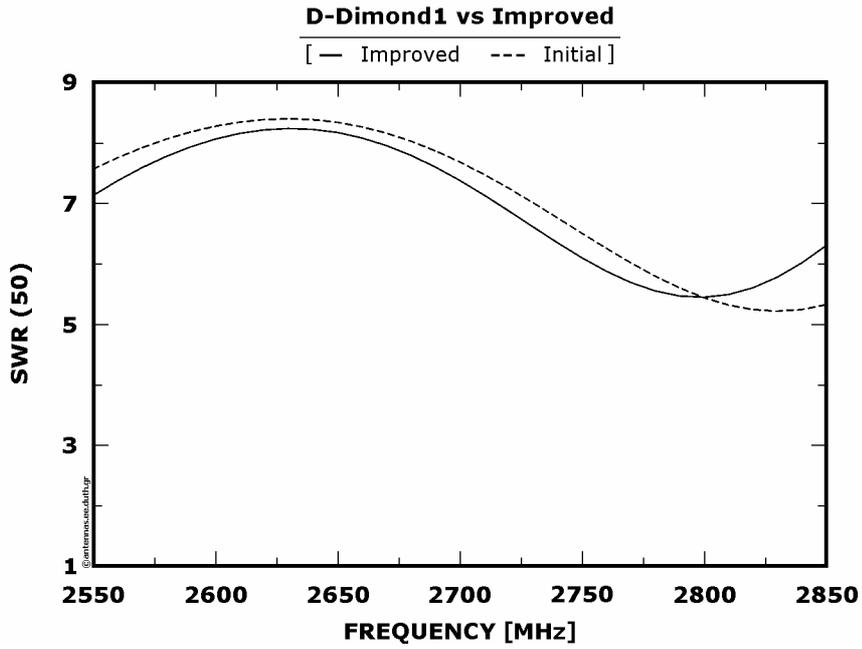

<u>Figure 3</u> : SWR (with reference to 50 [Ω]) at the initial and the improved disposition
as a function of frequency

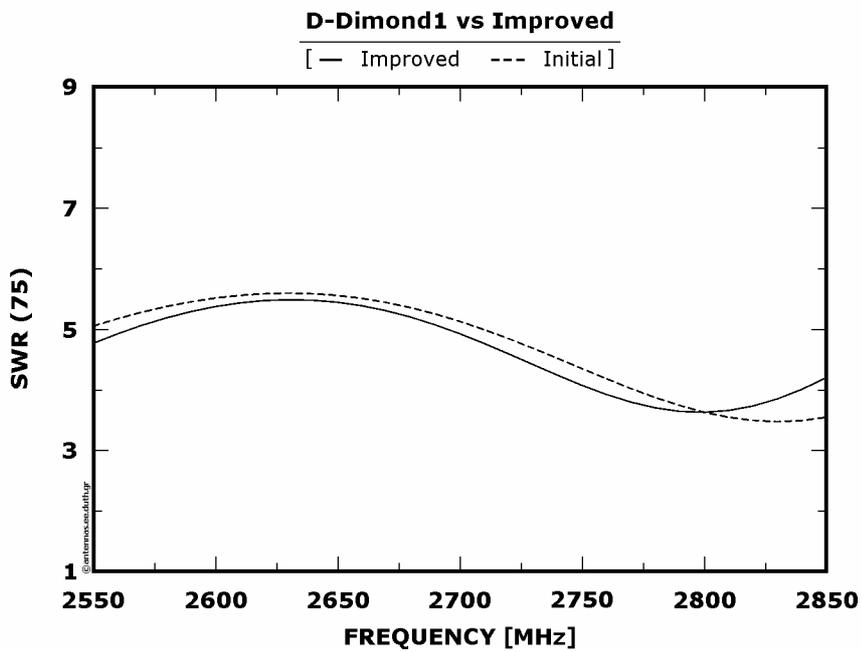

<u>Figure 4</u> : SWR (with reference to 75 [Ω]) at the initial and the improved disposition
as a function of frequency

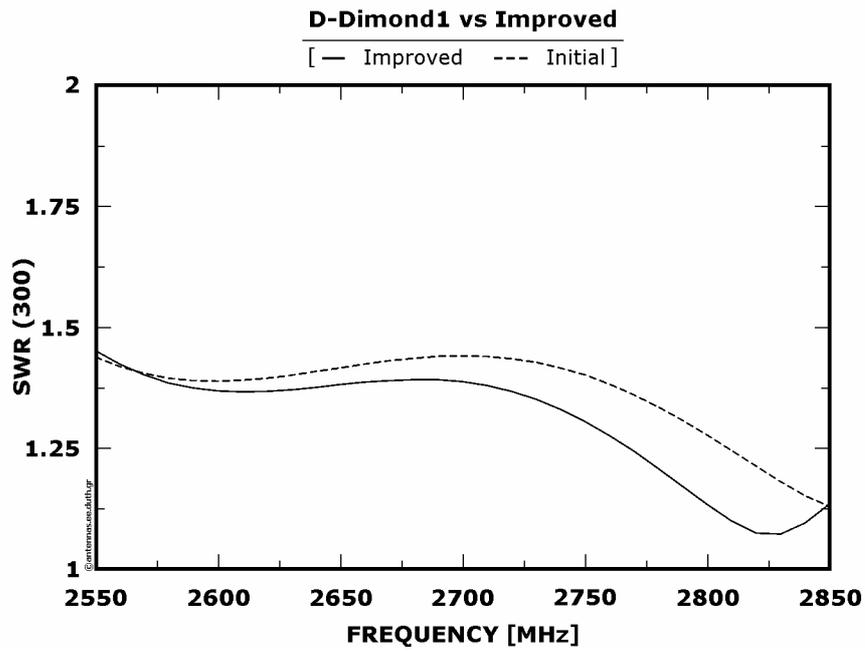

<u>Figure 5</u> : SWR (with reference to 300 [Ω]) at the initial and the improved disposition as a function of frequency

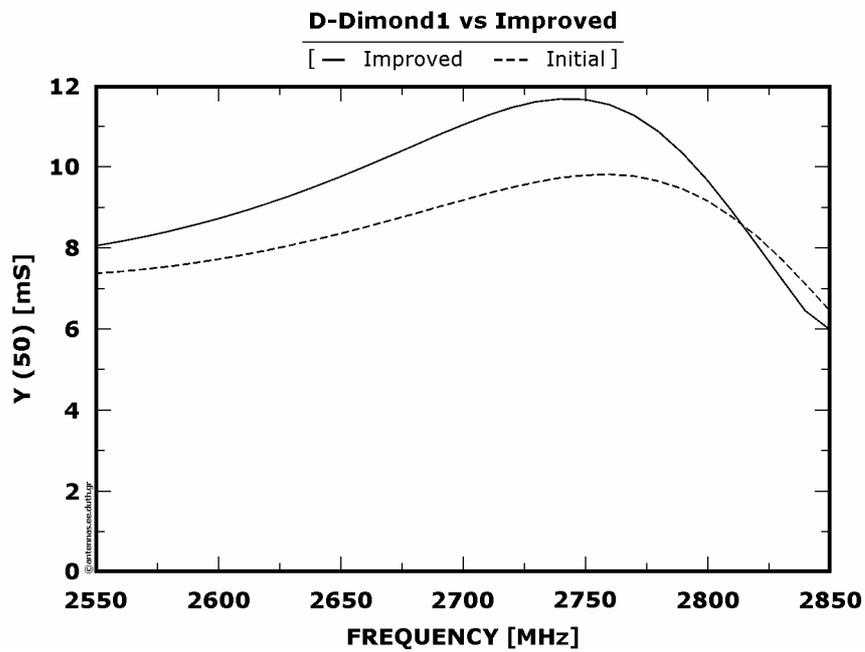

<u>Figure 6</u> : Normalized radiation intensity (with reference to 50 [Ω]) at the initial and the improved disposition as a function of frequency

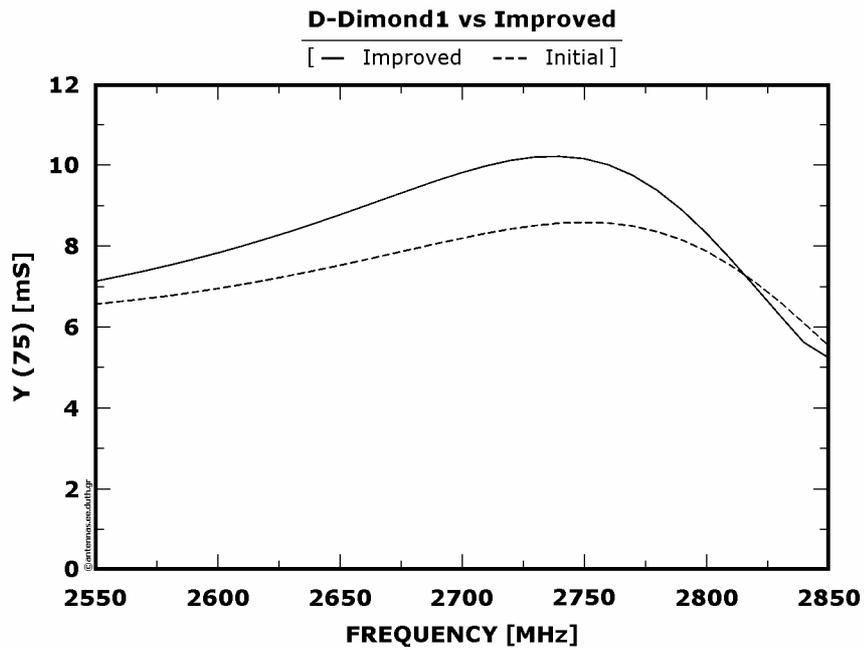

<u>Figure 7</u> : Normalized radiation intensity (with reference to 75 [$\Omega$]) at the initial and the improved disposition as a function of frequency

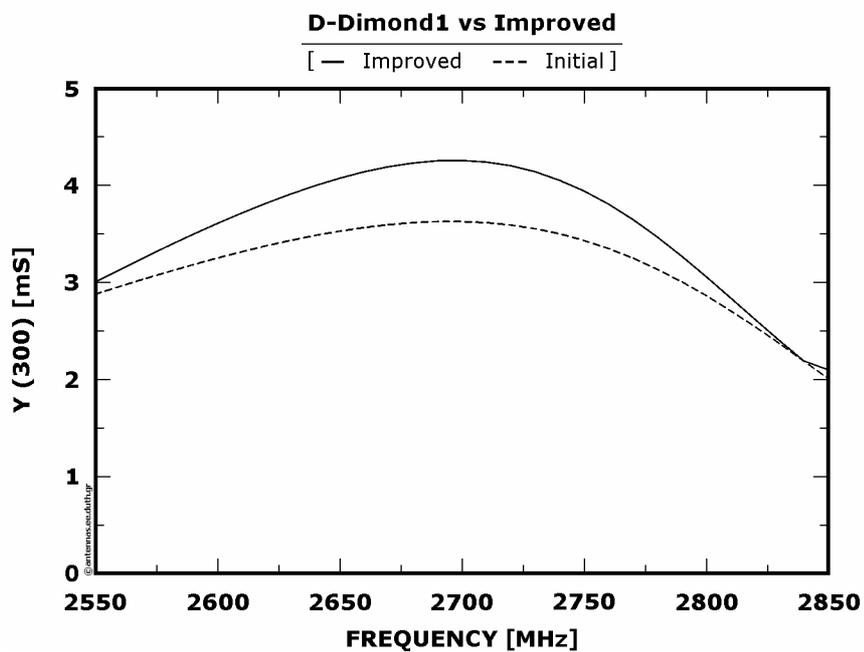

<u>Figure 8</u> : Normalized radiation intensity (with reference to 300 [$\Omega$]) at the initial and the improved disposition as a function of frequency

## 5.6 : Conclusions

Summing all the above, some useful conclusions are deduced regarding the operation of D-Dimond1's improved dispositions in comparison with the initial one at a frequency range from 200 to 2850 [MHz]. Below the appropriate tables are cited.

Comparing Table 5.6.1 and Table 4.3.4 of the improved and original modxdel respectively it infers that for a maximum acceptable SWR equals to 2 the antenna's connection with both a 50 [Ω] and 75 [Ω] transmission line leads to reduced bandwidth at the improved model comparing with the original model, that is 20 compared to 50 [MHz] and 80 compared to 100 [MHz] respectively. On the other hand, improved model's connection with a 300 [Ω] transmission line leads to a significant broader bandwidth, that is 180 compared to 60 [MHz] of the original model. Moreover, directivity's values are considerably high at the improved model compared to the original. On the other hand, there are greater variations which can be taken into consideration only for connection of a 75 [Ω] transmission line. For a design requiring a maximum SWR equals to 3 the connection of the improved antenna system with a 50 [Ω] transmission line continues to offer narrower bandwidth than the original disposition, that is 90 compared to 110 [MHz]. On the contrary, the utilization of a 75 [Ω] transmission line offers a broader bandwidth, which is 180 [MHz]. The usage of a 300 [Ω] transmission line keeps offering significantly greater bandwidth when connected to the improved model than to the original, that is 230 compared to 140 [MHz]. Even if directivity's values are appeared to be quite high at the improved model however they tend to fluctuate more. The usage of a 300 [Ω] transmission line could be an exception to the above statement which even if it offers great bandwidth, (230 [MHz]), causes the minimum directivity's fluctuation at the same time.

Comparing Table 5.6.2 with Table 4.3.5 does not arise practically any difference between the initial and the improved model. Even after the improving process there is no capability of connecting a transmission line with a different characteristic resistance in spite of a 300 [Ω]. The only differentiation from the above condition rests when operating the improved model at higher frequencies, where directivity takes higher values which however present greater fluctuations.

Table 1 : Frequency range from 2200 to 2500 [MHz]

| Acceptable SWR | SWR(50) | | SWR(75) | | SWR(300) | |
|---|---|---|---|---|---|---|
| | BW [MHz] | D [dBd] | BW [MHz] | D [dBd] | BW [MHz] | D [dBd] |
| 2 | 2260 − 2240 = 20 | 3.16 έως 3.50 | 2320 − 2240 = 80 | 3.16 έως 4.33 | 2500 − 2320 = 180 | 4.45 έως 4.16 |
| 3 | 2320 − 2230 = 90 | 3.26 έως 4.45 | 2410 − 2230 = 180 | 2.99 έως 4.71 | 2500 − 2270 = 230 | 3.62 έως 4.04 |

Table 2 : Frequency range from 2550 to 2850 [MHz]

| Acceptable SWR | SWR(50) | | SWR(75) | | SWR(300) | |
|---|---|---|---|---|---|---|
| | BW [MHz] | D [dBd] | BW [MHz] | D [dBd] | BW [MHz] | D [dBd] |
| 2 | -- | --- | --- | --- | 2850 − 2550 = 300 | 3.55 έως 1.87 |
| 3 | --- | --- | --- | --- | 2850 − 2550 = 300 | 3.55 έως 1.87 |

**Final conclusions**

Taking all the above into consideration some quite useful conclusions can be derived:

1. The improving models that have been applied do not ameliorate D-Dimond1's operation at the following frequency ranges: from 400 to 800 [MHz], from 1100 to 1500 [MHz] and from 1800 to 2200 [MHz]. Namely, the best operation at the above frequency ranges is achieved with the conservation of the antenna's initial geometrical shape.

2. At the frequency range from 2200 to 2500 [MHz] the best operation is achieved according to the modifications that the presented models produce. The improvement that was achieved consists in the remarkably greater value of the directivity but also in the broader bandwidth when a 300 [Ω] transmission line is used, as shown in Table 5.6.1. On the contrary, the bandwidth lessens when a connection of both a 50 [Ω] and 75 [Ω] transmission lines is utilized while directivity's values significantly increase however along with their fluctuations.

3. At the frequency range from 2550 to 2850 [MHz] the applied models were inefficient to produce remarkable practical improvements as there is still the incapacity of connecting any other transmission lines except a 300 [Ω] one, as shown in Table 5.6.2. Simultaneously an increase of directivity's values is achieved along with their greater fluctuations, inside the produced operating bandwidth.

4. The choice of the normalized radiation intensities (Y) as the determinant for the selection of the frequency ranges is judged as ineffective regarding the bandwidth improvement but is considered satisfactory regarding the remarkable increase of the antenna's directivity. This is the case because Y is strongly depended from directivity's value for the reason that Y is directly proportional to it. At the same time, due to Y's more complex relationship with SWR's value, which is not directly proportional, Y is less strongly depended.

# Annex A :      Radiation patterns in space

The below computer program that is written in language C carries out an estimation of the Radiation Pattern at space per $5°$. The program's output creates the file OUT3D.TXT, which is the input file of the computer program RADPAT. The following procedure is executed by the program: firstly the magnitude of $\overline{\mathbb{E}}$ is calculated, afterwards its maximum value is discovered and $\overline{\mathbb{E}}$ is normalized by this value, finally $\mathcal{E}_x$, $\mathcal{E}_y$, $\mathcal{E}_z$ are computed and printed at the output file.

```
#include <stdio.h>
#include <float.h>
#include <limits.h>
#include <math.h>
#include <stdlib.h>
#define pi 3.1415926535897932384626433832795
#define sq3 1.7320508075688772935274463415059

main()
{
 int i,j;
 double E[37][73],E_k[37][73],Ea1,Ea2,Ea3,Ea4,E1,E2,E3,E4,E5,E6,E7;
 double Eolth,Eolph a1,a2,a3,b1,b2,b3,c1,c2,c3,max,Ex,Ey,Ez,th,ph;
 FILE *diagramma;
 diagramma=fopen("out3d.txt","w");
 for (i=0;i<=36;++i)
 {
        for (j=0;j<=72;++j)
        {
                th=(i*5)*(pi/180.0);
                ph=(j*5)*(pi/180.0);
                if (th==0)
                {
                        th=0.001*(pi/180);
                }
                if (th==180*(pi/180))
                {
                th=179.999*(pi/180);
                }
```

```
if ((th==30*(pi/180))&&(ph==90*(pi/180))||
   (th==30*(pi/180))&&(ph==270*(pi/180)))
{
        th=29.999*(pi/180);
}
if ((th==90*(pi/180))&&(ph==90*(pi/180))||
   (th==90*(pi/180))&&(ph==270*(pi/180)))
{
        th=89.999*(pi/180);
}
if ((th==150*(pi/180))&&(ph==90*(pi/180))||
   (th==150*(pi/180))&&(ph==270*(pi/180)))
{
        th=149.999*(pi/180);
}

a1=(pi/2)*(sin(th)*sin(ph)-cos(th));
a2=(pi/2)*(sin(th)*sin(ph)+cos(th));
a3=(pi/2)*((1+sq3)*cos(th));
b1=(1.0/2)*(sin(th)*sin(ph)+sq3*cos(th));
c1=cos(th)*sin(ph)-sq3*sin(th);
b2=(1.0/2)*(sq3*cos(th)-sin(th)*sin(ph));
c2=cos(th)*sin(ph)+sq3*sin(th);
b3=sin(th)*sin(ph);
c3=cos(th)*sin(ph);

Ea1=2.0/sin(th);
Ea2=1.0/(1-pow(b1,2));
Ea3=1.0/(1-pow(b2,2));
Ea4=1.0/(1-pow(b3,2));

E1=cos((pi/2)*cos(th))*(cos(a2)-cos(a1));
E2=cos((pi/2)*b1)*(cos(a1)-cos(a3));
E3=cos((pi/2)*b2)*(cos(a2)-cos(a3));
E4=2*cos((pi/2)*b3);
E5=(cos(th)-sin((pi/2)*cos(th)))*(sin(a2)+sin(a1));
E6=(b1-sin((pi/2)*b1))*(sin(a3)-sin(a1));
E7=(b2-sin((pi/2)*b2))*(sin(a2)+sin(a3));

Eolth=(Ea1*E1)+(E2*Ea2*c1)+(E3*Ea3*c2)+
(E4*Ea4*c3)-(Ea1*E5)+(E6*Ea2*c1)+(E7*Ea3*c2);
Eolph=(cos(ph))*((E2*Ea2)+(E3*Ea3)+(E4*Ea4)+(E6*Ea2)+
(E7*Ea3));
```

```
                E[i][j]=sqrt(pow(Eolth,2)+pow(Eolph,2));
        }
}
max=0.0;
for (i=0;i<=36;++i)
{
        for (j=0;j<=72;++j)
        {
                if (E[i][j]>=max)
                {
                        max=E[i][j];
                }
        }
}
printf("%f \n",max);
system("pause");

for (i=0;i<=36;++i)
{
        for (j=0;j<=72;++j)
        {
                E_k[i][j]=E[i][j]/max;
        }
}
for (i=0;i<=36;++i)
{
        for (j=0;j<=72;++j)
        {
                th=(i*5.0)*(pi/180.0);
                ph=(j*5.0)*(pi/180.0);
                Ex=E_k[i][j]*sin(th)*cos(ph);
                Ey=E_k[i][j]*sin(th)*sin(ph);
                Ez=E_k[i][j]*cos(th);
                fprintf(diagramma,"%f \t %f \t %f \n ",Ex,Ey,Ez);
        }
}
fclose(diagramma);
}
```

## Annex B :   Radiation patterns at the three main planes

The below computer program that is written in language C carries out an estimation of the Radiation Pattern at the three main planes per $1°$. The program's output creates the file OUT2D.TXT, which is the input file of the computer program RADPAT. The following procedure is executed by the program: firstly the magnitude of $\overline{\overline{\mathbb{E}}}$ is calculated, afterwards its maximum value is discovered and $\overline{\overline{\mathbb{E}}}$ is normalized by this value, finally $\mathcal{E}_{xOy}$, $\mathcal{E}_{yOz}$, $\mathcal{E}_{zOx}$ are computed and printed at the output file.

```
#include <stdio.h>
#include <float.h>
#include <limits.h>
#include <math.h>
#include <stdlib.h>
#define pi 3.14159265358979323846426433832795
#define sq3 1.73205080756887729352742463415059
double En(double th, double ph);

main()
{
 int i,j;
 double th,ph,E,Ek,max1,max2,max3,max;
 FILE *da2d;
 da2d=fopen("out2d.txt","w");
 //**********MAXIMUM EVALUATION AT XOY*********************
 //**************TH=90 & PH=[0,360]****************
 max1=0.0;
 for (j=0;j<=360;++j)
        {
                ph=j*(pi/180.0);
                th=90*(pi/180.0);

                if ((ph==90*(pi/180))||(ph==270*(pi/180)))
                {
                        th=89.986*(pi/180);
                }
                E=En(th,ph);
                if (E>=max1)
                {
```

```c
                        max1=E;
                }
        }
}
printf("%f",max1);
system("pause");
//**********MAXIMUM EVALUATION AT ZOY***********************
//**************TH=[0,180] & PH=90,270***************
max2=0.0;
for (j=90;j<=270;j=j+180)
{
        for(i=0;i<=180;++i)
        {
                ph=j*(pi/180.0);
                th=i*(pi/180.0);
                if (th==0)
                {
                        th=0.001*(pi/180);
                }
                if (th==180*(pi/180))
                {
                        th=179.999*(pi/180);
                }
                if ((th==30*(pi/180))&&(ph==90*(pi/180))||
                (th==30*(pi/180))&&(ph==270*(pi/180)))
                {
                        th=29.999*(pi/180);
                }
                if ((th==90*(pi/180))&&(ph==90*(pi/180))||
                   (th==90*(pi/180))&&(ph==270*(pi/180)))
                {
                        th=89.999*(pi/180);
                }
                if ((th==150*(pi/180))&&(ph==90*(pi/180))||
                (th==150*(pi/180))&&(ph==270*(pi/180)))
                {
                        th=149.999*(pi/180);
                }
                E=En(th,ph);
                if(E>=max2)
                {
                        max2=E;
                }
        }
}
```

```c
printf("%f",max2);
system("pause");
//**********MAXIMUM EVALUATION AT XOZ***********************
//**************TH=[0,180] & PH=0,180***************
max3=0.0;
for (j=0;j<=180;j=j+180)
{
        for(i=0;i<=180;++i)
        {
                ph=j*(pi/180.0);
                th=i*(pi/180.0);
                if (th==0)
                {
                        th=0.001*(pi/180);
                }
                if (th==180*(pi/180))
                {
                        th=179.999*(pi/180);
                }
                E=En(th,ph);
                if(E>=max3)
                {
                        max3=E;
                }
        }
}
printf("%f",max3);
system("pause");
//**************MAXIMUM VALUE EVALUATION *************
max=max1;
if(max<=max2)
{
        max=max2;
}
if(max<=max3)
{
        max=max3;
}
printf("%f",max);
system("pause");
//*********************** XOY PLANE ************************
th=90*(pi/180.0);
for (j=0;j<=360;++j)
{
```

```
        ph=j*(pi/180.0);
        if ((ph==90*(pi/180))||(ph==270*(pi/180)))
        {
                th=89.986*(pi/180);
        }
        E=En(th,ph);
        Ek=E/max;
        fprintf(da2d,"%f \n ",Ek);
}
//*********************** YOZ PLANE ****************************
ph=90*(pi/180.0);
for(i=0;i<=180;++i)
{
        th=i*(pi/180.0);
        if (th==0)
        {
                th=0.001*(pi/180);
        }
        if (th==180*(pi/180))
        {
                th=179.999*(pi/180);
        }
        if ((th==30*(pi/180)))
        {
                th=29.999*(pi/180);
        }
        if ((th==90*(pi/180)))
        {
                th=89.999*(pi/180);
        }
        if ((th==150*(pi/180)))
        {
                th=149.999*(pi/180);
        }
        E=En(th,ph);
        Ek=E/max;
        fprintf(da2d,"%f \n", Ek);
}
ph=270*(pi/180.0);
for (i=179;i>=0;--i)
{
        th=i*(pi/180.0);
        if (th==0)
        {
```

```
                th=0.001*(pi/180);
        }
        if ((th==30*(pi/180)))
        {
                th=29.999*(pi/180);
        }
        if ((th==90*(pi/180)))
        {
                th=89.999*(pi/180);
        }
        if ((th==150*(pi/180)))
        {
                th=149.999*(pi/180);
        }
        E=En(th,ph);
        Ek=E/max;
        fprintf(da2d,"%f \n", Ek);
}
//******************** ZOX PLANE ****************************
ph=0;
for (i=0;i<=180;++i)
{
        th=i*(pi/180.0);
        if (th==0)
        {
                th=0.001*(pi/180);
        }
        if (th==180*(pi/180))
        {
                th=179.999*(pi/180);
        }
        E=En(th,ph);
        Ek=E/max;
        fprintf(da2d,"%f \n", Ek);
}
ph=180*(pi/180.0);
for (i=179;i>=0;--i)
{
        th=i*(pi/180.0);
        if (th==0)
        {
                th=0.001*(pi/180);
        }
        E=En(th,ph);
```

```c
        Ek=E/max;
        fprintf(da2d,"%f \n", Ek);
}
fclose(da2d);
}

double En(double th, double ph)
{
 double E1,E2,E3,E4,E5,E6,E7,Ea1,Ea2,Ea3,Ea4,Ex,Ey,Ez;
 double a1,a2,a3,b1,c1,b2,a5,c2,a6,b3,c3,Eolth,Eolph,En;

 a1=(pi/2)*(sin(th)*sin(ph)-cos(th));
 a2=(pi/2)*(sin(th)*sin(ph)+cos(th));
 a3=(pi/2)*((1+sq3)*cos(th));
 b1=(1.0/2)*(sin(th)*sin(ph)+sq3*cos(th));
 c1=cos(th)*sin(ph)-sq3*sin(th);
 b2=(1.0/2)*(sq3*cos(th)-sin(th)*sin(ph));
 c2=cos(th)*sin(ph)+sq3*sin(th);
 b3=sin(th)*sin(ph);
 c3=cos(th)*sin(ph);

 Ea1=2.0/sin(th);
 Ea2=1.0/(1-pow(b1,2));
 Ea3=1.0/(1-pow(b2,2));
 Ea4=1.0/(1-pow(b3,2));

 E1=cos((pi/2)*cos(th))*(cos(a2)-cos(a1));
 E2=cos((pi/2)*b1)*(cos(a1)-cos(a3));
 E3=cos((pi/2)*b2)*(cos(a2)-cos(a3));
 E4=2*cos((pi/2)*b3);
 E5=(cos(th)-sin((pi/2)*cos(th)))*(sin(a2)+sin(a1));
 E6=(b1-sin((pi/2)*b1))*(sin(a3)-sin(a1));
 E7=(b2-sin((pi/2)*b2))*(sin(a2)+sin(a3));

 Eolth=(Ea1*E1)+(E2*Ea2*c1)+(E3*Ea3*c2)+(E4*Ea4*c3)-(Ea1*E5)+(E6*Ea2*c1)
+(E7*Ea3*c2);
 Eolph=(cos(ph))*((E2*Ea2)+(E3*Ea3)+(E4*Ea4)+(E6*Ea2)+(E7*Ea3));

 En=sqrt(pow(Eolth,2)+pow(Eolph,2));
 return(En);
}
```

## Annex C :   Division of the antenna

The below computer program is written in language C, creates an output file called INPUT for every different number of segments that the antenna is being divided which is the input file at RICHWIRE.

```c
#include <stdio.h>
#include <float.h>
#include <limits.h>
#include <math.h>
#include <stdlib.h>
#define pi 3.14159265358979323846264433832795
#define sq3 1.7320508075688772935274463415059

main()
{
 int k,i,j,m,r,Z;
 float f;
 double x1,y1,z1,x2,y2,z2,x3,y3,z3,x4,y4,z4,x5,y5,z5,x6,y6,z6,x7,y7,z7;
 double x8,y8,z8,x9,y9,z9,x10,y10,z10,x11,y11,z11,x12,y12,z12,x13,y13,
 double orio,x,y,z,Lq,z13,min_L;
 FILE *coor;
 coor=fopen("input.txt","w");
 printf("\n GIVE THE NUMBER OF SEGMENT'S SUB-SEGMENT l/4     [>=2]: ");
 scanf("%d",&m);
 if(m<2)
        {
        printf("WRONG VALUE! \n");
        system("pause");
        exit;
        }
 printf("GIVE THE NUMBER OF THE SUB-SEGMENTS' DIVISION : %d \n",m);
 system("pause");
 f=1111.0;
 Lq=(1.0/4)*(300.0/1111.0);            //CALCULATION OF THE LENGTH
Lamda/4
 orio=0.002;            //MINIMUM ACCEPTABLE SEGMENT LENGTH LIMIT
 min_L=Lq/m;            //MINIMUM ACCEPTABLE SEGMENT LENGTH
 Z=14*m;
//***CHECK OF THE MINIMUM SEGMENT LENGTH******
 if(min_L<=orio)
 {
        printf("DECREASE NUMBER OF DIVISION - LIMIT EXCEEDED!** \n");
        system("pause");
```

```
        exit;
    }
//***COORDINATE DISCOVERY OF THE STABLE POINTS***
    //SEGMENT 1
x1=0.0;
y1=0.0;
z1=0.0;
    //SEGMENT 2
x2=0.0;
y2=-Lq;
z2=0.0;
    //SEGMENT 3
x3=0.0;
y3=-Lq;
z3=Lq;
    //SEGMENT 4
x4=0.0;
y4=-Lq/2;
z4=Lq+sin(pi/3)*Lq;
    //SEGMENT 5
x5=0.0;
y5=0.0;
z5=Lq+sin(pi/3)*2*Lq;
    //SEGMENT 6
x6=0.0;
y6=Lq/2;
z6=Lq+sin(pi/3)*Lq;
    //SEGMENT 7
x7=0.0;
y7=Lq;
z7=Lq;
    //SEGMENT 8
x8=0.0;
y8=Lq;
z8=0.0;
    //SEGMENT 9
x9=0.0;
y9=Lq;
z9=-Lq;
    //SEGMENT 10
x10=0.0;
y10=Lq/2;
z10=-Lq-sin(pi/3)*Lq;
    //SEGMENT 11

x11=0.0;
y11=0.0;
z11=-Lq-sin(pi/3)*2*Lq;
    //SEGMENT 12
x12=0.0;
y12=-Lq/2;
z12=-Lq-sin(pi/3)*Lq;
    //SEGMENT 13
x13=0.0;
y13=-Lq;
z13=-Lq;

fprintf(coor,"%4.1f \n",f);
fprintf(coor,"-1.0 0.001 \n");
fprintf(coor,"%d \n",Z-1);

fprintf(coor,"1    %1.7f    %1.7f    %1.7f \n",x1,y1,z1);
fprintf(coor,"2    %1.7f    %1.7f    %1.7f \n",x2,y2,z2);
fprintf(coor,"3    %1.7f    %1.7f    %1.7f \n",x3,y3,z3);
fprintf(coor,"4    %1.7f    %1.7f    %1.7f \n",x4,y4,z4);
fprintf(coor,"5    %1.7f    %1.7f    %1.7f \n",x5,y5,z5);
fprintf(coor,"6    %1.7f    %1.7f    %1.7f \n",x6,y6,z6);
fprintf(coor,"7    %1.7f    %1.7f    %1.7f \n",x7,y7,z7);
fprintf(coor,"8    %1.7f    %1.7f    %1.7f \n",x8,y8,z8);
fprintf(coor,"9    %1.7f    %1.7f    %1.7f \n",x9,y9,z9);
fprintf(coor,"10   %1.7f    %1.7f    %1.7f \n",x11,y10,z10);
fprintf(coor,"11   %1.7f    %1.7f    %1.7f \n",x11,y11,z11);
fprintf(coor,"12   %1.7f    %1.7f    %1.7f \n",x12,y12,z12);
fprintf(coor,"13   %1.7f    %1.7f    %1.7f \n",x13,y13,z13);
//***COORDINATE ESTIMATION OF THE NON-STABLE POINTS***
r=14;        //CONSECUTIVE NUMBER KOMBWN
    //SEGMENT 1
for (i=1;i<=m-1;++i)
{
        x=0.0;
        y=y1+((i*Lq)/m);
        z=0.0;
        fprintf(coor,"%d  %1.7f    %1.7f    %1.7f \n",r,x,y,z);
        r=r+1;
}
    //SEGMENT 2
for (i=1;i<=m-1;++i)
{
x=0.0;
```

```
y=y7;
z=z7-((i*Lq)/m);
fprintf(coor,"%d  %1.7f   %1.7f   %1.7f \n",r,x,y,z);
r=r+1;
}
//SEGMENT 3
for (i=1;i<=m-1;++i)
{
     x=0.0;
     y=y6+((i*Lq)/m)*sin(pi/6);
     z=z6-((i*Lq)/m)*cos(pi/6);
     fprintf(coor,"%d  %1.7f   %1.7f   %1.7f \n",r,x,y,z);
     r=r+1;
}
//SEGMENT 4
for (i=1;i<=m-1;++i)
{
     x=0.0;
     y=y6-((i*Lq)/m)*sin(pi/6);
     z=z6+((i*Lq)/m)*cos(pi/6);
     fprintf(coor,"%d  %1.7f   %1.7f   %1.7f \n",r,x,y,z);
     r=r+1;
}
//SEGMENT 5
for (i=1;i<=m-1;++i)
{
     x=0.0;
     y=y5-((i*Lq)/m)*cos(pi/3);
     z=z5-((i*Lq)/m)*sin(pi/3);
     fprintf(coor,"%d  %1.7f   %1.7f   %1.7f \n",r,x,y,z);
     r=r+1;
}
//SEGMENT 6
for (i=1;i<=m-1;++i)
{
x=0.0;
y=y3+((i*Lq)/m)*cos(pi/3);
z=z3+((i*Lq)/m)*sin(pi/3);
fprintf(coor,"%d  %1.7f   %1.7f   %1.7f \n",r,x,y,z);
r=r+1;
}
//SEGMENT 7
for (i=1;i<=m-1;++i)
{
```

```
     x=0.0;
     y=y2;
     z=z2+(i*Lq)/m;
     fprintf(coor,"%d  %1.7f   %1.7f   %1.7f \n",r,x,y,z);
     r=r+1;
}
//SEGMENT 8
for (i=1;i<=m-1;++i)
{
     x=0.0;
     y=y2;
     z=z2-((i*Lq)/m);
     fprintf(coor,"%d  %1.7f   %1.7f   %1.7f \n",r,x,y,z);
     r=r+1;
}
//SEGMENT 9
for (i=1;i<=m-1;++i)
{
     x=0.0;
     y=y13+((i*Lq)/m)*cos(pi/3);
     z=z13-((i*Lq)/m)*sin(pi/3);
     fprintf(coor,"%d  %1.7f   %1.7f   %1.7f \n",r,x,y,z);
     r=r+1;
}
//SEGMENT 10
for (i=1;i<=m-1;++i)
{
     x=0.0;
     y=y11-((i*Lq)/m)*cos(pi/3);
     z=z11+((i*Lq)/m)*sin(pi/3);
     fprintf(coor,"%d  %1.7f   %1.7f   %1.7f \n",r,x,y,z);
     r=r+1;
}
//SEGMENT 11
for (i=1;i<=m-1;++i)
{
     x=0.0;
     y=y10-((i*Lq)/m)*cos(pi/3);
     z=z10-((i*Lq)/m)*sin(pi/3);
     fprintf(coor,"%d  %1.7f   %1.7f   %1.7f \n",r,x,y,z);
     r=r+1;
}
//SEGMENT 12
for (i=1;i<=m-1;++i)
```

```
{
        x=0.0;
        y=y10+((i*Lq)/m)*cos(pi/3);
        z=z10+((i*Lq)/m)*sin(pi/3);
        fprintf(coor,"%d  %1.7f    %1.7f    %1.7f \n",r,x,y,z);
        r=r+1;
}
//SEGMENT 13
for (i=1;i<=m-1;++i)
{
        x=0.0;
        y=y9;
        z=z9+((i*Lq)/m);
        fprintf(coor,"%d  %1.7f    %1.7f    %1.7f \n",r,x,y,z);
        r=r+1;
}
//SEGMENT 14
for (i=1;i<=m-1;++i)
{
        x=0.0;
        y=y2+((i*Lq)/m);
        z=0.0;
        fprintf(coor,"%d  %1.7f    %1.7f    %1.7f \n",r,x,y,z);
        r=r+1;
}
fprintf(coor,"%d \n",Z);
//****SEGMENTS DECLARATION***
j=1;                    //CONSECUTIVE NUMBER
k=13;          //SEGMENTS NUMERATION
//SEGMENT 1
fprintf(coor,"1   1    %d \n",k+1);
k=k+1;
if(m>2)
{
        for (i=1;i<=m-2;++i)
        {
                j=j+1;
                fprintf(coor,"%d    %d    %d  \n",j,k,k+1);
                k=k+1;
        }
}
fprintf(coor,"%d    %d    8 \n",j+1,k);
//SEGMENT 2
fprintf(coor,"%d    7    %d \n",j+2,k+1);
```

```
j=j+2;
k=k+1;
if (m>2)
{
        for (i=1;i<=m-2;++i)
        {
                j=j+1;
                fprintf(coor,"%d    %d    %d  \n",j,k,k+1);
                k=k+1;
        }
}
fprintf(coor,"%d    %d    8 \n",j+1,k);
//SEGMENT 3
fprintf(coor,"%d    6    %d \n",j+2,k+1);
j=j+2;
k=k+1;
if (m>2)
{
        for (i=1;i<=m-2;++i)
        {
                j=j+1;
                fprintf(coor,"%d    %d    %d  \n",j,k,k+1);
                k=k+1;
        }
}
fprintf(coor,"%d    %d    7\n",j+1,k);
//SEGMENT 4
fprintf(coor,"%d    6    %d \n",j+2,k+1);
j=j+2;
k=k+1;
if (m>2)
{
        for (i=1;i<=m-2;++i)
        {
                j=j+1;
                fprintf(coor,"%d    %d    %d  \n",j,k,k+1);
                k=k+1;
        }
}
fprintf(coor,"%d    %d    5\n",j+1,k);
//SEGMENT 5
fprintf(coor,"%d    5    %d \n",j+2,k+1);
j=j+2;
k=k+1;
```

```c
if(m>2)
{
        for (i=1;i<=m-2;++i)
        {
                j=j+1;
                fprintf(coor,"%d   %d   %d \n",j,k,k+1);
                k=k+1;
        }
}
fprintf(coor,"%d   %d   4 \n",j+1,k);
//SEGMENT 6
fprintf(coor,"%d   3   %d \n",j+2,k+1);
j=j+2;
k=k+1;
if(m>2)
{
        for (i=1;i<=m-2;++i)
        {
                j=j+1;
                fprintf(coor,"%d   %d   %d \n",j,k,k+1);
                k=k+1;
        }
}
fprintf(coor,"%d   %d   4 \n",j+1,k);
//SEGMENT 7
fprintf(coor,"%d   2   %d \n",j+2,k+1);
j=j+2;
k=k+1;
if (m>2)
{
        for (i=1;i<=m-2;++i)
        {
                j=j+1;
                fprintf(coor,"%d   %d   %d \n",j,k,k+1);
                k=k+1;
        }
}
fprintf(coor,"%d   %d   3\n",j+1,k);
//SEGMENT 8
fprintf(coor,"%d   2   %d \n",j+2,k+1);
j=j+2;
k=k+1;
if(m>2)
{
```

```c
        for (i=1;i<=m-2;++i)
        {
                j=j+1;
                fprintf(coor,"%d   %d   %d \n",j,k,k+1);
                k=k+1;
        }
}
fprintf(coor,"%d   %d   13\n",j+1,k);
//SEGMENT 9
fprintf(coor,"%d   13   %d \n",j+2,k+1);
j=j+2;
k=k+1;
if(m>2)
{
        for (i=1;i<=m-2;++i)
        {
                j=j+1;
                fprintf(coor,"%d   %d   %d \n",j,k,k+1);
                k=k+1;
        }
}
fprintf(coor,"%d   %d   12 \n",j+1,k);
//SEGMENT 10
fprintf(coor,"%d   11   %d \n",j+2,k+1);
j=j+2;
k=k+1;
if(m>2)
{
        for (i=1;i<=m-2;++i)
        {
                j=j+1;
                fprintf(coor,"%d   %d   %d \n",j,k,k+1);
                k=k+1;
        }
}
fprintf(coor,"%d   %d   12 \n",j+1,k);
//SEGMENT 11
fprintf(coor,"%d   10   %d \n",j+2,k+1);
j=j+2;
k=k+1;
if(m>2)
{
        for (i=1;i<=m-2;++i)
        {
```

```
        j=j+1;
        fprintf(coor,"%d   %d   \n",j,k,k+1);
        k=k+1;
    }
}
fprintf(coor,"%d   %d   11\n",j+1,k);
//SEGMENT 12
fprintf(coor,"%d   10   %d \n",j+2,k+1);
j=j+2;
k=k+1;
if(m>2)
{
    for (i=1;i<=m-2;++i)
    {
        j=j+1;
        fprintf(coor,"%d   %d   %d \n",j,k,k+1);
        k=k+1;
    }
}
fprintf(coor,"%d   %d   9\n",j+1,k);
//SEGMENT 13
fprintf(coor,"%d   9   %d \n",j+2,k+1);
j=j+2;
k=k+1;
if(m>2)
{
    for (i=1;i<=m-2;++i)
    {
        j=j+1;
        fprintf(coor,"%d   %d   %d \n",j,k,k+1);
        k=k+1;
    }
}
fprintf(coor,"%d   %d   8 \n",j+1,k);
//SEGMENT 14
fprintf(coor,"%d   2   %d \n",j+2,k+1);
j=j+2;
k=k+1;
if(m>2)
{
    for (i=1;i<=m-2;++i)
    {
        j=j+1;
        fprintf(coor,"%d   %d   %d \n",j,k,k+1);
```

```
        k=k+1;
    }
}
fprintf(coor,"%d   %d   1 \n",j+1,k);
//*******************************//
fprintf(coor, "%d \n",1);
fprintf (coor,"0 \n");
fprintf(coor,"1  1.0  1.0  1.0 \n");
fprintf(coor,"1 90.  0.  90. 180.");
}
```

# Annex D : Antenna's improvement (1<sup>st</sup> model)

The below computer program written in C creates an output file called INPUT which is RICHWIRE's input file. The current program results to the displacement of the active dipole towards the positive z axis. A detailed description of the model is given at <u>Section 5.1</u>.

```
#include <stdio.h>
#include <float.h>
#include <limits.h>
#include <math.h>
#include <stdlib.h>
#define pi 3.1415926535897932384626433832795
#define sq3 1.7320508075688772935274463415059
main()
{
 int k,i,j,m,r,Z;
 long int h;
 float f;
 double x1,y1,z1,x2,y2,z2,x3,y3,z3,x4,y4,z4,x5,y5,z5,x6,y6,z6,x7,y7;
 double x8,y8,z8,x9,y9,z9,x10,y10,z10,x11,y11,z11,x12,y12, z7,x,y,z,Lq;
 double xa1,xa2,ya1,ya2,za1,za2,z12,x13,y13,z13;
 FILE *coor;
 coor=fopen("input.txt","w");
 f=2330.0;
 m=4;
 Lq=(1.0/4)*(300.0/1111.0);        //CALCULATION OF THE LENGTH Lamda/4
 Z=56;
 printf("GIVE THE DISPLACEMENT GRADE OF ACTIVE DIPOLE [0 32]: ");
 scanf("%d",&h);
 if(h<0||h>32)
```

```
{                                          //NODE 9
      printf("WRONG VALUE! \n");           x9=0.0;
      system("pause");                     y9=Lq;
      exit;                                z9=-Lq;
}                                          //NODE 10
else                                       x10=0.0;
{                                          y10=Lq/2;
      printf("GIVE INPUT VALUE : %d",h);   z10=-Lq-sin(pi/3)*Lq;
      system("pause");                     //NODE 11
      exit;                                x11=0.0;
}                                          y11=0.0;
//***COORDINATE DISCOVERY OF THE STABLE POINTS***   z11=-Lq-sin(pi/3)*2*Lq;
//NODE 1                                   //NODE 12
x1=0.0;                                    x12=0.0;
y1=0.0;                                    y12=-Lq/2;
z1=0.0;                                    z12=-Lq-sin(pi/3)*Lq;
//NODE 2                                   //NODE 13
x2=0.0;                                    x13=0.0;
y2=-Lq;                                    y13=-Lq;
z2=0.0;                                    z13=-Lq;
//NODE 3                                   //**MOVING NODE1**
x3=0.0;                                    xa1=0.0;
y3=-Lq;                                    ya1=Lq;
z3=Lq;                                     za1=h*(Lq/32.0);
//NODE 4                                   //**MOVING NODE2**
x4=0.0;                                    xa2=0.0;
y4=-Lq/2;                                  ya2=-Lq;
z4=Lq+sin(pi/3)*Lq;                        za2=h*(Lq/32.0);
//NODE 5
x5=0.0;                                    switch(h)
y5=0.0;                                    {
z5=Lq+sin(pi/3)*2*Lq;                      case 0:
//NODE 6                                   case 8:
x6=0.0;                                    case 16:
y6=Lq/2;                                   case 24:
z6=Lq+sin(pi/3)*Lq;                        case 32:
//NODE 7                                          fprintf(coor,"%4.1f \n",f);
x7=0.0;                                           fprintf(coor,"-1.0 0.001 \n");
y7=Lq;                                            fprintf(coor,"%d \n",Z-1);
z7=Lq;                                            break;
//NODE 8                                   default:
x8=0.0;                                           fprintf(coor,"%4.1f \n",f);
y8=Lq;                                            fprintf(coor,"-1.0 0.001 \n");
z8=0.0;                                           fprintf(coor,"%d \n",Z+1);
```

```c
        break;
}
fprintf(coor,"1    %1.7f    %1.7f    %1.7f \n",x1,y1,za1);
fprintf(coor,"2    %1.7f    %1.7f    %1.7f \n",x2,y2,z2);
fprintf(coor,"3    %1.7f    %1.7f    %1.7f \n",x3,y3,z3);
fprintf(coor,"4    %1.7f    %1.7f    %1.7f \n",x4,y4,z4);
fprintf(coor,"5    %1.7f    %1.7f    %1.7f \n",x5,y5,z5);
fprintf(coor,"6    %1.7f    %1.7f    %1.7f \n",x6,y6,z6);
fprintf(coor,"7    %1.7f    %1.7f    %1.7f \n",x7,y7,z7);
fprintf(coor,"8    %1.7f    %1.7f    %1.7f \n",x8,y8,z8);
fprintf(coor,"9    %1.7f    %1.7f    %1.7f \n",x9,y9,z9);
fprintf(coor,"10   %1.7f    %1.7f    %1.7f \n",x10,y10,z10);
fprintf(coor,"11   %1.7f    %1.7f    %1.7f \n",x11,y11,z11);
fprintf(coor,"12   %1.7f    %1.7f    %1.7f \n",x12,y12,z12);
fprintf(coor,"13   %1.7f    %1.7f    %1.7f \n",x13,y13,z13);
//***COORDINATE ESTIMATION OF THE NON-STABLE POINTS***
r=14;        //CONSECUTIVE NUMBER KOMBWN
//SEGMENT 1
for (i=1;i<=m-1;++i)
{
        x=0.0;
        y=y1+((i*Lq)/m);
        z=za1;
        fprintf(coor,"%d %1.7f    %1.7f    %1.7f \n",r,x,y,z);
        r=r+1;
}
//SEGMENT 2
for (i=1;i<=m-1;++i)
{
        x=0.0;
        y=y7;
        z=Lq-((i*Lq)/m);
        fprintf(coor,"%d %1.7f    %1.7f    %1.7f \n",r,x,y,z);
        r=r+1;
}
//SEGMENT 3
for (i=1;i<=m-1;++i)
{
        x=0.0;
        y=y6+((i*Lq)/m)*sin(pi/6);
        z=z6-((i*Lq)/m)*cos(pi/6);
        fprintf(coor,"%d %1.7f    %1.7f    %1.7f \n",r,x,y,z);
        r=r+1;
}
//SEGMENT 4
for (i=1;i<=m-1;++i)
{
        x=0.0;
        y=y6-((i*Lq)/m)*sin(pi/6);
        z=z6+((i*Lq)/m)*cos(pi/6);
        fprintf(coor,"%d %1.7f    %1.7f    %1.7f \n",r,x,y,z);
        r=r+1;
}
//SEGMENT 5
for (i=1;i<=m-1;++i)
{
        x=0.0;
        y=y5-((i*Lq)/m)*cos(pi/3);
        z=z5-((i*Lq)/m)*sin(pi/3);
        fprintf(coor,"%d %1.7f    %1.7f    %1.7f \n",r,x,y,z);
        r=r+1;
}
//SEGMENT 6
for (i=1;i<=m-1;++i)
{
        x=0.0;
        y=y3+((i*Lq)/m)*cos(pi/3);
        z=z3+((i*Lq)/m)*sin(pi/3);
        fprintf(coor,"%d %1.7f    %1.7f    %1.7f \n",r,x,y,z);
        r=r+1;
}
//SEGMENT 7
for (i=1;i<=m-1;++i)
{
        x=0.0;
        y=y2;
        z=(i*Lq)/m;
        fprintf(coor,"%d %1.7f    %1.7f    %1.7f \n",r,x,y,z);
        r=r+1;
}
//SEGMENT 8
for (i=1;i<=m-1;++i)
{
        x=0.0;
        y=y2;
        z=-(i*Lq)/m;
        fprintf(coor,"%d %1.7f    %1.7f    %1.7f \n",r,x,y,z);
        r=r+1;
```

```c
}
//SEGMENT 9
for (i=1;i<=m-1;++i)
{
        x=0.0;
        y=y13+((i*Lq)/m)*cos(pi/3);
        z=z13-((i*Lq)/m)*sin(pi/3);
        fprintf(coor,"%d  %1.7f    %1.7f    %1.7f \n",r,x,y,z);
        r=r+1;
}
//SEGMENT 10
for (i=1;i<=m-1;++i)
{
        x=0.0;
        y=y11-((i*Lq)/m)*cos(pi/3);
        z=z11+((i*Lq)/m)*sin(pi/3);
        fprintf(coor,"%d  %1.7f    %1.7f    %1.7f \n",r,x,y,z);
        r=r+1;
}
//SEGMENT 11
for (i=1;i<=m-1;++i)
{
        x=0.0;
        y=y10-((i*Lq)/m)*cos(pi/3);
        z=z10-((i*Lq)/m)*sin(pi/3);
        fprintf(coor,"%d  %1.7f    %1.7f    %1.7f \n",r,x,y,z);
        r=r+1;
}
//SEGMENT 12
for (i=1;i<=m-1;++i)
{
        x=0.0;
        y=y10+((i*Lq)/m)*cos(pi/3);
        z=z10+((i*Lq)/m)*sin(pi/3);
        fprintf(coor,"%d  %1.7f    %1.7f    %1.7f \n",r,x,y,z);
        r=r+1;
}
//SEGMENT 13
for (i=1;i<=m-1;++i)
{
        x=0.0;
        y=y9;
        z=z9+((i*Lq)/m);
        fprintf(coor,"%d  %1.7f    %1.7f    %1.7f \n",r,x,y,z);
```

```c
        r=r+1;
}
//SEGMENT 14
for (i=1;i<=m-1;++i)
{
        x=0.0;
        y=y2+((i*Lq)/m);
        z=za2;
        fprintf(coor,"%d  %1.7f    %1.7f    %1.7f \n",r,x,y,z);
        r=r+1;
}
switch(h)
{
case 0:
case 8:
case 16:
case 24:
case 32:
        fprintf(coor,"%d \n",Z);
        break;
default:
        fprintf(coor,"56  %1.7f    %1.7f    %1.7f \n",xa1,ya1,za1);
        fprintf(coor,"57  %1.7f    %1.7f    %1.7f \n",xa2,ya2,za2);
        fprintf(coor,"%d \n",Z+2);
        break;
}
//****SEGMENTS DECLARATION***

j=1;                //CONSECUTIVE NUMBER
k=13;        //SEGMENTS NUMERATION
//SEGMENT 1
fprintf(coor,"1   1    %d \n",k+1);
k=k+1;
if(m>2)
{
        for (i=1;i<=m-2;++i)
        {
                j=j+1;
                fprintf(coor,"%d    %d    %d  \n",j,k,k+1);
                k=k+1;
        }
}
switch (h)
{
```

Π-13

```
case 0:
        fprintf(coor,"%d    %d    8 \n",j+1,k);
        break;
case 8:
        fprintf(coor,"%d    %d    19 \n",j+1,k);
        break;
case 16:
        fprintf(coor,"%d    %d    18 \n",j+1,k);
        break;
case 24:
        fprintf(coor,"%d    %d    17 \n",j+1,k);
        break;
case 32:
        fprintf(coor,"%d    %d    7 \n",j+1,k);
        break;
default:
        fprintf(coor,"%d    %d    56 \n",j+1,k);
        break;
}
//SEGMENT 2
switch(h)
{
case 1:
case 2:
case 3:
case 4:
case 5:
case 6:
case 7:
        fprintf(coor,"%d    7    17 \n",j+2);
        fprintf(coor,"%d    17    18 \n",j+3);
        fprintf(coor,"%d    18    19 \n",j+4);
        fprintf(coor,"%d    19    56 \n",j+5);
        fprintf(coor,"%d    56    8 \n",j+6);
        j=j+5;
        k=k+3;
        break;
case 9:
case 10:
case 11:
case 12:
case 13:
case 14:
case 15:

        fprintf(coor,"%d    7    17 \n",j+2);
        fprintf(coor,"%d    17    18 \n",j+3);
        fprintf(coor,"%d    18    56 \n",j+4);
        fprintf(coor,"%d    56    19 \n",j+5);
        fprintf(coor,"%d    19    8 \n",j+6);
        j=j+5;
        k=k+3;
        break;
case 17:
case 18:
case 19:
case 20:
case 21:
case 22:
case 23:
        fprintf(coor,"%d    7    17 \n",j+2);
        fprintf(coor,"%d    17    56 \n",j+3);
        fprintf(coor,"%d    56    18 \n",j+4);
        fprintf(coor,"%d    18    19 \n",j+5);
        fprintf(coor,"%d    19    8 \n",j+6);
        j=j+5;
        k=k+3;
        break;
case 25:
case 26:
case 27:
case 28:
case 29:
case 30:
case 31:
        fprintf(coor,"%d    7    56 \n",j+2);
        fprintf(coor,"%d    56    17 \n",j+3);
        fprintf(coor,"%d    17    18 \n",j+4);
        fprintf(coor,"%d    18    19 \n",j+5);
        fprintf(coor,"%d    19    8 \n",j+6);
        j=j+5;
        k=k+3;
        break;
        default:
        fprintf(coor,"%d    7    17 \n",j+2);
        fprintf(coor,"%d    17    18 \n",j+3);
        fprintf(coor,"%d    18    19 \n",j+4);
        fprintf(coor,"%d    19    8 \n",j+5);
        j=j+4;
```

```c
        k=k+3;
        break;
}
//SEGMENT 3
fprintf(coor,"%d   6    %d \n",j+2,k+1);
j=j+2;
k=k+1;
if (m>2)
{
        for (i=1;i<=m-2;++i)
                {
                        j=j+1;
                        fprintf(coor,"%d   %d    %d  \n",j,k,k+1);
                        k=k+1;
                }
}
fprintf(coor,"%d   %d   7\n",j+1,k);
//SEGMENT 4
fprintf(coor,"%d   6    %d \n",j+2,k+1);
j=j+2;
k=k+1;
if (m>2)
{
        for (i=1;i<=m-2;++i)
                {
                        j=j+1;
                        fprintf(coor,"%d   %d    %d  \n",j,k,k+1);
                        k=k+1;
                }
}
fprintf(coor,"%d   %d   5\n",j+1,k);
//SEGMENT 5
fprintf(coor,"%d   5    %d \n",j+2,k+1);
j=j+2;
k=k+1;
if(m>2)
{
        for (i=1;i<=m-2;++i)
                {
                        j=j+1;
                        fprintf(coor,"%d   %d    %d  \n",j,k,k+1);
                        k=k+1;
                }
}
```

```c
fprintf(coor,"%d   %d    4 \n",j+1,k);
//SEGMENT 6
fprintf(coor,"%d   3    %d \n",j+2,k+1);
j=j+2;
k=k+1;
if(m>2)
{
        for (i=1;i<=m-2;++i)
                {
                        j=j+1;
                        fprintf(coor,"%d   %d    %d  \n",j,k,k+1);
                        k=k+1;
                }
}
fprintf(coor,"%d   %d    4 \n",j+1,k);
//SEGMENT 7
switch(h)
{
case 1:
case 2:
case 3:
case 4:
case 5:
case 6:
case 7:
        fprintf(coor,"%d   2  57 \n",j+2);
        fprintf(coor,"%d  57  32 \n",j+3);
        fprintf(coor,"%d  32  33 \n",j+4);
        fprintf(coor,"%d  33  34 \n",j+5);
        fprintf(coor,"%d  34   3 \n",j+6);
        j=j+5;
        k=k+3;
        break;
case 9:
case 10:
case 11:
case 12:
case 13:
case 14:
case 15:
        fprintf(coor,"%d   2  32 \n",j+2);
        fprintf(coor,"%d  32  57 \n",j+3);
        fprintf(coor,"%d  57  33 \n",j+4);
        fprintf(coor,"%d  33  34 \n",j+5);
```

```c
        fprintf(coor,"%d   34    3 \n",j+6);
        j=j+5;
        k=k+3;
        break;
case 17:
case 18:
case 19:
case 20:
case 21:
case 22:
case 23:
        fprintf(coor,"%d    2  32 \n",j+2);
        fprintf(coor,"%d  32  33 \n",j+3);
        fprintf(coor,"%d  33  57 \n",j+4);
        fprintf(coor,"%d  57  34 \n",j+5);
        fprintf(coor,"%d  34    3 \n",j+6);
        j=j+5;
        k=k+3;
        break;
case 25:
case 26:
case 27:
case 28:
case 29:
case 30:
case 31:
        fprintf(coor,"%d    2  32 \n",j+2);
        fprintf(coor,"%d  32  33 \n",j+3);
        fprintf(coor,"%d  33  34 \n",j+4);
        fprintf(coor,"%d  34  57 \n",j+5);
        fprintf(coor,"%d  57   3 \n",j+6);
        j=j+5;
        k=k+3;
        break;
default:
        fprintf(coor,"%d    2  32 \n",j+2);
        fprintf(coor,"%d  32  33 \n",j+3);
        fprintf(coor,"%d  33  34 \n",j+4);
        fprintf(coor,"%d  34    3 \n",j+5);
        j=j+4;
        k=k+3;
        break;
}
//SEGMENT 8
```

```c
switch(h)
{
default:
        fprintf(coor,"%d    2  35 \n",j+2);
        fprintf(coor,"%d  35  36 \n",j+3);
        fprintf(coor,"%d  36  37 \n",j+4);
        fprintf(coor,"%d  37  13 \n",j+5);
        j=j+4;
        k=k+3;
        break;
}
//SEGMENT 9
fprintf(coor,"%d   13    %d \n",j+2,k+1);
j=j+2;
k=k+1;
if(m>2)
{
        for (i=1;i<=m-2;++i)
        {
                j=j+1;
                fprintf(coor,"%d    %d    %d  \n",j,k,k+1);
                k=k+1;
        }
}
fprintf(coor,"%d    %d   12 \n",j+1,k);
//SEGMENT 10
fprintf(coor,"%d   11    %d \n",j+2,k+1);
j=j+2;
k=k+1;
if(m>2)
{
        for (i=1;i<=m-2;++i)
        {
        j=j+1;
        fprintf(coor,"%d    %d    %d  \n",j,k,k+1);
        k=k+1;
        }
}
fprintf(coor,"%d    %d   12 \n",j+1,k);
//SEGMENT 11
fprintf(coor,"%d   10    %d \n",j+2,k+1);
j=j+2;
k=k+1;
if(m>2)
```

```
{
        for (i=1;i<=m-2;++i)
        {
                j=j+1;
                fprintf(coor,"%d   %d    %d \n",j,k,k+1);
                k=k+1;
        }
}
fprintf(coor,"%d   %d   11\n",j+1,k);
//SEGMENT 12
fprintf(coor,"%d   10   %d \n",j+2,k+1);
j=j+2;
k=k+1;
if(m>2)
{
        for (i=1;i<=m-2;++i)
        {
                j=j+1;
                fprintf(coor,"%d   %d   %d \n",j,k,k+1);
                k=k+1;
        }
}
fprintf(coor,"%d   %d   9\n",j+1,k);
//SEGMENT 13
switch(h)
{
default:
        fprintf(coor,"%d   9   50 \n",j+2);
        fprintf(coor,"%d   50   51 \n",j+3);
        fprintf(coor,"%d   51   52 \n",j+4);
        fprintf(coor,"%d   52   8 \n",j+5);
        j=j+4;
        k=k+3;
        break;
}
//SEGMENT 14
switch (h)
{
case 0:
        fprintf(coor,"%d   2   %d \n",j+2,k+1);
        break;
case 8:
        fprintf(coor,"%d   32   %d \n",j+2,k+1);
        break;
```

```
case 16:
        fprintf(coor,"%d   33   %d \n",j+2,k+1);
        break;
case 24:
        fprintf(coor,"%d   34   %d \n",j+2,k+1);
        break;
case 32:
        fprintf(coor,"%d   3   %d \n",j+2,k+1);
        break;
default:
        fprintf(coor,"%d   57   %d \n",j+2,k+1);
        break;
}
j=j+2;
k=k+1;
if(m>2)
{
        for (i=1;i<=m-2;++i)
        {
                j=j+1;
                fprintf(coor,"%d   %d   %d \n",j,k,k+1);
                k=k+1;
        }
}
fprintf(coor,"%d   55   1 \n",j+1);
//*****************************//
fprintf(coor, "%d \n",1);
fprintf (coor,"0 \n");
fprintf(coor,"1 1.0  1.0  1.0 \n");
fprintf(coor,"1 90.  0.   90.  180.");
}
```

# Annex E :   Antenna's improvement (2<sup>nd</sup> model)

The below computer program written in C creates an output file called INPUT which is RICHWIRE's input file. The current program results to the elongation of the two vertical monopoles situated at the positive z axis. A detailed description of the model is given at <u>Section 5.1</u>.

```
#include <stdio.h>
#include <float.h>
#include <limits.h>
```

```c
#include <math.h>
#include <stdlib.h>
#define pi 3.141592653589793238462643383279S
#define sq3 1.7320508075688772935274463415059

main()
{
 int k,i,j,m,r,Z,h,f;
 double x1,y1,z1,x2,y2,z2,x3,y3,z3,x4,y4,z4,x5,y5,z5,x6,y6,z6,x7,y7,z7;
 double x,y,z,Lq, x8,y8,z8,x9,y9,z9,x10,y10,z10,x11,y11,z11;
 double x12,y12,z12,x13,y13,z13,min_L, orio,xa1a,xa2a,xa3a,ya1a,ya2a;
 double ya3a,za1a,za2a,za3a,xa1b,ya1b,za1b,xa2b,ya2b,za2b,xa3b,ya3b;
 double za3b,xopt1,yopt1,zopt1,xopt2,yopt2,zopt2;
FILE *coor;
coor=fopen("input.txt","w");
f=1867;
m=4;
Lq=(1.0/4)*(300.0/1111.0);      //CALCULATION OF THE LENGTH Lamda/4
orio=0.002;                //MINIMUM ACCEPTABLE SEGMENT LENGTH LIMIT
min_L=Lq/m;                //MINIMUM ACCEPTABLE SEGMENT LENGTH
Z=14*m;
printf("GIVE THE DISPLACEMENT GRADE OF THE DIPOLE AT Z AXIS [0-32]: ");
scanf("%d",&h);
if(h<0||h>32)
{
        printf("WRONG VALUE! \n");
        system("pause");
        exit;
}
else
{
        printf("GIVE INPUT VALUE : %d",h);
        system("pause");
        exit;
}
//***COORDINATE DISCOVERY OF THE STABLE POINTS***
//NODE 1
x1=0.0;
y1=0.0;
z1=0.0;
//NODE 2
x2=0.0;
y2=-Lq;
z2=0.0;

//OPTIMAZATION NODE 1
xopt1=0.0;
yopt1=-Lq;
zopt1=Lq+h*(Lq/32);
//NODE 3
x3=0.0;
y3=-Lq;
z3=Lq;
//NODE 4
x4=0.0;
y4=-Lq/2;
z4=z3+h*(Lq/32)+sin(pi/3)*Lq;
//NODE 5
x5=0.0;
y5=0.0;
z5=z3+h*(Lq/32)+sin(pi/3)*2*Lq;
//NODE 6
x6=0.0;
y6=Lq/2;
z6=z4;
//OPTIMIZATION NODE 2
xopt2=0.0;
yopt2=Lq;
zopt2=zopt1;
//NODE 7
x7=0.0;
y7=Lq;
z7=z3;
//NODE 8
x8=0.0;
y8=Lq;
z8=0.0;
//NODE 9
x9=0.0;
y9=Lq;
z9=-Lq;
//NODE 10
x10=0.0;
y10=Lq/2;
z10=-Lq-sin(pi/3)*Lq;
//NODE 11
x11=0.0;
y11=0.0;
z11=-Lq-sin(pi/3)*2*Lq;
```

```c
//NODE 12
x12=0.0;
y12=-Lq/2;
z12=-Lq-sin(pi/3)*Lq;
//NODE 13
x13=0.0;
y13=-Lq;
z13=-Lq;
//ADDITIONAL NODE 1a
xa1a=0.0;
ya1a=Lq;
za1a=(5*Lq)/4;
//ADDITIONAL NODE 1b
xa1b=0.0;
ya1b=-Lq;
za1b=(5*Lq)/4;
//ADDITIONAL NODE 2a
xa2a=0.0;
ya2a=Lq;
za2a=(3*Lq/2);
//ADDITIONAL NODE 2b
xa2b=0.0;
ya2b=-Lq;
za2b=(3*Lq/2);
//ADDITIONAL NODE 3a
xa3a=0.0;
ya3a=Lq;
za3a=(7*Lq)/4;
//ADDITIONAL NODE 3b
xa3b=0.0;
ya3b=-Lq;
za3b=(7*Lq)/4;
switch(h)
{
case 1:
case 2:
case 3:
case 4:
case 5:
case 6:
case 7:
case 8:
        fprintf(coor,"%4d \n",f);
        fprintf(coor,"-1.0 0.001 \n");
        fprintf(coor,"%d \n",Z+1);
        break;
case 9:
case 10:
case 11:
case 12:
case 13:
case 14:
case 15:
case 16:
        fprintf(coor,"%4d \n",f);
        fprintf(coor,"-1.0 0.001 \n");
        fprintf(coor,"%d \n",Z+3);
        break;
case 17:
case 18:
case 19:
case 20:
case 21:
case 22:
case 23:
case 24:
        fprintf(coor,"%4d \n",f);
        fprintf(coor,"-1.0 0.001 \n");
        fprintf(coor,"%d \n",Z+5);
        break;
case 25:
case 26:
case 27:
case 28:
case 29:
case 30:
case 31:
case 32:
        fprintf(coor,"%4d \n",f);
        fprintf(coor,"-1.0 0.001 \n");
        fprintf(coor,"%d \n",Z+7);
        break;
default:
        fprintf(coor,"%4d \n",f);
        fprintf(coor,"-1.0 0.001 \n");
        fprintf(coor,"%d \n",Z-1);
        break;
}
```

```
fprintf(coor,"1    %1.7f    %1.7f    %1.7f \n",x1,y1,z1);
fprintf(coor,"2    %1.7f    %1.7f    %1.7f \n",x2,y2,z2);
fprintf(coor,"3    %1.7f    %1.7f    %1.7f \n",x3,y3,z3);
fprintf(coor,"4    %1.7f    %1.7f    %1.7f \n",x4,y4,z4);
fprintf(coor,"5    %1.7f    %1.7f    %1.7f \n",x5,y5,z5);
fprintf(coor,"6    %1.7f    %1.7f    %1.7f \n",x6,y6,z6);
fprintf(coor,"7    %1.7f    %1.7f    %1.7f \n",x7,y7,z7);
fprintf(coor,"8    %1.7f    %1.7f    %1.7f \n",x8,y8,z8);
fprintf(coor,"9    %1.7f    %1.7f    %1.7f \n",x9,y9,z9);
fprintf(coor,"10   %1.7f    %1.7f    %1.7f \n",x10,y10,z10);
fprintf(coor,"11   %1.7f    %1.7f    %1.7f \n",x11,y11,z11);
fprintf(coor,"12   %1.7f    %1.7f    %1.7f \n",x12,y12,z12);
fprintf(coor,"13   %1.7f    %1.7f    %1.7f \n",x13,y13,z13);
//***COORDINATE ESTIMATION OF THE NON-STABLE POINTS***
r=14;          //CONSECUTIVE NUMBER KOMBWN
//SEGMENT 1
for (i=1;i<=m-1;++i)
{
        x=0.0;
        y=y1+((i*Lq)/m);
        z=0.0;
        fprintf(coor,"%d  %1.7f    %1.7f    %1.7f \n",r,x,y,z);
        r=r+1;
}
//SEGMENT 2
switch(h)
{
case 1:
case 2:
case 3:
case 4:
case 5:
case 6:
case 7:
case 8:
        fprintf(coor,"%d  %1.7f    %1.7f    %1.7f \n",r,xopt2,yopt2,zopt2);
        r=r+1;
        break;
case 9:
case 10:
case 11:
case 12:
case 13:
case 14:
```

```
case 15:
case 16:
        fprintf(coor,"%d  %1.7f    %1.7f    %1.7f \n",r,xopt2,yopt2,zopt2);
        fprintf(coor,"%d  %1.7f    %1.7f    %1.7f \n",r+1,xa1a,ya1a,za1a);
        r=r+2;
        break;
case 17:
case 18:
case 19:
case 20:
case 21:
case 22:
case 23:
case 24:
        fprintf(coor,"%d  %1.7f    %1.7f    %1.7f \n",r,xopt2,yopt2,zopt2);
        fprintf(coor,"%d  %1.7f    %1.7f    %1.7f \n",r+1,xa2a,ya2a,za2a);
        fprintf(coor,"%d  %1.7f    %1.7f    %1.7f \n",r+2,xa1a,ya1a,za1a);
        r=r+3;
        break;
case 25:
case 26:
case 27:
case 28:
case 29:
case 30:
case 31:
case 32:
        fprintf(coor,"%d  %1.7f    %1.7f    %1.7f \n",r,xopt2,yopt2,zopt2);
        fprintf(coor,"%d  %1.7f    %1.7f    %1.7f \n",r+1,xa3a,ya3a,za3a);
        fprintf(coor,"%d  %1.7f    %1.7f    %1.7f \n",r+2,xa2a,ya2a,za2a);
        fprintf(coor,"%d  %1.7f    %1.7f    %1.7f \n",r+3,xa1a,ya1a,za1a);
        r=r+4;
        break;
default:
        break;
}
for (i=1;i<=m-1;++i)
{
        x=0.0;
        y=y7;
        z=z7-((i*Lq)/m);
        fprintf(coor,"%d  %1.7f    %1.7f    %1.7f \n",r,x,y,z);
        r=r+1;
}
```

```c
//SEGMENT 3
for (i=1;i<=m-1;++i)
{
        x=0.0;
        y=y6+((i*Lq)/m)*sin(pi/6);
        z=z6-((i*Lq)/m)*cos(pi/6);
        fprintf(coor,"%d  %1.7f   %1.7f   %1.7f \n",r,x,y,z);
        r=r+1;
}

//SEGMENT 4
for (i=1;i<=m-1;++i)
{
        x=0.0;
        y=y6-((i*Lq)/m)*sin(pi/6);
        z=z6+((i*Lq)/m)*cos(pi/6);
        fprintf(coor,"%d  %1.7f   %1.7f   %1.7f \n",r,x,y,z);
        r=r+1;
}
//SEGMENT 5
for (i=1;i<=m-1;++i)
{
        x=0.0;
        y=y5-((i*Lq)/m)*cos(pi/3);
        z=z5-((i*Lq)/m)*sin(pi/3);
        fprintf(coor,"%d  %1.7f   %1.7f   %1.7f \n",r,x,y,z);
        r=r+1;
}
//SEGMENT 6
for (i=1;i<=m-1;++i)
{
        x=0.0;
        y=y3+((i*Lq)/m)*cos(pi/3);
        switch(h)
        {
        case 0:
                z=z3+((i*Lq)/m)*sin(pi/3);
                break;
        default:
                z=zopt1+((i*Lq)/m)*sin(pi/3);
                break;
        }
        fprintf(coor,"%d  %1.7f   %1.7f   %1.7f \n",r,x,y,z);
        r=r+1;
}
```

```c
}
//SEGMENT 7
for (i=1;i<=m-1;++i)
{
        x=0.0;
        y=y2;
        z=(i*Lq)/m;
        fprintf(coor,"%d  %1.7f   %1.7f   %1.7f \n",r,x,y,z);
        r=r+1;
}
switch(h)
{
case 1:
case 2:
case 3:
case 4:
case 5:
case 6:
case 7:
case 8:
        fprintf(coor,"%d  %1.7f   %1.7f   %1.7f \n",r,xopt1,yopt1,zopt1);
        r=r+1;
        break;
case 9:
case 10:
case 11:
case 12:
case 13:
case 14:
case 15:
case 16:
        fprintf(coor,"%d  %1.7f   %1.7f   %1.7f \n",r,xa1b,ya1b,za1b);
        fprintf(coor,"%d  %1.7f  %1.7f   %1.7f \n",r+1,xopt1,yopt1,zopt1);
        r=r+2;
        break;
case 17:
case 18:
case 19:
case 20:
case 21:
case 22:
case 23:
case 24:
        fprintf(coor,"%d  %1.7f   %1.7f   %1.7f \n",r,xa1b,ya1b,za1b);
```

```c
        fprintf(coor,"%d  %1.7f    %1.7f    %1.7f \n",r+1,xa2b,ya2b,za2b);
        fprintf(coor,"%d %1.7f  %1.7f  %1.7f \n",r+2,xopt1,yopt1,zopt1);
        r=r+3;
        break;
        case 25:
case 26:
case 27:
case 28:
case 29:
case 30:
case 31:
case 32:
        fprintf(coor,"%d %1.7f   %1.7f    %1.7f \n",r,xa1b,ya1b,za1b);
        fprintf(coor,"%d %1.7f   %1.7f    %1.7f \n",r+1,xa2b,ya2b,za2b);
        fprintf(coor,"%d %1.7f   %1.7f    %1.7f \n",r+2,xa3b,ya3b,za3b);
        fprintf(coor,"%d %1.7f  %1.7f %1.7f \n",r+3,xopt1,yopt1,zopt1);
        r=r+4;
        break;
default:
        break;
}
//SEGMENT 8
for (i=1;i<=m-1;++i)
{
        x=0.0;
        y=y2;
        z=z2-((i*Lq)/m);
        fprintf(coor,"%d  %1.7f    %1.7f    %1.7f \n",r,x,y,z);
        r=r+1;
}
//SEGMENT 9
for (i=1;i<=m-1;++i)
{
        x=0.0;
        y=y13+((i*Lq)/m)*cos(pi/3);
        z=z13-((i*Lq)/m)*sin(pi/3);
        fprintf(coor,"%d  %1.7f    %1.7f    %1.7f \n",r,x,y,z);
        r=r+1;
}
//SEGMENT 10
for (i=1;i<=m-1;++i)
{
        x=0.0;
        y=y11-((i*Lq)/m)*cos(pi/3);
```

```c
        z=z11+((i*Lq)/m)*sin(pi/3);
        fprintf(coor,"%d  %1.7f    %1.7f    %1.7f \n",r,x,y,z);
        r=r+1;
}
//SEGMENT 11
for (i=1;i<=m-1;++i)
{
        x=0.0;
        y=y10-((i*Lq)/m)*cos(pi/3);
        z=z10-((i*Lq)/m)*sin(pi/3);
        fprintf(coor,"%d  %1.7f    %1.7f    %1.7f \n",r,x,y,z);
        r=r+1;
}
//SEGMENT 12
for (i=1;i<=m-1;++i)
{
        x=0.0;
        y=y10+((i*Lq)/m)*cos(pi/3);
        z=z10+((i*Lq)/m)*sin(pi/3);
        fprintf(coor,"%d  %1.7f    %1.7f    %1.7f \n",r,x,y,z);
        r=r+1;
}
//SEGMENT 13
for (i=1;i<=m-1;++i)
{
        x=0.0;
        y=y9;
        z=z9+((i*Lq)/m);
        fprintf(coor,"%d  %1.7f    %1.7f    %1.7f \n",r,x,y,z);
        r=r+1;
}
//SEGMENT 14
for (i=1;i<=m-1;++i)
{
        x=0.0;
        y=y2+((i*Lq)/m);
        z=0.0;
        fprintf(coor,"%d  %1.7f    %1.7f    %1.7f \n",r,x,y,z);
        r=r+1;
}
switch(h)
{
case 1:
case 2:
```

```
case 3:
case 4:
case 5:
case 6:
case 7:
case 8:
        fprintf(coor,"%d \n",Z+2);
        break;
case 9:
case 10:
case 11:
case 12:
case 13:
case 14:
case 15:
case 16:
        fprintf(coor,"%d \n",Z+4);
        break;
case 17:
case 18:
case 19:
case 20:
case 21:
case 22:
case 23:
case 24:
        fprintf(coor,"%d \n",Z+6);
        break;
case 25:
case 26:
case 27:
case 28:
case 29:
case 30:
case 31:
case 32:
        fprintf(coor,"%d \n",Z+8);
        break;
        default:
        fprintf(coor,"%d \n",Z);
        break;
}
//****SEGMENTS DECLARATION***
j=1;                    //CONSECUTIVE NUMBER

k=13;                   //SEGMENTS NUMERATION
//SEGMENT 1
fprintf(coor,"1   1   %d \n",k+1);
k=k+1;
if(m>2)
{
        for (i=1;i<=m-2;++i)
        {
                j=j+1;
                fprintf(coor,"%d   %d   %d \n",j,k,k+1);
                k=k+1;
        }
}
fprintf(coor,"%d   %d   8 \n",j+1,k);
//SEGMENT 2
switch(h)
{
case 1:
case 2:
case 3:
case 4:
case 5:
case 6:
case 7:
case 8:
        fprintf(coor,"%d   %d   7 \n",j+2,k+1);
        j=j+1;
        k=k+1;
        break;
case 9:
case 10:
case 11:
case 12:
case 13:
case 14:
case 15:
case 16:
        fprintf(coor,"%d   %d   %d \n",j+2,k+1,k+2);
        fprintf(coor,"%d   %d   7 \n",j+3,k+2);
        j=j+2;
        k=k+2;
        break;
case 17:
case 18:
```

```
case 19:
case 20:
case 21:
case 22:
case 23:
case 24:
        fprintf(coor,"%d   %d    %d \n",j+2,k+1,k+2);
        fprintf(coor,"%d   %d    %d \n",j+3,k+2,k+3);
        fprintf(coor,"%d   %d    7 \n",j+4,k+3);
        j=j+3;
        k=k+3;
        break;
case 25:
case 26:
case 27:
case 28:
case 29:
case 30:
case 31:
case 32:
        fprintf(coor,"%d   %d    %d \n",j+2,k+1,k+2);
        fprintf(coor,"%d   %d    %d \n",j+3,k+2,k+3);
        fprintf(coor,"%d   %d    %d \n",j+4,k+3,k+4);
        fprintf(coor,"%d   %d    7 \n",j+5,k+4);
        j=j+4;
        k=k+4;
        break;
default:
        break;
}
fprintf(coor,"%d   7   %d \n",j+2,k+1);
j=j+2;
k=k+1;
for (i=1;i<=m-2;++i)
{
        j=j+1;
        fprintf(coor,"%d   %d   %d  \n",j,k,k+1);
        k=k+1;
}
fprintf(coor,"%d   %d    8 \n",j+1,k);
//SEGMENT 3
fprintf(coor,"%d   6   %d \n",j+2,k+1);
j=j+2;
k=k+1;

if (m>2)
{
        for (i=1;i<=m-2;++i)
        {
                j=j+1;
                fprintf(coor,"%d   %d    %d  \n",j,k,k+1);
                k=k+1;
        }
}
switch(h)
{
case 0:
        fprintf(coor,"%d   %d    7\n",j+1,k);
        break;
default:
        fprintf(coor,"%d   %d    17\n",j+1,k);
        break;
}
//SEGMENT 4
fprintf(coor,"%d   6   %d \n",j+2,k+1);
j=j+2;
k=k+1;
if (m>2)
{
        for (i=1;i<=m-2;++i)
        {
                j=j+1;
                fprintf(coor,"%d   %d    %d  \n",j,k,k+1);
                k=k+1;
        }
}
fprintf(coor,"%d   %d    5\n",j+1,k);
//SEGMENT 5
fprintf(coor,"%d   5   %d \n",j+2,k+1);
j=j+2;
k=k+1;
if(m>2)
{
        for (i=1;i<=m-2;++i)
        {
                j=j+1;
                fprintf(coor,"%d   %d    %d  \n",j,k,k+1);
                k=k+1;
        }
```

```c
}
fprintf(coor,"%d   %d   4 \n",j+1,k);
//SEGMENT 6
switch(h)
{
case 1:
case 2:
case 3:
case 4:
case 5:
case 6:
case 7:
case 8:
        fprintf(coor,"%d   36   %d \n",j+2,k+1);
        j=j+2;
        k=k+1;
        break;
case 9:
case 10:
case 11:
case 12:
case 13:
case 14:
case 15:
case 16:
        fprintf(coor,"%d   38   %d \n",j+2,k+1);
        j=j+2;
        k=k+1;
        break;
case 17:
case 18:
case 19:
case 20:
case 21:
case 22:
case 23:
case 24:
        fprintf(coor,"%d   40   %d \n",j+2,k+1);
        j=j+2;
        k=k+1;
        break;
case 25:
case 26:
case 27:
```

```c
case 28:
case 29:
case 30:
case 31:
case 32:
        fprintf(coor,"%d   42   %d \n",j+2,k+1);
        j=j+2;
        k=k+1;
        break;
default:
        fprintf(coor,"%d   3   %d \n",j+2,k+1);
        j=j+2;
        k=k+1;
        break;
}
if(m>2)
{
        for (i=1;i<=m-2;++i)
        {
                j=j+1;
                fprintf(coor,"%d   %d   %d \n",j,k,k+1);
                k=k+1;
        }
}
fprintf(coor,"%d   %d   4 \n",j+1,k);
//SEGMENT 7
fprintf(coor,"%d   2   %d \n",j+2,k+1);
j=j+2;
k=k+1;
if (m>2)
{
        for (i=1;i<=m-2;++i)
        {
                j=j+1;
                fprintf(coor,"%d   %d   %d \n",j,k,k+1);
                k=k+1;
        }
}
fprintf(coor,"%d   %d   3\n",j+1,k);
switch(h)
{
case 1:
case 2:
case 3:
```

```c
case 4:
case 5:
case 6:
case 7:
case 8:
        fprintf(coor,"%d   3   %d \n",j+2,k+1);
        j=j+1;
        k=k+1;
        break;
case 9:
case 10:
case 11:
case 12:
case 13:
case 14:
case 15:
case 16:
        fprintf(coor,"%d   3   %d \n",j+2,k+1);
        fprintf(coor,"%d   %d   %d \n",j+3,k+1,k+2);
        j=j+2;
        k=k+2;
        break;
case 17:
case 18:
case 19:
case 20:
case 21:
case 22:
case 23:
case 24:
        fprintf(coor,"%d   3   %d \n",j+2,k+1);
        fprintf(coor,"%d   %d   %d \n",j+3,k+1,k+2);
        fprintf(coor,"%d   %d   %d \n",j+4,k+2,k+3);
        j=j+3;
        k=k+3;
        break;
case 25:
case 26:
case 27:
case 28:
case 29:
case 30:
case 31:
case 32:
```

```c
        fprintf(coor,"%d   3   %d \n",j+2,k+1);
        fprintf(coor,"%d   %d   %d \n",j+3,k+1,k+2);
        fprintf(coor,"%d   %d   %d \n",j+4,k+2,k+3);
        fprintf(coor,"%d   %d   %d \n",j+5,k+3,k+4);
        j=j+4;
        k=k+4;
        break;
default:
        break;
}
//SEGMENT 8
fprintf(coor,"%d   2   %d \n",j+2,k+1);
j=j+2;
k=k+1;
if(m>2)
{
        for (i=1;i<=m-2;++i)
        {
                j=j+1;
                fprintf(coor,"%d   %d   %d  \n",j,k,k+1);
                k=k+1;
        }
}
fprintf(coor,"%d   %d   13\n",j+1,k);
//SEGMENT 9
fprintf(coor,"%d   13   %d \n",j+2,k+1);
j=j+2;
k=k+1;
if(m>2)
{
        for (i=1;i<=m-2;++i)
        {
                j=j+1;
                fprintf(coor,"%d   %d   %d  \n",j,k,k+1);
                k=k+1;
        }
}
fprintf(coor,"%d   %d   12 \n",j+1,k);
//SEGMENT 10
fprintf(coor,"%d   11   %d \n",j+2,k+1);
j=j+2;
k=k+1;
if(m>2)
{
```

```c
        for (i=1;i<=m-2;++i)
        {
                j=j+1;
                fprintf(coor,"%d   %d   %d  \n",j,k,k+1);
                k=k+1;
        }
}
fprintf(coor,"%d   %d   12 \n",j+1,k);
//SEGMENT 11
fprintf(coor,"%d   10   %d \n",j+2,k+1);
j=j+2;
k=k+1;
if(m>2)
{
        for (i=1;i<=m-2;++i)
        {
                j=j+1;
                fprintf(coor,"%d   %d   %d  \n",j,k,k+1);
                k=k+1;
        }
}
fprintf(coor,"%d   %d   11\n",j+1,k);
//SEGMENT 12
fprintf(coor,"%d   10   %d \n",j+2,k+1);
j=j+2;
k=k+1;
if(m>2)
{
        for (i=1;i<=m-2;++i)
        {
                j=j+1;
                fprintf(coor,"%d   %d   %d  \n",j,k,k+1);
                k=k+1;
        }
}
fprintf(coor,"%d   %d   9\n",j+1,k);
//SEGMENT 13
fprintf(coor,"%d   9   %d \n",j+2,k+1);
j=j+2;
k=k+1;
if(m>2)
{
        for (i=1;i<=m-2;++i)
        {
```

```c
        j=j+1;
        fprintf(coor,"%d   %d   %d  \n",j,k,k+1);
        k=k+1;
    }
}
fprintf(coor,"%d   %d   8 \n",j+1,k);
//SEGMENT 14
fprintf(coor,"%d   2   %d \n",j+2,k+1);
j=j+2;
k=k+1;
if(m>2)
{
        for (i=1;i<=m-2;++i)
        {
                j=j+1;
                fprintf(coor,"%d   %d   %d  \n",j,k,k+1);
                k=k+1;
        }
}
switch(h)
{
case 1:
case 2:
case 3:
case 4:
case 5:
case 6:
case 7:
case 8:
        fprintf(coor,"%d   %d   1 \n",j+1,Z+1);
        break;
case 9:
case 10:
case 11:
case 12:
case 13:
case 14:
case 15:
case 16:
        fprintf(coor,"%d   %d   1 \n",j+1,Z+3);
        break;
case 17:
case 18:
case 19:
```

```
case 20:
case 21:
case 22:
case 23:
case 24:
        fprintf(coor,"%d    %d    1 \n",j+1,Z+5);
        break;
case 25:
case 26:
case 27:
case 28:
case 29:
case 30:
case 31:
case 32:
        fprintf(coor,"%d    %d    1 \n",j+1,Z+7);
        break;
default:
        fprintf(coor,"%d    %d    1 \n",j+1,Z-1);
        break;
}
//*********************************//
fprintf(coor, "%d \n",1);
fprintf (coor,"0 \n");
fprintf(coor,"1  1.0   1.0   1.0 \n");
fprintf(coor,"1  90.  0.   90.  180.");
}
```

## Annex F : Antenna's improvement (3ʳᵈ model)

The below computer program written in C creates an output file called INPUT which is RICHWIRE's input file. The current program results to the elongation of the active dipole on either side of its edges towards to the y axis. A detailed description of the model is given at <u>Section 5.1</u>.

```
#include <stdio.h>
#include <float.h>
#include <limits.h>
#include <math.h>
#include <stdlib.h>
#define pi 3.14159265358979323846264338327950
#define sq3 1.73205080756887729352744463415059
```

```
main()
{
 int k,i,j,m,r,Z;
 long int h;
 float f;
 double x1,y1,z1,x2,y2,z2,x3,y3,z3,x4,y4,z4,x5,y5,z5,x6,y6,z6,x7,y7,z7;
 double x8,y8,z8,x9,y9,z9,x10,y10,z10,x11,y11,z11,x12,y12,z12,x,y,z,Lq;
 double xa1,xa2,ya1,ya2,za1,za2,xb1,xb2,yb1,yb2,zb1,zb2,x13,y13,z13;
 FILE *coor;
 coor=fopen("input.txt","w");
 f=1867.0;
 m=4;
 Lq=(1.0/4)*(300.0/1111.0);         //CALCULATION OF THE LENGTH Lamda/4
 Z=56;
 printf("DWSTE TON BA8MO EPIMHKYNSHS TOY ENERGOY DIPOLOY [0 32]: ");
 scanf("%d",&h);
 if(h<0||h>32)
 {
        printf("WRONG VALUE! \n");
        system("pause");
        exit;
 }
 else
 {
        printf("GIVE INPUT VALUE : %d",h);
        system("pause");
        exit;
 }
 //***COORDINATE DISCOVERY OF THE STABLE POINTS***
 //NODE 1
 x1=0.0;
 y1=0.0;
 z1=0.0;
 //NODE 2
 x2=0.0;
 y2=-Lq;
 z2=0.0;
 //NODE 3
 x3=0.0;
 y3=-Lq-h*(Lq/32.0);
 z3=Lq;
 //NODE 4
 x4=0.0;
 y4=-Lq*(0.5+(h/64.0));
```

```c
z4=Lq+(Lq/4)*sqrt((2.0-(h/16.0))*(6.0+(h/16.0)));
//NODE 5
x5=0.0;
y5=0.0;
z5=Lq+Lq*sqrt((1.0-(h/32.0))*(3.0+(h/32.0)));
//NODE 6
x6=0.0;
y6=-y4;
z6=z4;
//NODE 7
x7=0.0;
y7=Lq+h*(Lq/32.0);
z7=Lq;
//NODE 8
x8=0.0;
y8=Lq;
z8=0.0;
//NODE 9
x9=0.0;
y9=Lq+h*(Lq/32.0);
z9=-Lq;
//NODE 10
x10=0.0;
y10=y6;
z10=-z6;
//NODE 11
x11=0.0;
y11=0.0;
z11=-z5;
//NODE 12
x12=0.0;
y12=-y10;
z12=z10;
//NODE 13
x13=0.0;
y13=-Lq-h*(Lq/32.0);
z13=-Lq;
//**MOVING NODE1**
xa1=0.0;
ya1=Lq+h*(Lq/32.0);
za1=0.0;
//**MOVING NODE2**
xa2=0.0;
ya2=-Lq-h*(Lq/32.0);

za2=0.0;
//**STABLE NODE1**
xb1=0.0;
yb1=(3*Lq)/2.0;
zb1=0.0;
//**STABLE NODE2**
xb2=0.0;
yb2=-yb1;
zb2=0.0;
switch(h)
{
case 0:
        fprintf(coor,"%4.1f \n",f);
        fprintf(coor,"-1.0 0.001 \n");
        fprintf(coor,"%d \n",Z-1);
        break;
case 17:
case 18:
case 19:
case 20:
case 21:
case 22:
case 23:
case 24:
case 25:
case 26:
case 27:
case 28:
case 29:
case 30:
case 31:
case 32:
        fprintf(coor,"%4.1f \n",f);
        fprintf(coor,"-1.0 0.001 \n");
        fprintf(coor,"%d \n",Z+3);
        break;
default:
        fprintf(coor,"%4.1f \n",f);
        fprintf(coor,"-1.0 0.001 \n");
        fprintf(coor,"%d \n",Z+1);
        break;
}
fprintf(coor,"1   %1.7f   %1.7f   %1.7f \n",x1,y1,za1);
fprintf(coor,"2   %1.7f   %1.7f   %1.7f \n",x2,y2,z2);
```

```c
        fprintf(coor,"3     %1.7f     %1.7f     %1.7f \n",x3,y3,z3);
        fprintf(coor,"4     %1.7f     %1.7f     %1.7f \n",x4,y4,z4);
        fprintf(coor,"5     %1.7f     %1.7f     %1.7f \n",x5,y5,z5);
        fprintf(coor,"6     %1.7f     %1.7f     %1.7f \n",x6,y6,z6);
        fprintf(coor,"7     %1.7f     %1.7f     %1.7f \n",x7,y7,z7);
        fprintf(coor,"8     %1.7f     %1.7f     %1.7f \n",x8,y8,z8);
        fprintf(coor,"9     %1.7f     %1.7f     %1.7f \n",x9,y9,z9);
        fprintf(coor,"10    %1.7f     %1.7f     %1.7f \n",x10,y10,z10);
        fprintf(coor,"11    %1.7f     %1.7f     %1.7f \n",x11,y11,z11);
        fprintf(coor,"12    %1.7f     %1.7f     %1.7f \n",x12,y12,z12);
        fprintf(coor,"13    %1.7f     %1.7f     %1.7f \n",x13,y13,z13);
//***COORDINATE ESTIMATION OF THE NON-STABLE POINTS***
r=14;          //CONSECUTIVE NUMBER KOMBWN
//SEGMENT 1
for (i=1;i<=m-1;++i)
{
        x=0.0;
        y=y1+((i*Lq)/m);
        z=za1;
        fprintf(coor,"%d  %1.7f    %1.7f    %1.7f \n",r,x,y,z);
        r=r+1;
}
//SEGMENT 2
for (i=1;i<=m-1;++i)
{
        x=0.0;
        y=y7;
        z=Lq-((i*Lq)/m);
        fprintf(coor,"%d  %1.7f    %1.7f    %1.7f \n",r,x,y,z);
        r=r+1;
}
//SEGMENT 3
for (i=1;i<=m-1;++i)
{
        x=0.0;
        y=y6+((i*Lq)/m)*(0.5*(1.0+(h/32.0)));
        z=z6-((i*Lq)/m)*(0.5*sqrt((1.0-(h/32.0))*(3.0+(h/32.0))));
        fprintf(coor,"%d  %1.7f    %1.7f    %1.7f \n",r,x,y,z);
        r=r+1;
)
//SEGMENT 4
for (i=1;i<=m-1;++i)
{
        x=0.0;
```

```c
        y=y6-((i*Lq)/m)*(0.5*(1+(h/32.0)));
        z=z6+((i*Lq)/m)*(0.5*sqrt((1-(h/32.0))*(3+(h/32.0))));
        fprintf(coor,"%d  %1.7f    %1.7f    %1.7f \n",r,x,y,z);
        r=r+1;
}
//SEGMENT 5
for (i=1;i<=m-1;++i)
{
        x=0.0;
        y=y5-((i*Lq)/m)*(0.5*(1+(h/32.0)));
        z=z5-((i*Lq)/m)*(0.5*sqrt((1-(h/32.0))*(3+(h/32.0))));
        fprintf(coor,"%d  %1.7f    %1.7f    %1.7f \n",r,x,y,z);
        r=r+1;
}
//SEGMENT 6
for (i=1;i<=m-1;++i)
{
        x=0.0;
        y=y3+((i*Lq)/m)*(0.5*(1+(h/32.0)));
        z=z3+((i*Lq)/m)*(0.5*sqrt((1-(h/32.0))*(3+(h/32.0))));
        fprintf(coor,"%d  %1.7f    %1.7f    %1.7f \n",r,x,y,z);
        r=r+1;
}
//SEGMENT 7
for (i=1;i<=m-1;++i)
{
        x=0.0;
        y=ya2;
        z=(i*Lq)/m;
        fprintf(coor,"%d  %1.7f    %1.7f    %1.7f \n",r,x,y,z);
        r=r+1;

}
//SEGMENT 8
for (i=1;i<=m-1;++i)
(
        x=0.0;
        y=ya2;
        z=-(i*Lq)/m;
        fprintf(coor,"%d  %1.7f    %1.7f    %1.7f \n",r,x,y,z);
        r=r+1;
)
//SEGMENT 9
for (i=1;i<=m-1;++i)
{
```

```
        x=0.0;
        y=y13+((i*Lq)/m)*(0.5*(1+(h/32.0)));
        z=z13-((i*Lq)/m)*(0.5*sqrt((1-(h/32.0))*(3+(h/32.0))));
        fprintf(coor,"%d  %1.7f   %1.7f   %1.7f \n",r,x,y,z);
        r=r+1;
}
//SEGMENT 10
for (i=1;i<=m-1;++i)
{
        x=0.0;
        y=y11-((i*Lq)/m)*(0.5*(1+(h/32.0)));
        z=z11+((i*Lq)/m)*(0.5*sqrt((1-(h/32.0))*(3+(h/32.0))));
        fprintf(coor,"%d  %1.7f   %1.7f   %1.7f \n",r,x,y,z);
        r=r+1;
}
//SEGMENT 11
for (i=1;i<=m-1;++i)
{
        x=0.0;
        y=y10-((i*Lq)/m)*(0.5*(1+(h/32.0)));
        z=z10-((i*Lq)/m)*(0.5*sqrt((1-(h/32.0))*(3+(h/32.0))));
        fprintf(coor,"%d  %1.7f   %1.7f   %1.7f \n",r,x,y,z);
        r=r+1;
)
//SEGMENT 12
for (i=1;i<=m-1;++i)
{
        x=0.0;
        y=y10+((i*Lq)/m)*(0.5*(1+(h/32.0)));
        z=z10+((i*Lq)/m)*(0.5*sqrt((1-(h/32.0))*(3+(h/32.0))));
        fprintf(coor,"%d  %1.7f   %1.7f   %1.7f \n",r,x,y,z);
        r=r+1;
}
//SEGMENT 13
for (i=1;i<=m-1;++i)
{
        x=0.0;
        y=y9;
        z=z9+((i*Lq)/m);
        fprintf(coor,"%d  %1.7f   %1.7f   %1.7f \n",r,x,y,z);
        r=r+1;
}
//SEGMENT 14
for (i=1;i<=m-1;++i)
```

```
{
        x=0.0;
        y=y2+((i*Lq)/m);
        z=0.0;
        fprintf(coor,"%d  %1.7f   %1.7f   %1.7f \n",r,x,y,z);
        r=r+1;
}
switch(h)
{
case 0:
        fprintf(coor,"%d \n",Z);
        break;
case 17:
case 18:
case 19:
case 20:
case 21:
case 22:
case 23:
case 24:
case 25:
case 26:
case 27:
case 28:
case 29:
case 30:
case 31:
case 32:
        fprintf(coor,"56  %1.7f   %1.7f   %1.7f \n",xa1,ya1,za1);
        fprintf(coor,"57  %1.7f   %1.7f   %1.7f \n",xa2,ya2,za2);
        fprintf(coor,"58  %1.7f   %1.7f   %1.7f \n",xb1,yb1,zb1);
        fprintf(coor,"59  %1.7f   %1.7f   %1.7f \n",xb2,yb2,zb2);
        fprintf(coor,"%d \n",Z+4);
        break;
default:
        fprintf(coor,"56  %1.7f   %1.7f   %1.7f \n",xa1,ya1,za1);
        fprintf(coor,"57  %1.7f   %1.7f   %1.7f \n",xa2,ya2,za2);
        fprintf(coor,"%d \n",Z+2);
        break;
}
//****SEGMENTS DECLARATION***
j=1;            //CONSECUTIVE NUMBER
k=13;           //SEGMENTS NUMERATION
//SEGMENT 1
```

```
fprintf(coor,"1    1    %d \n",k+1);
k=k+1;
if(m>2)
{
        for (i=1;i<=m-2;++i)
        {
                j=j+1;
                fprintf(coor,"%d    %d    %d  \n",j,k,k+1);
                k=k+1;
        }
}
switch (h)
{
case 0:
        fprintf(coor,"%d    %d    8 \n",j+1,k);
        break;
case 17:
case 18:
case 19:
case 20:
case 21:
case 22:
case 23:
case 24:
case 25:
case 26:
case 27:
case 28:
case 29:
case 30:
case 31:
case 32:
        fprintf(coor,"%d    %d    8 \n",j+1,k);
        fprintf(coor,"%d    8    58 \n",j+2);
        fprintf(coor,"%d    58    56 \n",j+3);
        j=j+2;
        break;
default:
        fprintf(coor,"%d    %d    8 \n",j+1,k);
        fprintf(coor,"%d    8    56 \n",j+2);
        j=j+1;
        break;
}
//SEGMENT 2
```

```
switch(h)
{
default:
        fprintf(coor,"%d    7    17 \n",j+2);
        fprintf(coor,"%d    17    18 \n",j+3);
        fprintf(coor,"%d    18    19 \n",j+4);
        fprintf(coor,"%d    19    56 \n",j+5);
        j=j+4;
        k=k+3;
        break;
case 0:
        fprintf(coor,"%d    7    17 \n",j+2);
        fprintf(coor,"%d    17    18 \n",j+3);
        fprintf(coor,"%d    18    19 \n",j+4);
        fprintf(coor,"%d    19    8 \n",j+5);
        j=j+4;
        k=k+3;
        break;
}
//SEGMENT 3
fprintf(coor,"%d    6    %d \n",j+2,k+1);
j=j+2;
k=k+1;
if (m>2)
{
        for (i=1;i<=m-2;++i)
        {
                j=j+1;
                fprintf(coor,"%d    %d    %d  \n",j,k,k+1);
                k=k+1;
        }
}
fprintf(coor,"%d    %d    7\n",j+1,k);
//SEGMENT 4
fprintf(coor,"%d    6    %d \n",j+2,k+1);
j=j+2;
k=k+1;
if (m>2)
{
        for (i=1;i<=m-2;++i)
        {
                j=j+1;
                fprintf(coor,"%d    %d    %d  \n",j,k,k+1);
                k=k+1;
```

```
        }
}
fprintf(coor,"%d   %d    5\n",j+1,k);
//SEGMENT 5
fprintf(coor,"%d   5    %d \n",j+2,k+1);
j=j+2;
k=k+1;
if(m>2)
{
        for (i=1;i<=m-2;++i)
        {
                j=j+1;
                fprintf(coor,"%d    %d    %d  \n",j,k,k+1);
                k=k+1;
        }
}
fprintf(coor,"%d   %d    4 \n",j+1,k);
//SEGMENT 6
fprintf(coor,"%d   3    %d \n",j+2,k+1);
j=j+2;
k=k+1;
if(m>2)
{
        for (i=1;i<=m-2;++i)
        {
                j=j+1;
                fprintf(coor,"%d    %d    %d  \n",j,k,k+1);
                k=k+1;
        }
}
fprintf(coor,"%d   %d    4 \n",j+1,k);
//SEGMENT 7
switch(h)
{
default:
        fprintf(coor,"%d  57   32 \n",j+2);
        fprintf(coor,"%d  32   33 \n",j+3);
        fprintf(coor,"%d  33   34 \n",j+4);
        fprintf(coor,"%d  34    3 \n",j+5);
        j=j+4;
        k=k+3;
        break;
case 0:
        fprintf(coor,"%d   2   32 \n",j+2);
```

```
        fprintf(coor,"%d  32   33 \n",j+3);
        fprintf(coor,"%d  33   34 \n",j+4);
        fprintf(coor,"%d  34    3 \n",j+5);
        j=j+4;
        k=k+3;
        break;
}
//SEGMENT 8
switch(h)
{
default:
        fprintf(coor,"%d  57   35 \n",j+2);
        fprintf(coor,"%d  35   36 \n",j+3);
        fprintf(coor,"%d  36   37 \n",j+4);
        fprintf(coor,"%d  37   13 \n",j+5);
        j=j+4;
        k=k+3;
        break;
case 0:
        fprintf(coor,"%d   2   35 \n",j+2);
        fprintf(coor,"%d  35   36 \n",j+3);
        fprintf(coor,"%d  36   37 \n",j+4);
        fprintf(coor,"%d  37   13 \n",j+5);
        j=j+4;
        k=k+3;
        break;
}
//SEGMENT 9
fprintf(coor,"%d  13    %d \n",j+2,k+1);
j=j+2;
k=k+1;
if(m>2)
{
        for (i=1;i<=m-2;++i)
        {
                j=j+1;
                fprintf(coor,"%d    %d    %d  \n",j,k,k+1);
                k=k+1;
        }
}
fprintf(coor,"%d   %d   12 \n",j+1,k);
//SEGMENT 10
fprintf(coor,"%d  11    %d \n",j+2,k+1);
j=j+2;
```

```c
k=k+1;
if(m>2)
{
        for (i=1;i<=m-2;++i)
        {
                j=j+1;
                fprintf(coor,"%d    %d    %d  \n",j,k,k+1);
                k=k+1;
        }
}
fprintf(coor,"%d    %d    12 \n",j+1,k);
//SEGMENT 11
fprintf(coor,"%d    10    %d \n",j+2,k+1);
j=j+2;
k=k+1;
if(m>2)
{
        for (i=1;i<=m-2;++i)
        {
                j=j+1;
                fprintf(coor,"%d    %d    %d  \n",j,k,k+1);
                k=k+1;
        }
}
fprintf(coor,"%d    %d    11\n",j+1,k);
//SEGMENT 12
fprintf(coor,"%d    10    %d \n",j+2,k+1);
j=j+2;
k=k+1;
if(m>2)
{
        for (i=1;i<=m-2;++i)
        {
                j=j+1;
                fprintf(coor,"%d    %d    %d  \n",j,k,k+1);
                k=k+1;
        }
}
fprintf(coor,"%d    %d    9\n",j+1,k);
//SEGMENT 13
switch(h)
{
default:
        fprintf(coor,"%d    9    50\n",j+2);
```

```c
        fprintf(coor,"%d    50    51 \n",j+3);
        fprintf(coor,"%d    51    52 \n",j+4);
        fprintf(coor,"%d    52    56 \n",j+5);
        j=j+4;
        k=k+3;
        break;
case 0:
        fprintf(coor,"%d    9    50 \n",j+2);
        fprintf(coor,"%d    50    51 \n",j+3);
        fprintf(coor,"%d    51    52 \n",j+4);
        fprintf(coor,"%d    52    8 \n",j+5);
        j=j+4;
        k=k+3;
        break;
}
//SEGMENT 14
switch (h)
{
case 0:
fprintf(coor,"%d    2    %d \n",j+2,k+1);
break;
case 17:
case 18:
case 19:
case 20:
case 21:
case 22:
case 23:
case 24:
case 25:
case 26:
case 27:
case 28:
case 29:
case 30:
case 31:
case 32:
        fprintf(coor,"%d    57    59 \n",j+2);
        fprintf(coor,"%d    59    2 \n",j+3);
        fprintf(coor,"%d    2    %d \n",j+4,k+1);
        j=j+2;
        break;
default:
        fprintf(coor,"%d    57    2 \n",j+2);
```

```
        fprintf(coor,"%d   2   %d \n",j+3,k+1);
        j=j+1;
        break;
}
j=j+2;
k=k+1;
if(m>2)
{
        for (i=1;i<=m-2;++i)
        {
                j=j+1;
                fprintf(coor,"%d   %d   %d  \n",j,k,k+1);
                k=k+1;
        }
}
fprintf(coor,"%d   55   1 \n",j+1);
//*******************************//
fprintf(coor, "%d \n",1);
fprintf (coor,"0 \n");
fprintf(coor,"1 1.0  1.0  1.0 \n");
fprintf(coor,"1 90.  0.   90.  180.");
}
```

# **Annex G : Mathematical formulation of $U_{max}$**

At the present annex a mathematical analysis will be attempted in order the formula of the quantity of maximum radiation intensity, $U_{max}$ , of an antenna to be extracted. This particular quantity is a useful indicator of the antenna's behavior as it depends on, as it will be shown, the value of SWR and directivity (D). Therefore this particular indicator acts subserviently to the extraction of useful conclusions during the antenna's study.

**Electromagnetic Theory**

The analysis will commence from the area of the electromagnetism where the following equations are in effect:

(1) : $U_{max} = D \, U_{ave}$

where $D$ : the directivity, $U_{ave}$ : the average radiation intensity [W/sterad]

However, the average radiation intensity is given from the formula:

(2) : $U_{ave} = \dfrac{P}{4\pi}$ και $\underline{P} = \varepsilon P$

where: $\underline{P}$ : the –far field- power radiation [W], $\varepsilon$ : antenna's efficiency

Assuming, without significant error, that the antenna is ideal, that is its whole input electrical power is converted into radiation, it implies:

(3) : $\varepsilon \simeq 1 \Rightarrow \underline{P} \simeq P$

where $P$ : the antenna's power input [W]

Thus <u>Z.(1)</u> becomes:

(4) : $U_{max} = \dfrac{1}{4\pi} D \, P$

However, the antenna's power input is given by the formula:

(5) : $P = \mathrm{Re}\{\dot{W}\}$

where: $\dot{W}$ : the complex input power which is:

(6) : $\dot{W} = \dfrac{1}{2}\dot{V}\dot{I}^* = \dfrac{1}{2}\left|\dot{V}\right|^2 \dfrac{\dot{Z}}{\left|\dot{Z}\right|^2}$

where: $\dot{V}$, $\dot{I}$ and $\dot{Z}$ are the voltage, the current and the antenna's input impedance respectively and in general these are complex quantities.

Hence Z.(5) due to Z.(6) becomes:

(7) : $P = \dfrac{1}{2}\left|\dot{V}\right|^2 \dfrac{R}{\left|\dot{Z}\right|^2}$

Utilizing the expressions Z.(7) and Z.(2), Z.(1) gives:

(8) : $U_{max} = \dfrac{1}{8\pi}D\left|\dot{V}\right|^2 \dfrac{R}{\left|\dot{Z}\right|^2}$

**Theory of transmission lines**

At the present point the mathematical formula of the antenna's input voltage will be outlined using the basic principles of the transmission line theory. In general, a complex source $\dot{V}_s$ is assumed with an impedance $\dot{Z}_s$. The source feeds a complex load $\dot{Z}_L$ which is, at the present case, the antenna's input impedance through a transmission line with a complex characteristic resistance $\dot{Z}_0$, complex propagation constant $\dot{\gamma}$ and length $\ell$. All the above are sufficiently clarified from Figure Z.1.

Following David M. Pozar's analysis [8] it is assumed that the source's position is at the point $z = -\ell$ and the load's position at $z = 0$. Therefore, it results that the definition scope of the variable z is the closed range $[-\ell, 0]$. The following equations describe the voltage, the current and the impedance respectively at every point on the transmission line. It implies:

(9) : $\dot{V}(z) = \dot{V}_1\,e^{-\dot{\gamma}z} + \dot{V}_2\,e^{\dot{\gamma}z}$

(10) : $\dot{I}(z) = \dfrac{1}{\dot{Z}_0}\left[\dot{V}_1\,e^{-\dot{\gamma}z} - \dot{V}_2\,e^{\dot{\gamma}z}\right]$

(11) : $\dot{Z}(z) = \dfrac{\dot{V}(z)}{\dot{I}(z)}$

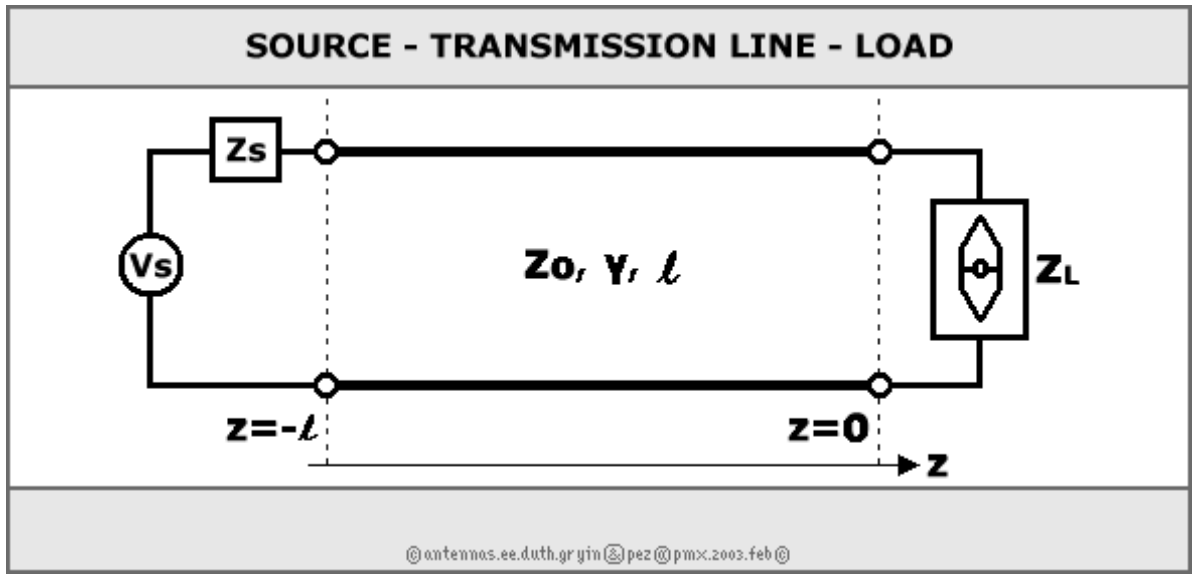

Figure 1 : Source-transmission line-load

More specifically, the quantity $\dot{V}_1\,e^{-\dot{\gamma}z}$ describes a propagating wave towards the positive direction of the $z$ axis. Equivalently, the quantity $\dot{V}_2\,e^{\dot{\gamma}z}$ describes a reflecting wave moving towards the negative direction of the $z$ axis.

The below equations are written:

$$\dot{V}(z) = \dot{V}_1 e^{-\dot{\gamma} z}\left[1 + \left(\frac{\dot{V}_2 e^{\dot{\gamma} z}}{\dot{V}_1 e^{-\dot{\gamma} z}}\right)\right]$$

$$\dot{I}(z) = \frac{\dot{V}_1}{\dot{Z}_0} e^{-\dot{\gamma} z}\left[1 - \left(\frac{\dot{V}_2 e^{\dot{\gamma} z}}{\dot{V}_1 e^{-\dot{\gamma} z}}\right)\right]$$

$$\dot{Z}(z) = \dot{Z}_0 \frac{1 + \left(\dfrac{\dot{V}_2 e^{\dot{\gamma} z}}{\dot{V}_1 e^{-\dot{\gamma} z}}\right)}{1 - \left(\dfrac{\dot{V}_2 e^{\dot{\gamma} z}}{\dot{V}_1 e^{-\dot{\gamma} z}}\right)}$$

ή

$$(12): \ \dot{V}(z) = \dot{V}_1 e^{-\dot{\gamma} z}\left[1 + \left(\frac{\dot{V}_2}{\dot{V}_1} e^{2\dot{\gamma} z}\right)\right]$$

$$(13): \ \dot{I}(z) = \frac{\dot{V}_1}{\dot{Z}_0} e^{-\dot{\gamma} z}\left[1 - \left(\frac{\dot{V}_2}{\dot{V}_1} e^{2\dot{\gamma} z}\right)\right]$$

$$(14): \ \dot{Z}(z) = \dot{Z}_0 \frac{1 + \left(\dfrac{\dot{V}_2}{\dot{V}_1} e^{2\dot{\gamma} z}\right)}{1 - \left(\dfrac{\dot{V}_2}{\dot{V}_1} e^{2\dot{\gamma} z}\right)}$$

The voltage reflection coefficient, which in general is a complex quantity, is defined as the ratio of a voltage reflecting wave to the counterpart traveling wave. Specifically, the voltage reflection coefficient at every point on the transmission line is given by the formula:

$$(15): \ \dot{\rho}(z) = \frac{\dot{V}_2 e^{\dot{\gamma} z}}{\dot{V}_1 e^{-\dot{\gamma} z}} = \frac{\dot{V}_2}{\dot{V}_1} e^{2\dot{\gamma} z}$$

$$(16): \ \dot{\rho}(0) = \frac{\dot{V}_2}{\dot{V}_1} \Rightarrow \dot{\rho}(z) = \dot{\rho}(0) e^{2\dot{\gamma} z}$$

where $\dot{\rho}(0)$ is the reflection coefficient at the point $z = 0$, that is at the position of the antenna's load. For the point $z = -\ell$ Z.(15) implies:

$$\dot{\rho}(-\ell) = \dot{\rho}(0) e^{-2\dot{\gamma}\ell}$$

$$\dot{\rho}(0) = \dot{\rho}(-\ell) e^{2\dot{\gamma}\ell}$$

Thus, the equations Z.(12), Z.(13) and Z.(14) can be expressed as a function of the reflection coefficient as follows:

$$(17) : \dot{V}(z) = \dot{V}_1 \, e^{-\dot{\gamma} z} \left[ 1 + \dot{\rho}(z) \right]$$

$$(18) : \dot{I}(z) = \frac{\dot{V}_1}{\dot{Z}_0} e^{-\dot{\gamma} z} \left[ 1 - \dot{\rho}(z) \right]$$

$$(19) : \dot{Z}(z) = \dot{Z}_0 \, \frac{1 + \dot{\rho}(z)}{1 - \dot{\rho}(z)}$$

From Z.(19) it implies that:

$$(20) : \dot{\rho}(z) = \dot{Z}_0 \, \frac{\dot{Z}(z) - \dot{Z}_0}{\dot{Z}(z) + \dot{Z}_0} \quad \text{or otherwise:}$$

$$(21) : \dot{\rho}(z) = \frac{\dot{z}(z) - 1}{\dot{z}(z) + 1} \quad \text{where:}$$

$$(22) : \dot{z}(z) = \frac{\dot{Z}(z)}{\dot{Z}_0}$$

From Z.(21) it implies:

$$(23) : \dot{z}(z) = \frac{1 + \dot{\rho}(z)}{1 - \dot{\rho}(z)}$$

At the point $z = -\ell$ at the source, using the formulas Z.(17), Z.(18) and Z.(23) it implies:

$$\dot{V}(s) = \dot{I}(-\ell) \dot{Z}_s + \dot{V}(-\ell) \Rightarrow$$

$$\dot{V}(s) = \frac{\dot{Z}_s}{\dot{Z}_0} \dot{V}_1 e^{\dot{\gamma} \ell} \left[ 1 - \dot{\rho}(-\ell) \right] + \dot{V}_1 e^{\dot{\gamma} \ell} \left[ 1 + \dot{\rho}(-\ell) \right] \Rightarrow$$

$$\dot{V}(s) = \dot{z}_s \, \dot{V}_1 e^{\dot{\gamma} \ell} \left[ 1 - \dot{\rho}(-\ell) \right] + \dot{V}_1 e^{\dot{\gamma} \ell} \left[ 1 + \dot{\rho}(-\ell) \right] \Rightarrow$$

$$\dot{V}(s) = \frac{1 + \dot{\rho}_s}{1 - \dot{\rho}_s} \dot{V}_1 e^{\dot{\gamma} \ell} \left[ 1 - \dot{\rho}(-\ell) \right] + \dot{V}_1 e^{\dot{\gamma} \ell} \left[ 1 + \dot{\rho}(-\ell) \right] \Rightarrow$$

$$(24) : \dot{V}(s)(1 - \dot{\rho}_s) = \dot{V}_1 e^{\dot{\gamma} \ell} \left[ (1 + \dot{\rho}_s)(1 - \dot{\rho}(-\ell)) + (1 + \dot{\rho}_s)(1 + \dot{\rho}(-\ell)) \right]$$

where: $(1 + \dot{\rho}_s)[1 - \dot{\rho}(-\ell)] + (1 + \dot{\rho}_s)[1 + \dot{\rho}(-\ell)] = 2[1 - \dot{\rho}_s\,\dot{\rho}(-\ell)]$

Substituting Z.(24):

$(25): \dot{V}_l = \dfrac{1}{2}\dot{V}_s\,e^{-\dot{\gamma}\ell}\,\dfrac{1 - \dot{\rho}_s}{1 - \dot{\rho}_s\,\dot{\rho}(-\ell)}$

For $z = -\ell$ Z.(17) gives:

$(26): \dot{V}(-\ell) = \dot{V}_l\,e^{\dot{\gamma}\ell}[1 + \dot{\rho}(-\ell)]$

Substituting Z.(25) to Z.(26):

$(27): \dot{V}(-\ell) = \dfrac{1}{2}\dot{V}_s\,e^{-\dot{\gamma}\ell}\,\dfrac{1 - \dot{\rho}_s}{1 - \dot{\rho}_s\,\dot{\rho}(-\ell)}\,e^{\dot{\gamma}\ell}[1 + \dot{\rho}(-\ell)]$

In general, the adaptation between the source's internal impedance and the transmission line's characteristic resistance is postulated in order maximum transfer power from the source to the transmission line to be achieved, hence it implies:

$\dot{\rho}_s = 0$

Hence, Z.(27) becomes:

$(28): \dot{V}(-\ell) = \dfrac{1}{2}\dot{V}_s[1 + \dot{\rho}(-\ell)] \Rightarrow$

$(29): \left|\dot{V}(-\ell)\right|^2 = \dfrac{1}{4}\left|\dot{V}_s\right|^2\left|1 + \dot{\rho}(-\ell)\right|^2$

where: $\left|1 + \dot{\rho}(-\ell)\right|^2 = \left|1 + \dfrac{\dot{Z}_L - \dot{Z}_0}{\dot{Z}_L + \dot{Z}_0}\right|^2 = 4\dfrac{\left|\dot{Z}_L\right|^2}{\left|\dot{Z}_L + \dot{Z}_0\right|^2}$

Therefore Z.(29) becomes:

$$(30): \left| \dot{V}(-\ell) \right|^2 = \left| \dot{V}_s \right|^2 \frac{\left| \dot{Z}_L \right|^2}{\left| \dot{Z}_L + \dot{Z}_0 \right|^2}$$

Substituting Z.(30) to Z.(8) the final mathematical expression for the maximum radiation intensity is extracted:

$$(31): \mathrm{U}_{max} = \frac{1}{8\pi} D \left| \dot{V}_s \right|^2 \frac{R_L}{\left| \dot{Z}_L + \dot{Z}_0 \right|^2}$$

**Importance of U$_{max}$**

Observing the equation Z.(31) it is deduced that it associates the antenna's electromagnetic characteristics ($D$, $\mathrm{U}_{max}$) with the equivalent circuitry characteristics, which are those of the feeding disposition ($\left| \dot{V}_s \right|$), of the transmission line ($\dot{Z}_0$) and of the antenna itself ($\dot{Z}_L$, $\dot{R}_L$). During the essay it is desirable the maximization of $\mathrm{U}_{max}$ which implies that $D$ will be maximized too, as these two quantities are directly proportional. Moreover, the maximization of the mathematical term $\frac{R_L}{\left| \dot{Z}_L + \dot{Z}_0 \right|^2}$ is happening. If it is assumed that : $\dot{Z}_0 = Z_0$, which is a real number and : $\dot{Z}_L = R_L + i X_L$ then the above term becomes: $\frac{R_L}{(R_L + Z_0)^2 + X_L{}^2}$. For an independent variation of $X_L$ the pre-mentioned term is maximized when $X_L = 0$. Similarly, for an independent variation of $R_L$ the term is maximized when : $R_L = \pm Z_0$. Consequently, the maximization occurs when the antenna is matched to the transmission line. At this situation the reflection coefficient at the load $\dot{\rho}(0)$ is nullified and SWR (standing wave ratio) obtains its minimal unitary value. Therefore, the maximization of the general expression Z.(31) is also associated with SWR minimization. However, the précised dependence of these two quantities cannot be derived directly from their mathematical expressions.

At the present essay the computational program RICHWIRE assumes that $\left|\dot{V}_s\right| = 1$. Also, Z.(32) is multiplied with the factor 1000 so that its resultant values will be more legible. Consequently, the quantity that arises is called normalized radiation intensity and is symbolized as Y with dimensions of [mS].

# Annex H : Evaluation of the total radiation pattern

For the sake of simplification and better inspection of the results the complex vectorial quantities of 2.3.(7) up to 2.3.(20) will be grouped at separated sums. Thus, each radiation pattern will be expressed in terms of either of its components $\vec{\theta}$ και $\vec{\phi}$ and its pseudo-real ($\dot{\mathbb{E}}_R$) and pseudo-imaginary parts ($\dot{\mathbb{E}}_I$). Every quantity will be analyzed according to the expression:

$$(1): \overline{\mathbb{E}} = \begin{bmatrix} \dot{\mathbb{E}}_{R\,\theta}\ \vec{\theta} \\ \dot{\mathbb{E}}_{R\,\phi}\ \vec{\phi} \end{bmatrix} + i \begin{bmatrix} \dot{\mathbb{E}}_{I\,\theta}\ \vec{\theta} \\ \dot{\mathbb{E}}_{I\,\phi}\ \vec{\phi} \end{bmatrix}$$

$1^{st}$ grouping: $\overline{\mathbb{E}}_1$, $\overline{\mathbb{E}}_6$, $\overline{\mathbb{E}}_7$, $\overline{\mathbb{E}}_{12}$

$$\dot{\mathbb{E}}_{1R_\theta} = -\left(\frac{\dot{I}_\delta}{2\pi}\right)\left(\frac{1}{\sin\theta}\right)\left[\cos\left(\frac{\pi}{2}\cos\theta\right)\right]e^{-i\frac{\pi}{2}(\sin\theta\sin\phi - \cos\theta)}$$

$$\dot{\mathbb{E}}_{6R_\theta} = \left(\frac{\dot{I}_\delta}{2\pi}\right)\left(\frac{1}{\sin\theta}\right)\left[\cos\left(\frac{\pi}{2}\cos\theta\right)\right]e^{i\frac{\pi}{2}(\sin\theta\sin\phi + \cos\theta)}$$

$$\dot{\mathbb{E}}_{7R_\theta} = -\left(\frac{\dot{I}_\delta}{2\pi}\right)\left(\frac{1}{\sin\theta}\right)\left[\cos\left(\frac{\pi}{2}\cos\theta\right)\right]e^{\frac{\pi}{2}(\sin\theta\sin\phi - \cos\theta)}$$

$$\dot{\mathbb{E}}_{12R_\theta} = \left(\frac{\dot{I}_\delta}{2\pi}\right)\left(\frac{1}{\sin\theta}\right)\left[\cos\left(\frac{\pi}{2}\cos\theta\right)\right]e^{-\frac{\pi}{2}(\sin\theta\sin\phi + \cos\theta)}$$

Again for the sake of simplicity of the mathematical operations the following quantities are posed:

$$a_1 = \frac{\pi}{2}(\sin\theta\sin\phi - \cos\theta)$$
$$a_2 = \frac{\pi}{2}(\sin\theta\sin\phi + \cos\theta)$$

Summing all the above components it is obtained:

$$\dot{\mathbb{E}}_{1_{R_\theta}} + \dot{\mathbb{E}}_{6_{R_\theta}} + \dot{\mathbb{E}}_{7_{R_\theta}} + \dot{\mathbb{E}}_{12_{R_\theta}} = \dot{\mathbb{E}}_{1,6,7,12_{R_\theta}} \Rightarrow$$

$$\dot{\mathbb{E}}_{1,6,7,12_{R_\theta}} = (\frac{\dot{I_\delta}}{2\pi})(\frac{1}{\sin\theta})[\cos(\frac{\pi}{2}\cos\theta)][-e^{-ia_1} + e^{ia_2} - e^{ia_1} + e^{-ia_2}]$$

Resolving the complex exponentials according to Euler's expression it is obtained:

$$(2): \dot{\mathbb{E}}_{1,6,7,12_{R_\theta}} = (\frac{\dot{I_\delta}}{\pi})(\frac{1}{\sin\theta})[\cos(\frac{\pi}{2}\cos\theta)][\cos(a_2) - \cos(a_1)]$$

Afterwards, the pseudo-imaginary components are written:

$$\dot{\mathbb{E}}_{1_{I_\theta}} = -(\frac{\dot{I_\delta}}{2\pi})(\frac{1}{\sin\theta})[\cos\theta - \sin(\frac{\pi}{2}\cos\theta)]e^{-ia_1}$$

$$\dot{\mathbb{E}}_{6_{I_\theta}} = (\frac{\dot{I_\delta}}{2\pi})(\frac{1}{\sin\theta})[\cos\theta - \sin(\frac{\pi}{2}\cos\theta)]e^{ia_2}$$

$$\mathbb{E}_{7_{I_\theta}} = (\frac{\dot{I_\delta}}{2\pi})(\frac{1}{\sin\theta})[\cos\theta - \sin(\frac{\pi}{2}\cos\theta)]e^{ia_1}$$

$$\dot{\mathbb{E}}_{12_{I_\theta}} = -(\frac{\dot{I_\delta}}{2\pi})(\frac{1}{\sin\theta})[\cos\theta - \sin(\frac{\pi}{2}\cos\theta)]e^{-ia_2}$$

Summing all the above, it is obtained:

$$\dot{\mathbb{E}}_{1,6,7,12_{I_\theta}} = (\frac{\dot{I_\delta}}{2\pi})(\frac{1}{\sin\theta})[\cos\theta - \sin(\frac{\pi}{2}\cos\theta)][-e^{-ia_1} + e^{ia_2} + e^{ia_1} - e^{-ia_2}] \Rightarrow$$

$$(3): \dot{\mathbb{E}}_{1,6,7,12_{I_\theta}} = i(\frac{\dot{I_\delta}}{\pi})(\frac{1}{\sin\theta})[\cos\theta - \sin(\frac{\pi}{2}\cos\theta)][\sin(a_2) + \sin(a_1)]$$

The particular radiation patterns don't have components at the $\vec{\phi}_\kappa$ direction.

2$^{nd}$ grouping: $\bar{\bar{\mathbb{E}}}_2, \bar{\bar{\mathbb{E}}}_8$

The following mathematical quantities are posed:

$$b_1 = \frac{1}{2}(\sin\theta\sin\phi + \sqrt{3}\cos\theta)$$

$$c_1 = \cos\theta\sin\phi - \sqrt{3}\sin\theta$$

For the pseudo-real parts at $\vec{\theta}$ direction it is obtained:

$$\dot{\mathbb{E}}_{2R_\theta} = (\frac{\dot{I}_\delta}{4\pi})\frac{\cos(b_1\frac{\pi}{2})}{1-b_1{}^2}e^{-ia_1}\,c_1$$

$$\dot{\mathbb{E}}_{8R_\theta} = (\frac{\dot{I}_\delta}{4\pi})\frac{\cos(b_1\frac{\pi}{2})}{1-b_1{}^2}e^{ia_1}\,c_1$$

For the equivalent pseudo-imaginary parts it is obtained:

$$\dot{\mathbb{E}}_{2I_\theta} = -(\frac{\dot{I}_\delta}{4\pi})\frac{b_1-\sin(b_1\frac{\pi}{2})}{1-b_1{}^2}e^{-ia_1}\,c_1$$

$$\dot{\mathbb{E}}_{8I_\theta} = (\frac{\dot{I}_\delta}{4\pi})\frac{b_1-\sin(b_1\frac{\pi}{2})}{1-b_1{}^2}e^{ia_1}\,c_1$$

Summing all the above:

$$(4): \dot{\mathbb{E}}_{2,8R_\theta} = (\frac{\dot{I}_\delta}{2\pi})\frac{\cos(b_1\frac{\pi}{2})}{1-b_1{}^2}c_1\cos a_1$$

$$(5): \dot{\mathbb{E}}_{2,8I_\theta} = i(\frac{\dot{I}_\delta}{2\pi})\frac{b_1-\sin(b_1\frac{\pi}{2})}{1-b_1{}^2}c_1\sin a_1$$

Expressing the corresponding components at $\vec{\phi}$ direction it is obtained:

$$\dot{\mathbb{E}}_{2R_\phi} = (\frac{\dot{I}_\delta}{4\pi}) \frac{\cos(b_1 \frac{\pi}{2})}{1 - b_1^2} e^{-i a_1} \cos\phi$$

$$\dot{\mathbb{E}}_{8R_\phi} = (\frac{\dot{I}_\delta}{4\pi}) \frac{\cos(b_1 \frac{\pi}{2})}{1 - b_1^2} e^{i a_1} \cos\phi$$

$$\dot{\mathbb{E}}_{2I_\phi} = -(\frac{\dot{I}_\delta}{4\pi}) \frac{b_1 - \sin(b_1 \frac{\pi}{2})}{1 - b_1^2} e^{-i a_1} \cos\phi$$

$$\dot{\mathbb{E}}_{8I_\phi} = (\frac{\dot{I}_\delta}{4\pi}) \frac{b_1 - \sin(b_1 \frac{\pi}{2})}{1 - b_1^2} e^{i a_1} \cos\phi$$

Summing all the above:

$$(6): \dot{\mathbb{E}}_{2,8R_\phi} = (\frac{\dot{I}_\delta}{2\pi}) \frac{\cos(b_1 \frac{\pi}{2})}{1 - b_1^2} \cos\phi \cos a_1$$

$$(7): \dot{\mathbb{E}}_{2,8I_\phi} = i (\frac{\dot{I}_\delta}{2\pi}) \frac{b_1 - \sin(b_1 \frac{\pi}{2})}{1 - b_1^2} \cos\phi \sin a_1$$

3$^{rd}$ grouping: $\overline{\overline{\mathbb{E}}}_3$, $\overline{\overline{\mathbb{E}}}_9$

The following mathematical quantity is posed:

$$a_3 = \frac{\pi}{2} [(1 + \sqrt{3}) \cos\theta]$$

For the pseudo-real and pseudo-imaginary parts at $\vec{\theta}$ direction it is obtained:

$$\dot{\mathbb{E}}_{3R_\theta} = -\left(\frac{\dot{I}_\delta}{4\pi}\right) \frac{\cos(b_1 \frac{\pi}{2})}{1 - b_1^2} e^{i a_3} c_1$$

$$\dot{\mathbb{E}}_{9R_\theta} = -\left(\frac{\dot{I}_\delta}{4\pi}\right) \frac{\cos(b_1 \frac{\pi}{2})}{1 - b_1^2} e^{-i a_3} c_1$$

$$\dot{\mathbb{E}}_{3I_\theta} = -\left(\frac{\dot{I}_\delta}{4\pi}\right) \frac{b_1 - \sin(b_1 \frac{\pi}{2})}{1 - b_1^2} e^{i a_3} c_1$$

$$\dot{\mathbb{E}}_{9I_\theta} = \left(\frac{\dot{I}_\delta}{4\pi}\right) \frac{b_1 - \sin(b_1 \frac{\pi}{2})}{1 - b_1^2} e^{-i a_3} c_1$$

Summing all the equivalents parts:

$$(8): \dot{\mathbb{E}}_{3,9R_\theta} = -\left(\frac{\dot{I}_\delta}{2\pi}\right) \frac{\cos(b_1 \frac{\pi}{2})}{1 - b_1^2} c_1 \cos a_3$$

$$(9): \dot{\mathbb{E}}_{3,9I_\theta} = -i\left(\frac{\dot{I}_\delta}{2\pi}\right) \frac{b_1 - \sin(b_1 \frac{\pi}{2})}{1 - b_1^2} c_1 \sin a_3$$

Taking all the equivalent sums at $\bar{\varphi}$ direction it is obtained:

$$(10): \dot{\mathbb{E}}_{3,9R_\phi} = -\left(\frac{\dot{I}_\delta}{2\pi}\right) \frac{\cos(b_1 \frac{\pi}{2})}{1 - b_1^2} \cos\phi \cos a_3$$

$$(11): \dot{\mathbb{E}}_{3,9I_\phi} = -i\left(\frac{\dot{I}_\delta}{2\pi}\right) \frac{b_1 - \sin(b_1 \frac{\pi}{2})}{1 - b_1^2} \cos\phi \sin a_3$$

4$^{th}$ grouping: $\bar{\bar{\mathbb{E}}}_4, \bar{\bar{\mathbb{E}}}_{10}$

The following mathematical quantities are posed:

$$b_2 = \frac{1}{2}\left(\sqrt{3}\cos\theta - \sin\theta\sin\phi\right)$$
$$c_2 = \cos\theta\sin\phi + \sqrt{3}\sin\theta$$

For the pseudo-real and pseudo-imaginary parts at $\vec{\theta}$ direction it is obtained:

$$\dot{\mathbb{E}}_{4_{R_\theta}} = -\left(\frac{\dot{I}_\delta}{4\pi}\right)\frac{\cos\left(b_2\frac{\pi}{2}\right)}{1 - b_2^{\,2}}\,e^{i\,a_3}\,c_2$$

$$\dot{\mathbb{E}}_{10_{R_\theta}} = -\left(\frac{\dot{I}_\delta}{4\pi}\right)\frac{\cos\left(b_2\frac{\pi}{2}\right)}{1 - b_2^{\,2}}\,e^{-i\,a_3}\,c_2$$

$$\dot{\mathbb{E}}_{4_{I_\theta}} = -\left(\frac{\dot{I}_\delta}{4\pi}\right)\frac{b_2 - \sin\left(b_2\frac{\pi}{2}\right)}{1 - b_2^{\,2}}\,e^{i\,a_3}\,c_2$$

$$\dot{\mathbb{E}}_{10_{I_\theta}} = \left(\frac{\dot{I}_\delta}{4\pi}\right)\frac{b_2 - \sin\left(b_2\frac{\pi}{2}\right)}{1 - b_2^{\,2}}\,e^{-i\,a_3}\,c_2$$

Summing all the corresponding parts:

$$(12):\ \dot{\mathbb{E}}_{4,10_{R_\theta}} = -\left(\frac{\dot{I}_\delta}{2\pi}\right)\frac{\cos\left(b_2\frac{\pi}{2}\right)}{1 - b_2^{\,2}}\,c_2\,\cos a_3$$

$$(13):\ \dot{\mathbb{E}}_{4,10_{I_\theta}} = -i\left(\frac{\dot{I}_\delta}{2\pi}\right)\frac{b_2 - \sin\left(b_2\frac{\pi}{2}\right)}{1 - b_2^{\,2}}\,c_2\,\sin a_3$$

Taking the equivalent sums at $\vec{\varphi}$ direction it is obtained:

$$(14):\ \dot{\mathbb{E}}_{4,10_{R_\phi}} = -\left(\frac{\dot{I}_\delta}{2\pi}\right)\frac{\cos\left(b_2\frac{\pi}{2}\right)}{1 - b_2^{\,2}}\,\cos\phi\,\cos a_3$$

$$(15):\ \dot{\mathbb{E}}_{4,10_{I_\phi}} = -i\left(\frac{\dot{I}_\delta}{2\pi}\right)\frac{b_2 - \sin\left(b_2\frac{\pi}{2}\right)}{1 - b_2^{\,2}}\,\cos\phi\,\sin a_3$$

5th grouping: $\overline{\mathbb{E}}_5$, $\overline{\mathbb{E}}_{11}$

For the pseudo-real and pseudo-imaginary parts at $\vec{\theta}$ direction it is obtained:

$$\dot{\mathbb{E}}_{5R_\theta} = (\frac{\dot{I}\delta}{4\pi}) \frac{\cos(b_2 \frac{\pi}{2})}{1 - b_2{}^2} e^{ia_2} c_2$$

$$\dot{\mathbb{E}}_{11R_\theta} = (\frac{\dot{I}\delta}{4\pi}) \frac{\cos(b_2 \frac{\pi}{2})}{1 - b_2{}^2} e^{-ia_2} c_2$$

$$\dot{\mathbb{E}}_{5I_\theta} = -(\frac{\dot{I}\delta}{4\pi}) \frac{b_2 - \sin(b_2 \frac{\pi}{2})}{1 - b_2{}^2} e^{ia_2} c_2$$

$$\dot{\mathbb{E}}_{11I_\theta} = (\frac{\dot{I}\delta}{4\pi}) \frac{b_2 - \sin(b_2 \frac{\pi}{2})}{1 - b_2{}^2} e^{-ia_2} c_2$$

Summing all the equivalent parts:

$$(16): \ \dot{\mathbb{E}}_{5,11R_\theta} = (\frac{\dot{I}\delta}{2\pi}) \frac{\cos(b_2 \frac{\pi}{2})}{1 - b_2{}^2} c_2 \cos a_2$$

$$(17): \ \dot{\mathbb{E}}_{5,11I_\theta} = -i(\frac{\dot{I}\delta}{2\pi}) \frac{b_2 - \sin(b_2 \frac{\pi}{2})}{1 - b_2{}^2} c_2 \sin a_2$$

Taking the corresponding sums at $\vec{\phi}$ direction it is obtained:

$$(18): \ \dot{\mathbb{E}}_{5,11R_\phi} = (\frac{\dot{I}\delta}{2\pi}) \frac{\cos(b_2 \frac{\pi}{2})}{1 - b_2{}^2} \cos\phi \cos a_2$$

$$(19): \ \dot{\mathbb{E}}_{5,11I_\phi} = -i(\frac{\dot{I}\delta}{2\pi}) \frac{b_2 - \sin(b_2 \frac{\pi}{2})}{1 - b_2{}^2} \cos\phi \sin a_2$$

$6^{th}$ grouping: $\overline{\overline{\mathbb{E}}}_{13}, \overline{\overline{\mathbb{E}}}_{14}$

The following mathematical quantities are posed:

$b_3 = \sin\theta \sin\phi$
$c_3 = \cos\theta \sin\phi$

For the pseudo-real and pseudo-imaginary parts at $\vec{\theta}$ direction it is obtained:

$$\mathbb{\dot{E}}_{13_{R_\theta}} = (\frac{\dot{I}_\delta}{2\pi}) \frac{\cos(\frac{\pi}{2}b_3)}{1 - b_3^2} c_3$$

$$\mathbb{\dot{E}}_{13_{I_\theta}} = (\frac{\dot{I}_\delta}{2\pi}) \frac{b_3 - \sin(\frac{\pi}{2}b_3)}{1 - b_3^2} c_3$$

$$\mathbb{\dot{E}}_{14_{R_\theta}} = (\frac{\dot{I}_\delta}{2\pi}) \frac{\cos(\frac{\pi}{2}b_3)}{1 - b_3^2} c_3$$

$$\mathbb{\dot{E}}_{14_{I_\theta}} = - (\frac{\dot{I}_\delta}{2\pi}) \frac{b_3 - \sin(\frac{\pi}{2}b_3)}{1 - b_3^2} c_3$$

Summing all the corresponding parts:

$$(20): \mathbb{\dot{E}}_{13,14_{R_\theta}} = (\frac{\dot{I}_\delta}{\pi}) \frac{\cos(b_3 \frac{\pi}{2})}{1 - b_3^2} c_3$$

$$(21): \mathbb{\dot{E}}_{13,14_{I_\theta}} = 0$$

Taking the equivalent sums at $\vec{\phi}$ direction it is obtained:

$$(22): \mathbb{\dot{E}}_{13,14_{R_\phi}} = (\frac{\dot{I}_\delta}{\pi}) \frac{\cos(b_3 \frac{\pi}{2})}{1 - b_3^2} \cos\phi$$

$$(23): \mathbb{\dot{E}}_{13,14_{I_\phi}} = 0$$

Taking all the separated sums and making use of equations 2.3.(2) up to 2.3.(23), the final formulas of the components of pseudo-real and pseudo-imaginary parts are taken at $\vec{\theta}$ and $\vec{\phi}$ directions of the total radiation pattern.

$$(24): \mathbb{\dot{E}}_{o\lambda_{R_\theta}} = (\frac{\dot{I}_\delta}{2\pi}) [(\frac{2}{\sin\theta}) [\cos(\frac{\pi}{2}\cos\theta)] [\cos(a_2) \quad \cos(a_1)]$$

$$+ \frac{\cos(\frac{\pi}{2}b_1)}{1 - b_1{}^2} c_1 \left[\cos(a_1) - \cos(a_3)\right] + \frac{\cos(\frac{\pi}{2}b_2)}{1 - b_2{}^2} c_2 \left[\cos(a_2) - \cos(a_3)\right] + \frac{2\cos(\frac{\pi}{2}b_3)}{1 - b_3{}^2} c_3 \Bigr]$$

$$(25): \ \dot{\mathbb{E}}_{o\lambda_{I_\theta}} = (i\frac{\dot{I}_\delta}{2\pi}) \Bigl[ (\frac{2}{\sin\theta})\left[\cos\theta - \sin(\frac{\pi}{2}\cos\theta)\right]\left[\sin(a_2) + \sin(a_1)\right]$$

$$- \frac{b_1 - \sin(\frac{\pi}{2}b_1)}{1 - b_1{}^2} c_1 \left[\sin(a_3) - \sin(a_1)\right] - \frac{b_2 - \sin(\frac{\pi}{2}b_2)}{1 - b_2{}^2} c_2 \left[\sin(a_2) + \sin(a_3)\right] \Bigr]$$

$$(26): \ \dot{\mathbb{E}}_{o\lambda_{R_\phi}} = (\frac{\dot{I}_\delta}{2\pi}\cos\phi) \Bigl[ \frac{\cos(\frac{\pi}{2}b_1)}{1 - b_1{}^2} \left[\cos(a_1) - \cos(a_3)\right] + \frac{\cos(\frac{\pi}{2}b_2)}{1 - b_2{}^2} \left[\cos(a_2) - \cos(a_3)\right]$$

$$+ \frac{2\cos(\frac{\pi}{2}b_3)}{1 - b_3{}^2} \Bigr]$$

$$(27): \ \dot{\mathbb{E}}_{o\lambda_{I_\phi}} = (i\frac{\dot{I}_\delta}{2\pi}\cos\phi) \Bigl[ - \frac{b_1 - \sin(\frac{\pi}{2}b_1)}{1 - b_1{}^2} \left[\sin(a_3) - \sin(a_1)\right]$$

$$- \frac{b_2 - \sin(\frac{\pi}{2}b_2)}{1 - b_2{}^2} \left[\sin(a_2) + \sin(a_3)\right] \Bigr]$$

It is observed that equations 2.3.(25) and 2.3.(27) are made up of purely imaginary components, except the complex value of current $\dot{I}_\delta$. Therefore, it follows from 2.3.(1) that the total radiation pattern will finally have only real part at $\vec{\theta}$ and $\vec{\phi}$ directions.

Consequently, the final formulas of both components of the total radiation pattern will be:

$(28): \dot{\mathbb{E}}_{o\lambda_\theta} = (\frac{\dot{I}_\delta}{2\pi})\left[(\frac{2}{\sin\theta})[\cos(\frac{\pi}{2}\cos\theta)][\cos(a_2) - \cos(a_1)]\right.$

$$+ \frac{\cos(\frac{\pi}{2}b_1)}{1 - b_1{}^2}c_1[\cos(a_1) - \cos(a_3)] + \frac{\cos(\frac{\pi}{2}b_2)}{1 - b_2{}^2}c_2[\cos(a_2) - \cos(a_3)]$$

$$+ \frac{2\cos(\frac{\pi}{2}b_3)}{1 - b_3{}^2}c_3 - (\frac{2}{\sin\theta})[\cos\theta - \sin(\frac{\pi}{2}\cos\theta)][\sin(a_2) + \sin(a_1)]$$

$$+ \frac{b_1 - \sin(\frac{\pi}{2}b_1)}{1 - b_1{}^2}c_1[\sin(a_3) - \sin(a_1)]$$

$$+ \left.\frac{b_2 - \sin(\frac{\pi}{2}b_2)}{1 - b_2{}^2}c_2[\sin(a_2) + \sin(a_3)]\right]$$

$(29): \dot{\mathbb{E}}_{o\lambda_\phi} = (\frac{\dot{I}_\delta}{2\pi}\cos\phi)\left[\frac{\cos(\frac{\pi}{2}b_1)}{1 - b_1{}^2}[\cos(a_1) - \cos(a_3)] + \frac{\cos(\frac{\pi}{2}b_2)}{1 - b_2{}^2}[\cos(a_2) - \cos(a_3)]\right.$

$$+ \frac{2\cos(\frac{\pi}{2}b_3)}{1 - b_3{}^2} + \frac{b_1 - \sin(\frac{\pi}{2}b_1)}{1 - b_1{}^2}[\sin(a_3) - \sin(a_1)]$$

$$+ \left.\frac{b_2 - \sin(\frac{\pi}{2}b_2)}{1 - b_2{}^2}[\sin(a_2) + \sin(a_3)]\right]$$

# Bibliography

[1]     Zimourtopoulos P.E.,
        "Antennas I/II : Analysis",
        Antennas-DUTH, Xanthi, 1999-2005,
        http://antennas.ee.duth.gr

[2]     Giannopoulou N.I., Zimourtopoulos P.E.,
        "RADPAT4W : Antenna Radiation Pattern for Windows, ver.4.3.7",
        Antennas-DUTH, Xanthi, 1999-2005,
        http://antennas.ee.duth.gr

[3]     Giannopoulou N.I., Zimourtopoulos P.E.,
        "Radiation Patterns Maps",
        Antennas-DUTH, Xanthi, 2002,
        http://antennas.ee.duth.gr

[4]     Zimourtopoulos P.E.,
        "RICHWIRE : A corrected version of THINWIRE, ver.1.2",
        Antennas-DUTH, Xanthi, 1996-2000,
        http://antennas.ee.duth.gr

[5]     Papadaniel Glykeria,
        "Thesis by elaboration",
        Antennas-DUTH,
        Thesis #40, Xanthi 2006

[6]     Katebenis Athanasios,
        "Tantamount – Prototype Antenna-02",
        Antennas-DUTH,
        Thesis #36, Xanthi 2006

[7]     Rec radio amateur antenna,
        "http://tinyurl.com/gy8u8"

[8]     Pozar M. David
        "Microwave Engineering, 2$^{nd}$ Edition",
        John Wiley & Sons, Inc.,
        page 65